*Università degli Studi di Firenze*

Scuola di Dottorato in Elettronica per l'Ingegneria dell'Informazione

Dottorato di Ricerca Internazionale in RF, Microonde ed Elettromagnetismo

Ciclo XXV

Coordinatore: Prof. Gianfranco Manes

# Low Noise Readout Circuits for Particle and Radiation Sensors

Settore Scientifico Disciplinare ING-INF/01

**Dottorando**
Dott. Claudio Gotti

**Tutore**
Prof. Gianfranco Manes

**Co-Tutore**
Prof. Gianluigi Pessina

Anni 2010 - 2012

*He who sleeps in continual noise is wakened by silence.*

William Dean Howells

# Summary

The present thesis follows a three years' work in design, realization and operation of electronic circuits for the readout of particle and radiation sensors, carried out in close collaboration with the Istituto Nazionale di Fisica Nucleare (INFN), sezione di Milano Bicocca. The work was mainly focused to applications in particle physics experiments which are currently in the construction phase, or to existing experiments which planned major hardware upgrades in the next years, involving the design of new front-end circuits. The circuits developed are in principle applicable also outside the field of pure science research, for applications in nuclear instrumentation, medical imaging, security and industrial scanners, and others.

Chapter 1 provides an overview on particle and radiation sensors and their readout electronics. The main issues related to the design of typical readout circuits are reviewed, with particular emphasis on electronic noise and its impact on energy and timing resolution.

Chapter 2 describes the CLARO, a low power, wide-bandwidth integrated circuit for fast single photon counting with pixellated photon sensors. It was primarily designed to readout multi-anode photomultiplier tubes (Ma-PMTs) in the upgraded LHCb RICH detector at the Large Hadron Collider at CERN, but may find application also in the readout of other photon sensors such as microchannel plates (MCP-PMTs) and silicon photomultipliers (SiPMs). The first prototype of the CLARO was realized in a 0.35 µm CMOS technology. The design proved adequate to sustain the high event rate foreseen in the upgraded LHCb RICH, up to $10^7$ hits per second for the pixels in the hottest regions, with a power consumption of 1 mW per channel. The low power consumption is a crucial requirement due to the large number of closely packed channels to be readout in the upgraded LHCb RICH, of the order of $10^5$. The timing resolution of the circuit was demonstrated down to 10 ps RMS, an outstanding result considering its low power consumption.

Chapter 3 describes a charge sensitive amplifier named GeFRO developed and proposed for the phase II of the GERDA experiment at Laboratori Nazionali del Gran Sasso (LNGS). The design introduces a new approach to the readout of Broad Energy Germanium (BEGe) detectors and of ionization sensors in general, minimizing the number of components near the sensors. The first stage of the readout circuit is located near the sensors, submerged in liquid Argon at a temperature of 87 K, and features a minimal number of discrete components. The signals are driven through a transmission line to a second stage located more than ten meters away, which also provides slow feedback through a second line. This allows to obtain a high degree of radiopurity without compromising resolution and bandwidth, which can extend up to a few tens of MHz, improving the capability to discriminate the rare nuclear events of interest from the background. If accepted by the GERDA collaboration, about 50 channels are foreseen to be deployed in the upgraded experiment.

Chapter 4 describes the working principles of bolometric sensors and the requirements of three bolometric experiments, named CUORE, LUCIFER and MARE, for what concerns the front-end amplifiers. CUORE and LUCIFER are cryogenic detectors for rare event searches currently under construction at LNGS. Despite the similarity with the physics goals of GERDA the overall design of the experiments is very different, as CUORE and LUCIFER are based on macrobolometers, that is crystals held at very low temperature, of the order of a few tens of mK, where nuclear events of interest cause temperature variations which are measured by thermistors. In the case of LUCIFER a dual readout technique is exploited to detect the scintillation light produced in the crystals, improving the capability to discriminate the rare events of interest from background. CUORE will be composed of about 1000 channels, while about 50 channels are foreseen for LUCIFER. The MARE experiment is instead based on microbolometers, which give faster thermal signals and are more suitable to higher event rates, and aims at a direct kinematic measurement of the neutrino mass. In the case of CUORE and LUCIFER the electronics are located at room temperature, while in the case of MARE the first stage is located inside the cryostat at around $120 - 130$ K in order to preserve the signal bandwidth. The main requirements for the front-end electronics of these experiments are low noise at low frequency, high precision and high stability.

# Contents

# 1 Readout of particle and radiation sensors

## 1.1 Introduction

There is no real difference between particle sensors and radiation sensors, for there is no real difference between particles and radiation. The term radiation is used to denote an energy flow through vacuum or a medium which is not required for its propagation, and is associated at the quantum scale with travelling atomic or subatomic particles. Electromagnetic radiation, for instance, is ultimately made of photons, massless particles travelling at the speed of light. Photons can travel in vacuum, and the presence of matter can only interfere with their propagation. On the other side, it is known that photons exhibit wave-like properties on the quantum scale. Their description in terms of travelling waves, or electromagnetic radiation, is sometimes preferred. The same holds true for every other kind of radiation, from rare, spontaneous nuclear decays to high intensity, high energy accelerated particle beams: particles and radiation are in general just two faces of the same coin, and will be treated as one in the following. In this sense, radiation is radically different from other kinds of energy transfer which do require matter as a substrate for propagation and are not necessarily associated with the presence of travelling particles, such as mechanical or sound waves.

Particle and radiation sensors are employed in a broad range of applications, from fundamental research in physics and astronomy to commercial instrumentation and devices.[1] The current generation of high energy particle physics experiments dominates the first category. These experiments detect and analyze the products of the collisions of accelerated particle beams, probing the validity of the current physical models and looking for deviations. The extremely high collision rate, of the order of $10^8$ proton-proton collisions per second in the case of the Large Hadron Collider (LHC) accelerator at CERN, allows to collect a huge amout of statistics necessary for precise measurements. The four large detectors - ATLAS, CMS, LHCb, ALICE - deployed and taking data at the LHC are among the most demanding systems ever built for what concerns the total number of particle and radiation sensors and corresponding readout channels. The challenges faced by the designers of the front-end circuits for the LHC experiments, aside from the obvious requirements for signal to noise ratio, are mainly related with the close packing of the sensors employed and their readout speed, needed to assure the adequate space and time resolution in detecting the products of particle collisions. In most cases the electronics needs to be able to tolerate the large amount of radiation present in the accelerator environment.

[1] Sensors are often also called "detectors". In this chapter we prefer to denote by "sensors" the individual sensing devices, and by "detectors" the complex systems built from individual sensors. This distinction is somewhat artificial, and will be dropped in the subsequent chapters.

Another active field of research is that of neutrino physics. Neutrinos are elusive particles, rarely interacting with matter, and their detection requires experiments of very large size and mass to maximize the probability of interaction. Since the events of interest are usually very rare, all kinds of background from natural radioactivity need to be minimized, which leads to experiments being built deep underground where the overlying rock shields from cosmic radiation. Moreover, their construction requires great care in the selection of radiopure detector materials, especially those which are closer to the particle sensors, in order to keep background from the natural radioactivity to a negligible level. The readout electronics makes no exception: low mass and radiopure materials are key requirements if the front-end circuits are placed near the sensors.

The particle physics experiments presented in this thesis belong to the categories just mentioned. Other active experiments exist in the field of pure research in particle physics, but will not be considered here.

As in other fields, pure research drives the development of new technology. New particle and radiation sensors are conceived in physics research facilities, answering to the requirements and needs of physics experiments. Prototypes are built. At some point, when a new sensor concept is solid and promising, it can wake the interest of private companies. After a further engineering effort the sensor may be commercialized and produced on large scales. Experiments with a very large number of channels can take great advantage from this technology transfer, as the sensors become available for large scale deployment. Moreover, there are many commercial applications which can benefit from new sensor technology. The most reknown applications whose requirements are closely intertwined with those of particle physics experiments are medical imaging, where X or gamma rays are used to probe tissues and provide non-invasive information on biological processes, and security and industrial scanners, where particle sensors are employed to scan objects in order to gather information on their composition and structure. Thus, despite the relative abstractness of goals of the particle physics experiments above mentioned, their impact on new technology development in the field of particle and radiation sensor development is huge, and broad is the scope of use of such sensors in the fields of applied science and industry.

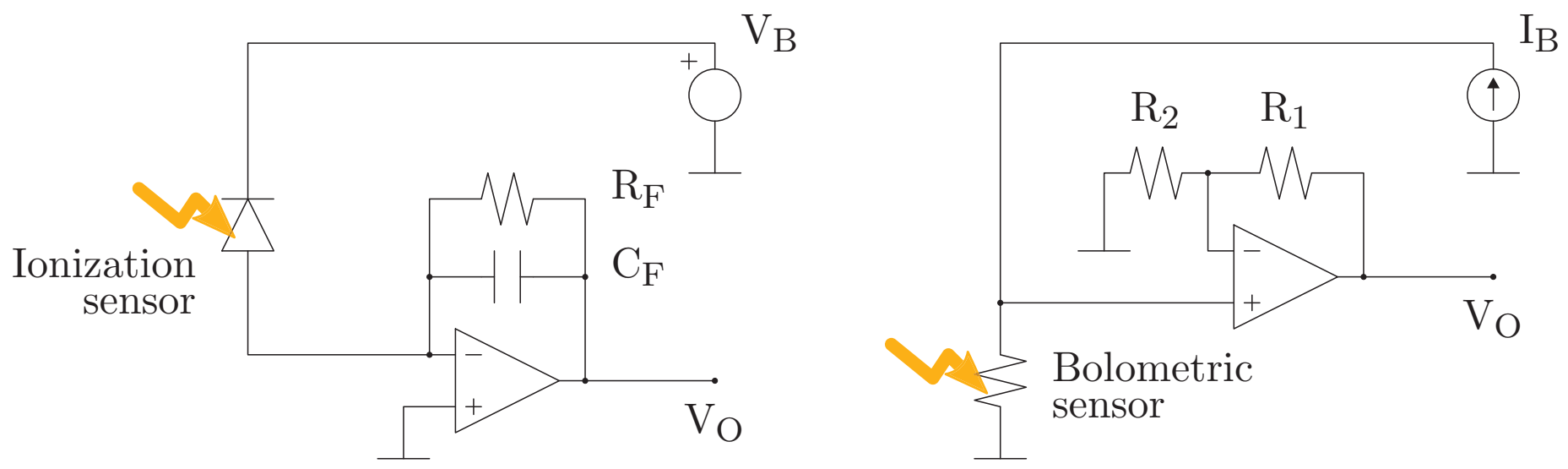

Figure 1.1: Two examples of particle and radiation sensors, including typical biasing and readout circuits. On the left, an ionization sensor readout with a charge amplifier. On the right, a bolometric sensor readout with a voltage amplifier.

## 1.2 Typical sensors and their readout circuits

In a broad sense, a sensor is a device which converts a physical information into another. In the case of particle sensors the primary quantity to be measured may be the energy of particles, or their position, their time of arrival, or other quantities leading to particle identification such as electric charge, velocity, mass, etc [1]. In the case of energy measurement the particle is usually absorbed or destroyed in the process, while in other cases the primary particles can survive the measurement without being significantly perturbed. The physical mechanism underlying the working principle of different sensors may vary, and depends ultimately on the physics of the sensor itself, which is optimized to obtain the highest sensitivity to the quantity of interest. The quantity of interest is converted by the sensor to another physical quantity, usually an electric signal which is then managed by the front-end readout circuitry. In many cases the signal from the sensor consists of a current pulse carrying a given amount of electric charge. In other cases the signal consists of the modulation of the sensor impedance. In most cases the sensor needs to be biased, meaning that it needs an external power source for proper operation.

Figure 1.1 depicts two typical cases. On the left the case of a solid-state ionization sensor is shown. The sensor is based on a junction of a properly doped semiconductor (Silicon, Germanium, . . . ) and is basically a diode. The diode is reverse biased with a voltage source $V_B$, and its depletion region is the active area of the sensor. The bias voltage $V_B$ can range from a few V to several kV, depending on the characteristics and dimensions of the sensor. Impinging particles interact with the atoms in the junction, creating electron-hole pairs which are accelerated in opposite directions by the electric field. For each detected event charges of opposite polarity reach the two electrodes of the sensor giving current signals. In the case depicted in the left side of figure 1.1 positive current pulses are collected at the sensor anode by the front-end amplifier. If the sensor is large enough to stop the particle, all the energy is converted to ionization inside the sensor and the output signal is generally proportional to the particle energy. This is the typical case for calorimetric or spectroscopic detectors, meant to measure the energy of the incoming particles or radiation. If the sensor is small only a part of the energy of the particles is converted to ionization, in which case the sensor is used to obtain information on the presence, position and timing of the impinging particles with a minimal perturbation of their motion. If such small sensors are used to fill a space region, the trajectories of the particles can be reconstructed. This is the typical case of tracking detectors (or trackers).

The charge signal from an ionization sensor can be modelled as a current generator $I_S$. The current pulse corresponding to a particle event coming from the sensor at $t = 0$ can be modelled as

$$I_S(t) = Qf(t) \tag{1.1}$$

where Q is the total charge collected and $f(t)$ is a function describing the pulse shape. Since the integral of the current pulse over time corresponds to the total collected charge, the function $f(t)$ must satisfy the constraint

$$\int_{-\infty}^{\infty} f(t)dt = 1 \tag{1.2}$$

The shape of the current pulse depends on the physics of the sensor, but it is usually very localized in time. The current pulse at the anode is canonically readout with a charge amplifier, that is essentially a current integrator. The basic concept is that of an inverting amplifier with very high open loop gain and capacitive feedback, such as shown in the left of figure 1.1. Due to the negative feedback the input node of the circuit is a virtual ground. Let us neglect $R_F$ for a moment, and let us assume the amplifier to have infinite bandwidth. The charge amplifier integrates the current pulse from the sensor on the feedback capacitor $C_F$, giving at the output

$$V_O(t) = -\frac{Q}{C_F}\int_{-\infty}^{t} f(t')dt' = -\frac{Q}{C_F}F(t) \tag{1.3}$$

On the time scale where the response of the sensor can be considered to be instantaneous, $f(t)$ can be approximated by the Dirac delta distribution $\delta(t)$, and we have that $F(t) \simeq \vartheta(t)$, where $\vartheta(t)$ is the step function whose value is 0 for $t < 0$ and 1 otherwise. In this case the output pulse takes the form

$$V_O(t) = -\frac{Q}{C_F}\vartheta(t) \tag{1.4}$$

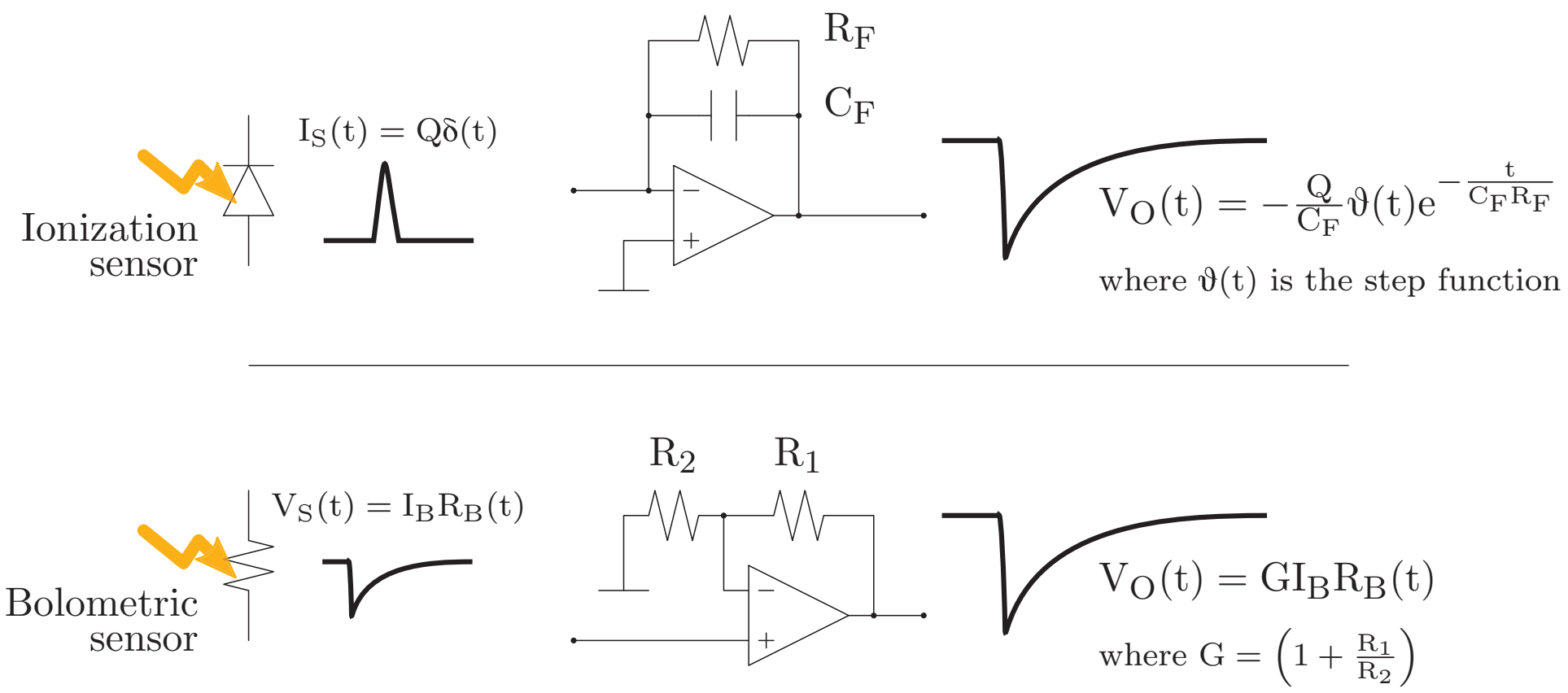

Figure 1.2: Typical signal shape for an ionization sensor readout with a charge amplifier (top) and for a bolometric sensor readout with a voltage amplifier (bottom).

In a real circuit the ideal voltage step will be anyway smoothed by the finite bandwidth of the readout amplifier.

Usually a large value resistor or a switch is connected in parallel with $C_F$ to discharge the capacitor after each event. In the case shown in figure 1.1 the large value resistor $R_F$ is used. The discharge is then exponential with time constant $C_F R_F$ which is chosen to be much larger than the rise time. The output pulse under these assumptions has the form

$$V_O(t) = -\frac{Q}{C_F}\vartheta(t)e^{-\frac{t}{C_F R_F}} \tag{1.5}$$

This case is summarized at the top of figure 1.2.

Figure 1.1 on the right shows the case of a bolometric sensor. In this case the quantity of interest is the temperature of a crystal, which is converted to a variation of the impedance of the sensor. To a first order approximation the temperature rise in the crystal is proportional to the energy deposited by impinging particles. After each event the crystal relaxes to the base temperature, restoring the initial conditions. A thermistor is connected to the crystal and biased with a constant current $I_B$. A thermal signal on the termistor induces a resistance variation $R_B(t)$ which becomes a voltage signal

$$V_S(t) = I_B R_B(t) \tag{1.6}$$

Since normally in semiconductor sensors a larger temperature corresponds to smaller resistance values, for positive values of $I_B$ the signal polarity results to be negative. If the resistance value is large, as in most cases, the low pass effect on the signals due to the parasitic capacitance of the electrical links to the front-end amplifier must be considered. The signal at the output of the amplifier is given by

$$V_O(t) = \left(1 + \frac{R_1}{R_2}\right) I_B R_B(t) = G I_B R_B(t) \tag{1.7}$$

where G is the gain of the amplifier. Here the thermistor signal is faithfully amplified by the front-end circuit by a factor G, without any shaping. The shape of the pulse in this case depends on the thermal excitation and relaxation characteristics of the bolometric sensor, which are contained in $R_B(t)$. The transitions usually show exponential profiles, with the time constant of the temperature increase being smaller than that of the relaxation to the base temperature. This case is summarized at the bottom of figure 1.2. As already mentioned, the speed of the thermal signal may be anyway affected by the effect of the parasitic capacitance of the connecting links from the sensor to the front-end amplifier, or by the finite bandwidth of the amplifier.

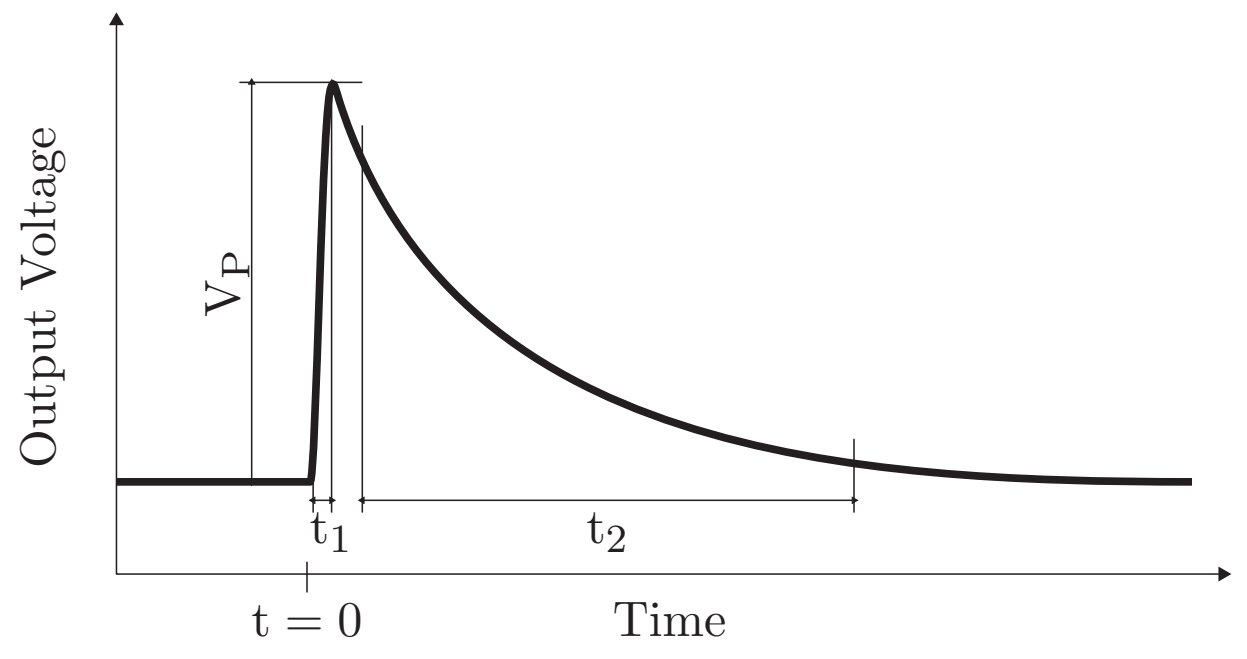

Figure 1.3: A typical signal from a particle or radiation sensor at the output of the readout circuit, with amplitude $V_P$, rise time $t_1$ and fall time $t_2$.

## 1.3 Typical signal shape

Even if based on very different physical mechanisms and readout circuit topologies, the shapes of the signals at the output of the readout circuits in the two cases presented in section 1.2 are very similar and can be traced back to the general case shown in figure 1.3. In this figure the polarity of the pulse is reversed with respect to those of figure 1.2 to better guide the eyes. The generic signal shown is characterized by its amplitude, rise time and fall time, with the latter being usually larger than the former of at least one order of magnitude. As already mentioned, the signal amplitude $V_P$ is usually proportional to the energy deposited in the sensor by a particle event. The rise time $t_1$ is related to the speed of response of the sensor or to the limited bandwidth of the readout circuit, depending on which of the two is slower. The fall time $t_2$ is related to the feedback time constant of the charge amplifier $C_F R_F$ in the case of the ionization sensor, and to the thermal relaxation time in the case of the bolometric sensor.

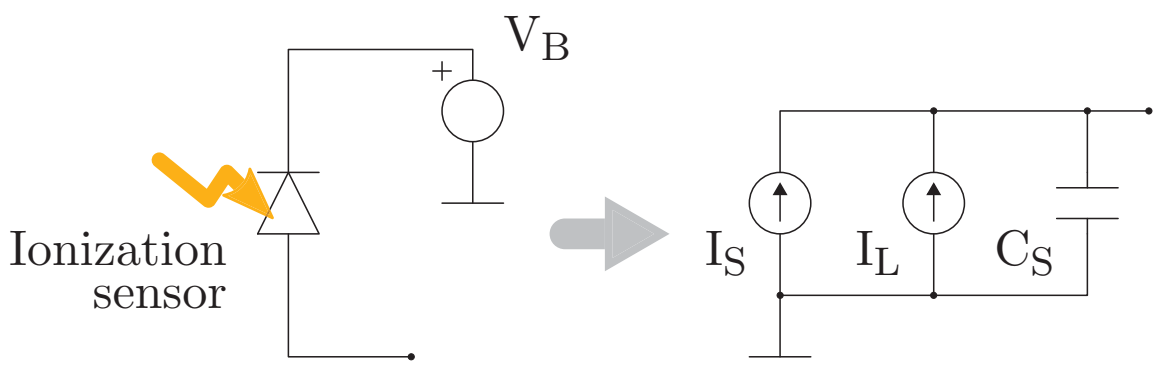

Figure 1.4: Electrical model of a typical ionization sensor.

In most cases the signals have exponential rising and falling edges, and the pulse starting from $t = 0$ shown in figure 1.3 can be expressed as

$$V_O(t) = V'_P \left( e^{-\frac{t}{\tau_2}} - e^{-\frac{t}{\tau_1}} \right) \tag{1.8}$$

For such signal, assuming $\tau_1 \ll \tau_2$, the 10% to 90% rise time $t_1$ and the 90% to 10% fall time $t_2$ are given by $\ln(9) \times \tau_1$ and $\ln(9) \times \tau_2$ respectively, with $\ln(9) \simeq 2.2$. The peak amplitude $V_P$ is given by

$$V_P = V'_P \left( \left( \frac{\tau_1}{\tau_2} \right)^{\frac{\tau_1}{\tau_2 - \tau_1}} - \left( \frac{\tau_1}{\tau_2} \right)^{\frac{\tau_2}{\tau_2 - \tau_1}} \right) \tag{1.9}$$

which is equal to the pulse parameter $V'_P$ of equation (1.8) only if $\tau_1 \ll \tau_2$. In many cases the amplitude is the only quantity of interest. In others a fast rise time can be used for precise timing. In some cases the rise and fall times or other shape parameters can be used to some extent for particle identification. Some practical applications of all the above cases will be presented in this thesis.

## 1.4 Sensor modelling

In section 1.2 two typical sensors were considered. From the electrical point of view, the sensors together with their bias circuits can be represented by equivalent circuits. The case of the ionization sensor is depicted in figure 1.4. The signal is represented by the current generator $I_S$, already considered in section 1.2, whose shape in time domain can be expressed by equation (1.1). As already mentioned, the shape of the charge collection profile $f(t)$ can change widely between different sensors, depending on their working principles, technology and dimensions. The rising edge of f(t) is due to the charge collection time in the detector. The falling edge is related to the restoration of equlibrium conditions in the sensor after a pulse. For faster sensors, or anyway if the readout circuit is slower than the sensor, the signals can be considered instantaneous and f(t) can be approximated by the Dirac delta. The current generator $I_L$ shown in figure 1.4 acconts for the bias current of the sensor. In the case of solid-state ionization detectors (diodes) $I_L$ is the reverse leakage current. The source impedance of the sensor is usually capacitive, and is represented in figure 1.4 by $C_S$. Since the sensor is biased by an ideal voltage source, $C_S$ and $I_L$ can be considered to be referred to ground. This model covers not only the case of semiconductor-based sensors, but is commonly used at least as a first approximation to model most of the sensors which give current pulses as their output signals, including all ionization sensors (gaseous, liquid and solid-state) and photomultipliers. This class of sensors covers the largest part of particle and radiation sensors, which are sometimes referred to as “capacitive” sensors. An ideal sensor would have negligible values for $C_S$ and $I_L$. In most real cases $C_S$ ranges from a few pF to a few nF and $I_L$ ranges from a few pA to a few nA. Photomultipliers will be dealt with in chapter 2. The case of a solid-state Germanium ionization sensor will be described in chapter 3.

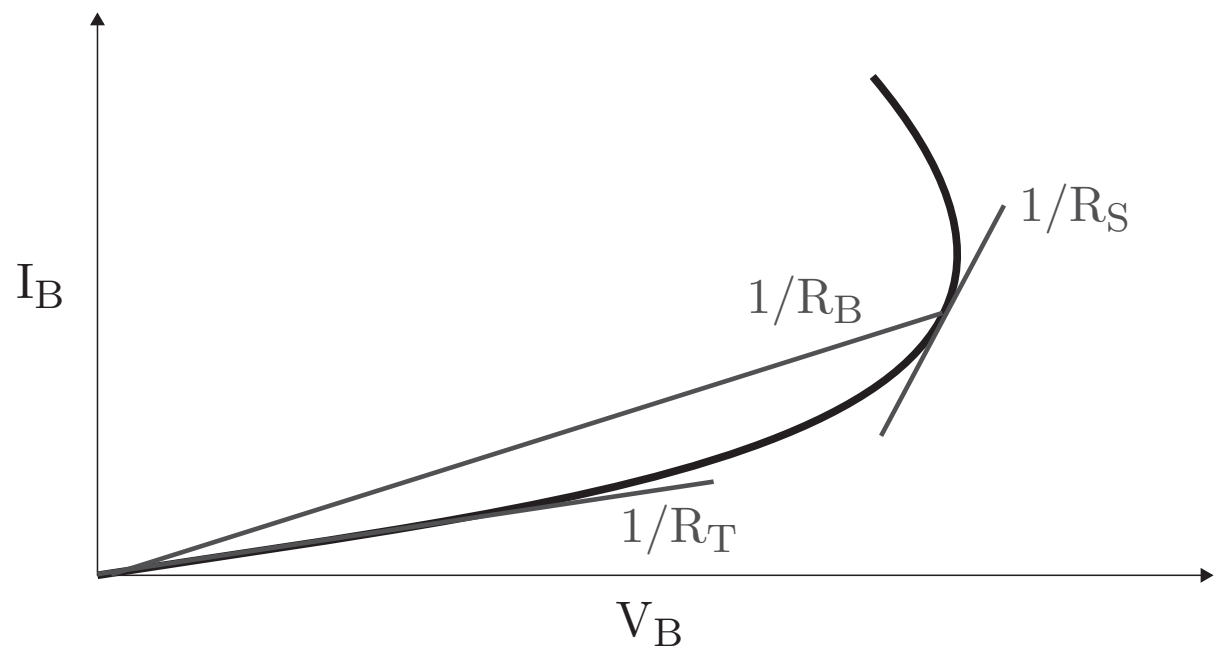

Figure 1.5: Impedance curve of a typical bolometric sensor.

Let us now consider the case of a bolometric sensor. In this case, the sensing element is a thermistor, that is a resistor whose value changes with temperature allowing to measure the thermal pulses in the bolometer. The static impedance of the thermistor has a positive real part, whose value depends on temperature. Since biasing the thermistor with a current dissipates power and thus induces a temperature variation in the thermistor, the static ratio $V_B/I_B$ is not constant unless for very small values of $I_B$, that is for a negligible power dissipated in the thermistor by the bias current. This is shown in figure 1.5, which illustrates the typical $I_B$ versus $V_B$ curve of a bolometer. The value of $R_B = V_B/I_B$ changes along the curve, and is equal to the resistance of the thermistor $R_T$ only for small values of $I_B$. The electrical and thermal properties of the bolometer are thus closely linked, a mechanism known as "electrothermal feedback". If a thermistor has a negative thermal coefficient, a stable working point is reached by biasing it with a constant current, since a larger value for $R_B$ would result in a larger power dissipation $R_B I_B^2$ by the bias current, forcing a smaller value for $R_B$. If on the other side the thermistor is biased with a constant voltage, the sign of the electrothermal feedback results to be positive, making the working point unstable. It can be shown that bolometers with a negative thermal coefficient show an inductive component, and can be represented by the equivalent circuit shown in figure 1.6. The physical interpretation of $L_P$ can be understood as follows. If the bias current is static, the heat capacity of the bolometer can be neglected, and the power dissipated by the bias current flows through the thermal conductance between the bolometer and the heat sink, that is the temperature inside the cryostat where the bolometer is held, causing the temperature of the bolometer to increase. This gives the behaviour of the curve shown in figure 1.5. If instead the power dissipated by the bias current changes abruptly, the transient is absorbed in the heat capacitance, which is equivalent to a thermal short to the heat sink. On the transient the temperature of the bolometer is then constant, and the impedance of the bolometer is equal to $R_T$. The description of this model will be resumed with more detail in chapter 4, section 4.2. For the moment, let us only denote the dynamic DC resistance of the sensor as

$$R_S = R_T || R_P \tag{1.10}$$

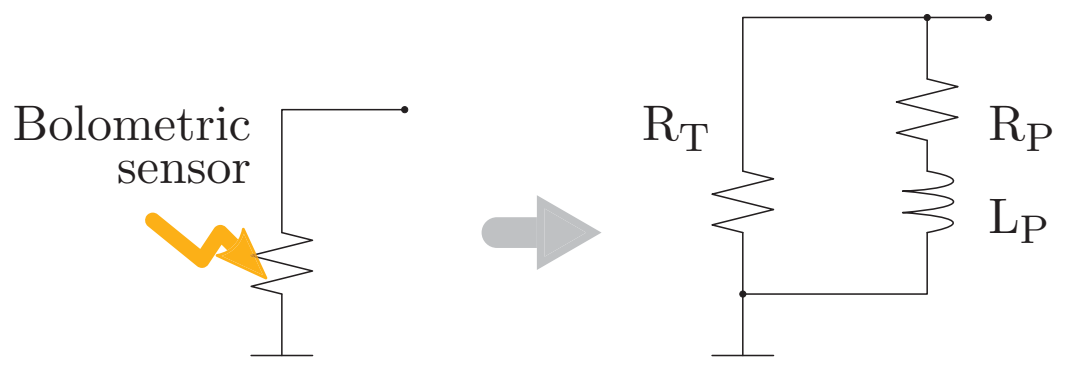

Figure 1.6: Electrical model of a typical bolometric sensor.

The dynamic resistance is given by the slope of the $V_B$ versus $I_B$ curve, which for large $I_B$ is different from $R_B$, as clearly shown in figure 1.5. Due to the electrothermal feedback $R_S$ can assume positive, zero and negative values, depending on the bias current $I_B$. Stable working points usually chosen for operation are those where $R_S$ is small and positive.

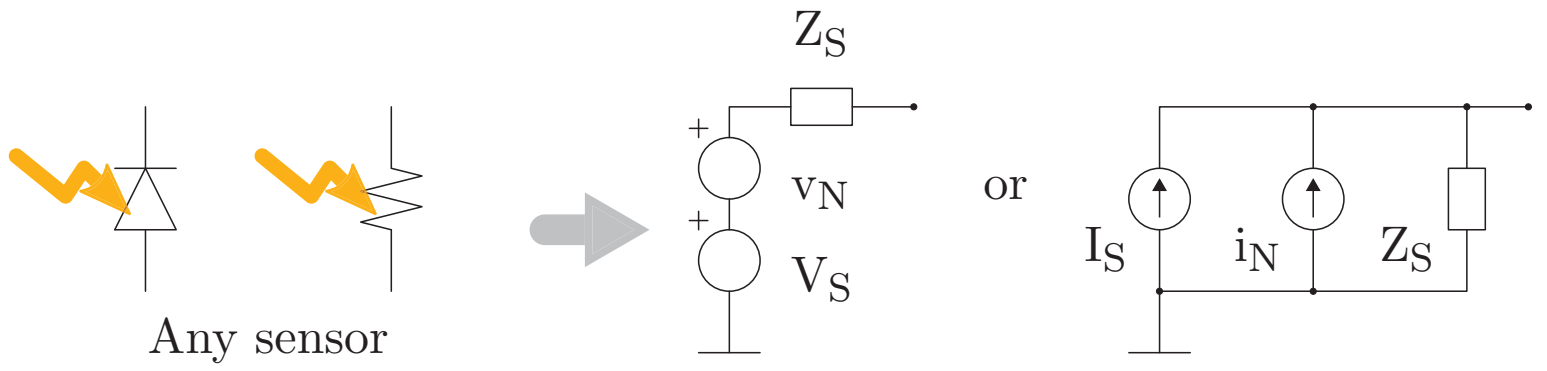

Figure 1.7: General sensor model. For any given sensor, the choice between the two electrical representations on the right is arbitrary.

Both of the cases presented above can be described in a general way, which will be useful in the next sections when discussing the signal to noise ratio of the circuits which can be used for sensor readout. The general model is that of a current source in parallel with the impedance $Z_S$, or equivalently a voltage source in series with the impedance $Z_S$. The voltage and current generators can be splitted between the signal, represented by $V_S$ or $I_S$, and noise, represented by $v_N$ or $i_N$. This is shown in figure 1.7. This is an application of the well known property of any given circuit to be represented as a Thevenin voltage source or a Norton current source. The choice of one representation or the other is arbitrary. These two representations are linked by the relations

$$\begin{aligned} V_S &= I_S Z_S \\ v_N &= i_N Z_S \end{aligned} \tag{1.11}$$

The sensor bias circuits are included in the model. If the readout circuit is a voltage amplifier with high input impedance the Thevenin representation is usually chosen. If instead the circuit is a current amplifier with low input impedance the Norton representation is usually chosen.

When considering the resolution of a sensor, three different factors need to be considered. The first is the intrinsic resolution of the sensor, which is related to fundamental fluctuations of the physical quantities on which the working principles of the sensor rely. This contribution depends only on the sensor class and material, and not on the individual devices. For the case of ionization sensors, this fundamental limit is dictated by the statistics of electron-hole pairs creation, usually referred to as Fano limit, and will be discussed with more detail when considering a Germanium detector in chapter 3. A similar fundamental limit due to phonon fluctuations can be found in bolometric sensors, and will be discussed in chapter 4. For any given sensor this resolution limit cannot be exceeded. Real sensors in the best cases can reach this theoretical limit, but in some cases nonidealities in their realization or operation can lead to a worsening in the resolution. This is the second factor, which is a property of each given sensor in real working conditions. The third factor to be considered is electronic noise. It can be originated in the sensor or in the readout circuit, but in any case its weight on resolution depends on the characteristics of the readout circuit. In the design of front-end amplifiers for sensor readout the third factor is the main concern, and will be treated in the remainder of this chapter. The other factors will be considered in the next chapters, when introducing the sensors to be readout in the applications considered in this thesis.

## 1.5 Electronic noise sources

The electronic noise of any given circuit in stationary conditions is a random Gaussian fluctuation in voltage or current, which can be described in terms of its frequency-dependent power spectral density. The power spectral density expresses how much electrical power is fluctuating at any given frequency. The term "noise spectrum" is commonly used to refer to the square root of the power spectral density. The description in terms of voltage or current fluctuations is arbitrary for a given circuit since the two are connected by Ohm's law. The voltage noise spectrum is expressed in $\mathrm{V}/\sqrt{\mathrm{Hz}}$, while the current noise spectrum is expressed in $\mathrm{A}/\sqrt{\mathrm{Hz}}$. Since the fluctuations are random, the values of noise generators are commonly indicated as root-mean-squared values. However the notation is usually simplified. For instance, a current noise generator is often indicated as

$$\sqrt{\overline{\mathrm{i}_\mathrm{N}^2(\mathrm{f})}} = \mathrm{i}_\mathrm{N}(\mathrm{f}) \tag{1.12}$$

where f is the frequency. Since noise is a fluctuation superimposed on the "true", noiseless signals, its mean value is zero by definition and its spectrum is zero at DC. Its AC content instead can vary depending on the characteristics of the circuits considered. The overall root mean square (RMS) fluctuation is given by the square root of the integral of the squared noise spectrum over the whole frequency range, that is

$$\sigma_\mathrm{i} = \sqrt{\int_0^\infty \mathrm{i}_\mathrm{N}^2(\mathrm{f})\mathrm{df}} \tag{1.13}$$

The above expression follows from Parseval's theorem, which states that energy is conserved from time domain to frequency domain. The same expressions can be written for a voltage noise generator.

The physical mechanisms behind noise are known. Dissipative elements, that is resistors, exhibit foundamental fluctuations which depend on temperature and on the resistance value [2, 3]. A resistor of value R gives a thermal current noise spectrum which is represented by a current source $\mathrm{i}_\mathrm{th}$ in parallel with the resistor. The current noise spectrum is the same at every frequency and its value is

$$\mathrm{i}_\mathrm{th} = \sqrt{\frac{4\mathrm{k}_\mathrm{B}\mathrm{T}}{\mathrm{R}}} \tag{1.14}$$

where $k_B$ is the Boltzmann constant. The noise spectrum can equivalently be represented by a voltage source in series with the resistor:

$$v_{th} = Ri_{th} = \sqrt{4k_B TR} \tag{1.15}$$

As already mentioned, choosing one description or the other is arbitrary. Generally in this thesis the representation as a current source will be preferred.

Diodes and other semiconductor junctions exhibit the so called "shot" noise due to the presence of a unidirectional bottleneck for charge conduction, that is the potential barrier on which their operation relies [4]. The discretization of charge transfer in such devices results in a white current noise source of value

$$i_{sh} = \sqrt{2qI} \tag{1.16}$$

where q is the elementary charge (the absolute value of the electron charge) and I is the current which passes through the device. The two noise meachanisms presented so far, thermal and shot noise, are related to fundamental physical properties of the circuit elements and do not depend on technology. Moreover, they are physically related, as the thermal noise can be seen as a bidirectional shot noise due to the thermal agitation of electrons in a resistive material, where the current is zero in average and depends on temperature. They both give a frequency independent, or "white", spectrum, which is related to the fact that these noise mechanisms do not exhibit an appreciable self-correlation over time.

On the contrary, there are other noise mechanisms which are strongly dependent on the technology adopted, and usually exhibit a measurable self correlation over time, which gives the spectra a more pronounced presence at lower frequency. Their power spectral density often goes as the reciprocal of frequency, and for this reason these noise sources are usually filed under "$1/f$ noise" [5–8].[2] In most cases $1/f$ noise is related to the presence of traps for charge carriers in the bulk or surface of a semiconductor material. The traps can be due to impurities, introducing spurious states in the energy gap of the semiconductor, or to oxide, where the charge carriers can be trapped via tunnelling. This trapping mechanism gives rise to a slow fluctuation in the number of charge carriers (electrons and holes) available for conduction. If a dominant impurity type is present, there is a dominant time constant in the process, and the noise spectrum takes the form of a Lorentzian function (the transfer function of a single pole lowpass filter). This mechanism goes under the name of Generation-Recombination (GR) noise [9, 10]. The ratio between the position of the traps in the energy gap of the semiconductor and its thermal energy $k_B T$ determines the time constant of the process. Given a frequency range of interest there is usually a precise temperature where the effect of an individual trap type on noise is maximum. Deep traps show their effect at room temperature, while shallow traps become evident only at low temperature. This can be clearly seen in figure 1.8, which shows the noise spectra of two high quality JFET samples at different temperatures. The effect of individual impurities can be seen from the behaviour of the curves, as in definite temperature ranges the individual trapping centers are activated.

[2] Noise with power spectra going as $1/f^a$ with $a \simeq 1$ are found in many different fields, and the mechanisms behind it are not in all cases fully understood. At least for the case of electronic devices its sources are quite well known.

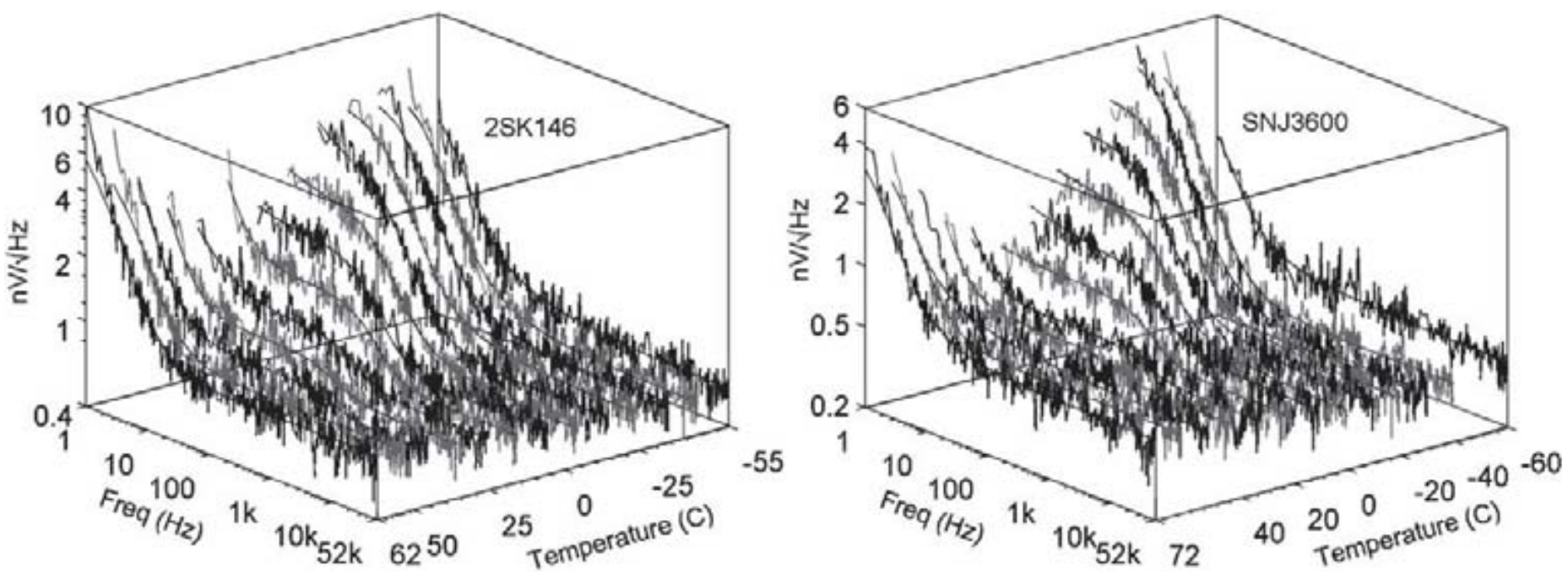

Figure 1.8: Examples of temperature dependence of the noise spectra in high quality JFET transistors, taken from [10].

If a variety of traps is present, their individual GR contributions generally combine to give a $1/f$ power spectral density. At room temperature it is not common to observe the effect of GR noise from individual impurities, aside from devices of very good quality. In most cases the effects of many impurities combine to give $1/f$ noise. The $1/f$ noise level due to impurities in semiconductors is thus very technology dependent, and often also wafer dependent, as the amount and type of impurities can vary from a production batch to another. By using very pure materials and manifacturing processes the presence of impurities and hence $1/f$ noise can be kept under control, the major exception being MOS technology. Even without impurities, in the case of MOS technology charge trapping due to the presence of the gate oxide (amorphous $SiO_2$) is a major source of $1/f$ noise at any temperature. This makes $1/f$ noise generally much higher for MOS transistors compared to bipolar transistors and JFETs.

The noise at the input of any given transistor is modelled with two noise generators, a voltage generator in series with the input and a current generator in parallel with it. For this reason the voltage and current noise are often referred to as “series” and “parallel” noise respectively. The two contributions cannot be merged unless the value of impedance connected at the input of the transistor is fixed, which is not a convenient thing to do at this stage.

In the case of bipolar transistors, the voltage noise is contributed by a $1/f$ term, by the shot noise of the junctions, which is referred to the input by dividing it by the transconductance $g_m$, and by the thermal noise of the series base spreading resistance $R_{BB'}$, whose value depends on the design of the transistor. Uncorrelated noise sources are summed together at the level of their power spectral density, so the corresponding noise generators are summed in quadrature. The quadratic sum of the three contributions gives

$$v_{BJT} = \sqrt{\frac{A_f}{f} + \frac{2k_B T}{g_m} + 4k_B T R_{BB'}} \tag{1.17}$$

where $A_f$ is the device-dependent $1/f$ voltage noise coefficient, which is usually higher for heterojunction transistors such as SiGe than for purely Si transistors. The second term is due to shot noise, but through the known equations which rule the operation of bipolar transistors it could be expressed as the thermal noise of a resistor of value $0.5\ g_m^{-1}$.

In the case of MOS transistors and JFETs in strong inversion the voltage noise spectrum is contributed by a $1/f$ term and by the thermal resistance of the drain-source resistive channel:

$$v_{FET} = \sqrt{\frac{A_f}{f} + \frac{8k_B T}{3g_m}} \tag{1.18}$$

where $A_f$ is again the device-dependent $1/f$ voltage noise coefficient, which is usually much higher for MOS transistors than for JFETs. For MOS transistors and JFETs in a given technology the value of $A_f$ results to be inversely proportional to the transistor area. In the case of MOS transistors this noise is due to trapping in the gate oxide. The second term is due to the thermal noise in the channel, but it behaves as if due to a resistance of value $0.7\ g_m^{-1}$. The transconductance $g_m$ is generally lower for MOS transistors and JFETs than for bipolar transistors, since its dependance on the input voltage is quadratic in MOS transistors and JFETs whereas it is exponential in bipolar transistors. This leads to larger white noise. An exception is given by MOS transistors operated in weak inversion, where the transconductance depends exponentially on the input voltage and MOS transistors behave like bipolar transistors. MOS transistors and JFETs with very short gate or operated in weak inversion may show deviations from equation (1.18). Nevertheless in most cases equation (1.18) gives anyway a satisfactory approximation.

The current noise at the transistor input is related to the shot noise of the current entering the transistor. The shot noise is high for bipolar transistors, and is much lower (usually zero) for MOS transistors and JFETs, where the current entering the gate can be neglected. For bipolar transistors an additional $1/f$ term is usually present, and we can write

$$i_{BJT} = \sqrt{\frac{A_B I_B^2}{f} + 2qI_B} \tag{1.19}$$

where $I_B$ is the base current. The $1/f$ noise coefficient is proportional to $I_B^2$, so it becomes dominant at higher base currents. In the case of the work presented in this thesis $A_B$ could be always neglected, so it will not be considered in the following. Also the noise correlation between input and output nodes can be generally neglected in the readout of particle and radiation sensors.

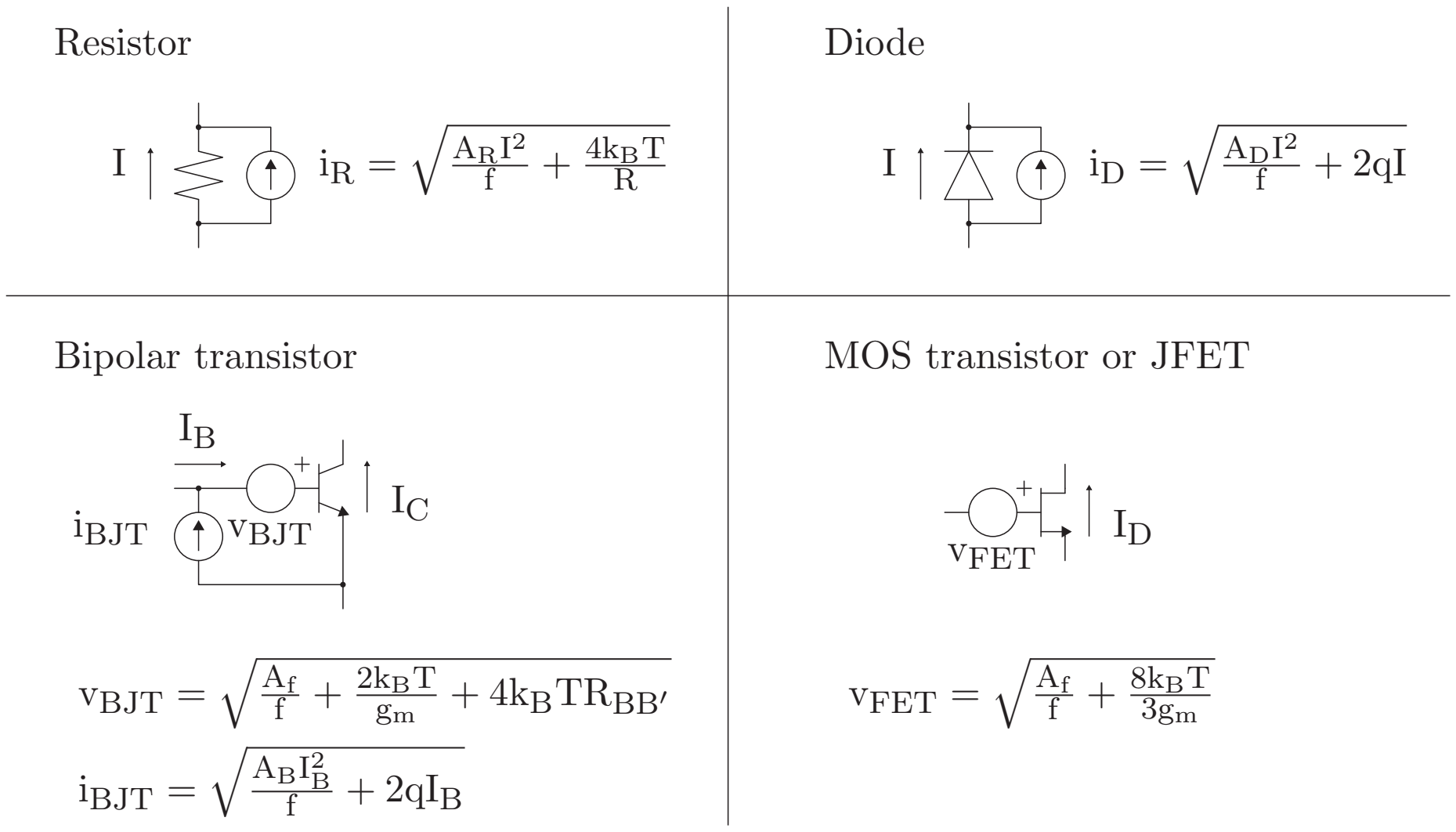

Figure 1.9: Summary of noise sources in electronic devices.

If resistors or diodes are made from a semiconductor material, aside from thermal noise they will usually show also $1/f$ noise. As in the case of the base current of the BJT, the amount of noise is usually found to be proportional to the current flowing in the device, and can be kept under control with pure materials and manufacturing processes. Considering the sum of the $1/f$ and white contributions, with the former being proportional to the current I flowing through the device, the noise of a real resistor can be expressed as

$$i_R = \sqrt{\frac{A_R I^2}{f} + \frac{4k_B T}{R}} \tag{1.20}$$

Similarly the noise of a real diode can be expressed as

$$i_D = \sqrt{\frac{A_D I^2}{f} + 2qI} \tag{1.21}$$

In the case of resistors the $1/f$ noise coefficient $A_R$ is generally found to be proportional to the resistance value R, and at a given fixed bias current I its contribution is thus more significant for large value resistors. If instead the voltage RI is fixed, then clearly the $1/f$ noise contribution results to be larger for smaller resistance values.

The noise sources of resitors, diodes and transistors mentioned above are summarized in figure 1.9. A charge amplifier or a voltage amplifier for sensor readout are made of many passive and active components. Each contributes to the total output noise of the amplifier. Anyway for most designs the dominant noise source is the first stage of amplification and the feedback elements, which are directly connected to the input. Subsequent stages may (and usually do) have a higher noise, but their effect on the signal to noise ratio is made negligible by the gain of the first stage. In the first stage the main noise source is usually the input transistor, whose choice is crucial for the design of front-end amplifiers for particle and radiation sensors. In most cases where optimal signal to noise ratio is a major requirement JFETs are used, since they do not have parallel noise and their $1/f$ voltage noise spectrum is much lower than that of MOS transistors. In other cases, where the sensor impedance is low, or wide bandwidth is more important than low noise, bipolar transistors may be used. Finally for integrated readout circuits in CMOS technologies the only available choice may be that of MOS transistors, whose feature sizes can anyway be taylored for the sensor to be readout, sometimes obtaining better results than with discrete components due to the lower impact of parasitics. There is no general rule to fit all sensors. Each application with its characteristics and requirements must be individually studied if the optimal signal to noise ratio is to be obtained.

## 1.6 Current amplifiers and voltage amplifiers

As was shown in section 1.4 any sensor together with its bias circuit can be arbitrarily represented as a current or a voltage source. The representation as a current source naturally fits the case where the readout circuit is a current amplifier. This case is represented in figure 1.10. In this circuit the signal current is converted to a voltage through the feedback impedance $Z_F$. The noise of the resistive part of $Z_F$ is represented as a current generator $i_F$. The input impedance of the amplifier $Z_A$ is also shown, together with its voltage and current noise sources $v_A$ an $i_A$.

In the domain of the Laplace complex frequency the transfer function of the circuit is given by

$$T_C(s) = \frac{V_O}{I_S} = -\frac{Z_F(s)}{1 + s\tau_B} \tag{1.22}$$

where $\tau_B$ is the time constant associated with the bandwidth of the amplifier. As can be clearly seen by equation (1.22) the transfer function does not explicitly depend on the open loop input impedance, that is given by the parallel combination of $Z_S$ and $Z_A$. However the input impedance generally affects the loop gain and thus $\tau_B$.

Let us now consider the noise sources, starting from the parallel contributions. One end of the current noise source $i_F$ is connected to the output of

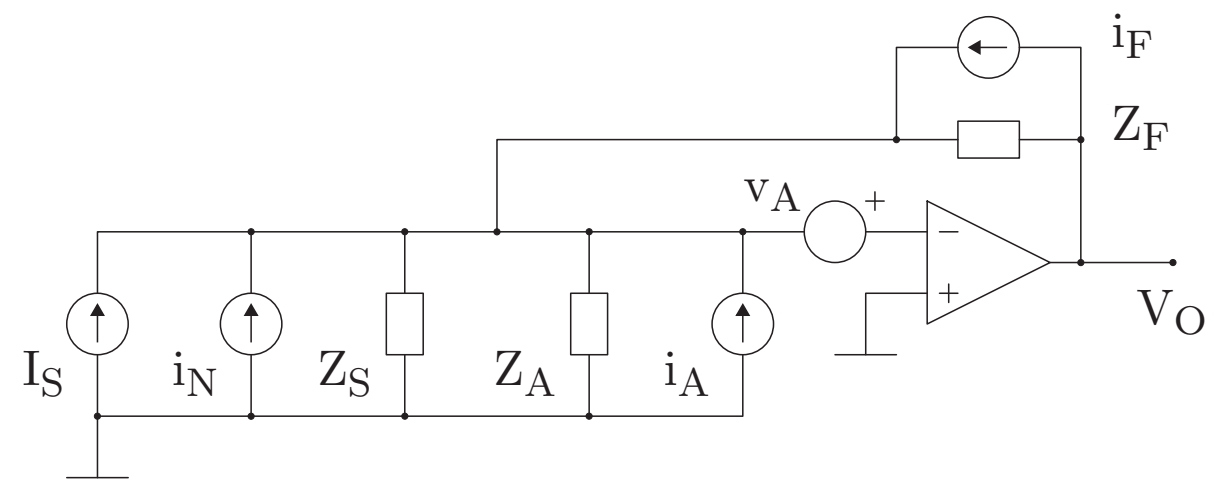

Figure 1.10: General scheme for a sensor readout with a current amplifier, including all the noise sources.

---

the amplifier, which is ideally a low impedance node. That end can be considered at ground, so that $i_F$ becomes a current generator between ground and the input, joining the parallel combination of $i_N$ and $i_A$. These current noise generators share the same transfer function as the signal, equation (1.22), so if we define the quadratic sum of these current noise sources as

$$i_T = \sqrt{\overline{i_N^2} + \overline{i_A^2} + \overline{i_F^2}} = i_N \oplus i_A \oplus i_F \tag{1.23}$$

where the $\oplus$ operator was used to denote the quadratic sum of uncorrelated noise spectra, then we have at the output

$$v_{Oi}(s) = i_T T_C(s) = i_T \frac{Z_F(s)}{1 + s\tau_B} \tag{1.24}$$

where the sign of the transfer function was dropped, since it is irrelevant for noise calculations. Let us now consider the series noise. The transfer function for the voltage noise source $v_A$ differs from that of the signal. The feedback makes the inverting input node a virtual ground. By Ohm's law the voltage noise source $v_A$ can then be considered as a current noise source of value $v_A/Z_T$, where $Z_T$ is the total impedance seen at the input, and is given by the parallel combination of $Z_S$, $Z_A$ and $Z_F$:

$$Z_T(s) = Z_S(s) \;||\; Z_A(s) \;||\; Z_F(s) \tag{1.25}$$

From here, the transfer function (1.22) can be used to bring the contribution to the output, obtaining

$$v_{Ov}(s) = \frac{v_A}{Z_T(s)} T_C(s) = \frac{v_A}{Z_T(s)} \frac{Z_F(s)}{1 + s\tau_B} \tag{1.26}$$

If we sum together quadratically the noise contributions at the output we obtain

$$v_O(s) = \left( i_T \oplus \frac{v_A}{Z_T(s)} \right) \frac{Z_F(s)}{1 + s\tau_B} \tag{1.27}$$

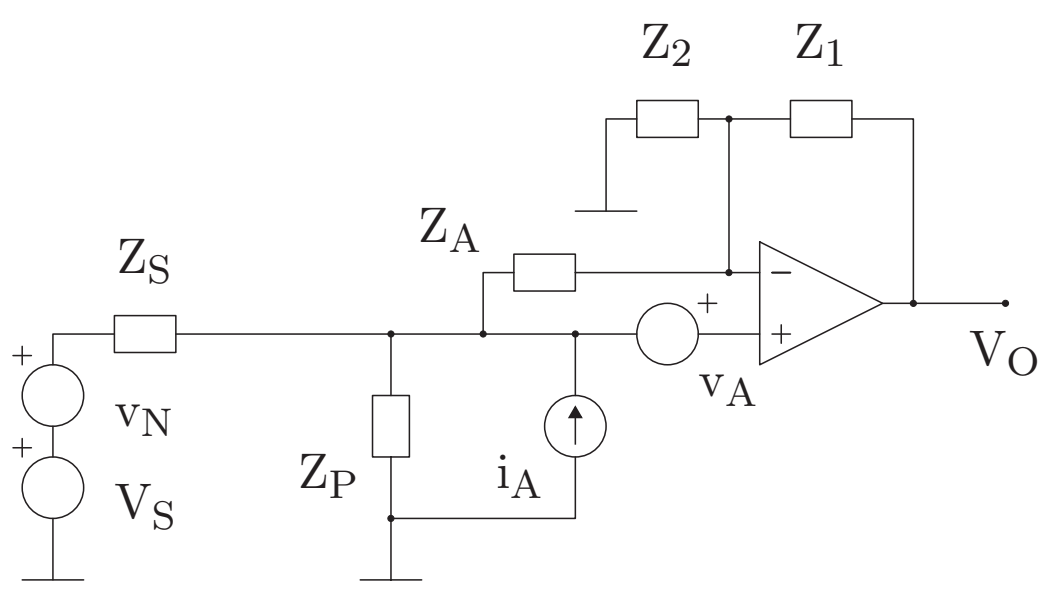

Figure 1.11: General scheme for a sensor readout with a voltage amplifier, including all the noise sources.

By dividing this expression by the transfer function (1.22) we can refer it to the input as a current noise generator in parallel with the signal generator $I_S$, obtaining

$$i_I = i_T \oplus \frac{v_A}{Z_T(s)} = i_N \oplus i_A \oplus i_F \oplus \frac{v_A}{Z_T(s)} \tag{1.28}$$

Equivalently by equation (1.11) this current noise source can be converted to a voltage noise source obtaining

$$v_I = i_I Z_S(s) = i_N Z_S(s) \oplus i_A Z_S(s) \oplus i_F Z_S(s) \oplus v_A \frac{Z_S(s)}{Z_T(s)} \tag{1.29}$$

Let us now consider the case of a voltage amplifier, which naturally fits the case where the sensor is represented as a voltage source. This case is shown in figure 1.11. Again the noise sources of the amplifier are represented as the voltage and current generators $v_A$ and $i_A$. The impedance of the amplifier $Z_A$ is connected between the inputs and it is now bootstrapped. For generality, the parasitic impedance from the input to ground $Z_P$ is also shown in the figure. The gain of the amplifier is given by

$$G(s) = 1 + \frac{Z_1(s)}{Z_2(s)} \tag{1.30}$$

The noise due to $Z_1$ and $Z_2$ is considered to be contained in $v_A$. The transfer function for the signal is given by

$$T_V(s) = \frac{V_O}{V_S} = \frac{Z_P(s)}{Z_P(s) + Z_S(s)} \frac{G(s)}{1 + s\tau_B} \tag{1.31}$$

where $\tau_B$ as before takes into account the bandwidth of the closed loop amplifier, which in this case depends on $Z_1$ and $Z_2$. By defining

$$Z_Q(s) = Z_S(s) \mid\mid Z_P(s) \tag{1.32}$$

then equation (1.31) can be written as

$$T_V(s) = \frac{Z_Q(s)}{Z_S(s)} \frac{G(s)}{1 + s\tau_B} \tag{1.33}$$

The voltage noise source $v_N$ obviously shares the same transfer function of the signal. The other series noise source $v_A$ does not go through the partition between $Z_P$ and $Z_S$, and is instead affected by $Z_A$. The sum of these contributions at the output is given by

$$v_{Ov}(s) = \left( v_N \frac{Z_Q(s)}{Z_S(s)} \oplus v_A \frac{Z_Q(s) + Z_A(s)}{Z_A(s)} \right) \frac{G(s)}{1 + s\tau_B} \tag{1.34}$$

Let us now consider the current noise source $i_A$. This current noise source becomes a voltage $i_A Z_Q$ at the input. From here it goes to the output of the amplifier as

$$v_{Oi}(s) = i_A Z_Q(s) \frac{G(s)}{1 + s\tau_B} \tag{1.35}$$

If all these noise sources are summed at the output, the resulting total noise is

$$v_O(s) = \left( i_A Z_Q(s) \oplus v_N \frac{Z_Q(s)}{Z_S(s)} \oplus v_A \frac{Z_Q(s) + Z_A(s)}{Z_A(s)} \right) \frac{G(s)}{1 + s\tau_B} \tag{1.36}$$

which can be divided by the transfer function (1.33) to refer it to the input as a voltage noise source in series with the signal generator $V_S$:

$$v_I = i_A Z_S(s) \oplus v_N \oplus v_A \frac{Z_S(s)}{Z_T(s)} \tag{1.37}$$

where $Z_T$ is the total input impedance, defined as

$$Z_T(s) = Z_Q(s) \parallel Z_A(s) = Z_S(s) \parallel Z_P(s) \parallel Z_A(s) \tag{1.38}$$

Now we can also convert this voltage noise source to a current noise source by dividing it by $Z_S$, obtaining

$$i_I = \frac{v_I}{Z_S(s)} = i_A \oplus \frac{v_N}{Z_S(s)} \oplus \frac{v_A}{Z_T(s)} = i_A \oplus i_N \oplus \frac{v_A}{Z_T(s)} \tag{1.39}$$

where equation (1.11) was used to convert $v_N$ to $i_N$. As can be clearly seen equation (1.37) is equivalent to equation (1.29), and equation (1.39) is equivalent to equation (1.28). This proves that the signal to noise ratio is independent of the choice of the readout configuration between a current amplifier and a voltage amplifier [11]. The choice of the readout circuit can be made arbitrarily based on other considerations. For instance, there is an important difference which makes the current amplifier preferrable in the case of closely packed channels, that is the fact that being the input at virtual ground its voltage is fixed, and crosstalk due to stray capacitance between neighbouring channels is miminized. On the other side, voltage amplifiers can be more naturally arranged in differential configurations, reducing the sensitivity to common mode disturbances, and are thus usually preferred when there is a long connection between the sensor and the readout circuit. In the case of bolometric sensors the voltage readout is favoured also by electrothermal feedback, which makes the operating point stable in case of constant current biasing. Considering the voltage noise contribution to the input-referred current noise expressed by equations (1.28) and (1.39), there is an important aspect to be noted, namely the fact that the weight of the voltage noise is inversely proportional to the total input impedance $Z_T$. If $Z_T$ is capacitive, then $Z_T = 1/sC_T$, and the voltage noise results to be directly proportional to the total input capacitance, a result which will be discussed with more detail in section 1.8.

## 1.7 Integrating amplifiers and flat gain amplifiers

The transfer functions of the readout circuits considered in the previous section are based on the value of $Z_F$ for the current amplifier and on the value of $Z_T$ and G for the voltage amplifier. For both cases, two configurations are usually found.

The first is the case of integrators, where the output signal is proportional to the integral of the sensor signal. For the current amplifier this happens if the feedback impedance is a capacitor $C_F$, so that

$$Z_F(s) = \frac{1}{sC_F} \tag{1.40}$$

that is the transfer function of an integrator. Since the integral of the current signal from the sensor gives the total collected charge Q, the readout circuit in this case is a charge amplifier. If instead the feedback impedance is a resistor $R_F$ then the signal at the output of the circuit is proportional to the input current from the sensor, without any integration. In this case the circuit has a flat current to voltage conversion gain. If $Z_F$ is the parallel combination of a capacitor $C_F$ and a resistor $R_F$, then

$$Z_F(s) = \frac{R_F}{1 + sC_F R_F} \tag{1.41}$$

In this case the circuit is an integrator for frequencies well above $\tau_F = C_F R_F$, where (1.41) can be approximated as (1.40). For frequencies below $\tau_F$ the circuit behaves instead as a flat gain amplifier. Charge amplifiers are usually implemented with a large value feedback resistor $R_F$, such as that discussed in section 1.2 for the readout of an ionization sensor. The signal current pulses from the sensor are integrated on the capacitor $C_F$, and the output signal is proportional to the total collected charge Q. On a longer timescale, the capacitor is then discharged with time constant $\tau_F$. This case will be discussed with more detail in section 1.8.

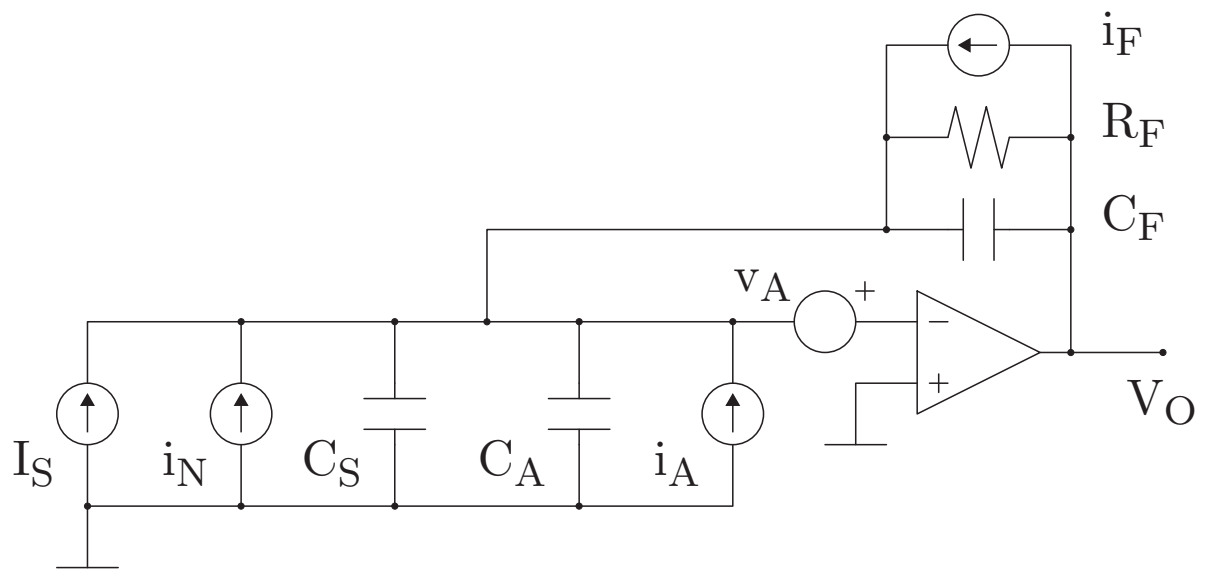

Figure 1.12: Noise sources in the typical readout chain of a capacitive sensor.

The same considerations can be carried out for the voltage amplifier. In this case the transfer function depends on the input impedance $Z_T$ and on the gain G. Let us consider only the case where the amplifier has a flat gain G, independent of frequency. If the input impedance is purely capacitive, that is if

$$Z_Q(s) = Z_S(s) \mid\mid Z_P(s) = \frac{1}{s(C_S + C_P)} = \frac{1}{sC_Q} \tag{1.42}$$

where $C_S$ and $C_P$ are respectively the sensor capacitance and the parasitic capacitance, then the overall transfer function is that of an integrator. The input pulse is integrated on the capacitance $C_Q$, and from there it is amplified to the output with gain G. If instead the input impedance is resistive, then the overall transfer function has a flat gain, and the signals from the sensor are brought to the output without any integration. This is the case of the second circuit discussed in section 1.2. If the resistance at the input is very large, as often happens with bolometric sensors, then the voltage amplifier has a flat gain only up to the frequency where the capacitance $C_Q$ starts to smooth the signals. From that frequency up, the transfer function can be considered that of an integrator. This case will be discussed with more detail in section 1.9.

## 1.8 Noise in a charge amplifier for capacitive sensor readout

In this section the transfer function of a charge amplifier for capacitive sensor readout will be reviewed with more detail, and the impact of the noise sources on the signal to noise ratio will be calculated. Figure 1.12 shows the equivalent schematic of the capacitive sensor connected to the charge amplifier, where all the noise sources from the sensor and the amplifier can be seen. The current source $i_N$ usually represents the shot noise of the sensor bias current. The current source $i_F$ gives the noise of the feedback resistor $R_F$. The voltage and current sources $v_A$ and $i_A$ represent the noise coming from the amplifier. The schematic shows also the detector capacitance $C_S$ and the input capacitance of the amplifier $C_A$. From equations (1.22) and (1.41) its transfer function is given by

$$T_C(s) = \frac{V_O}{I_S} = -\frac{1}{C_F}\frac{\tau_F}{(1+s\tau_F)(1+s\tau_B)} \tag{1.43}$$

where $\tau_F = C_F R_F$. By taking the inverse Laplace transform of equation (1.43) one obtains the response of the amplifier to a Dirac delta $\delta(t)$, that is

$$V_{O\delta}(t) = -\frac{1}{C_F}\frac{\tau_F}{\tau_F - \tau_B}\left(e^{-\frac{t}{\tau_F}} - e^{-\frac{t}{\tau_B}}\right) \tag{1.44}$$

Given an input signal $Qf(t)$, the signal at the output of the amplifier is given by

$$V_O(t) = Qf(t) \star V_{O\delta}(t) \tag{1.45}$$

where the $\star$ operator denotes the convolution.

If the frequency content of $f(t)$ is all inside the amplifier bandwidth, and assuming the input signal $f(t)$ to be much faster than $\tau_F$, then the output signal can be approximated for $\tau_B \simeq 0$ as

$$V_O(t) = -\frac{Q}{C_F}F(t)e^{-\frac{t}{C_F R_F}} \tag{1.46}$$

where $F(t) = \int_{-\infty}^{t} f(t')dt'$. This can be seen as follows. If $\tau_B \simeq 0$ then equation (1.43) multiplied by the frequency content of the input signal $Q\tilde{f}(s)$ becomes

$$V_O(s) \simeq -\frac{Q}{C_F}\frac{\tau_F}{1+s\tau_F}\tilde{f}(s) \tag{1.47}$$

which on short time scales, that is for large s, can be approximated as

$$V_O(s) \simeq -\frac{Q\tilde{f}(s)}{sC_F} \tag{1.48}$$

which in time domain gives

$$V_O(t) \simeq -\frac{Q}{C_F}\int_0^t f(t) = -\frac{Q}{C_F}F(t) \tag{1.49}$$

where the last equality is due to the fact that the input signal is zero by definition for $t < 0$. This proves the validity of equation (1.46) for small t. On longer time scales, that is for small s, equation (1.47) can be approximated by

$$V_O(s) \simeq -\frac{Q}{C_F}\frac{\tau_F}{1+s\tau_F}\tilde{f}(0) \tag{1.50}$$

whose inverse Laplace transform is

$$V_O(t) \simeq -\frac{Q}{C_F}\tilde{f}(0)e^{-\frac{t}{C_F R_F}} \tag{1.51}$$

and by the well known properties of the Laplace transform

$$\tilde{f}(0) = \int_0^\infty f(t')dt' = F(\infty) \tag{1.52}$$

therefore

$$V_O(t) \simeq -\frac{Q}{C_F}F(\infty)e^{-\frac{t}{C_F R_F}} \tag{1.53}$$

for large t, proving that equation (1.46) is valid also for large t.

On the other hand, if the input current pulse from the sensor is much faster than the amplifier it can be considered instantaneous, and modelled as $Q\delta(t)$. The resulting output signal is

$$V_O(t) = -\frac{Q}{C_F}\frac{\tau_F}{\tau_F-\tau_B}\left(e^{-\frac{t}{\tau_F}} - e^{-\frac{t}{\tau_B}}\right) \tag{1.54}$$

meaning that the rising edge is now limited by the bandwidth of the amplifier. Both (1.46) and (1.54) fall under the general case shown in section 1.3.

Let us consider first the effect of parallel noise, and let us denote the total current noise at the input by $i_T$, as in section 1.6. By substituting the actual value for $Z_F$ into equation (1.24) we obtain

$$v_{Oi}(s) = \frac{i_T}{C_F}\frac{\tau_F}{(1+s\tau_F)(1+s\tau_B)} \tag{1.55}$$

As argumented in section 1.5 the current noise spectra $i_N$, $i_A$ and $i_F$ are generally white, so they do not depend on s. For what concerns the noise of the feedback resistor $i_F$, the $1/f$ contribution is not considered since the DC current in the resistor is negligible. By taking the squared amplitude of equation (1.55) and integrating over frequency from 0 to $\infty$ we obtain the squared RMS contribution at the output

$$\begin{aligned}\sigma_{Oi}^2 &= \left(\frac{i_T}{C_F}\right)^2\int_0^\infty \frac{\tau_F^2}{(1+\omega^2\tau_F^2)(1+\omega^2\tau_B^2)}\frac{d\omega}{2\pi} \\ &= \left(\frac{i_T}{C_F}\right)^2\frac{\tau_F^2}{4(\tau_F+\tau_B)}\end{aligned} \tag{1.56}$$

The square root of equation (1.56) gives the RMS noise at the output of the amplifier due to the current noise alone:

$$\sigma_{Oi} = \frac{i_T}{C_F} \frac{\tau_F}{2\sqrt{\tau_F + \tau_B}} \tag{1.57}$$

If, as usually happens, $\tau_F$ is larger than $\tau_B$ of at least an order of magnitude, then the above expression simplifies to

$$\sigma_{Oi} \simeq \frac{i_T}{C_F} \frac{\sqrt{\tau_F}}{2} \qquad \text{for } \tau_B \ll \tau_F \tag{1.58}$$

From equation (1.58) it can be seen that the current noise is directly proportional to the square root of the fall time, regardless of the value of the rise time, as long as it is smaller than the fall time of at least an order of magnitude. This corresponds to the fact that, since the current noise is integrated, the current white noise is converted to $1/f^2$ noise, and thus a longer integration time results in a larger RMS fluctuation at the output.

Let us now consider the series noise. From equation (1.26) and (1.41) we have in this case

$$v_{Ov}(s) = v_A \frac{C_T}{C_F} \frac{\tau_F}{(1 + s\tau_F)(1 + s\tau_B)} \tag{1.59}$$

where

$$C_T = C_S + C_A + C_F \tag{1.60}$$

is the total capacitance seen at the input node. From equations (1.17) and (1.18) the voltage noise can be expressed as a superposition of white and $1/f$ contributions. We can then write

$$v_A = \sqrt{\frac{A_f}{f} + v_w^2} \tag{1.61}$$

where $A_f$ is the $1/f$ noise coefficient and $v_w$ is the purely white component. The weight of the two components on the RMS noise at the output can be evaluated separtely. For the $1/f$ part we have

$$\begin{aligned} \sigma^2_{Ov1/f} &= A_f \left(\frac{C_T}{C_F}\right)^2 \int_0^\infty \frac{2\pi\omega\tau_F^2}{(1 + \omega^2\tau_F^2)(1 + \omega^2\tau_B^2)} \frac{d\omega}{2\pi} \\ &= A_f \left(\frac{C_T}{C_F}\right)^2 \frac{\tau_F^2}{\tau_F^2 - \tau_B^2} \ln \frac{\tau_F}{\tau_B} \end{aligned} \tag{1.62}$$

so that the contribution of the $1/f$ voltage noise to the RMS fluctuation at the output is given by

$$\sigma_{Ov1/f} = \sqrt{A_f} \frac{C_T}{C_F} \frac{\tau_F}{\sqrt{\tau_F^2 - \tau_B^2}} \sqrt{\ln \frac{\tau_F}{\tau_B}} \tag{1.63}$$

If $\tau_F$ is much larger than $\tau_B$ the expression can be approximated to

$$\sigma_{Ov^{1/f}} \simeq \sqrt{A_f}\frac{C_T}{C_F}\sqrt{\ln\frac{\tau_F}{\tau_B}} \qquad \text{for } \tau_B \ll \tau_F \tag{1.64}$$

We can notice that in this approximation the contribution of the $^1/f$ voltage noise to the output RMS fluctuation depends on the ratio between $\tau_F$ and $\tau_B$ and not on their absolute values, and the dependence is weak since it involves a logarithm. For what concerns the white noise contribution we have

$$\begin{aligned}\sigma^2_{Ovw} &= v^2_w\left(\frac{C_T}{C_F}\right)^2\int_0^\infty\frac{\omega^2\tau_F^2}{(1+\omega^2\tau_F^2)(1+\omega^2\tau_B^2)}\frac{d\omega}{2\pi}\\ &= v^2_w\left(\frac{C_T}{C_F}\right)^2\frac{\tau_F^2}{4(\tau_F^2\tau_B+\tau_F\tau_B^2)}\end{aligned} \tag{1.65}$$

so that we have

$$\sigma_{Ovw} = v_w\frac{C_T}{C_F}\frac{\tau_F}{2\sqrt{\tau_F^2\tau_B+\tau_F\tau_B^2}} \tag{1.66}$$

which for $\tau_B$ much smaller than $\tau_F$ becomes

$$\sigma_{Ovw} \simeq v_w\frac{C_T}{C_F}\frac{1}{2\sqrt{\tau_B}} \qquad \text{for } \tau_B \ll \tau_F \tag{1.67}$$

From equation (1.67) the voltage white contribution appears to be inversely proportional to the square root of the rise time. Since $\tau_B$ is inversely proportional to the bandwidth of the amplifier, we have that the series white noise results to be directly proportional to the square root of the bandwidth of the amplifier.

By summing together in quadrature all the contributions considered in this section the overall RMS noise at the output of the amplifier can be calculated. In the limit for $\tau_B \ll \tau_F$ it is given by

$$\sigma_O \simeq \frac{1}{C_F}\sqrt{i_T^2\frac{\tau_F}{4}+A_fC_T^2\ln\frac{\tau_F}{\tau_B}+v_w^2C_T^2\frac{1}{4\tau_B}} \qquad \text{for } \tau_B \ll \tau_F \tag{1.68}$$

As already pointed out, the contributions belong to three distinct categories. By applying the approximation for $\tau_B \ll \tau_F$ we have that the parallel white noise goes as $\sqrt{\tau_F}$, the series white noise goes as $1/\sqrt{\tau_B}$, so goes as the square root of bandwidth, and the series $^1/f$ noise is almost constant, as it depends only (and weakly) from the ratio $\tau_F/\tau_B$.

It is common practice to refer the noise to the input of the charge amplifier as an equivalent noise charge $\sigma_Q$ by dividing the RMS noise at the output by the peak amplitude of the output signal in response to a unitary charge [12]. In other words, the equivalent noise charge is defined as

$$\sigma_Q = \frac{Q}{V_P(Q)}\sigma_O \tag{1.69}$$

where $V_P(Q)$ is the peak amplitude of the output signal in response to a charge Q. Assuming the signal to be a Dirac delta, by combining equations (1.8), (1.9) and (1.54) one can calculate the value of the output peak amplitude for a charge Q, that is

$$V_P(Q) = \frac{Q}{C_F}\frac{\tau_F}{\tau_F - \tau_B}\left(\left(\frac{\tau_B}{\tau_F}\right)^{\frac{\tau_B}{\tau_F-\tau_B}} - \left(\frac{\tau_B}{\tau_F}\right)^{\frac{\tau_F}{\tau_F-\tau_B}}\right) \tag{1.70}$$

which for $\tau_B \ll \tau_F$ becomes simply

$$V_P(Q) \simeq \frac{Q}{C_F} \qquad \text{for } \tau_B \ll \tau_F \tag{1.71}$$

Equation (1.71) holds even if the signal cannot be approximated by a Dirac delta, as long as $\tau_F$ is much longer that the rise time of the signal. In this approximation the equivalent noise charge can be written as

$$\sigma_Q \simeq C_F\sigma_O \qquad \text{for } \tau_B \ll \tau_F \tag{1.72}$$

obtaining

$$\sigma_Q = \sqrt{i_T^2\frac{\tau_F}{4} + A_f C_T^2 \ln\frac{\tau_F}{\tau_B} + v_w^2 C_T^2 \frac{1}{4\tau_B}} \qquad \text{for } \tau_B \ll \tau_F \tag{1.73}$$

Equation (1.73) gives the equivalent noise charge for a charge amplifier under the approximation that $\tau_B \ll \tau_F$. The ratio between the amount of charge Q contained in a current pulse at the input and the equivalent noise charge gives the signal to noise ratio of the amplifier. The total input capacitance $C_T$ is a fundamental parameter to be minimized to reduce the weight of the series noise. At a first glance, one could think to reduce the contribution of the current noise by reducing the value of the feedback resistor $R_F$, which is directly proportional to $\tau_F$. But this is not effective if $i_T^2$ is dominated by $i_F^2$, since the product $i_F^2\tau_F$ is independent of $R_F$. Moreover the circuit is a charge amplifier, that is an integrator, only above a frequency given by $1/2\pi\tau_F$. If $\tau_F$ becomes too small then the approximation fails. For frequencies below $1/2\pi\tau_F$ the circuit behaves as a flat gain amplifier, and changing the value of $R_F$ results in a change in the gain of the circuit, with no effect on the weight of the parallel noise of $R_F$. For this reasons it is better to leave $R_F$ as large as possible and to reduce the fall time constant with additional filtering, which is considered in section 1.10. Similarly, the weight of the series noise can be reduced by bandwidth reduction, that is with a larger $\tau_B$. The price to pay is the reduction of the timing resolution of the amplifier, to be discussed in section 1.11. The best compromise can then be obtained not by modifying the front-end circuit, but by applying additional filters to the signals its output. This can be done in real time with proper analog circuits or offline with digital signal processing, to be discussed in section 1.10.

## 1.9 Noise in a voltage amplifier for resistive sensor readout

In this section the transfer function and noise of a voltage amplifier for resistive sensor readout, such as for a bolometric application, will be considered with more detail. To some extent the derivations will be similar to those of section 1.8. Let us first clarify an important point. As shown in section 1.6, in the voltage amplifier configuration the input impedance contributes in defining the overall transfer function of the circuit. If the input impedance is capacitive, as happens with capacitive sensors, there is an integration at the input and the overall transfer function is that of an integrator. If the input impedance is resistive, as happens with resistive sensors, then the amplifier shows a flat gain at low frequency. However at higher frequency the effect of the input capacitance starts to dominate over the input resistance. The input capacitance is contributed by the capacitance of the front-end amplifier and of the parasitics of the connecting links. Since the series noise of the front-end amplifier becomes lower using larger input transistors, there is usually an advantage in employing transistors of large area, which feature a large transconductance but also a larger gate capacitance. The upper limit to transistor area is determined by the maximum value of input capacitance which can be accepted. If the signals are slow the front-end may be placed at a distance from the sensors, which could be advantageous in the overall design of an experiment. In this case also the capacitance of the connecting links must be considered, as was done in section 1.6. Thus the rising edge of the signals is generally integrated on the input capacitance. This is the case to be discussed in this section. However in some experiments the rise time of the signals may be of interest to discriminate different kinds of interactions in the sensor. For such applications, the capacitance at the input of the voltage amplifier should be minimized to preserve the bandwidth of the signals.

The typical pulse from a thermal sensor, as already mentioned, is usually characterized by exponential rise and fall profiles. The former is due to heat propagation in the bolometer lattice after a particle event, while the latter is due to its relaxation through the thermal link. Here we denote the associated time constants by $\tau_H$ and $\tau_K$. We can model the DC resistance of the sensor for a slow thermal event due to an impinging particle at $t = 0$ as

$$R_B(t) = R_B(0) - \Delta R_B \frac{\tau_K}{\tau_K - \tau_H} \left( e^{-\frac{t}{\tau_K}} - e^{-\frac{t}{\tau_H}} \right) \tag{1.74}$$

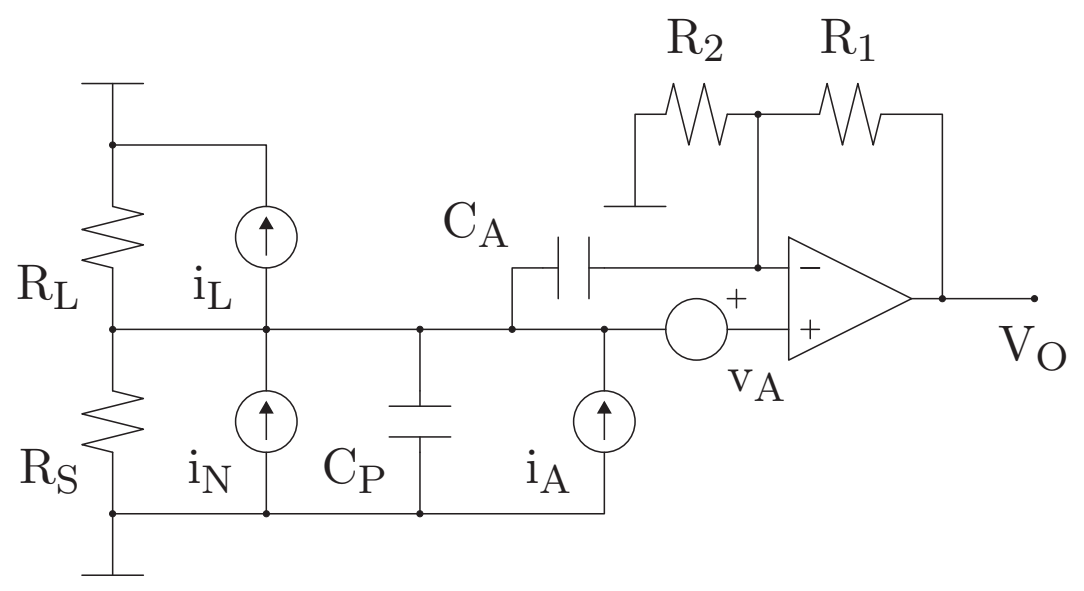

Figure 1.13: Noise sources in the typical readout chain of a resistive sensor.

where $R_B(0)$ is the dynamic resistance of the thermistor in thermal equilibrium with the rest of the system and $\Delta R_B$ is the momentary resistance variation induced by the slow thermal event. The magnitude of $\Delta R_B$ depends on the amount of energy released in the bolometer and on its thermal dynamics, as introduced in section 1.4 and explained with more detail in chapter 4. From the point of view of the thermal pulses the first term can be neglected, and since the thermistor is biased with a current $I_B$ the voltage signal at its ends is given by

$$\Delta V_B(t) = -I_B \Delta R_B \frac{\tau_K}{\tau_K - \tau_H}\left(e^{-\frac{t}{\tau_K}} - e^{-\frac{t}{\tau_H}}\right) \tag{1.75}$$

This signal is already of the kind presented in section 1.3. The frequency content of the signal is enclosed between a lower frequency $1/2\pi\tau_K$ and an upper frequency $1/2\pi\tau_H$, where the former is usually much lower than the second. Since $\tau_K \gg \tau_H$, let us now consider the case for $\tau_H \simeq 0$. The signal expressed by equation (1.75) becomes

$$\Delta V_B(t) = -I_B \Delta R_B e^{-\frac{t}{\tau_K}} \tag{1.76}$$

In the domain of the Laplace complex frequency, this gives

$$\Delta V_B(s) = -I_B \Delta R_B \frac{\tau_K}{1 + s\tau_K} \tag{1.77}$$

By neglecting $\tau_H$, the signal is approximated as a voltage step slowing discharging with time constant $\tau_K$.

Figure 1.13 shows the main noise sources in a typical readout chain for a bolometric sensor. The bias current $I_B$ is usually provided by means of a voltage generator in series with a large value resistor $R_L$, much larger than $R_B$. $R_L$ is usually referred to as the “load resistor”. The thermal current noise of $R_B$ and $R_L$ are shown, together with the voltage and current noise $v_A$ and $i_A$ from the amplifier. The noise of the sensor resistance $R_B$ was indicated by $i_N$. The voltage source $v_A$ also contains the thermal noise of the feedback resistors $R_1$ and $R_2$. The capacitance $C_P$ accounts for the parasitic capacitance of the connecting links from the sensor to the front-end circuit, while $C_A$ is the input capacitance of the amplifier. Since the value of $R_L$ is generally chosen to be much larger than $R_B$, at a first glance one would expect its noise to be negligible. But while $R_B$ is held in a cryostat at cryogenic temperatures, down to a few mK, the same is not generally true for $R_L$, which instead is often located at room temperature. So even if $R_L$ is chosen to be much larger than $R_B$ its current noise is not necessarily smaller due to their very different temperatures. As introduced in section 1.4 the bolometer is characterized by a dynamic impedance whose value depends on the thermal dynamics of the bolometer and on the characteristics of the thermistor. The impedance has an inductive component which can be neglected to a first order approximation. From the point of view of slow thermal signals, what matters is the dynamic resistance $R_S$ which depends on frequency and temperature of operation. From equation (1.33) the transfer function of the amplifier for voltage signals with a source impedance $R_S$ can be written as

$$T_V(s) = \frac{1}{1 + sC_P R_S} \frac{G}{1 + s\tau_B} = \frac{1}{1 + s\tau_P} \frac{G}{1 + s\tau_B} \tag{1.78}$$

where $\tau_P$ was used to indicate the time constant associated with the low pass effect at the input. For very small values of $R_S$ the dominant low pass effect is found to be related with the inductive contribution to the source impedance. In this case a damped oscillating behaviour may be visible on the signals due to the LC circuit formed at the input. The low pass effect related to $C_P$ would still be present, but $\tau_P$ in this case would be given by $\sqrt{L_P C_P}$, being $L_P$ the inductive component of the impedance of the bolometer.

The gain G is given by

$$G = 1 + \frac{R_1}{R_2} \tag{1.79}$$

and again $\tau_B$ indicates the bandwidth limit of the amplifier. As in the case of the charge amplifier, the signal at the output of the readout chain is obtained as the convolution between the thermal signal (1.75) and the response of the readout chain to a Dirac delta. If the bandwidth of the readout circuit extends over $1/2\pi\tau_H$, then no information is lost in the thermal signal. If the bandwidth of the circuit is smaller than $1/2\pi\tau_H$, then the higher frequency content of the thermal signal, that is related with its rise time, is lost. Of the three time constants which determine the rise time of the signal, $\tau_H$, $\tau_P$ and $\tau_B$, the first is often the smallest, at least for the cases where the front-end circuits are not close to the sensors, since the temperature rise of the bolometric sensor is usually fast even for large bolometers. The expression for the output signal can be obtained from equations (1.77) and (1.78) as

$$V_O(s) = \Delta V_B T_V = -I_B \Delta R_B \frac{\tau_K}{1+s\tau_K} \frac{1}{1+s\tau_P} \frac{G}{1+s\tau_B} \tag{1.80}$$

Let us consider the bandwidth of the amplifier to extend beyond the low pass frequency set at the input, as commonly happens. Then $\tau_P \gg \tau_B$ and the above expression can be simplified to

$$V_O(s) = -I_B \Delta R_B \frac{\tau_K}{1+s\tau_K} \frac{G}{1+s\tau_P} \tag{1.81}$$

which in time domain gives

$$V_O(t) = -G I_B \Delta R_B \frac{\tau_K}{\tau_K - \tau_P} \left( e^{-\frac{t}{\tau_K}} - e^{-\frac{t}{\tau_P}} \right) \tag{1.82}$$

Two cases may be considered. If the thermal relaxation is very slow, or if $\tau_P$ is small with respect to $\tau_K$, then there is no integration of the falling edge of the signal, and equation (1.82) for small t becomes

$$V_O(t) = -G I_B \Delta R_B \left( 1 - e^{-\frac{t}{\tau_P}} \right) \qquad \text{for } \tau_K \gg \tau_P \tag{1.83}$$

that is a voltage step smoothed by the input low pass with time constant $\tau_P$. The output peak in this case occurs for large t and is equal to

$$V_P(\Delta R_B) = -G I_B \Delta R_B \qquad \text{for } \tau_K \gg \tau_P \tag{1.84}$$

If instead $\tau_P$ is larger than $\tau_K$, then the signals are completely integrated by the parasitic capacitance at the input, and equation (1.82) can be written as

$$V_O(t) = -G I_B \Delta R_B \frac{\tau_K}{\tau_P} e^{-\frac{t}{\tau_P}} \qquad \text{for } \tau_K \ll \tau_P \tag{1.85}$$

whose peak value occurs for $t = 0$ and is equal to

$$V_P(\Delta R_B) = -G I_B \Delta R_B \frac{\tau_K}{\tau_P} \qquad \text{for } \tau_K \ll \tau_P \tag{1.86}$$

Equations (1.84) and (1.86) will be used later in this section.

From equation (1.35) the noise of the current sources $i_N$ and $i_L$ at the output is given by

$$v_{Oi}(s) = \frac{i_T R_S}{1+s\tau_P} \frac{G}{1+s\tau_B} \tag{1.87}$$

where

$$i_T = i_N \oplus i_L \oplus i_A \tag{1.88}$$

Under the assumption that $\tau_P \gg \tau_B$ we can drop the denominator of the last term and obtain

$$v_{Oi}(s) = \frac{i_T R_S G}{1 + s\tau_P} \qquad \text{for } \tau_B \ll \tau_P \tag{1.89}$$

By taking the squared amplitude of equation (1.55) and integrating over frequency from 0 to $\infty$ we obtain the squared RMS contribution at the output

$$\begin{aligned} \sigma_{Oi}^2 &= (i_T R_S G)^2 \int_0^\infty \frac{1}{1+\omega^2\tau_P^2} \frac{d\omega}{2\pi} \\ &= (i_T R_S G)^2 \frac{1}{4\tau_P} \end{aligned} \tag{1.90}$$

The square root of equation (1.90) gives the RMS noise at the output of the amplifier due to the current sources:

$$\sigma_{Oi} = \frac{i_T R_S G}{2\sqrt{\tau_P}} = \frac{i_T}{C_P} G \frac{\sqrt{\tau_P}}{2} \tag{1.91}$$

The contributions of the thermal noise of $R_B$ and $R_L$ and the current noise $i_A$ from the amplifier are thus directly proportional to the square root of the low pass frequency resulting from the parasitic capacitance $C_P$.

A similar derivation can be carried out for the amplifier voltage noise $v_A$, aside from the fact that this noise source does not suffer from the low pass effect due to $C_P$, being directly connected to the input of the amplifier. Its noise contribution at the output is

$$v_{Ov}(s) = \frac{v_A G}{1 + s\tau_B} \frac{C_A + C_P}{C_P} \tag{1.92}$$

Again the noise source $v_A$ can be split into a 1/f and a white components, as expressed by equation (1.61). For the $^1/_f$ component we have

$$\sigma_{Ov^1/f}^2 = A_f G^2 \left(\frac{C_A + C_P}{C_P}\right)^2 \int_0^\infty \frac{2\pi}{\omega\left(1+\omega^2\tau_B^2\right)} \frac{d\omega}{2\pi} \tag{1.93}$$

This integral does not converge in $\omega = 0$. We can consider the integral extending down to a lower frequency of interest $\omega_0$, and obtain

$$\begin{aligned} \sigma_{Ov^1/f}^2 &= A_f G^2 \left(\frac{C_A + C_P}{C_P}\right)^2 \int_{\omega_0}^\infty \frac{2\pi}{\omega\left(1+\omega^2\tau_B^2\right)} \frac{d\omega}{2\pi} \\ &= A_f G^2 \left(\frac{C_A + C_P}{C_P}\right)^2 \ln\left(\frac{1}{\omega_0\tau_B}\right) \end{aligned} \tag{1.94}$$

As the bandwidth of interest extends to lower frequency, the weight of the $^1/_f$ noise contribution becomes larger. The lower frequency of interest $\omega_0$ can be considered to be associated with the thermal relaxation of the sensor, that is $1/2\pi\tau_K$. The RMS noise at the output expressed by the square root of equation (1.94) then becomes

$$\sigma_{Ov^1/f} = \sqrt{A_f} G \frac{C_A + C_P}{C_P} \sqrt{\ln \frac{\tau_K}{\tau_B}} \tag{1.95}$$

In this approximation, the weight of the series $^1/f$ noise is directly proportional to the square root of the time constant associated with the lowest frequency of interest, which we assumed to be related to the thermal relaxation of the sensor. Concerning the white noise contribution we have

$$\begin{aligned} \sigma^2_{Ovw} &= \left( v_w G \frac{C_A + C_P}{C_P} \right)^2 \int_0^\infty \frac{1}{1 + \omega^2 \tau_B^2} \frac{d\omega}{2\pi} \\ &= \left( v_w G \frac{C_A + C_P}{C_P} \right)^2 \frac{1}{4\tau_B} \end{aligned} \tag{1.96}$$

which results in

$$\sigma_{Ovw} = \frac{v_w G}{2\sqrt{\tau_B}} \frac{C_A + C_P}{C_P} \tag{1.97}$$

So, as in the case of the charge amplifier, the weight of the series white noise on the output RMS noise is directly proportional to the square root of the bandwidth of the amplifier.

By summing together in quadrature all the contributions considered in this section the overall RMS noise at the output of the amplifier can be obtained:

$$\sigma_O = G \sqrt{i_T^2 \frac{\tau_P}{4C_P^2} + \left( A_f \ln \frac{\tau_K}{\tau_B} + v_w^2 \frac{1}{4\tau_B} \right) \left( \frac{C_A + C_P}{C_P} \right)^2} \tag{1.98}$$

Similarly to the case of the charge sensitive amplifier, the output noise can be referred to the input to be directly compared with the sensor signals. Let us consider to bring the noise contribution to the input as an equivalent fluctuation on the resistance, defined as

$$\sigma_R = \frac{\Delta R_B}{V_P(\Delta R_B)} \sigma_O \tag{1.99}$$

where $\Delta R_B$ is the maximum resistance variation and $V_P(\Delta R_B)$ is the peak amplitude of the corresponding output signal. Two cases can be considered. For small $\tau_P$ the resistance $R_S$ dominates the input impedance, the gain is flat, and equation (1.84) should be used for $V_P(\Delta R_B)$, obtaining

$$\sigma_R = \frac{1}{I_B} \sqrt{i_T^2 \frac{\tau_P}{4C_P^2} + \left( A_f \ln \frac{\tau_K}{\tau_B} + v_w^2 \frac{1}{4\tau_B} \right) \left( \frac{C_A + C_P}{C_P} \right)^2} \qquad \text{for } \tau_P \ll \tau_K$$

| | White current | $^1/_f$ voltage | White voltage |
|---|---|---|---|
| Signal to noise ratio | $\sim \sqrt{\tau_F}$ | $\sim \frac{1}{Z_T}$ | $\sim \frac{1}{Z_T\sqrt{\tau_B}}$ |

Figure 1.14: Summary of the weight of the noise sources on the signal to noise ratio. $Z_T$ is the total input impedance seen at the input of the front-end amplifier.

(1.100)

On the other side for large $\tau_P$ the input impedance is dominated by $C_P$, equation (1.86) should be used for $V_P(\Delta R_B)$, obtaining

$$\sigma_R = \frac{R_S}{I_B \tau_K}\sqrt{i_T^2 \frac{\tau_P}{4} + \left(A_f \ln\frac{\tau_K}{\tau_B} + v_w^2 \frac{1}{4\tau_B}\right)(C_A + C_P)^2} \qquad \text{for } \tau_P \gg \tau_K \tag{1.101}$$

Since $C_A + C_P$ is the total input capacitance $C_T$, equation (1.101) is equivalent to equation (1.73). Also in this case the contributions belong to three categories. The current white noise gives a contribution which goes as the square root of the time constant set at the input by $R_S$ and $C_P$, that is the same low pass which affects the signals. As in the case of the charge amplifier, the contribution of the voltage $^1/_f$ and white noise is inversely proportional to the input impedance. If the input capacitance dominates over the resistance, then the voltage $^1/_f$ and white noise contributions result to be directly proportional to its value. The white voltage noise of the amplifier goes as the square root of the bandwidth of the amplifier. Since the bandwidth of the signal is in any case limited by $\tau_P$, the additional noise should be filtered from the response, properly shaping the bandwidth for the best signal to noise ratio. Also in this case, as in the case of the charge amplifier, additional filtering can improve the noise performance. Filtering is either performed online with additional circuitry, or offline with digital signal processing, to be discussed in section 1.10.

## 1.10 Filters for energy resolution

In the previous sections the effect of various noise sources on the signal to noise ratio of two typical readout circuits were considered. As already pointed out, the voltage or current amplifiers give the same signal to noise ratio, summarized in figure 1.14. The weight of the parallel noise depends on the fall time of the signals $\tau_F$. The larger the time constant, the larger the impact of the parallel contribution on the total noise. The weight of the series $^1/_f$ noise is only weakly dependent on the values of the time constants involved. To a first order approximation its contribution can be considered to be independent on the values of the time constants. Its contribution is inversely proportional to the total input impedance $Z_T$. The weight of the series white noise is directly proportional to the square root of the bandwidth of the amplifier, given by $1/2\pi\tau_B$. Like for the series $1/f$ noise, its contribution is inversely proportional to the total input impedance $Z_T$.

These expressions are based on the calculation of the total RMS noise at the output of the readout circuit. As such, they directly give an error in the measurement of the value of the output signal at a given time. Since, as already pointed out, the energy of the particles detected by the sensor is directly proportional to the amplitude of the output signals, the electronic noise is directly related to an uncertainity in the energy measurement. In the cases where energy resolution is important there are known techniques to reduce the weight of electronic noise for a better signal to noise ratio at the output of the amplifier.

The first, simpler approach is to make all the time constants equal. This can be obtained by cascading the output of the readout circuit to a low pass and a high pass, both with the same time constant $\tau$, corresponding to a first order bandpass filter peaked at a frequency of $1/2\pi\tau$. This is known as $CR - RC$ shaping, and the time constant $\tau$ is equal to the RC of the high pass and low pass filters. Of course $\tau$ must be chosen so that $1/2\pi\tau$ falls inside the bandwidth of the signals. Moreover, the fall time constant of the unfiltered signal $\tau_F$ should be larger than $\tau$ of at least one order of magnitude, to allow the unfiltered signal to be approximated as a voltage step. If the fall time of the pulse is shorter than about 10 $\tau$ then the shaped signals will show an undershoot under the baseline. If $R_F$ cannot be made larger, then the undershoot can be corrected with pole-zero compensation by using a dedicated circuit. In the following, we will consider $\tau_F$ to be much larger than $\tau$, so that any effect related to a non-ideal integration of the pulses is negligible. After such filtering the table in figure 1.14 reduces to the one shown in figure 1.15.

At the output of the filter the equivalent noise charge takes the general form

$$\sigma_Q = \sqrt{i_T^2\beta\tau + A_f C_T^2\gamma + v_w^2 C_T^2\frac{\alpha}{\tau}} \tag{1.102}$$

The values of the coefficients $\alpha$, $\beta$ and $\gamma$ can be calculated by mimicking the derivations of section 1.8 after having multiplied the output signal by the transfer function of the filter. For instance, for the case of the charge amplifier, multiplying equation (1.55) by the transfer function of a $CR - RC$ filter with time constant $\tau$ yields

$$v_{Oi}(s) = \frac{i_T}{C_F}\frac{\tau_F}{(1+s\tau_F)(1+s\tau_B)}\frac{s\tau}{(1+s\tau)^2} \tag{1.103}$$

| | White current | $^1/_f$ voltage | White voltage |
|---|---|---|---|
| Signal to noise ratio | $\sim \sqrt{\tau}$ | $\sim \frac{1}{Z_T}$ | $\sim \frac{1}{Z_T\sqrt{\tau}}$ |

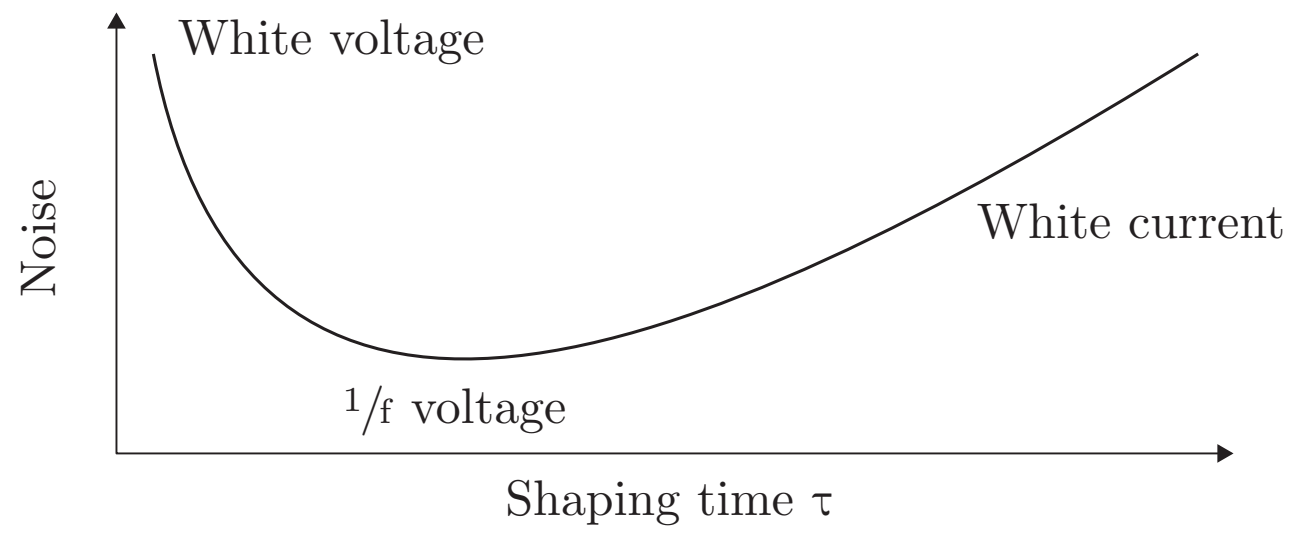

Figure 1.15: Summary of the weight of the noise sources on the signal to noise ratio after a $CR - RC^n$ filter with time constant $\tau$ is applied (on top). Qualitative trend of the noise sources versus shaping time (on bottom).

If $\tau_F \gg \tau \gg \tau_B$ the expression can be approximated for $\tau_F \to \infty$, $\tau_B \to 0$, giving

$$v_{Oi}(s) = \frac{i_T}{C_F}\frac{s\tau}{(1+s\tau)^2} \tag{1.104}$$

which is equivalent to equation (1.55) once the time constants are put equal to $\tau$. The calculations to obtain the equivalent noise charge can be carried out following the very same steps as in sections 1.8 and 1.9. The resulting values of the coefficients of equation (1.102) for the $CR - RC$ filter are

$$\begin{aligned} \alpha = \beta &= \frac{e^2}{8} \simeq 0.92 \\ \gamma &= \frac{e^2}{2} \simeq 3.69 \end{aligned} \tag{1.105}$$

Thus with respect to the unfiltered case of equation (1.73) the weight of the white current and voltage noise components can be changed by acting on the value of $\tau$. In the presence of all three noise sources, there is usually a precise value for $\tau$ which minimizes the equivalent noise charge, as can be seen at the bottom of figure 1.15. Similar considerations can be applied to the case of the voltage amplifier for resistive sensors.

Another kind of shaping traditionally used in spectroscopy is the Gaussian shaping, which gives a better signal to noise ratio with respect to the $CR-RC$ shaping. In this case the signals in time domain are filtered so that the output pulse becomes a Gaussian curve, expressed by

$$V_O = V_G e^{-\frac{1}{2}\left(\frac{t-t_D}{\tau_G}\right)^2} \tag{1.106}$$

where $V_G$ is the peak amplitude, $t_D$ is a delay term, and $\tau_G$ is the shaping time. In this case equation (1.102) is still valid, with $\tau_G$ in place of $\tau$. The coefficients $\alpha$, $\beta$ and $\gamma$ can be calculated, obtaining

$$\begin{aligned} \alpha &= \frac{\sqrt{\pi}}{4} \simeq 0.44 \\ \beta &= \frac{\sqrt{\pi}}{2} \simeq 0.89 \\ \gamma &= \pi \simeq 3.14 \end{aligned} \tag{1.107}$$

In this case the weight of the white voltage noise is reduced by about a factor of two with respect to a $CR-RC$ filter with the same time constant. The $1/f$ voltage contribution is reduced by about 20%. The parallel contribution is almost untouched, but since the weight of the series noise is reduced the optimum shaping time $\tau_G$ is likely to be smaller than in the case of the $CR-RC$ filter, so also the weight of the current noise can be reduced by choosing a smaller time constant. As can be demonstrated, the Gaussian shaping can be implemented by cascading one CR high pass filter to several RC low pass filters, obtaining a $CR-RC^n$ shaping. If the time constants of the individual CR and RC cells is chosen so that $RC = \tau_G/\sqrt{n}$, then for large n the shape of the signals approaches that of the Gaussian curve of equation (1.106). The delay in this case is given by $t_D = \sqrt{n}\tau_G = nRC$.

Figure 1.16 shows the unshaped signal, the signal after a $CR-RC$ shaping and the signal after a $CR-RC^{10}$ shaping. Axes are in arbitrary units. The rise time constant of the unshaped signal is $\tau_1 = 0.01$, its fall time constant $\tau_2 = 100$. The shaping time of the $CR-RC$ filter is $\tau = 1$. The shaping time of the $CR-RC^{10}$ filter is $\tau_G = 1$, meaning that the cutoff frequency of the individual CR and RC cells are $RC = \tau_G/\sqrt{10} \simeq 0.32$. As can be seen, even with a relatively small value of $n = 10$ the shape of the signal after the $CR-RC^{10}$ filter approximates very closely the Gaussian curve of equation (1.106). The delay of the Gaussian curve is $t_D = 3.2$ as expected. The shaped signals were multiplied by arbitrary factors to make the comparison easier to the eye.

Filtering by bandwidth limiting, that is with a long filter time constant, can be very beneficial to reduce noise and improve the resolution on amplitude measurements, especially in presence of dominating series noise. However if the filtered signals are too slow with respect to the average signal rate, the probability of overlapping signals or "pile-up" becomes larger. Pile-up signals of amplitudes $V_1$ and $V_2$ will be erroneously considered as a fake signal of amplitude up to $V_1+V_2$. Pile-up is minimized if short shaping time constants are chosen. This is depicted in figure 1.17, which shows two signals at t = 0 and t = 1. While the Gaussian shaping with $\tau = 0.2$ succeeds in resolving them, the same is not true for the Gaussian shaping with t = 1. Thus if on one side the best noise performance on single signals may be found at longer $\tau$, pile-up considerations can put an upper limit on its value. In cases where no compromise can be taken, a dual readout chain can be adopted: a slow readout for best amplitude resolution and a fast readout for pile-up rejection, at the price of increased complexity, space occupation and power consumption in the readout chain.

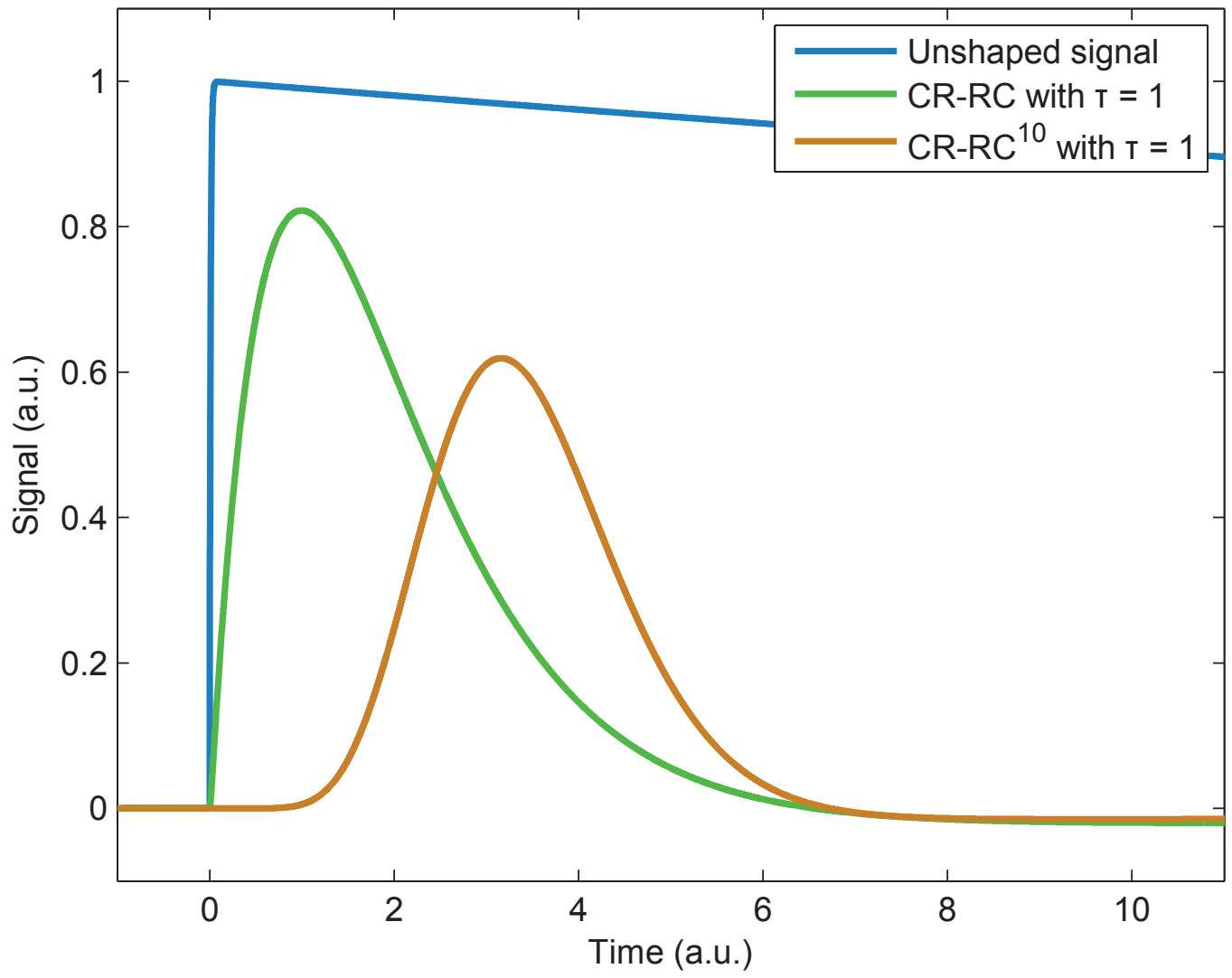

Figure 1.16: Comparison between the unshaped signal, the signal filtered with a $CR - RC$ filter and with a $CR - RC^{10}$, approximating a Gaussian shaping.

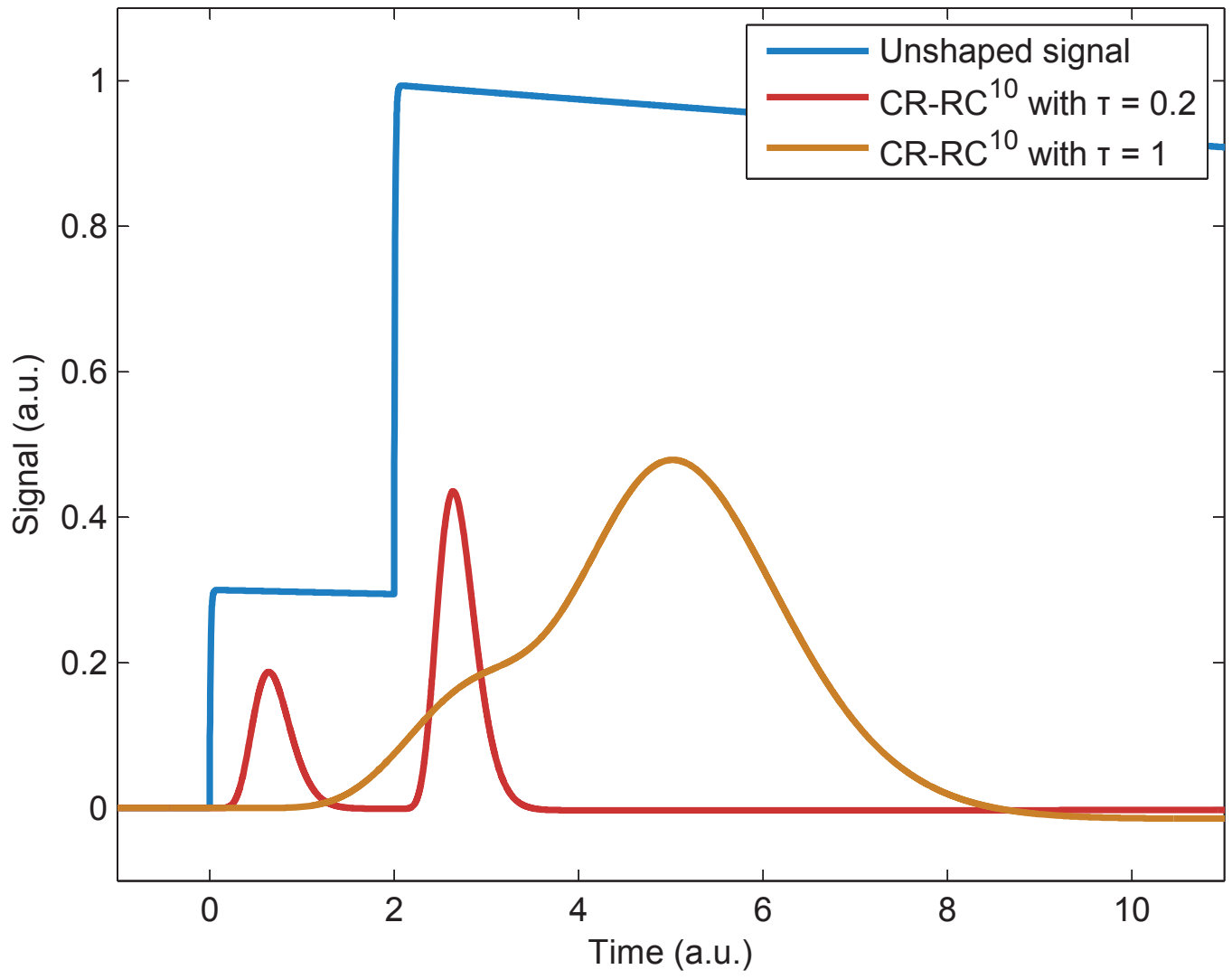

Figure 1.17: Comparison between the unshaped signals with pile-up, the signals shaped with a $CR - RC^{10}$ filter with $\tau = 0.2$ and with a $CR - RC^{10}$ filter with $\tau = 1$.

Another common shaping is the $CR^2 - RC^n$ shaping. In this case the filter has an additional high pass cell which gives the signals a bipolar shape, and can be advisable to avoid baseline shift due to pile-up at high rate. The noise performance of $CR^2 - RC^n$ shaping is generally worse than that of $CR - RC^n$ shaping.

The main advantage of $CR - RC^n$ and $CR^2 - RC^n$ filters is that they can be implemented with simple analog cells such as single pole low pass and high pass filters. For this reason Gaussian shaping, which among $CR - RC^n$ offers the best noise performance, has been extensively used in the past and is still used to this day in x and $\gamma$ spectroscopy applications. From the point on view of particle physics experiments it is considered somewhat obsolete. While $CR - RC$ and $CR - RC^2$ shapers are easily implemented with a few components, and for this reason they can also be used in integrated circuits, cascading many RC cells to obtain Gaussian shaping becomes unpractical in integrated circuits due to space contraints on Silicon. Moreover increasing the number of cells increases the power consumption per channel, which is not good if the channels are many or closely packed. So for experiments with high event rates such as the experiments at the LHC it is very uncommon to see $CR - RC^n$ shapers with $n > 2$. On the other side digital signal processing has made huge steps ahead in the last years. Complex filters can be nowadays more easily implemented in the digital domain than in the analog domain, offering more flexibility. Digital signal processing offers a broader choice of filters to be applied, since the filters do not need to be necessarily implementable in analog circuitry. There are several digital filters which may be used to improve the signal to noise ratio of acquired signals. For instance, a moving average with properly chosen weights can be used. If the size of the averaging window is small, the computation is relatively fast and the filter can effectively reduce the weight of noise at high frequency. For this reason in the cases where the signals can be processed digitally Gaussian shaping is generally not the best filter.

Moreover, since the speed of analog to digital converters and the amount of memory available for data storage has increased tremendously, it is often found convenient to sample all the signals as soon as possible, performing the largest possible part of signal processing offline where it can be better controlled or reprocessed. If signal filtering is performed offline, meaning that is not forced to operate in real time, then it is not constrained to be causal. This means that the output of the filter can depend not only on the value of the input signal in the past and present, but also in the future with respect to the filtered signal. If the shape of the signals is known, then a filter can be designed to maximally enhance the characteristics of the known signals from the acquired samples. Digital filters which are tailored for a given signal shape are usually called "optimum filters" [13]. For instance, let us assume that the frequency content of a noiseless signal is known, and let us denote it by S(s). Let us also denote its noise spectrum by N(s). Then one can define the transfer function of the optimum filter as

$$\mathrm{T_O(s)} = \frac{\mathrm{S^*(s)}}{|\mathrm{N(s)}|^2} \tag{1.108}$$

where S*(s) is the complex conjugate of the signal spectrum S(s), namely the Laplace transform of the specular image of the signal S(t) with respect to time. The basic concept behind such filter is that all fequencies in the spectrum are scaled proportionally to their expected signal to noise ratio. Other filters defined in similar ways as that of equation (1.108) exist (the Wiener filter, the matched filter, ...). The definition of the filter assumes that the frequency content of the signal and noise are known, which implies that the filter parameters must be optimized for the application. The optimum filter is generally non causal, meaning that the output of the filter at any given instant depends on the knowledge of the input signal over the entire time span. The filtered signals are obtained by multiplication of spectra in the frequency domain, or by convolution in time domain. In any case, the computations can be heavy in some cases, which makes the optimum filter practical only for applications with low event rates.

## 1.11 Filters for timing resolution

In the previous section, the problem of filtering the signals to reduce the relative RMS fluctuation in order to improve the signal to noise ratio was discussed. RMS noise directly affects the measurement of the output signal at a given instant, the most representative case being the measurement of the peak amplitude, which as mentioned in the previous sections is generally proportional to the energy deposited in the sensors by a particle event.

In other cases the main quantity of interest may be related with the time when the output signal assumes a given value. This case is schematically depicted in figure 1.18. The arrival time of a particle is determined by measuring the time $\mathrm{t_T}$ when the output signal crosses a given threshold voltage $\mathrm{V_T}$. Other more complex triggering algorithms may be implemented, but do not change the substance of the following discussion. The timing error for the signal $\mathrm{V_O(t)}$ crossing the threshold $\mathrm{V_T}$ is related to the RMS noise at the

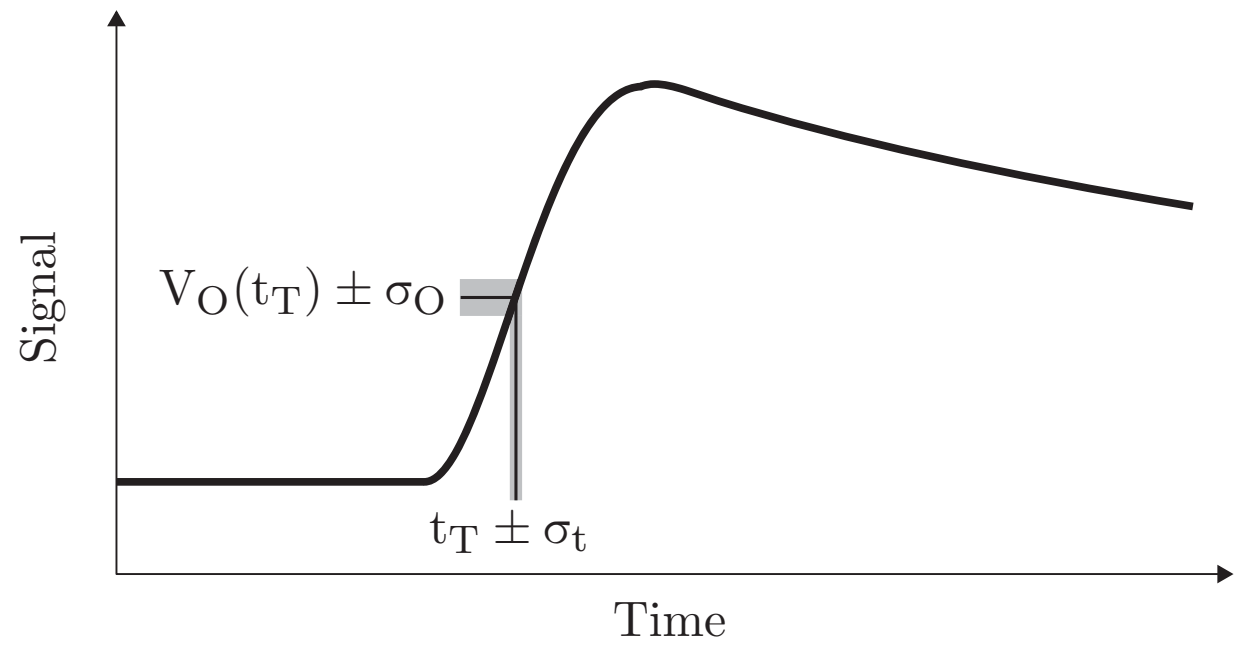

Figure 1.18: Noise in the output signal results directly in an amplitude error $\sigma_O$, or is divided by the slope of the signal at a given time to give a timing error $\sigma_t$.

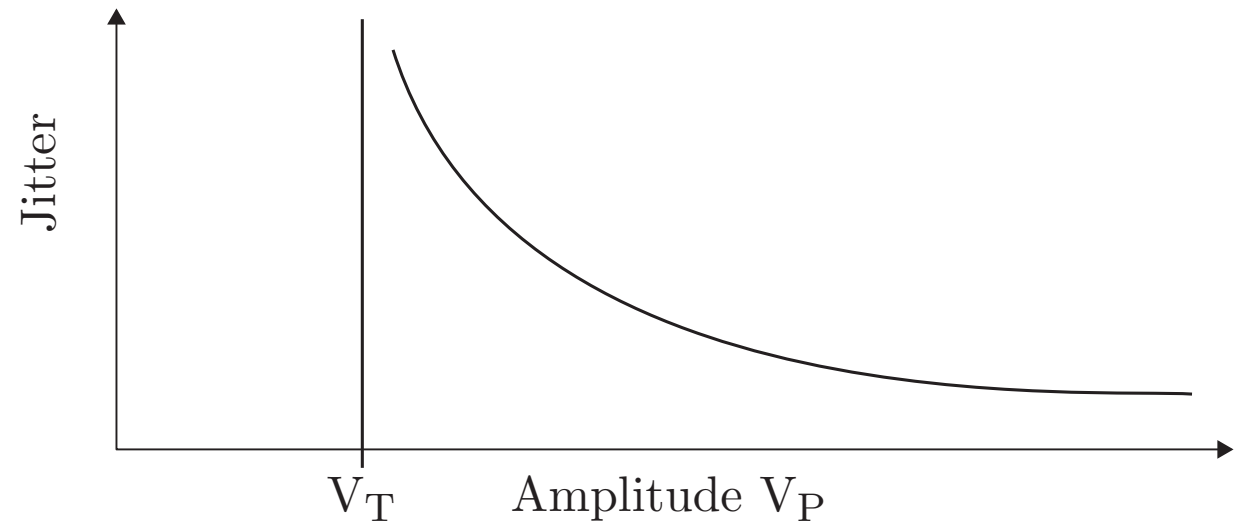

Figure 1.19: Typical trend of the jitter $\sigma_t$ versus signal amplitude.

output divided by the signal slope. In other words, the timing error is related to the output RMS noise by

$$\sigma_t = \frac{\sigma_O}{V'_O(t_T)} \tag{1.109}$$

where $\sigma_O$ is the RMS fluctuation at the output, as calculated in the previous sections, and

$$V'_O(t_T) = \left.\frac{dV_O(t)}{dt}\right|_{V_O(t)=V_T} \tag{1.110}$$

is the derivative of the output signal with respect to time, evaluated at $t_T$, that is the time when $V_O(t) = V_T$. Equation (1.109) can be simply derived from a first order expansion of the output signal for small fluctuations $\sigma_t$ around the threshold crossing time $t_T$.

As can be clearly seen from equation (1.109), in the case of timing measurements filtering should aim to reduce $\sigma_t$ by reducing $\sigma_O$ without sacrificing the signal slope $V'_O(t)$. Let us consider the case of a timing measurement on the rising edge of the signal, as depicted in figure 1.18, and let us approximate the signal expression (1.8) for $\tau_1 \ll \tau_2$, obtaining for small t

$$V_O(t) = V_P\left(1 - e^{-\frac{t}{\tau_1}}\right) \qquad \text{for } \tau_1 \ll \tau_2 \tag{1.111}$$

The time when this signal crosses the threshold $V_T$ is

$$t_T = \tau_1 \ln\left(\frac{V_P}{V_P - V_T}\right) \tag{1.112}$$

The larger the peak amplitude $V_P$ with respect to $V_T$, the smaller the time of threshold crossing $t_T$. The time derivative of the signal at $t_T$ gives

$$V'_O(t_T) = \frac{V_P}{\tau_1} e^{-\frac{t_T}{\tau_1}} = \frac{V_P - V_T}{\tau_1} \tag{1.113}$$

which is maximum and equal to $V_P/\tau_1$ for signals well above threshold, and approaches zero for $V_P$ approaching $V_T$. Thus, for a given value of $\sigma_O$, $\tau_1$ and $V_T$, the timing error $\sigma_T$ is expected to be large for $V_P$ just above threshold, and to become smaller for $V_P$ well above threshold. The qualitative trend of jitter versus signal amplitude is depicted in figure 1.19.

Let us now consider the case of the charge amplifier. By combining the expression for the equivalent noise charge for $\tau_B \ll \tau_F$, that is equation (1.68), with equations (1.71) and (1.113), into equation (1.109), we can write

$$\sigma_t = \frac{\tau_B}{Q - C_F V_T} \sqrt{i_T^2 \frac{\tau_F}{4} + A_f C_T^2 \ln \frac{\tau_F}{\tau_B} + v_w^2 C_T^2 \frac{1}{4\tau_B}} \tag{1.114}$$

under the assumption that the signal from the sensor is instantaneous, and that the speed of the output signal is limited by the bandwidth of the charge amplifier. A small $\tau_B$, or a wide bandwidth, is crucial to obtain a low jitter. It is also clear that fixed a threshold $V_T$ and the gain of the circuit, that is given by $C_F$, jitter decreases for large signals. If the signals from the sensor are slower than the readout circuit considered, the rise time of the signals should replace $\tau_B$ outside the square root in equation (1.114), but only outside the square root, as the noise term under the square root sign still depends on the amplifier bandwidth, regardless of the bandwidth of the signals. Thus for timing measurements filtering is only beneficial if it manages to reduce $\sigma_O$ while preserving the slope of the signals. Practically, this means that filtering can be used to remove the lower part of the frequency spectrum, which is more affected by parallel and $1/f$ noise. If the slope of the signals is limited by the bandwidth of the amplifier, filtering cannot be used to reduce the weight of series white noise, since series white noise is proportional to the square root of bandwidth while the slope of the signal is directly proportional to the bandwidth. Similar considerations apply also to other readout topologies.

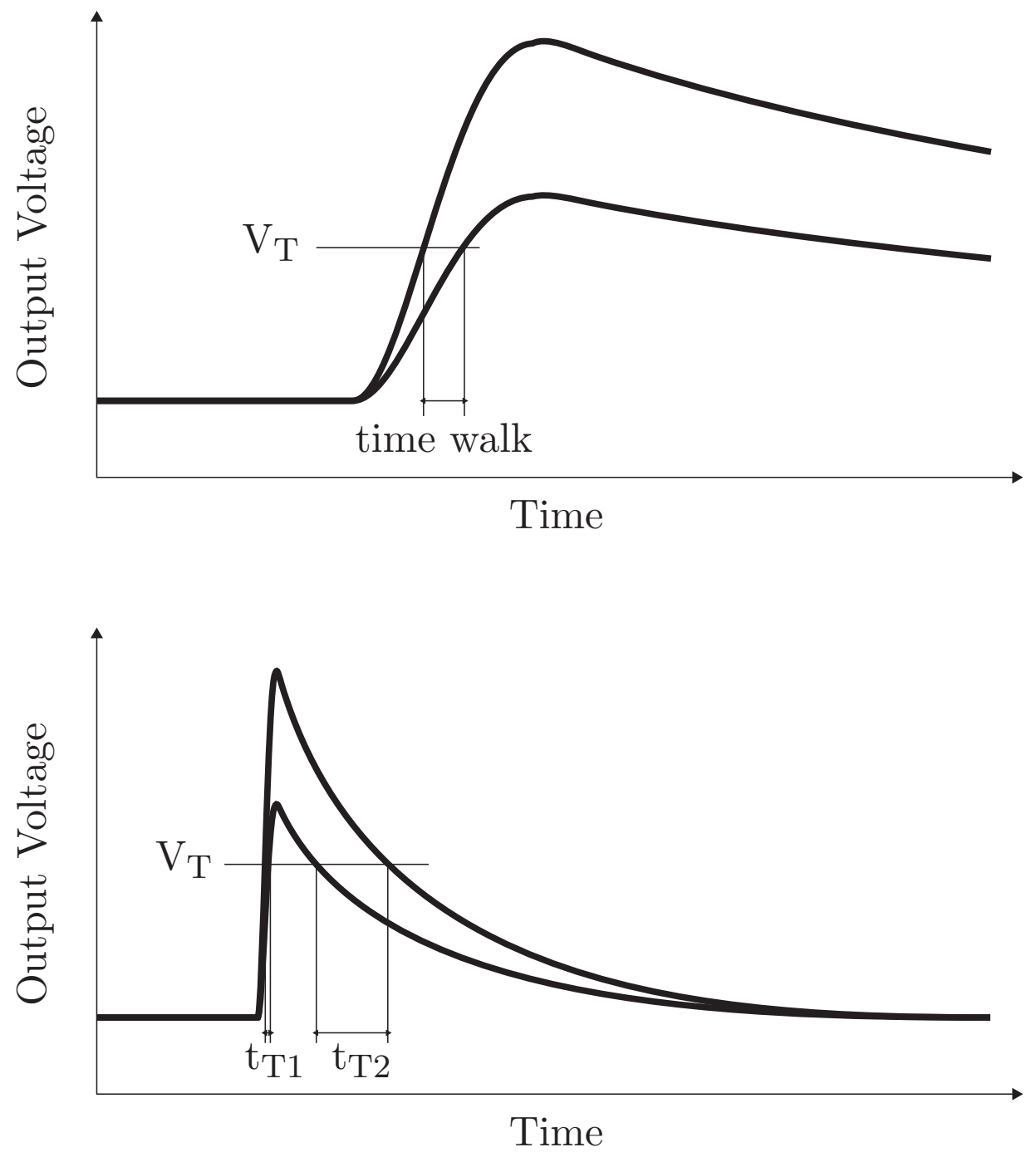

Figure 1.20: Definition of time walk between two signals of different amplitude (top) and its compensation with a time over threshold measurement (bottom).

There is another important effect to be considered affecting the resolution of a timing measurement, that is time walk. Since the amplitude of the pulses from the sensor is generally not constant, the time when the signals cross the threshold is different for different signal amplitudes. Figure 1.20 on top illustrates the definition. If the range of input signal amplitudes is large, as often happens, the value of the time walk is of the order of the rise time, completely spoiling the timing resolution unless a time walk compensation technique is employed. Time walk compensation is usually performed with two approaches. One is to perform a second timing measurement on the falling edge of the signals. In this way the total duration of each signal can be known, and since the duration of a signal is proportional to the logarithm of its amplitude, the amplitude can be calculated. Knowing the amplitude of each signal, time walk can be corrected. This technique is named "time over threshold" compensation, and is illustrated at the bottom of figure 1.20.

It needs little or no extra design effort in the analog circuitry, but requires two timing measurements to be performed instead of one. Another method to eliminate time walk is to shape the signals to make their threshold crossing time $t_T$ independent of their amplitude. This is commonly done with a constant fraction discriminator. This method requires the use of a delayed copy of the signal to be multiplied by a given scaling factor and summed to the main signal. This complicates the analog circuitry and makes this method generally not implementable in integrated circuits.

# 2 The CLARO ASIC for the LHCb RICH upgrade

## 2.1 The LHCb experiment

LHCb is an experiment devoted to heavy flavour physics deployed and taking data at the Large Hadron Collider (LHC) at CERN [14]. Its primary goal is the precise measurement of the CP violation parameters contained in the currently accepted particle physics model (the "Standard Model") and the search for new phenomena beyond it. CP violation gives a measure of the asymmetry between particles and antiparticles. While in the Big Bang equal amounts of matter and antimatter were originated, in the present day universe the situation is highly asymmetrical, as the first dominates over the second. The Standard Model contains CP violation terms but these are not enough to explain the asymmetry between matter and antimatter observed in the present day universe. The presence of new CP violation sources beyond the Standard Model is thus foreseen, and some answers are expected to come from the LHCb measurements.

LHCb is designed to detect the products of proton-proton collisions which contain Beauty and Charm hadrons. Most of these particles are produced with small angles with respect to the direction of the primary proton beams. For this reason LHCb was designed to cover only an angle from 15 mrad (limited by the presence of the beam pipe) to about 300 mrad with respect to the beam axis. A schematic side view of the LHCb detector is shown in figure 2.1. The collision vertex is on the left. Going outwards from the collision vertex LHCb is composed of various subsystems, which are the vertex locator (VELO), the RICH1, the trigger tracker (TT), the magnet, the trackers T1, T2 and T3, the RICH2, the first muon station M1, the electromagnetic and hadronic calorimeters (SPD/PS, ECAL and HCAL), and the remaining muon stations M2, M3, M4 and M5. The VELO is a silicon tracking detector with high spatial resolution, down to tens of µm, used to precisely locate the interaction vertices where the secondary particles are produced. The TT, T1, T2 and T3 are used to detect the trajectory of the secondary particles. Due to the presence of the large resistive magnet, able to generate an integrated field up to 4 Tm over 10 m, the trajectories of charged particles are bent, which allows to measure their momentum. The RICH detectors are used to measure the velocity of particles, which together with the momentum information can be used to determine their mass. The calorimeters are used to measure the energy of the secondary particles. The muon stations are used to detect muons produced in the collisions. Particle identification and precise event reconstruction is provided by combining the informations from all the above subsystems. The work presented in this chapter is related to the RICH subdetectors, which will be described with more detail in the following section.

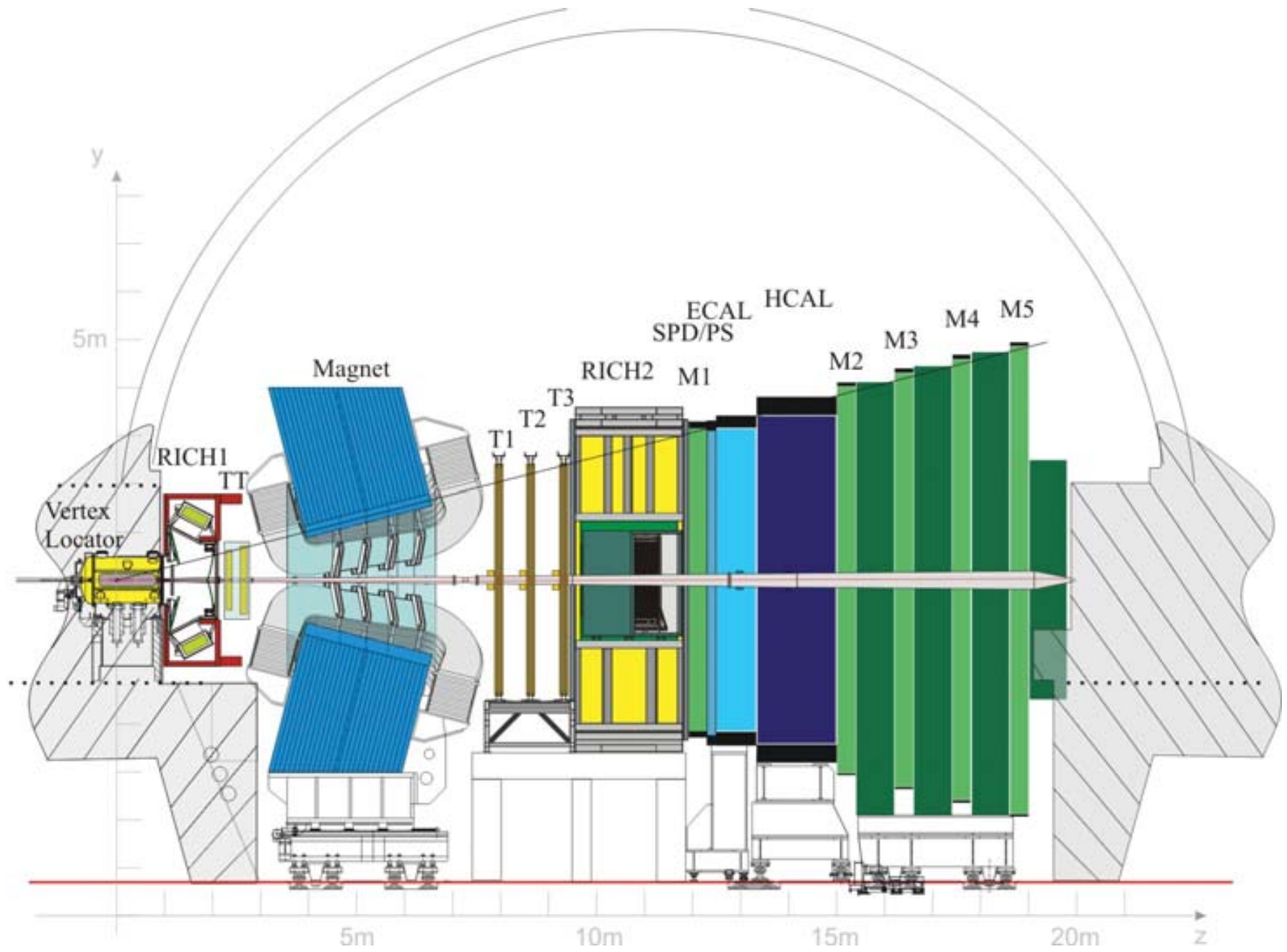

Figure 2.1: Schematic view of the LHCb Detector. The beam pipe runs from left to right in the figure.

The LHCb experiment is taking data since the end of 2009. The nominal design luminosity of $2 \times 10^{32}$ cm$^{-2}$s$^{-1}$ was reached at the end of 2010 at the energy of 7 TeV in the center of mass. In 2012 the energy of the LHC was increased to 8 TeV. At the end of 2012, after three years of operation, LHCb has accumulated more than 3 fb$^{-1}$ of data, corresponding to about $2 \times 10^{14}$ proton-proton collisions, of which more than half in 2012 alone. Figure 2.2 shows the plot of a LHCb event after offline reconstruction, showing the tracks of the detected particles originated in a proton-proton collision. The collision vertex is at the bottom left of the figure.

LHCb operates with a reduced luminosity with respect to the maximum luminosity which can be delivered by the LHC, in order to maximize the num-ber of single proton-proton collisions per bunch crossing. This makes the event analysis easier and limits the radiation damage in the detector components. The resulting mean event rate is 12 MHz, corresponding to one collision every 80 ns on average. The "Level 0" (L0) hardware trigger, largely contributed by the calorimeters and the muon stations, is used to reduce of a factor of 10 the amount of data to be trasmitted off detector. The design of the overall electronic architecture of the LHCb detector, from the front-end circuits to data acquisition, was driven by the resulting foreseen event rate of about 1 MHz. This was considered to be the maximum rate affordable considering the technology available at the time of the design. The front-end electronics of the subdetectors was designed to cope with such trigger rate. The data at 1 MHz is transmitted through optical links to an off-detector CPU farm, which triggers the events of interest more precisely in software and reduces to about 5 KHz the total event rate to be written to storage for offline physics analysis.

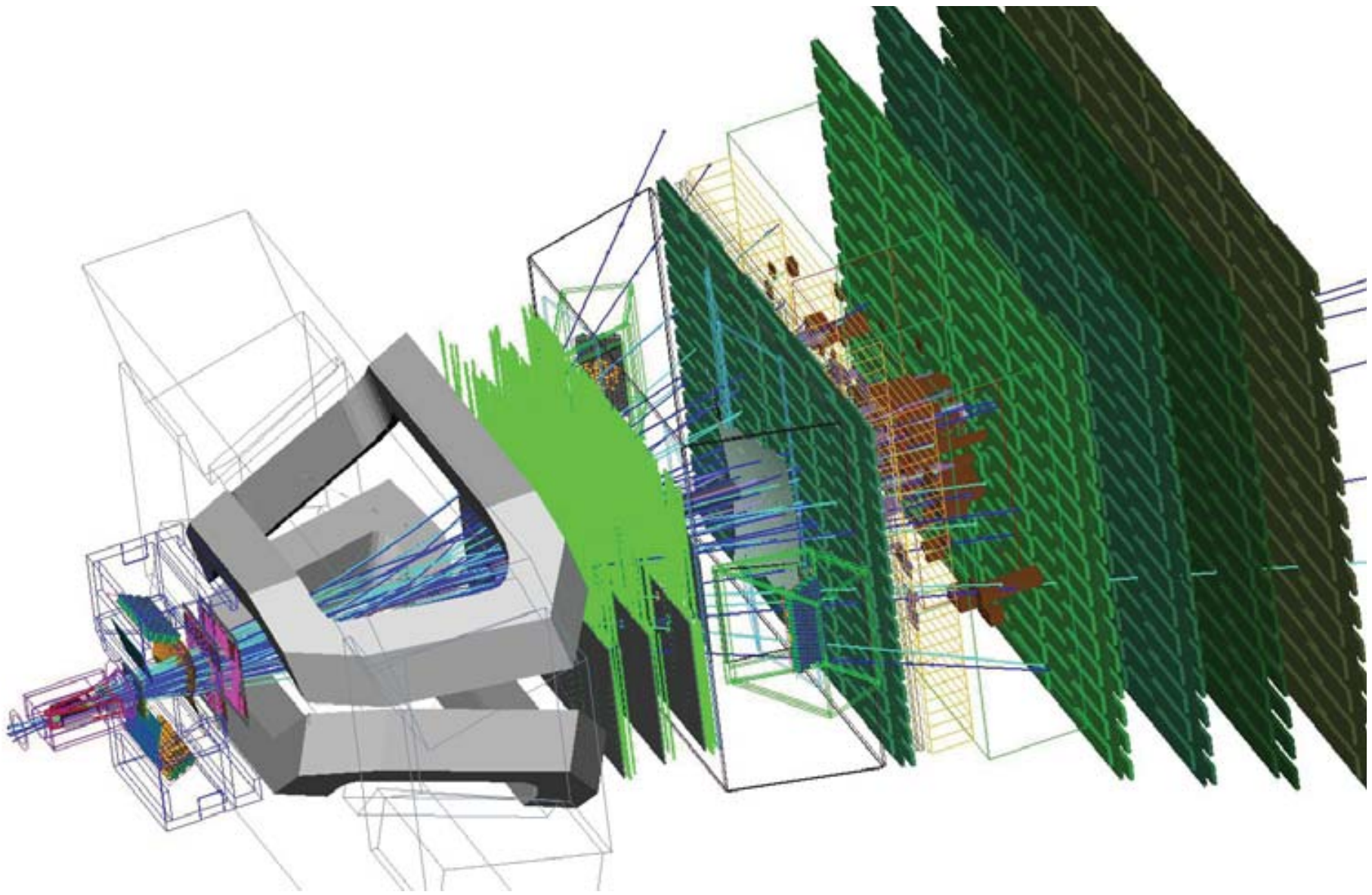

Figure 2.2: A proton-proton collision in LHCb. The tracks of the secondary particles produced are shown in blue.

By 2018 an upgrade of the whole detector is planned in order to increase the luminosity up to ten times the current design value [15,16]. The increased amount of data will allow to push the precision of LHCb measurements to a higher level. Moreover, the harware trigger is to be removed, and all triggering is to be done in software, increasing its flexibility and sensitivity to exotic decays and phenomena beyond the Standard Model. The upgraded LHCb will be able to accumulate 50 $fb^{-1}$ of data in ten years of operation. The requirement for a fully software trigger increases the event rate which needs to be readout from the current value of 1 MHz to 40 MHz. Most of the front-end electronic circuits were designed for the original data rate of 1 MHz, and thus need to be upgraded or re-designed to cope with the higher readout speed. In some of the subdetectors there are sensors which by 2018 will have suffered aging from long time operation and exposure to radiation, and will need in any case to be replaced. This gives more flexibility in the design of the upgraded subdetectors, since in some cases it is preferrable to replace the sensors together with the readout electronics to better match the 40 MHz specification. This is the case of the RICH1 and RICH2 subdetectors.

## 2.2 The LHCb RICH detectors

Ring imaging Cherenkov (RICH) detectors exploit the Cherenkov radiation produced by charged particles which cross a proper medium to measure their velocity. Their working principle is based on the Cherenkov effect. When a fast charged particle crosses a medium (the radiator) where the speed of light is lower than that of the particle, Cherenkov photons are produced. Their spectrum is peaked towards the blue and UV, ranging up to the soft X rays. The number of photons produced for a given radiator thickness depends on the refractive index n of the radiator, and is usually relatively small, typically of the order of a few tens to several hundreds. There is a lower velocity threshold below which no photons are emitted, whose value is c/n, where c is the speed of light. The higher the refractive index, the lower the threshold, and the higher the number of photons produced for a given particle velocity and radiator thickness. The total photon yield for a given particle velocity and refractive index is directly proportional to the thickness of the radiator. Cherenkov photons are emitted instantaneously and with a precise angle, forming a cone with respect to the trajectory of the primary particle. The Cherenkov angle $\vartheta_C$ is given by

$$\vartheta_C = \cos^{-1}\left(\frac{1}{n\beta}\right) \tag{2.1}$$

where $\beta = v/c$ is the ratio between the speed of the particle and the speed of light in vacuum. For very fast particles with $\beta \to 1$ the Cherenkov angle saturates to $\cos^{-1}(1/n)$. The Cherenkov effect is summarized in figure 2.3.

By measuring the Cherenkov photons on a plane properly equipped with photon sensors, rings are detected whose radius depends on the velocity of the primary particle. The spatial resolution, or granularity, of the photon sensors is a crucial parameter for precise measurements. The choice of the radiator defines the velocity range where the Cherenkov effect can be exploited, which is about a decade between the lower velocity threshold and the upper saturation limit, depending on the spatial resolution of the measurement.

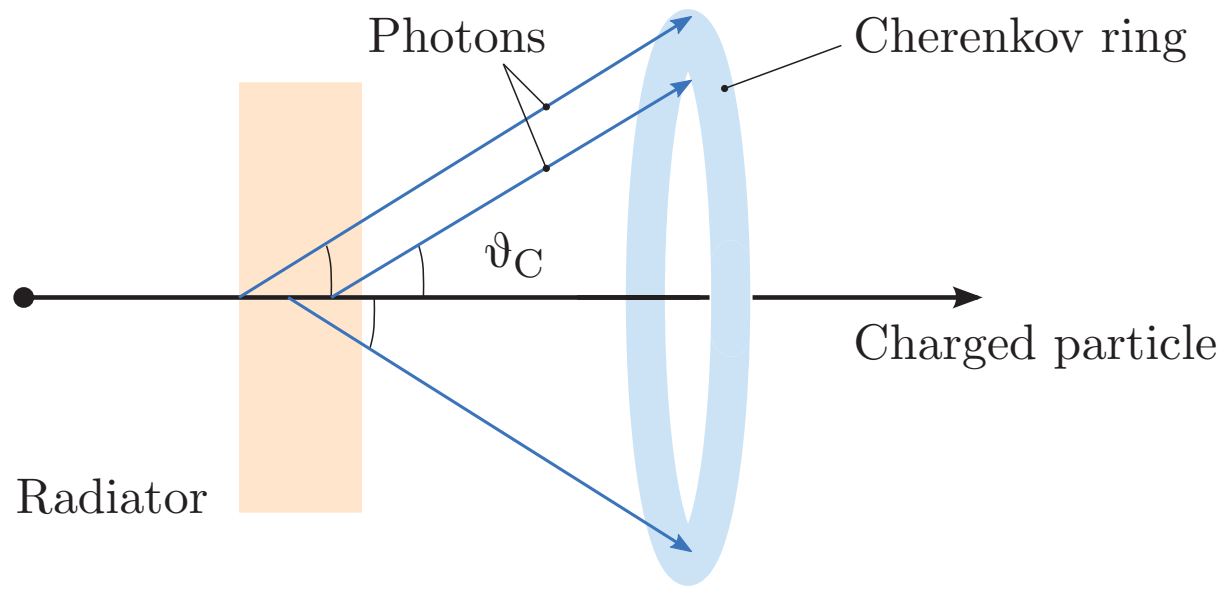

Figure 2.3: The Cherenkov effect, by which a charged particle which propagates through a given medium (the radiator) where the speed of light is lower than the speed of the particle produces photons which are emitted with a precise angle $\vartheta_C$ with respect to the direction of the particle.

In the simple case depicted in figure 2.3 the thickness of the ring is limited by the thickness of the radiator. To eliminate this dependence spherical mirrors can be used to focus the photons, improving the resolution. The use of mirrors offers also another advantage, since it allows to place the photon sensors farther from the beam, improving their effective granularity and reducing the radiation dose they have to tolerate.

In figure 2.3 the refraction of photons going out of the radiator was neglected, as commonly happens for low n radiators. If n is large the refraction must be considered, and internal reflection may occur. In some detector designs internal reflection is exploited to collect the Cherenkov photons at the sides of the radiator, as for the BaBar DIRC, which was operated at SLAC until 2008, and the LHCb TORCH, which is currently in the R&D phase.

In the case of LHCb, photons, electrons and neutral hadrons are identified through the measurement of their energy in the calorimeters. Charged hadrons are instead identified by the two RICH detectors (RICH1 and RICH2), which are oprimized to separate kaons, pions and protons in a broad momentum range, from 2 to 100 Gev/c. The RICH1 is located upstream with respect to the magnet, as close as possible to the interaction region, and is employed to detect lower momentum particles over the full 300 mrad detector acceptance. It employs two radiators: tiles of solid Silica Aerogel with n = 1.03 to cover momenta between 2 and 10 GeV/c, and gaseous $C_4F_{10}$ with n = 1.0014 to cover momenta up to about 40 GeV/c. The RICH2 is located downstream and has an acceptance of 120 mrad. It employs gaseous $CF_4$ with n = 1.0005 to cover momenta of the higher energy particles up to 100 GeV/c.

The average photon yield is about 20 photons for the Aerogel, and about 100 from the gaseous radiators, due to the larger thickness of the gas-filled

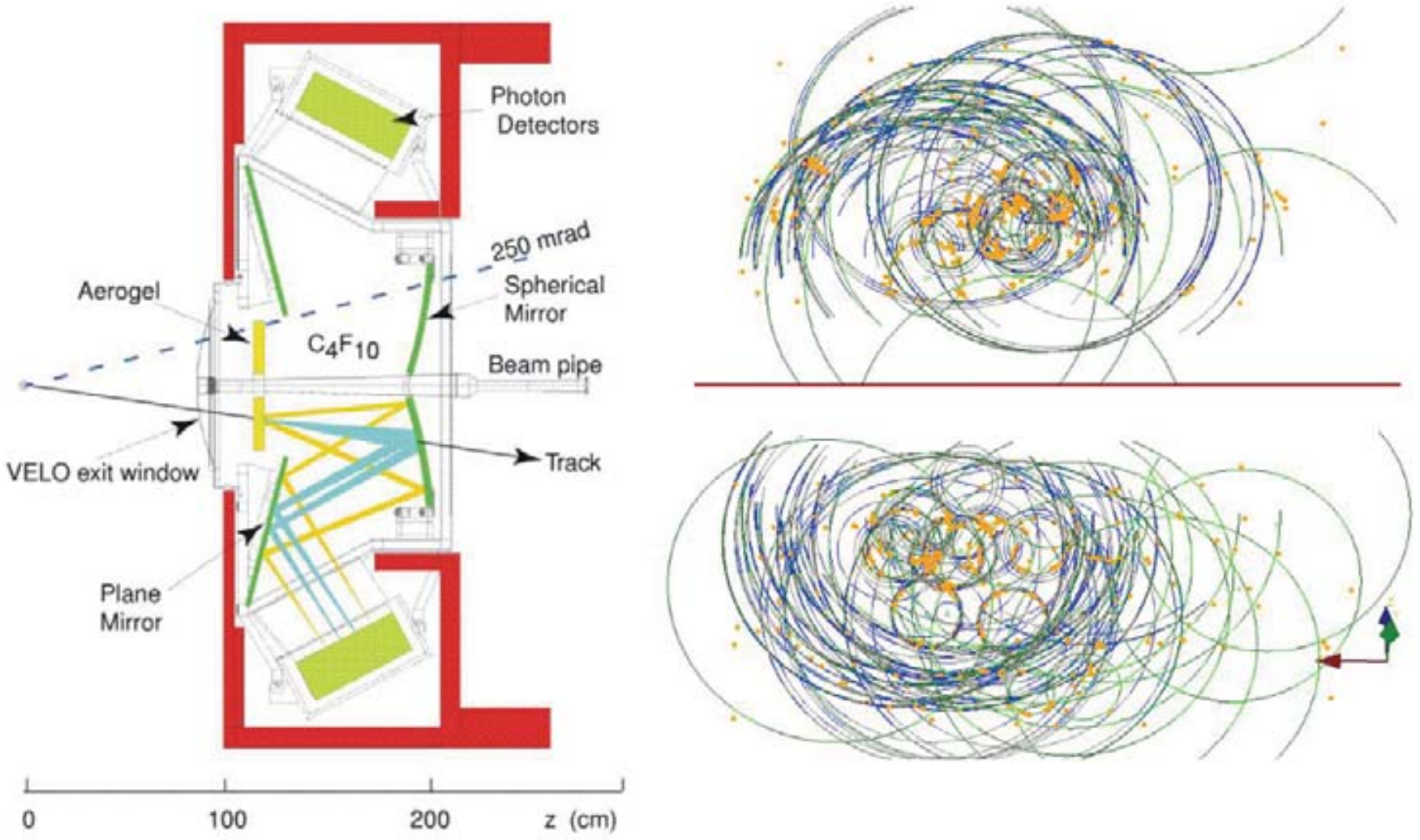

Figure 2.4: On the left, a side view schematic of the LHCb RICH1. On the right, Cherenkov rings detected in the upper and lower photosensitive planes of the LHCb RICH1.

---

volume. Both RICH detectors have similar optical systems, made of a tilted spherical focussing primary mirror and a secondary flat mirror, with reflectivity of about 90%. Each optical system is divided into two halves, placed vertically in the RICH1 and horizontally in the RICH2. Figure 2.4 shows on the left a side view schematic of the RICH1, and on the right the Cherenkov photons detected by the photon sensors.

In the current LHCb RICH detectors the Cherenkov photons are detected by hybrid photon detectors (HPD) [17], 196 in the RICH1 and 288 in the RICH2. Each HPD consists of a cylindrical vacuum tube with a 75 mm active diameter, equipped with a quartz window and a multialkali photocathode with a quantum efficiency up to 30% in the wavelength range $200 - 600$ nm. Each HPD has $32 \times 32$ pixels with $2.5 \times 2.5$ mm$^2$ effective size each. The incoming photons are converted to photoelectrons in the photocathode and accelerated by a voltage of $-16$ kV before striking a pixelated silicon sensor, that is a p-n junction in reverse bias, generating a signal of about 5000 e$^-$. The front-end electronics is a custom integrated circuit embedded in the vacuum envelope of the tube, with an equivalent noise charge close to 150 e$^-$. The timing resolution of the readout is better than 25 ns, that is the nominal bunch crossing rate of the LHC. The readout chip can transmit data at a maximum rate of 1 MHz, which matches the L0 trigger rate.

In 2018 the readout speed of the photon sensors will need to be increased to 40 MHz to match the upgrade specifications. This requires a redesign of the front-end electronics. Since currently the readout chip is embedded in the HPDs it cannot be removed without disassembling the devices. Moreover, by 2018 the HPDs will have to be replaced in any case due to aging after ten years of operation, and the fabrication of new HPDs in large numbers is not easy due to their custom nature. Replacing the HPDs with other photon sensors which are commercially available and are more naturally suited for a full 40 MHz readout is now considered a better option. The devices chosen for the upgraded RICH are multi-anode photomultiplier tubes. These photon sensors will be described in the next section.

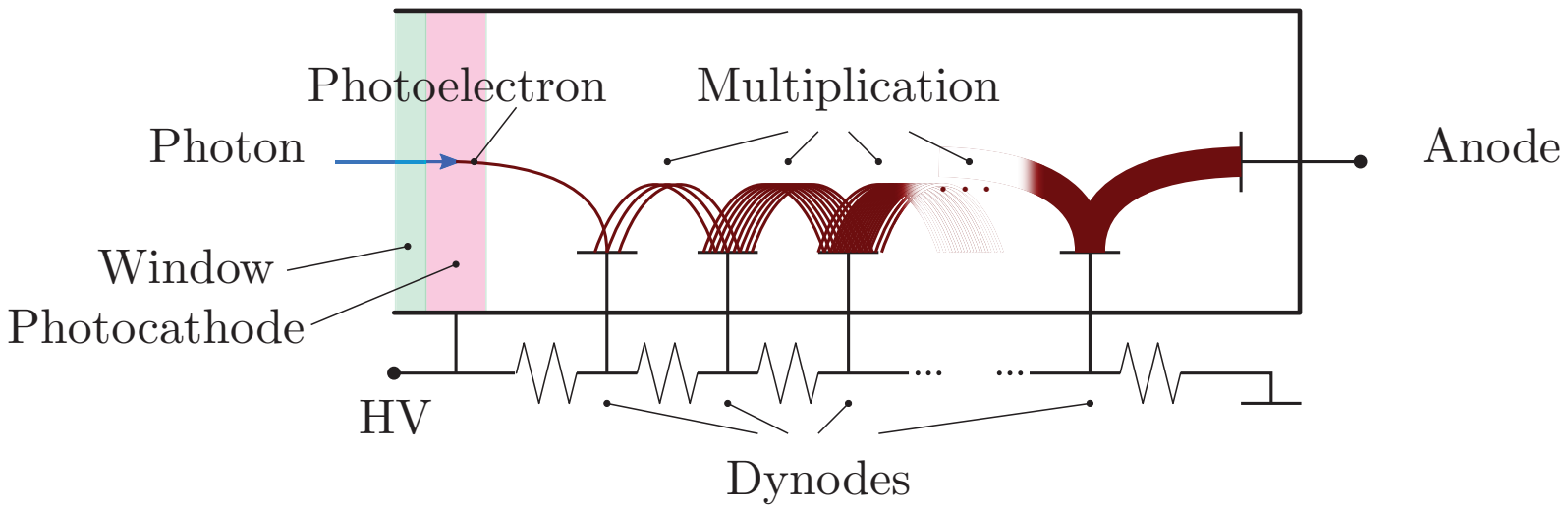

Figure 2.5: Drawing of the electron multiplication process in a photomultiplier tube.

## 2.3 Multi-anode photomultiplier tubes

Single channel photomultiplier tubes were invented about a century ago, and are employed since many decades to detect low light levels. As of now their use as general purpose light sensors is disfavoured due to their cost, which is much larger than that of Silicon light sensors. But thanks to their very low dark current, they still play a leading role when high sensitivity down to the single photon level is required.

A drawing of a single channel photomultiplier tube is shown in figure 2.5. Each tube is composed of an entrance window, a photocathode, a series of dynodes and a readout anode [18]. The window needs to be transparent to photons in the wavelength range of interest. For visible light detection Borosilicate glass is often used, while for increased sensitivity to UV photons other materials such as Quartz can be employed. The photocathode is a crucial element, as it provides the conversion between photons and electrons by photoelectric effect. For a good sensitivity, the photocathode material needs to have the lowest possible extraction potential. Alkali metals are most often used. The extraction probability is called quantum efficiency, and depends on the wavelength of incoming photons. The photocathode material should be chosen for the maximum sensitivity to the wavelength range of interest, which in the case of Cherenkov light extends towards the UV. Typical values of quantum efficiency are around 25%, while for some of the best cases it can range up to 40% peaking around 350 nm.

Photomultiplier tubes are biased with a negative high voltage close to −1 kV. The high voltage is usually applied to the photocathode, and the dynodes are biased with fractions of it in such a way that there is a large electric field between each dynode and the next. The readout anode is usually tied at virtual ground by the input of a charge amplifier. The working principle of these sensors is as follows. Photons hitting the entrance window are coverted to photoelectrons by photoelectric effect in the photocathode. Inside the vacuum tube the photoelectrons are accelerated by the electric field towards the first dynode. When a photoelectron strikes the first dynode it liberates a few secondary electrons. The process is a close relative of the photoelectric effect, and is called secondary emission. The secondary electrons are accelerated toward the second dynode, and the multiplication process is iterated, until a large signal after many stages of multiplication reaches the anode, which is the readout electrode.

Inefficiencies due to geometry may cause the loss of some photoelectrons before reaching the first dynode, determining the collection efficiency of the tube. Typical values in nominal bias conditions are close to 90%. Collection efficiency drops if the bias voltage between the photocathode and the first dynode is not large enough. The gain of each dynode $\mu$ is usually about 3 or 4, depending on the bias voltage. The total gain of the multiplication chain is then $\mu^d$, where d is the number of dynodes. Typical photomultiplier tubes have a number of dynodes between 8 and 12. For $d = 12$ the gain results in about $10^5 - 10^7$, thus the output pulse consists of million of electrons, which are fairly large signals compared to those of other photon sensors such as the HPDs mentioned above. Low noise is thus generally a minor concern for photomultiplier readout circuits, unless a very low jitter is required. Since $\mu$ depends almost linearly on the bias voltage $V_B$, the gain of the device follows a power law and goes approximately as $V_B^d$. The secondary emission from each dynode follows a Poisson distribution with mean $\mu$, whose standard deviation is $\sqrt{\mu}$. The fluctuation in $\mu$ is thus of the same order of magnitude as $\mu$ itself. The width of the output pulses varies consequently, being dominated by the fluctuation at the first dynodes. For this reason variations in signal amplitude up to about one order of magnitude around the mean value given by $\mu^d$ are commonplace. For small photomultipliers the signal is fast, with a rise time of the order of 1 ns and a timing resolution which ranges down to 200 ps for the fastest devices.

The dark current is the current coming out of the anode when no photons are hitting the photocathode. It is due to electrons which leave the photocathode or the dynodes due to thermal agitation and enter the multiplication chain giving signals at the output. Dark current events generated in the photocathode give signals which are indistinguishable from the real signals due to incoming photons, while those generated in the dynodes skip some multiplication steps and give smaller signals. By setting a threshold on the detected signals and increasing the gain at the first dynode the dark current from the dynodes can be kept under control. The dark current from the photocathode instead cannot be discriminated from the real signals, but its rate is usually low, being below a few Hz per mm$^2$ of photocathode area for most photocathodes at room temperature. It strongly depends on temperature, and large reduction in the dark current can be obtained by cooling the devices below room temperature.

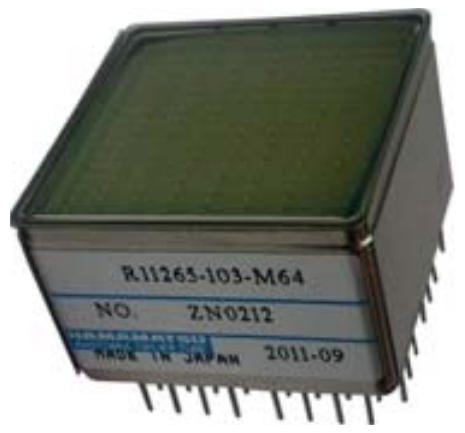

| | |
|---|---|
| Geometrical dimensions | $26.2 \times 26.2$ mm$^2$ |
| Window material / thickness | UV glass / 0.8 mm |
| Photocathode minimum effective area | $23 \times 23$ mm$^2$ ($> 80\%$) |
| Photocathode material | Super Bialkali |
| Spectral response range | $185 - 650$ nm |
| Number of pixels / dimensions | 64 / $2.9 \times 2.9$ mm$^2$ |
| Number of dynodes | 12 |
| Maximum supply voltage | $-1.1$ kV |
| Typical gain at $-1$ kV | $> 1 \times 10^6$ |
| Uniformity between pixels | $1 \div 3$ |
| Dark current (average per pixel) | 0.4 nA |
| Rise / transit time | 0.6 ns / 5.1 ns |

Figure 2.6: The Hamamatsu R11265 multi-anode photomultiplier tube.

In recent years, multichannel models appeared on the market. These are called multi-anode photomultiplier tubes (Ma-PMT). Most models offer pixellated anodes arranged in $4 \times 4$, $8 \times 8$ or $16 \times 16$ pixels. The pixel dimensions range from $2 \times 2$ mm$^2$ to about four times that value. The high voltage bias for the photocathode and dynodes are shared between channels. The industry leader for such devices is Hamamatsu Photonics. For the LHCb RICH upgrade, various devices were tested in Milano Bicocca: the H9500, with $16 \times 16$ pixels of $2.8 \times 2.8$ mm$^2$ size, the R7600, with $8 \times 8$ pixels of $2 \times 2$ mm$^2$ size, and more recently the R11265, with $8 \times 8$ pixels of $2.9 \times 2.9$ mm$^2$ size. Figure 2.6 shows a picture of the Hamamatsu R11265 and a table summarizing its main characteristics.

Being multichannel devices, additional performance parameters must be considered. One is the crosstalk, which is due to electrons jumping to neighbouring dynodes in the multiplication process, and may consitute a limit to space resolution. Another parameter to be considered is the gain uniformity between pixels. For the devices mentioned above, the uniformity is within a factor of three, meaning that at a given bias voltage one pixel may have a gain up to three times the gain of another pixel of the same device. Since in RICH detectors the photon sensors must be packed to obtain a large photosensitive area, the active area fraction of each device must be considered, and the presence of inactive borders near the sides of the device may become a limiting factor.

The H9500 offers the lowest cost for a given area coverage, but it is not designed for single photon counting applications. In laboratory measurements a crosstalk up to 30% was found at the single photon level, making these devices not optimal for a use in RICH detectors [19]. The R7600 offers excellent single photon counting performance and negligible crosstalk, but has an active area ratio of only about 50% due to the relatively large inactive border on the sides of the device. This devices could be used in a RICH detector if lenses were used to recover a larger active area [20]. The R11265, recently made available by Hamamatsu, performs similarly to the R7600 but overcomes its active area limitation making this device ideal for large area RICH detectors. The price of these devices is currently higher than that of the others, and may become a critical factor for large area deployment. Figure 2.7 shows the single photon spectra of some of the pixels of a R11265 Ma-PMT currently under test in Milano Bicocca. The shapes of the spectra reflect the expected large variations in signal amplitude due to the intrinsic random nature of the multiplication process.

In order to ascertain the compliance and good behaviour of these devices in the LHCb environment additional factors need to be considered. Intensive tests are being carried out on the various photomultiplier models mentioned above, and some critical aspects were already found. First, a residual magnetic field due to the large LHCb magnet is present in the region of the RICH subdetectors, especially in the RICH1. The residual field inside the iron enclosure of the RICH1, which already contributes to shielding, ranges up to about 30 G. In the case of the multi-anode photomultipliers tested, the magnetic field causes a gain and efficiency loss which is already noticeable at 30 G, especially critical for pixels near the edges of the devices. Additional shielding must then be considered at least for the RICH1 to operate these devices in the LHCb environment. Another important factor is the aging of the devices. The main effect related to aging is a gain reduction, which becomes evident in a few weeks of operation at the relatively high average anode current of 1 μA. In single photon counting applications with an average photon detection rate per pixel $\nu$, the average anode current is given by

$$I_{PMT} = qG\nu \tag{2.2}$$

where q is the electron charge and G is the gain of the device. For a typical gain of $1 \times 10^6$ an average current of 1 μA corresponds to an event rate $\nu$ of about 6 MHz, which is a fairly high value. Nevertheless aging should be considered in the high occupancy regions of the RICH1. Its effects can be mitigated by operating the devices with a low bias voltage, to keep the gain low, and then increasing the bias voltage later on to compensate for the gain loss. Another aspect which must be kept under control, and affects both the sensor and the readout electronics, is radiation damage. Radiation tests on these sensors have not yet been performed.

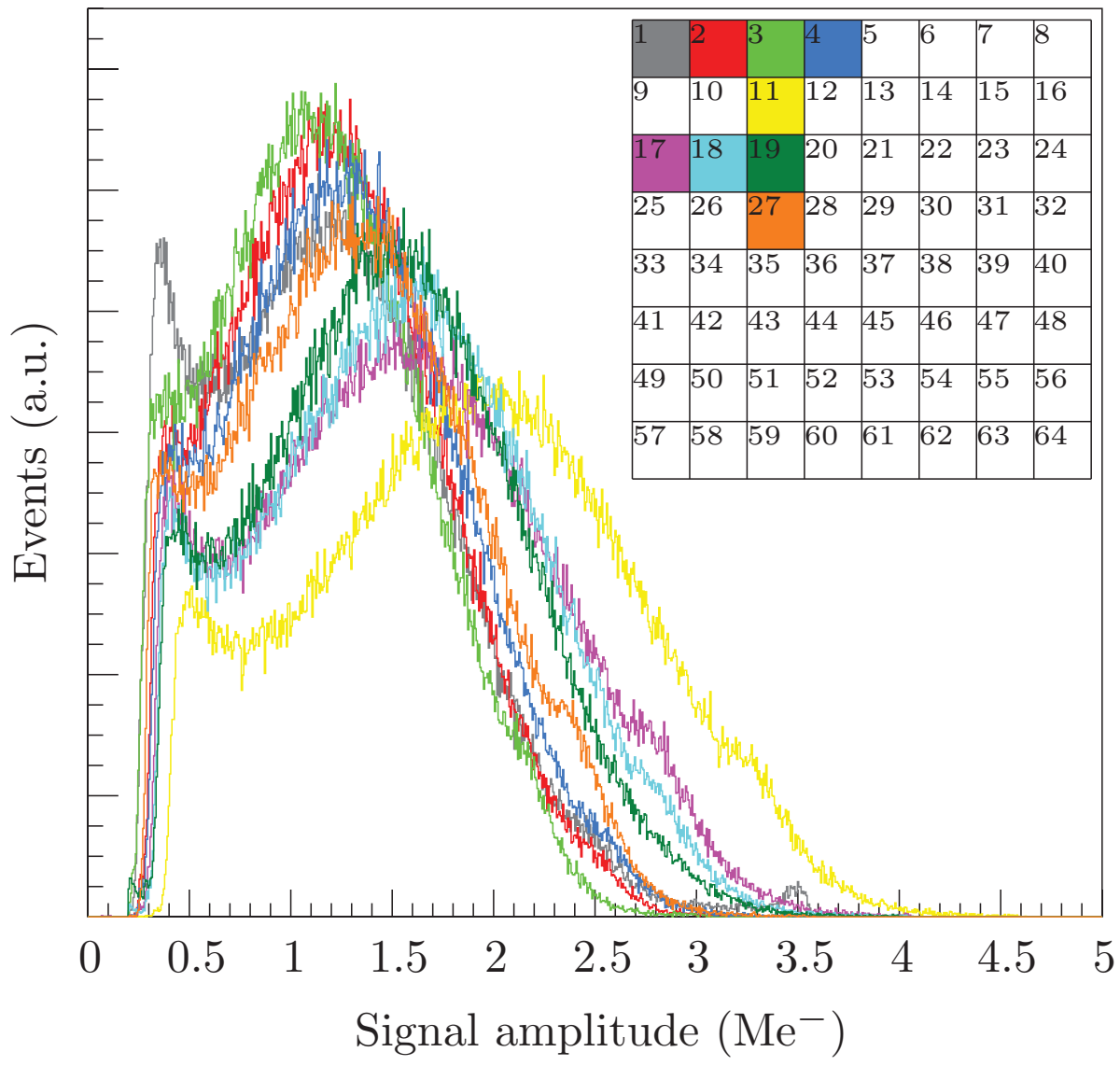

Figure 2.7: Single photon spectra of some pixels of a R11265 Ma-PMT biased at −850 V, with the threshold set at 300 Ke$^-$. The uniformity table provided by Hamamatsu is also shown, where the relative gain between pixels can be found. The colors of the spectra match those of the highlighted pixels in the uniformity table.

## 2.4 Readout circuit requirements

The anodes of R11265 photomultiplier tubes are modeled as current sources with capacitive impedance, falling under the case of the capacitive sensors discussed in chapter 1. In response to single photon excitations the pixels give charge signals of a few million electrons in nominal bias conditions, with the amplitudes randomly distributed according to the spectra shown in figure 2.7. The capacitance of each pixel is of the order of 1 pF and its leakage current is of the order of 1 nA, partially contributed by the dark counts.

The front-end circuit required for photon counting needs to read the current pulses from the pixel of the sensor on a virtual ground node and give at the output a binary information if the pixel was hit or not. The need for a virtual ground is due to the presence of a parasitic capacitance between the pixels of the photon sensor. If the input voltage is not fixed, crosstalk can be injected to neighbouring pixels through the parasitic capacitance. This source of crosstalk is minimized if the voltage at the input is fixed, which happens with a current or charge sensitive configuration. The readout circuit for each pixel can be considered to be composed of two main parts: a current or charge amplifier, which reads the input current signals on a virtual ground node, and a discriminator, that is a circuit which reads the analog signal from the amplifier, compares it to a given threshold, and gives a binary information at its output which indicates if the signal from the amplifier exceeds the threshold. From the binary output of each pixel the signal must be counted and time-tagged by purely digital circuitry, which can be implemented in a FPGA. Since as pointed out in section 2.3 the gain of the pixels of the photon sensor may vary of a factor of three in the same device, the gain of the amplifier or the threshold of the discriminator (or both) should be settable on a channel by channel basis.

The frequency of collisions of the proton beams (the bunch crossing rate) in the LHC accelerator is 40 MHz, which corresponds to collisions every 25 ns. As already pointed out the upgrade of the LHCb experiment aims at a tenfold increase in the number of proton-proton collisions per bunch crossing with respect to the current value, and to a full 40 MHz readout of the whole detector in order to perform all triggering in software. The main requirement of the readout circuit of the photon sensors of the upgraded RICH is the capability of handling such data rate. From the point of view of the individual pixels of the photon sensors to be employed in the RICH, this means that the readout circuit should be fast enough to avoid the superposition of events corresponding to different bunch crossings. The resulting hit rate per pixel depends on the layout of the detectors. For a given number of proton-proton collisions per bunch crossing, the number of photons hitting a pixel of the photon sensors depends on the radiator chosen, on the mirror geometry and position, and on the position and dimensions of the pixels of the photon sensors. These aspects may change in the design of the upgraded RICH detectors, and are not currently fixed. Assuming to keep everything as in the current configuration, simulations were carried out showing that in the central regions the pixel occupancy (that is the hit probability) can be as high as 20%. For a safe margin, this would force the readout circuit to guarantee that no dead time follows the detection of a photon, which means that the readout circuit after every hit would need to be ready to detect another hit in less than 25 ns. If some of the geometrical aspects of the RICH were changed in the upgrade, then the resulting occupancy in the hottest regions could be lower, possibly relaxing the readout speed requirements.

Due to the close packing of the photon sensors, the front-end circuitry of each pixel needs to be integrated on chip. This seems the only way of packing several photon sensors with pixels sizes of the order of a few $mm^2$ close to each other with no inactive area in between. In order to minimize cooling the power consumption on the front-end circuits should be low, of the order of a few mW per channel. Another important requirement of the readout circuits is their tolerance to the amount of radiation present in the LHCb environment. As is well known, deep submicron technologies can usually tolerate a larger amount of radiation with respect to older technologies, due to the lower thickness of the gate oxide and to the lower cross-section they offer to radiation. Older technologies on the contrary can suffer damage from ionizing radiation due to the build-up of charge trapped in the gate oxide. The widely used and relatively old 0.35 µm technology stands somewhat in between, as it can tolerate moderate amounts of radiation without serious damaging, but does not reach the level of radiation hardness of more scaled technologies such as 130 nm and below.

At the present state, there is no existing chip which satisfies all the above requirements. An ASIC (application-specific integrated circuit) purposely designed for multi-anode photomultipliers in AMS 0.35 µm SiGe-BiCMOS technology is the MAROC [21]. This chip has 64 channels, matching the number of channels of the Hamamatsu R11265, with a power consumption of about 3 mW per channel. Its main drawback is the readout speed, as the rise time of the binary signals at its outputs is close to 15 ns, and the recovery time is expected to be of the order of 100 ns. It could be used in the upgraded LHCb RICH only if significative changes were done in the geometry of the detector to reduce the occupancy in the hottest regions. Another aspect to consider is its radiation hardness, which still needs to be tested. Another ASIC developed for the ALICE experiment which fully satisfies the readout speed requirement is the NINO [22], designed in IBM 0.25 µm technology. In this case the output pulses are fully contained within 25 ns, guaranteeing no dead time at the LHC bunch crossing frequency. The main drawback in this case is power consumption, which is 27 mW per channel in the current version of the chip.

To match all the specifications outlined above, a new integrated circuit was designed during this PhD work. The circuit is named CLARO. Its design choices and the results obtained in the characterization of the first prototype are presented in the following sections. The first prototype named CLARO-CMOS was realized in AMS 0.35 µm CMOS technology, which is an ideal prototyping starting point thanks to its low cost. If the CLARO-CMOS will be found not to tolerate the amount of radiation foreseen in the upgraded LHCb RICH environment, the whole design could be ported to a deep submicron technology for a higher degree of radiation hardness.

## 2.5 The CLARO-CMOS prototype

The CLARO-CMOS is the first prototype of an ASIC designed for fast photon counting with multi-anode photomultiplier tubes [23, 24]. It features a very fast operation for full recovery before 25 ns, aiming to completely eliminate the dead time at the bunch crossing rate of the LHC accelerator. Moreover it consumes about 1 mW per channel, minimizing the cooling required for its operation in large arrays of photon sensors. As such it fully complies with the requirements for a readout ASIC for the upgraded RICH detectors of the LHCb experiment, also in the high occupancy regions. As will be discussed in the following sections, its features make it also useful to readout silicon photomultipliers (SiPM) and microchannel plates (MCP-PMT). In particular the outstandingly low jitter of the CLARO-CMOS, which will be illustrated in detail in section 2.10, allows to exploit the timing resolution of MCP-PMT devices without limitations, a feature which could be of use for new time of flight detector designs.

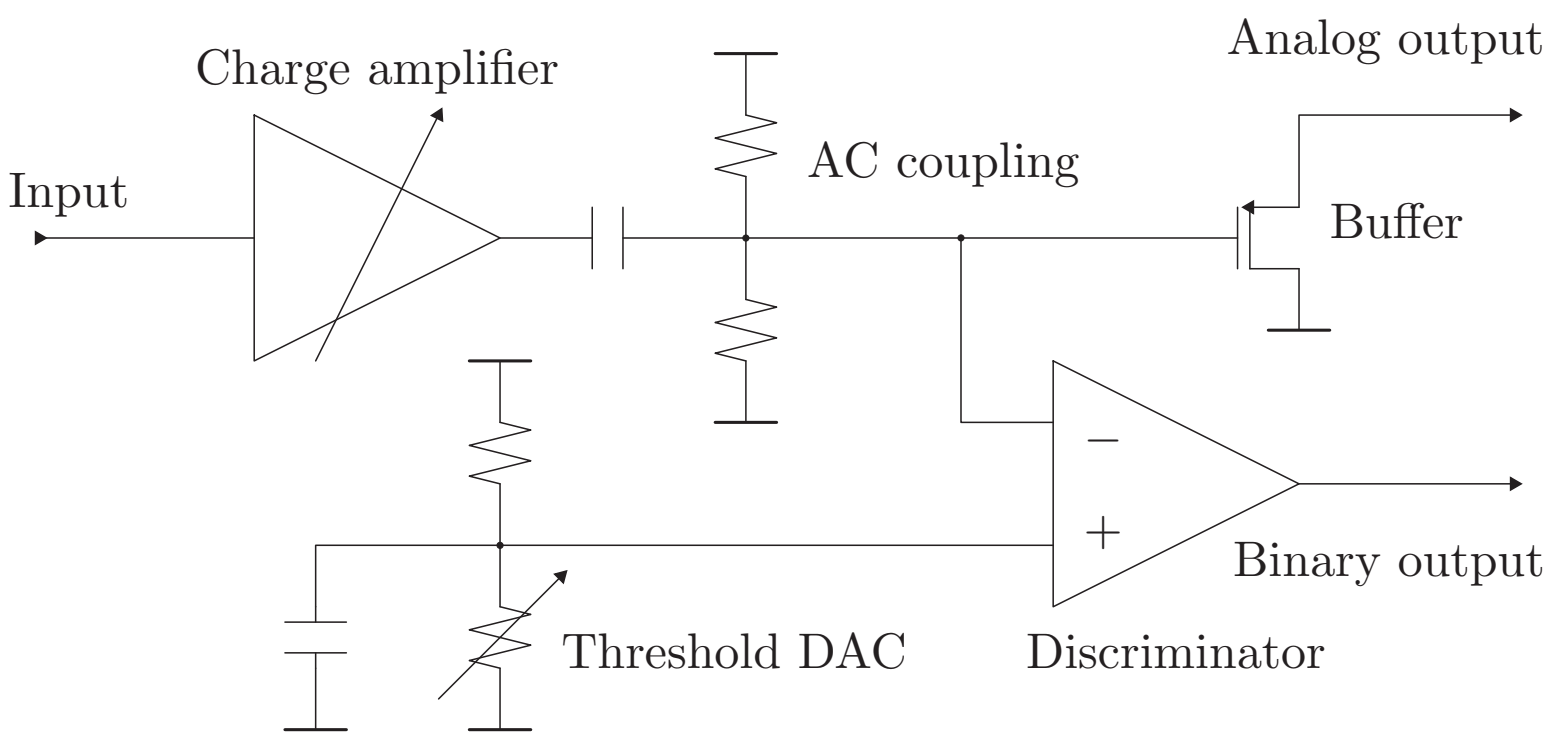

Figure 2.8: Block schematic of a channel of the CLARO-CMOS prototype.

The CLARO-CMOS has four channels, each made of a charge amplifier and a discriminator. Figure 2.8 shows the block schematic of a channel. The ASIC is designed to be operated between a positive 2.5 V supply rail and ground. The input is to be connected to a pixel of the photon sensor. The design is optimized for negative charge signals at the input, meaning that electrons are collected at the readout electrode, as happens for all vacuum-based photomultipliers. The charge amplifier reads the input current pulses on a virtual ground node, and gives at its output a voltage signal whose amplitude is proportional to the collected charge. The discriminator is a voltage comparator which compares the analog signals with a threshold, and whose output carries the binary information whether the amplitude of the input signal exceeds threshold or not. The charge amplifier is AC coupled to the discriminator in order to make the DC component of the analog signal at the input of the discriminator stable and equal to half the power supply voltage. The AC coupling time constant is 55 ns. Since as will be shown the signals at the output of the charge amplifier are fast, no noticeable baseline shift occurs unless the average rate of the input signals is larger than about 10 MHz. For single photon counting applications, as in the LHCb RICH detectors, the main output is the binary output. Two of the channels also feature a buffered analog output which gives the shaped analog signals at the output of the charge amplifier. The buffer is realized with a PMOS voltage follower of small area, with a small gate capacitance to avoid loading the output of the charge amplifier. The PMOS is to be biased with an external resistor tied to the positive voltage rail, and can be switched off when the readout of the analog signal is not required.

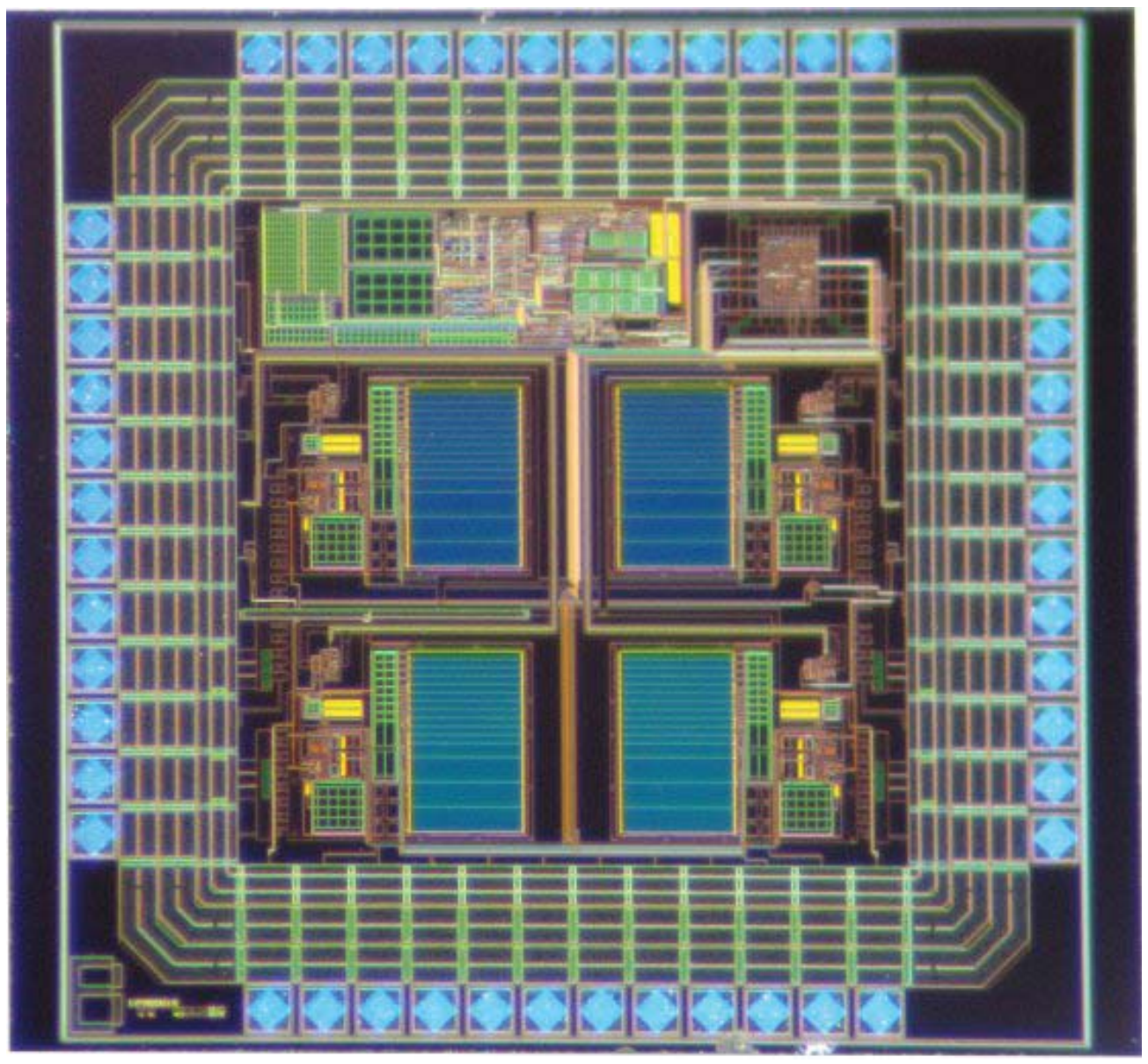

Figure 2.9: Photograph of the CLARO-CMOS. The die area is $2 \times 2$ mm$^2$.

The gain of the charge amplifier is programmable with a three bit resolution. Two bits are used to set a $1/4$ and $1/7$ attenuation. The setting of both bits results in an attenuation of $1/10$. The third bit gives a gain variation of a factor of 2, regardless of the attenuation setting. The threshold of the discriminator is set through a 5-bit DAC implemented as a simple resistive divider, which allows to choose the voltage at the non-inverting input of the discriminator between 32 values. The total number of setting bits for each channel is 8. In this prototype, the settings for channels B and D are copied from those of channels A and C, so that there are in total 16 bits to set. The settings are stored in a 16-bit shift register which can be programmed through a simple SPI interface.

Figure 2.9 shows a photograph of the CLARO-CMOS die at the microscope. The die area is $2 \times 2$ mm$^2$. The four identical structures corresponding to the four channels can be seen. The large devices near the center of the die are the resistors used for threshold setting. In a future version of the chip, the design of the DAC could be optimized to reduce the area required by each channel. The 16-bit shift register can be seen at the top of the die.

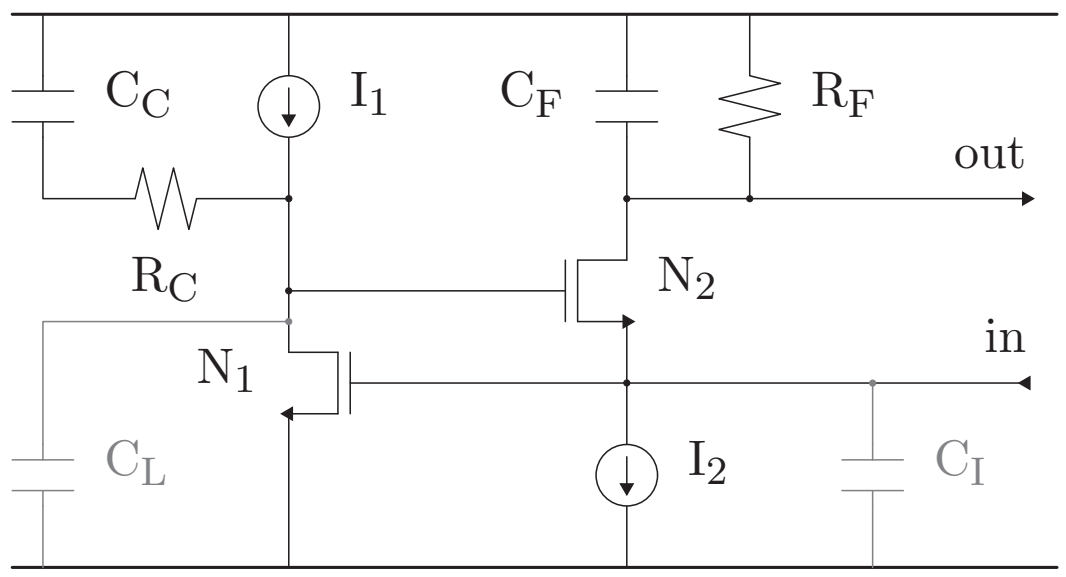

Figure 2.10: Simplified schematic of the charge amplifier.

## 2.6 Design of the charge amplifier

Figure 2.10 shows a simplified schematic of the charge amplifier, which includes the parasitic capacitance $C_L$ and the input capacitance $C_I$ for clarity. The input stage is an active cascode [25–27], a design widely used in the field of photosensor electronics due to its natural capability to readout large current pulses, also referred to as super common base [28–30]. This design uses a local feedback through $N_1$ to lower the impedance at the source of $N_2$, in order to read the input current pulses on a virtual ground node. The loop gain at intermediate frequency is $g_1 R_C$, where $g_1$ is the transconductance[1] of $N_1$. The current pulses are integrated by the capacitor $C_F$ at the drain of $N_2$, which discharges through the resistor $R_F$. The output signal in response to a (negative) charge Q injected at t = 0 is given by

$$V_O(t) = \frac{Q}{C_F}\frac{\tau_F}{\tau_F - \tau_R}\left(e^{-\frac{t}{\tau_F}} - e^{-\frac{t}{\tau_R}}\right) \tag{2.3}$$

where $\tau_R$ is the rise time constant given by the charge amplifier bandwidth and $\tau_F = C_F R_F$ is the fall time constant. The rise time constant $\tau_R$ is of the order of 1 ns, and is directly proportional to $C_I$ as will be shown. The ASIC is designed for fast photodetectors, where the input current pulse is short, of the order of 1 ns. The fall time constant $\tau_F$ was chosen to be 5 ns, large enough for an effective integration of fast pulses but small enough to sustain high rates without pile-up. The signal described by equation (2.3) is valid as long as the charge collection time is smaller than the 10% to 90% rise time of the circuit, that is 2.2 $\tau_R$. Otherwise the collection time should be considered in place of $\tau_R$ in modeling the signal at the output of the charge amplifier.

[1] The transconductance of MOS transistors is traditionally denoted in literature by $g_m$. However for a lighter notation we chose to denote the transconductance of transistor $N_1$ as $g_1$, that of transistor $N_2$ as $g_2$, and so on.

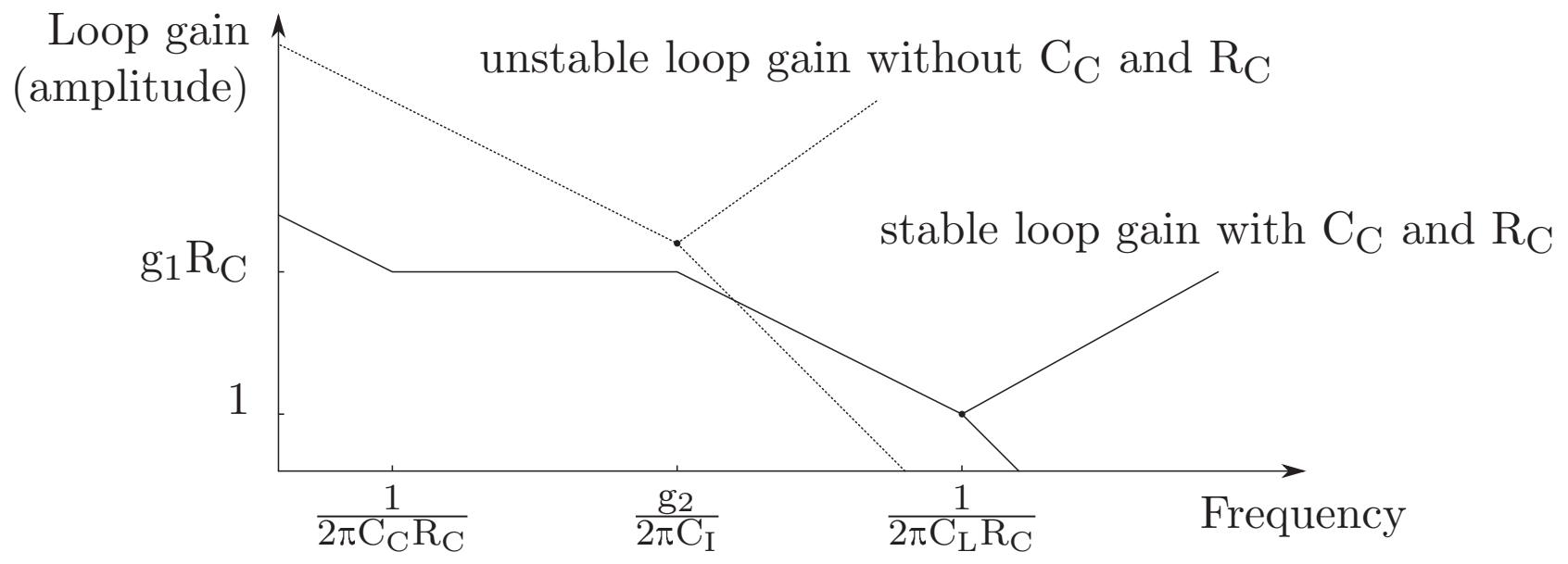

Figure 2.11: Bode diagram illustrating the stability of the input feedback loop.

In the simplified scheme of figure 2.10, the main voltage (or series) noise source is $N_1$ together with the bias circuit $I_1$, while the main current (or parallel) noise source is $R_F$ together with the bias circuit $I_2$. Transistor $N_2$ contributes to the series noise, but its contribution is divided by the loop gain and becomes negligible. The optimal noise performance corresponds to the case where $N_1$ is biased with a large current $I_1$ to keep its transconductance high and its series noise low. Since $R_F$ contributes to the parallel noise, its value cannot be too small, and this poses an upper limit to the bias current $I_2$ of $N_2$. With a low bias current, the transconductance $g_2$ of $N_2$ is low, and the input capacitance to ground $C_I$ due to the input bonding pad, the bonding wire, packaging, interconnects and to the sensor adds a pole to the input feedback loop at a frequency $g_2/2\pi C_I$. If $R_C$ and $C_C$ were not present, the load at the drain of $N_1$ would be purely capacitive, and there would be another pole at very low frequency due to $C_L$. This would be the lower frequency pole of the feedback loop. At the frequency of the second pole, that is $g_2/2\pi C_I$, the feedback loop would become unstable, unless it were already lower than 1, in which case it would be ineffective in lowering the input impedance at this frequency. This case is illustrated in the bode plot of figure 2.11, dashed line.

To compensate the pole due to $C_L$, $R_C$ and $C_C$ are used. This case is illustrated in the solid line of figure 2.11. The effect of compensation is to limit the loop gain to $g_1R_C$ at moderate frequency, higher than $1/2\pi C_C R_C$. This shifts the pole due to $C_L$ at a higher frequency given by $1/2\pi C_L R_C$. For this compensation to be effective, it is required that the value of $R_C$ is not too large and that $C_L$ is minimized with a proper layout. In particular, since the area of $C_C$ on silicon is larger than that of $R_C$, its parasitic capacitance to the substrate is larger. A much lower value for $C_L$ is obtained if $R_C$ is placed before $C_C$, as in figure 2.10. The relatively low value for $R_C$ strengthens the need to keep high the transconductance of $N_1$, while the transconductance of $N_2$ is less critical. As illustrated in the solid line of figure 2.11, the dominant pole of the input feedback loop is now at $g_2/2\pi C_I$. This ensures that the feedback loop is effective in lowering the input impedance up to a much higher frequency. The frequency where the loop gain becomes close to unity gives the bandwidth of the charge amplifier. The associated time constant gives the rise time of the output signal:

$$\tau_R = \frac{C_I}{g_2}\frac{1}{g_1 R_C + 1} \simeq \frac{C_I}{g_2 g_1 R_C} \tag{2.4}$$

The 10% to 90% rise time is given by 2.2 $\tau_R$. The rise time is thus directly proportional to the input capacitance $C_I$ and inversely proportional to the loop gain $g_1 R_C$. The stability of the feedback loop is ensured even if the sensor has a negligible capacitance, since the value of $C_I$ has a lower limit at a few pF due to the gate-drain capacitance of $N_1$, that is less than 100 fF but its contribution is multiplied by the loop gain, its gate-source and gate-bulk capacitance (about 0.5 pF in total) and the stray capacitance of the pads, the bonding wires, eccetera. Considering all the contributions from the circuit the input capacitance can be estimated to be about 1.5 pF, bonding pads excluded. With the CLARO-CMOS mounted in a small QFN48 package the total capacitance at the input (without the sensor) was measured to be about 3.3 pF.

Parasitic inductance in series with the input should also be kept under control to avoid adding phase shift which may compromise stability. However in simulations the effect of series inductance becomes noticeable only above about 100 nH: this is a fairly high value, much higher than the typical bonding wire inductance of a few nH, which can then be neglected.

The full schematic of the charge amplifier is shown in figure 2.12. To vary the gain, a set of MOS switches was included in the design. Two switches, $N_{S3}$ and $N_{S4}$, are used to attenuate the input signal: if the digital control signals $V_3$ or $V_4$ are set high, the switches are closed and a part of the input charge passes through $N_3$ or $N_4$ and is wasted on the positive rail. The amount of attenuation is set by choosing the dimensions of $N_3$ and $N_4$, which are 3 and 6 times larger than $N_2$ respectively, causing attenuations of $^1/_4$ and $^1/_7$. An attenuation of a factor of $^1/_{10}$ is obtained if both branches are enabled. The dummy switch $N_{S2}$ whose gate is tied to the positive rail was introduced to preserve the symmetry between the input branches.

Another switch $P_{SF}$ controlled by the digital control signal $V_F$ is used to change the value of $C_F$ and $R_F$, doubling $C_F$ and halving $R_F$, to change the gain by a factor of 2 while keeping the discharge time constant the same. The voltages $V_3$, $V_4$ and $V_F$ are the three control bits which allow gain setting on each channel. The reason why only one switch was used to change the values of $C_F$ and $R_F$ is related to the switch parasitics. If several switches were connected in series, their series resistance in the “on” state would have caused distortion in the shape of the output signal. If several of such switches

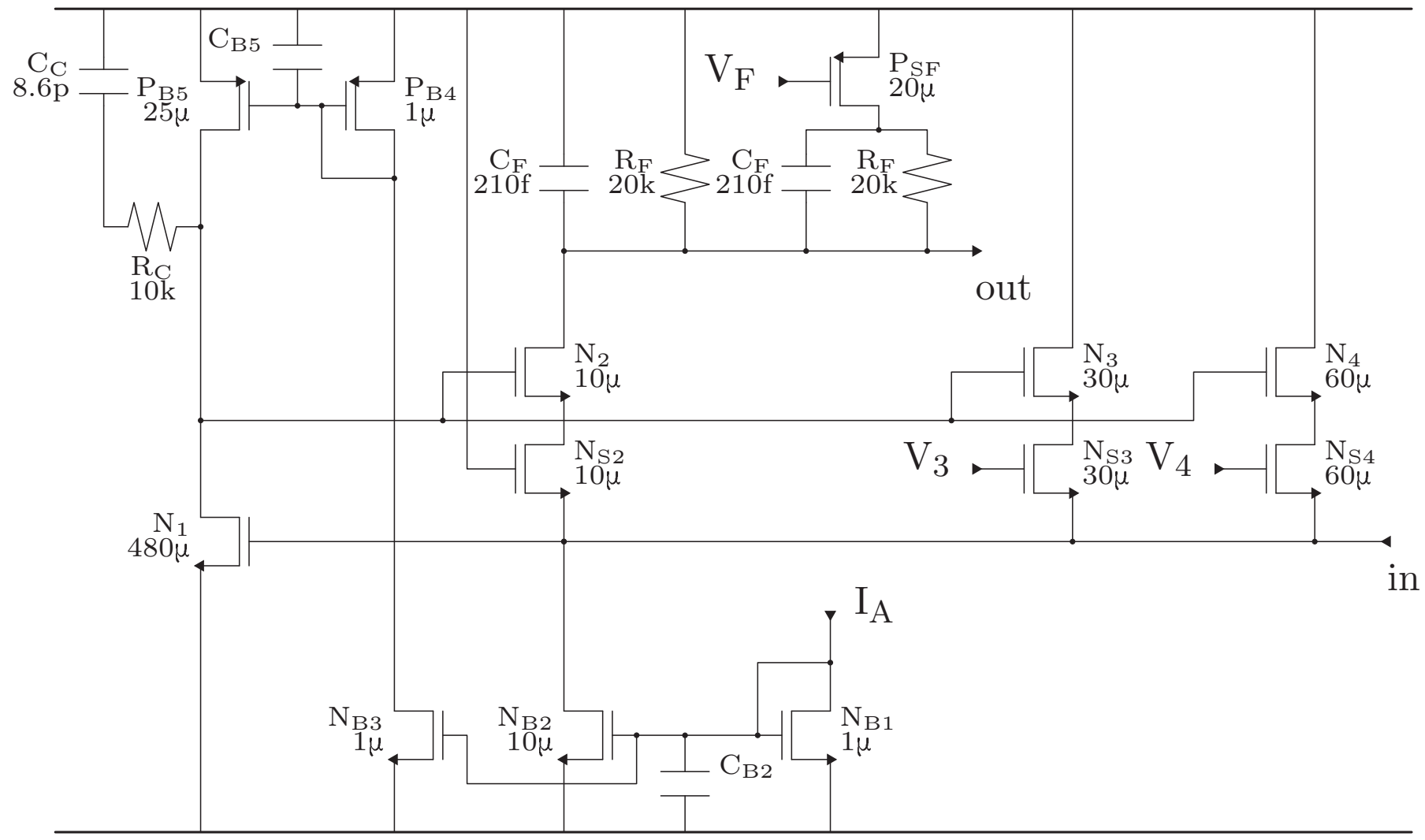

Figure 2.12: Full schematic of the charge amplifier. The width of the MOS transistors is shown. The gate length is 0.35 μm for all the transistors in the charge amplifier. The substrate of all NMOS (PMOS) transistors is tied to ground (to the positive rail).

---

were put in parallel, their capacitance in the "off" state would have been in parallel with $C_F$, reducing the maximum gain achievable.

The dimensions of the bias transistors $N_{B1} \ldots N_{B5}$ were chosen so that the bias current of $N_1$ is about 2.5 times larger than that of $N_2$. Transistor $N_1$ has a large area to obtain a high transconductance $g_1$, and is operated in the weak inversion region. In this prototype the bias current of the charge amplifier can be set by changing $I_A$ with an external resistor. Two operating modes were chosen: a "low power" mode, with $I_A = 2$ μA, and a "timing" mode, with $I_A = 5$ μA. In "low power" mode, $N_1$ is biased with 85 μA, resulting in $g_1 = 2$ mA/V. Since $R_C = 10$ kΩ, the low frequency gain of the input feedback loop is about 20. The input branches with $N_2$, $N_3$, $N_4$ are biased with a total current 25 μA. The total transconductance of $N_2$ in parallel with $N_3$ and $N_4$ is about 350 μA/V, depending on which of $N_3$ and $N_4$ are enabled. If the feedback loop were not present, the input impedance would be higher than 2 kΩ. The feedback loop lowers this value to about 130 Ω. From equation (2.4) the 10% to 90% rise time is expected to be about 1.2 ns for $C_I = 3.3$ pF, and 2.4 ns for $C_I = 6.5$ pF. In "timing" mode, $N_1$ is biased with 170 μA, and its transconductance becomes 3.8 mA/V, so that the loop gain roughly doubles. The total transconductance of $N_2$, $N_3$ and $N_4$ is about 500 μA/V. Thanks to the larger loop gain, the input impedance is now reduced to less than 100 Ω. The bandwidth of the charge amplifier is increased, and the loop gain at $1/2\pi C_L R_C$ becomes closer to unity, but stability of the feedback loop is still ensured even with a negligible sensor capacitance. The rise time of the signal at the output of the charge amplifier as given by equation (2.4) is roughly half than in "low power" mode thanks to the larger loop gain. The main consequence is a reduction of the time walk of the discriminator, as will be shown in the following.

The noise of the charge amplifier can be referred to the input as an equivalent noise charge $\sigma_Q$. The detailed noise calculations were given in chapter 1, section 1.8. Instead of approximating the expressions for $\tau_R \ll \tau_F$, as done in section 1.8, it is possible to obtain an expression valid for $\tau_R < 0.3\ \tau_F$, that is for $C_I < 10$ pF in "low power" mode. The equivalent noise charge results to be given by

$$\sigma_Q \simeq \sqrt{i_n^2 \frac{\tau_F + 3\tau_R}{4} + A_f C_I^2 \frac{\tau_F + 4\tau_R}{\tau_F} \ln \frac{\tau_F}{\tau_R} + e_n^2 C_I^2 \frac{\tau_F + 3\tau_R}{4\tau_F \tau_R}} \tag{2.5}$$

where $i_n$ is the total current noise density, $e_n$ is the total white voltage noise density and $A_f$ is the 1/f voltage noise coefficient, all referred to the input node of the charge amplifier. In addition to the noise from $N_1$ and $R_F$, it is necessary to consider the noise contributions coming from the bias transistor $N_{B2}$, whose current noise directly contributes to the parallel noise at the input, and $P_{B5}$, whose current noise is divided by the transconductance of $N_1$ and becomes a series noise contribution at the input. Moreover, if the values of the filtering capacitors $C_{B2}$ and $C_{B5}$ are not large enough, additional noise coming from $N_{B1}$, $N_{B3}$ and $P_{B4}$ can be injected through $N_{B2}$ and $P_{B5}$, contributing to the parallel and series noise respectively.

In this first CLARO-CMOS prototype, filter capacitors $C_{B2}$ and $C_{B5}$ are not present. The parallel noise is dominated by the channel current of $N_{B1}$ mirrored and multiplied by 10 by $N_{B2}$. Since in "low power" mode the transconductance of $N_{B1}$ is $g_{B1} = 35$ µA/V we have

$$i_{B1}^2 = 10^2 \times \frac{8}{3} kT g_{B1} \simeq \left(6.2\ \text{pA}/\sqrt{\text{Hz}}\right)^2 \tag{2.6}$$

Other contributions come from $N_{B2}$, about 2 pA/$\sqrt{\text{Hz}}$, and from $R_F$, about 0.9 pA/$\sqrt{\text{Hz}}$ if $V_F$ is high, 1.3 pA/$\sqrt{\text{Hz}}$ if $V_F$ is low, assuming the attenuation to be set to one. The weight of the noise generated by $R_F$ is directly proportional to the attenuation factor: at the maximum attenuation, that is a factor of 1/10, the noise from $R_F$ becomes the dominant parallel noise source with 9 pA/$\sqrt{\text{Hz}}$ if $V_F$ is high, 13 pA/$\sqrt{\text{Hz}}$ if $V_F$ is low. The other noise sources in the charge amplifier do not depend on the attenuation factor, since they share the same attenuation as the signal. Anyway the attenuation is meant to be used only when the signals are large; so in those cases the signal to noise ratio is expected to be anyway adequate. In the following, for all noise evaluations, we will consider the attenuation to be equal to one. The sum of all parallel noise is thus close to 7 pA/$\sqrt{\text{Hz}}$ in "low power" mode. In "timing" mode the parallel noise increases by about 20% due to the larger bias current which gives a larger transconductance to $N_{B1}$ and $N_{B2}$.

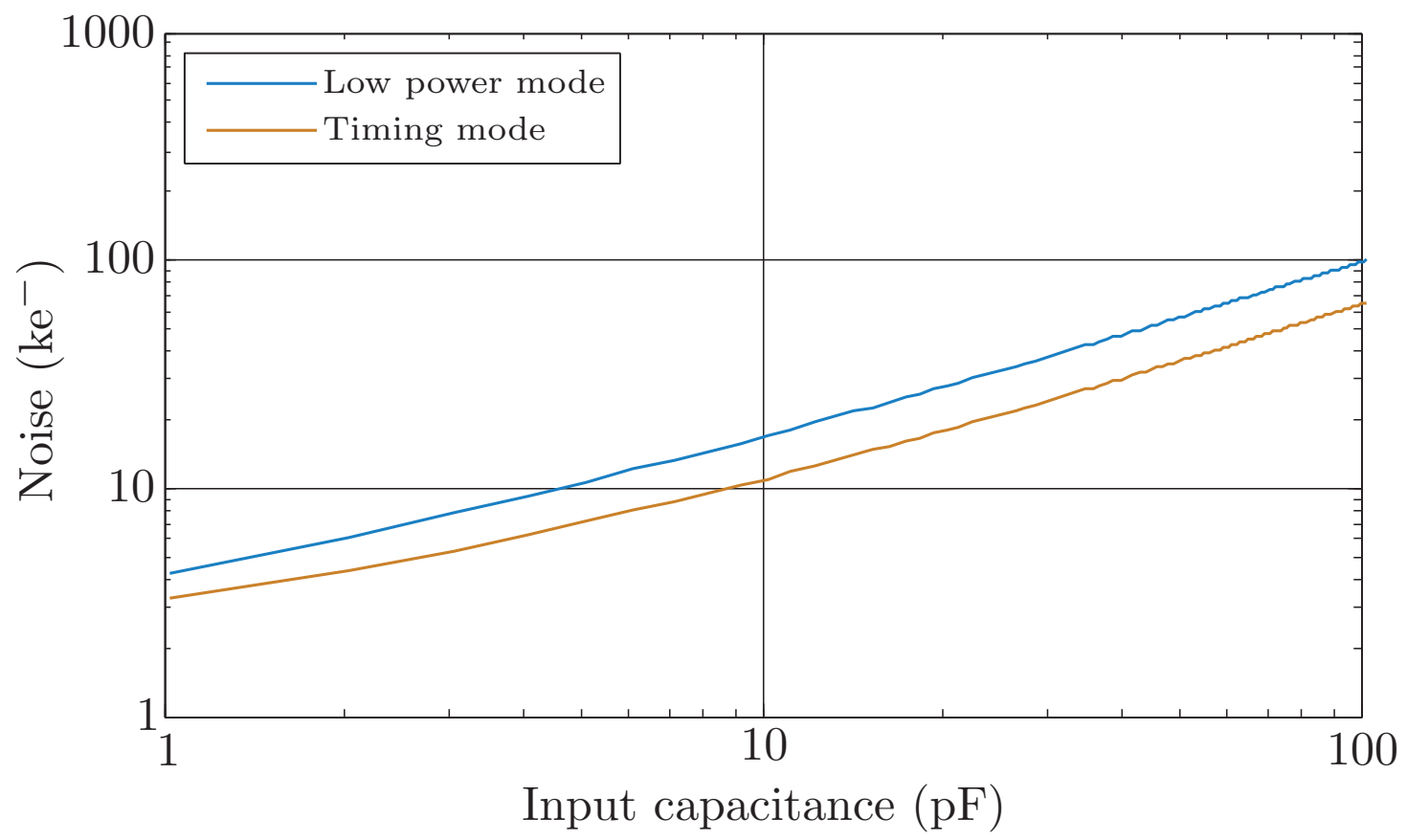

Figure 2.13: Calculated noise versus input capacitance in "low power" and "timing" modes.

The series noise is dominated by $N_1$ and $P_{B5}$. As already mentioned, additional noise from the other bias transistors is injected through $P_{B5}$ since its gate is not filtered. In "low power" mode, where $g_1 = 2$ mA/V, the series white noise is dominated by $N_{B1}$, $N_{B3}$ and $P_{B4}$, which all have a transconductance of about $g_{B1} = 35$ µA/V. The resulting white voltage noise at the input is

$$e^2_{B134} = 25^2 \times 3 \times \frac{8}{3} kT \frac{g_{B1}}{g_1^2} \simeq \left(13 \text{ nV}/\sqrt{\text{Hz}}\right)^2 \tag{2.7}$$

being 25 the area ratio between $P_{B5}$ and $P_{B4}$. Other contributions come from $N_1$, about 2.3 nV/$\sqrt{\text{Hz}}$, and from $P_{B5}$, about 1.6 nV/$\sqrt{\text{Hz}}$. The sum of all series white noise is about $e_n \simeq 14$ nV/$\sqrt{\text{Hz}}$. In "timing" mode the series noise reduces by almost a factor of 2, because of the larger transconductance of $N_1$ which gives a larger loop gain. Compared to the series white noise, the contribution of the 1/f component is expected to be negligible since from simulations it is possible to estimate $A_f < 10^{-9}$ V$^2$.

According to equations (2.4) and (2.5), the parallel noise contribution to the equivalent noise charge at the output of the charge amplifier in "low power" mode is expected to be about 1.8 ke$^-$ at $C_I = 3.3$ pF, and 2.0 ke$^-$ at $C_I = 6.5$ pF. The series noise contribution is expected to be about 7.5 ke$^-$ at $C_I = 3.3$ pF, and 12 ke$^-$ at $C_I = 6.5$ pF. The total noise of the charge amplifier in "low power" mode is thus expected to be 7.7 ke$^-$ at $C_I = 3.3$ pF, and 12 ke$^-$ at $C_I = 6.5$ pF. In "timing" mode the noise of the charge amplifier results about 50% higher, mostly because of the larger bandwidth. At the auxiliary output, the rise time is limited by the bandwidth of the analog buffer. In that case the weight of the series noise is expected to be smaller, while the weight of the parallel noise is expected to be larger, according to equation (2.5). For instance, assuming that the output buffer limits the output signals with time constants of $\tau_R = 1.3$ ns and $\tau_F = 7.2$ ns, equation (2.5) gives 5.6 ke$^-$ with an input capacitance of 3.3 pF, dominated by the series noise.

For larger values of input capacitance, the rise time becomes larger and the approximation $\tau_R < 0.3\ \tau_F$ that led to equation (2.5) is no longer valid. In this case the equivalent noise charge can be calculated without approximations starting from the same expressions presented in chapter 1, section 1.8. The resulting noise $\sigma_Q$ versus input capacitance up to $C_I = 100$ pF is plotted in figure 2.13. As can be seen, noise increases with larger input capacitance because of the larger series noise, reaching about 100 ke$^-$ for $C_I = 100$ pF. The increase is sublinear in $C_I$ because with larger input capacitance the bandwidth becomes smaller. The calculated curve closely matches simulations.

As already discussed, the filtering capacitors $C_{B2}$ and $C_{B5}$ can be used to improve the noise performance of the design, considerably reducing both the series and the parallel noise injected through the bias transistors, at the price of a larger layout area on silicon. Moreover, the noise coming from $N_{B2}$ and $P_{B5}$ can be reduced by decreasing their transconductance or by source degeneration. As an alternative, resistive biasing could be used in place of $N_{B2}$ and $P_{B5}$. These improvements will be considered for the next versions of the ASIC.

## 2.7 Design of the discriminator

Figure 2.14 shows the schematic of the voltage comparator used to discriminate the events above threshold. The input stage is a differential pair loaded with a current mirror. This is the only part of the comparator which dissipates a continuous current. Since $I_C$ is about 1 µA, the differential pair is biased with about 100 µA. The signal from the charge amplifier is connected to the inverting input of the comparator, while the noninverting input is held at a constant potential which defines the threshold. The threshold voltage at the inverting input of the discriminator can be set between 1.25 V (half the positive rail voltage) and 0.83 V (one third the positive rail voltage) in 32 steps, labelled from 0 to 31, thanks to a 5-bit DAC implemented as a simple voltage divider. Each step is about 13 mV. At the maximum gain, this corresponds to a threshold step of 150 ke$^-$.

In ready state, the output of the differential pair is low, and stays close to 0.5 V. This signal feeds the inverter made of $P_8$ and $N_8$. Transistor $N_8$ is small and has a large threshold, about 0.6 V. In this way $N_8$ is biased just below threshold: no current passes through the first inverter and its output is high.

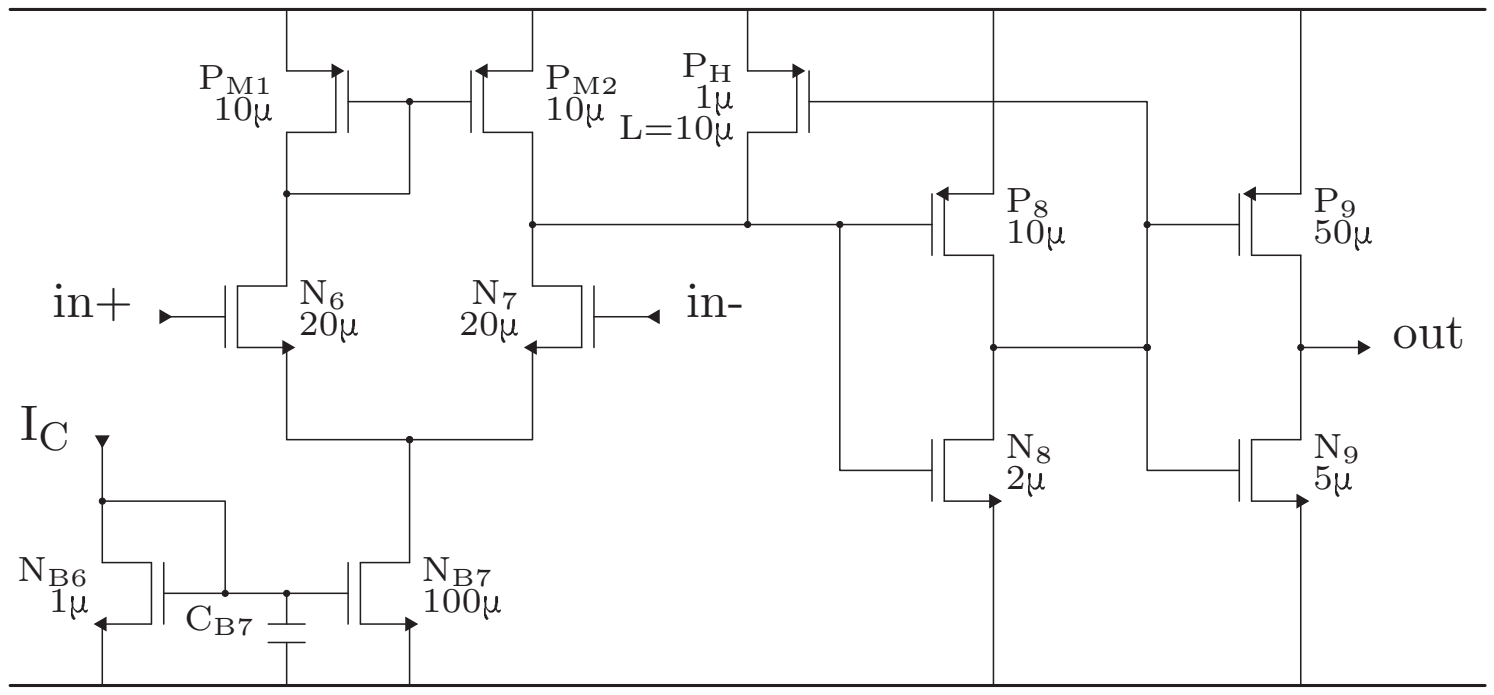

Figure 2.14: Full schematic of the comparator. The width of the MOS transistors is shown. The gate length is 0.35 µm for all the transistors except $P_H$. The substrate of all NMOS (PMOS) transistors is tied to ground (to the positive rail).

Transistor $P_H$ provides hysteresis, and since its gate is high it is switched off. The output of $P_8$ and $N_8$ is fed to the second inverter made of $P_9$ and $N_9$, which is also the output stage.

In response to a negative pulse from the charge amplifier, the output of the differential pair goes up, close to the positive rail. The output of first inverter goes to ground, closing the switch $P_H$, which draws current from the differential pair and holds up its output providing hysteresis. At the same time, the output of $P_9$ and $N_9$ swings to the positive rail. The gate length of $P_H$ is large: its "on" resistance is about 150 kΩ, so that only a fraction of the bias current of the differential pair passes through $P_H$, and after a few nanoseconds the output of the differential pair is able to get back to the initial condition. When the output of the differential pair goes down, the output of the first inverter goes up, transistor $P_H$ is opened and the output of the comparator goes down. After this the discriminator is ready to trigger another pulse from the charge amplifier. The width of the output pulses is proportional to the amplitude of the input signals, allowing to apply time over threshold algorithms to determine the input charge and compensate for time walk.

The gain of the input stage of the comparator is about 30 V/V for small signals around threshold at low frequency, with a pole at about 30 MHz. The corresponding time constant is $\tau_C \simeq 5$ ns, about the same as the fall time of the charge amplifier pulse $\tau_F$. The effect of hysteresis (which is essentially a small positive feedback) is to increase the gain to 600 V/V at low frequency. The gain of the inverters is about 20 V/V for each. The overall gain of the comparator at low frequency including hysteresis results in $24 \times 10^4$ V/V or 107 dB. Transistor $P_9$ is much larger than $N_9$, in order to obtain a very fast transition on the rising edge at the output. The rise and fall times of the output signal depend on the load at the output of the discriminator. The output stage was designed to drive only a short line to a digital processing circuit or to an external low impedance driver, located a few cm away on the same board. Thus a purely capacitive load of a few pF is expected. This was done in order to give the maximum flexibility in the design of a full system and to avoid unnecessary power consumption in the CLARO-CMOS. The output signal is limited by the slew rate of the output stage on the output load, that is $I_L/C_L$, where $I_L$ is the current from the output stage, and $C_L$ is the output load capacitance. The output current can be estimated to be $I_L \simeq 2.5\ V \times g_9$, where $g_9$ is the transconductance of the output stage. For small signals, the transconductance of $P_9$ is 2 mA/V, and that of $N_9$ is 0.8 mA/V, even if this values are largely non linear since the output stage swings from rail to rail. Anyway the rise time is expected to be about two times smaller than the fall time, since the rising edge is driven by $P_9$ while the falling edge is driven by $N_9$. With these numbers, the time required for the full swing from 0 V to 2.5 V at the output is about $2.5\ V/(I_L/C_L) \simeq C_L/g_9$. With a load capacitance of $C_L = 8$ pF, for instance, the output 0% to 100% swing takes 4 ns, which corresponds to a 10% to 90% rise time of 3.2 ns, and the output 100% to 0% swing takes 10 ns, which corresponds to a 90% to 10% fall time of 8 ns.

The input transistors $N_6$ and $N_7$ have a transconductance $g_C$ of about 700 μA/V, while $P_{M1}$ and $P_{M2}$ have a transconductance $g_M$ of about 300 μA/V. These are the main contributors to the noise of the comparator. Transistor $N_{B7}$ does not contribute because its noise is common mode while the input stage is differential. So in the case of the comparator the bias filtering capacitor $C_{B7}$ can be avoided. The input referred white voltage noise density can be expected to be

$$e_C^2 = 2 \times \frac{8}{3} kT \frac{1}{g_C} + 2 \times \frac{8}{3} kT \frac{g_M}{g_C^2} \simeq \left(6.7\ nV/\sqrt{Hz}\right)^2 \quad (2.8)$$

which together with the 1/f contributions corresponds to a voltage noise at the input of about 65 μV RMS. Compared with the RMS noise at the output of the charge amplifier, that is more than 1 mV RMS in the best case of a 3.3 pF input capacitance, this contribution is negligible, at least in the case of no attenuation. With larger attenuations the weight of the noise of the comparator grows accordingly, and at B = 10 it becomes significant. Since as already mentioned the attenuation is only meant to be used with very large signals, where the signal to noise ratio is a minor concern, we will anyway consider the case of no attenuation in the following. The jitter on the rising edge of the comparator for small values of the input capacitance $C_I$ is expressed by

$$\sigma_t \simeq \frac{\tau_C}{Q - Q_{TH}} \sqrt{i_n^2 \frac{\tau_C}{8} + A_f C_I^2 \frac{1}{2} + e_n^2 C_I^2 \frac{1}{8\tau_C}} \quad (2.9)$$

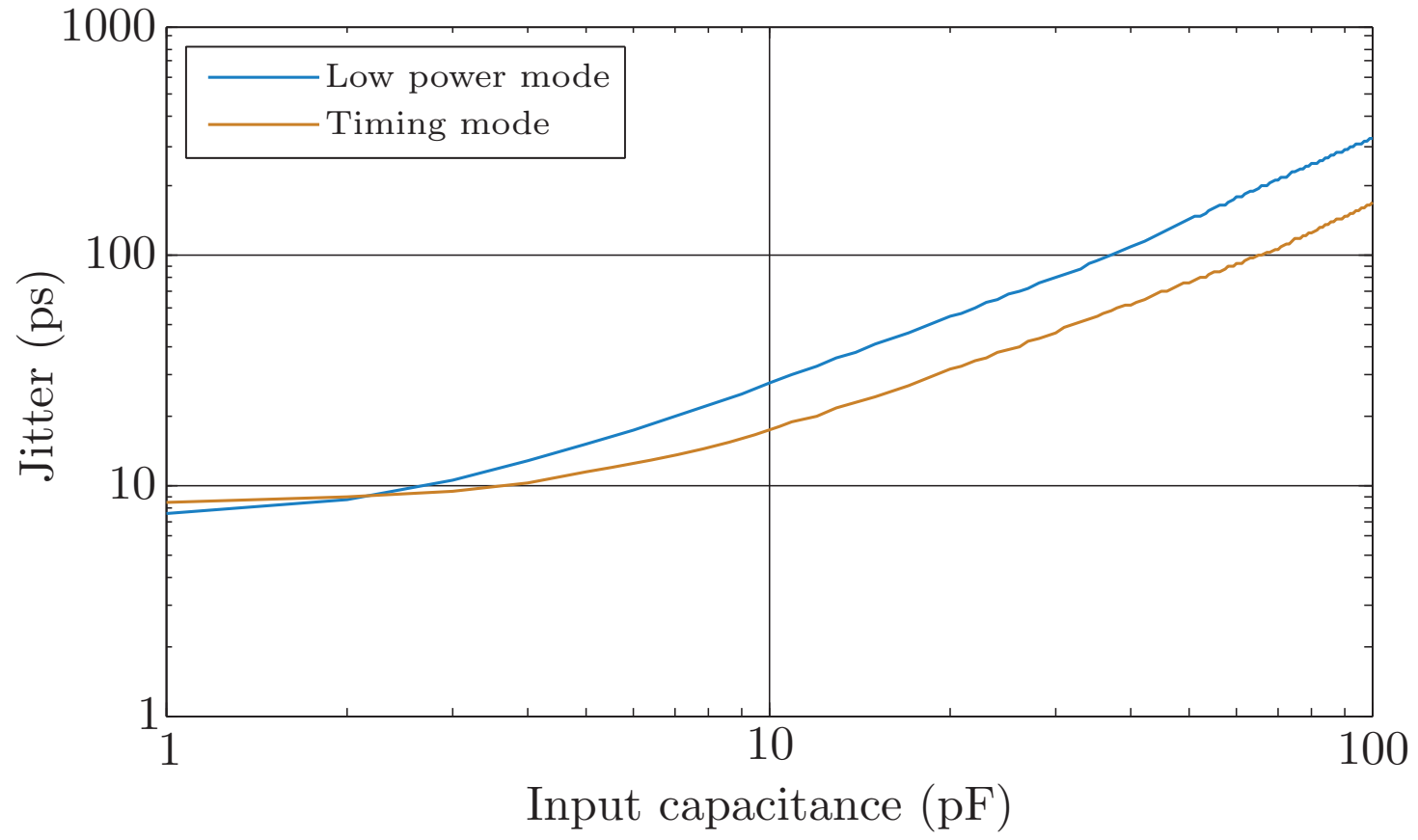

Figure 2.15: Calculated jitter versus input capacitance in "low power" and "timing" modes for 1 $Me^-$ signals when the threshold set at 100 $ke^-$.

The calculations to obtain equation (2.9) can be carried out following sections 1.8 and 1.11 approximating for $\tau_F \simeq \tau_B$, and indicating the time constant by $\tau_C$. The time constant $\tau_C \simeq 5$ ns is given by the bandwidth of the first stage of the comparator. When the threshold is set at 300 $ke^-$, equation (2.9) predicts a jitter of 32 ps for 600 $ke^-$ signals, of which 24 ps are due to the series noise, and 18 ps to the parallel noise. As for the case of the equivalent noise charge, the $^1/_f$ component is negligible. According to equation (2.9) the jitter is expected to decrease to 8 ps for 1.5 $Me^-$ signals. For larger signals, equation (2.9) predicts an unlimited improvement; in reality the slope of the signal at the first stage of the discriminator is also limited by slew rate. So, in contrast with equation (2.9), jitter is expected at some point to stop decreasing for larger signals, and to saturate to a constant value.

As for the case of the equivalent noise charge, for larger values of input capacitance the rise time becomes larger, and when $\tau_R$ becomes larger than $\tau_C$, $\tau_R$ should be considered in the calculations chapter 1 in place of $\tau_C$, and some approximations should be dropped accordingly. The resulting jitter versus input capacitance up to 100 pF is depicted in figure 2.15 for signals of 1 $Me^-$ when the threshold set at 100 $ke^-$. As can be seen, in the first part of the curves the increase in jitter with increasing capacitance is small, because the only effect is the increase in series noise. For values larger than about 10 pF, jitter starts to increase faster, because the effect of the slower rise time adds to the effect of larger series noise. In "low power" mode, the simulated jitter on 1 $Me^-$ signals with an input capacitance of 100 pF is about 320 ps RMS. In "timing" mode it is about 170 ps RMS. For larger signals, jitter is expected to reduce proportionally.

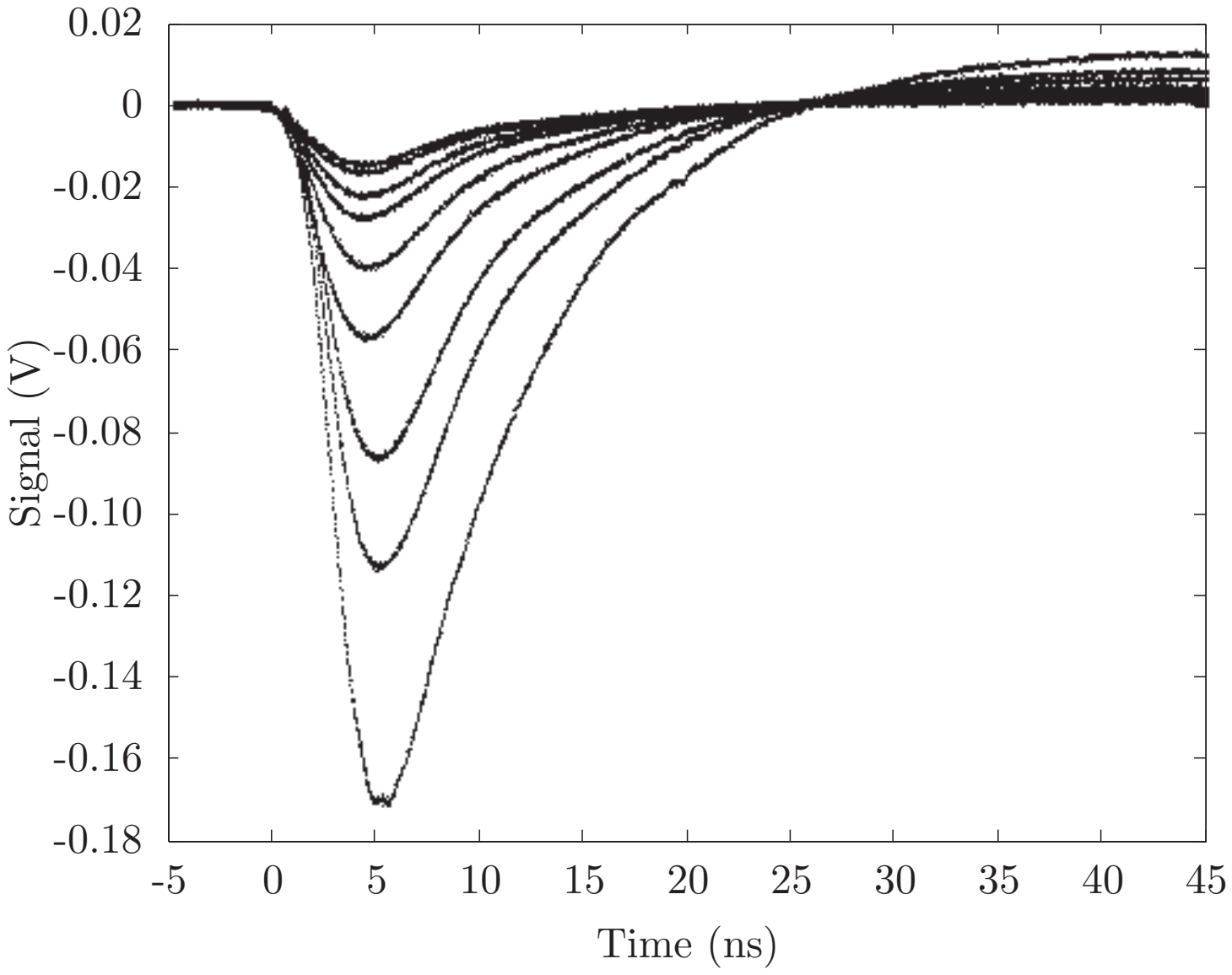

Figure 2.16: Signals at the output of the analog buffer (auxiliary output) in "low power" mode.

## 2.8 Output signals

Figure 2.16 shows the signal at the output of the charge amplifier in "low power" mode, read out at the auxiliary output through the PMOS follower biased with a 1 kΩ resistor to the positive rail. The gain was set to the maximum value ($V_3$ and $V_4$ were set low, $V_F$ was set high), and pulses from 330 ke$^-$ to 3.3 Me$^-$ were injected at the input by a Agilent 81130A 600 MHz step generator through a 0.5 pF test capacitance $C_T$. A block schematic of the measurement setup is presented in figure 2.17. The 10% to 90% rise time of the test signals is 0.6 ns, simulating the typical charge collection time of a fast photomultiplier. The output of the PMOS follower was buffered with a Texas Instruments LMH6703 fast opamp driving a terminated 50 Ω line. The signals were acquired with a Agilent DCA-X 86100D 20 GHz sampling scope with the bandwidth limited to 12 GHz in our measurements.

The leading edge of the measured analog signal in response to a 330 ke$^-$ pulse is 2.8 ns (10% to 90%), its trailing edge is 15.8 ns (90% to 10%), the pulse width at 50% is 8 ns. The corresponding time constants are $\tau_R = 1.3$ ns and $\tau_F = 7.2$ ns. Due to the finite bandwidth of the PMOS follower, the measured signal is slower than the signal at the output of the charge amplifier which feeds the input of the discriminator. Since the transconductance of the PMOS follower is less than 1 mA/V and its bias resistor is 1 kΩ, the amplitude of the buffered signal is smaller than at the output of the charge amplifier.

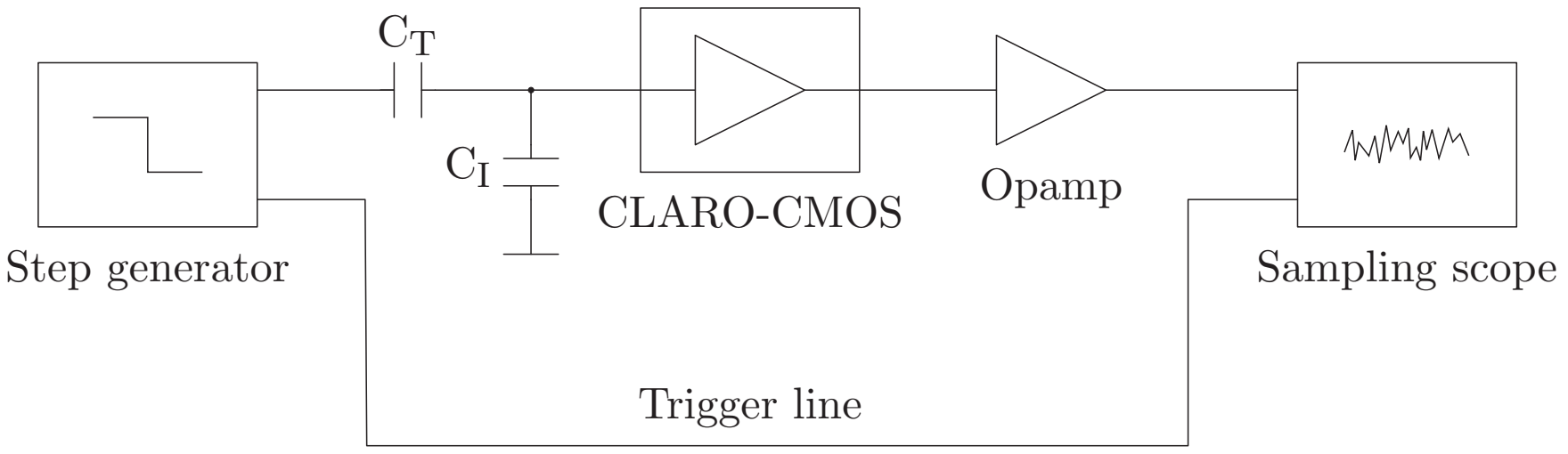

Figure 2.17: Setup used for the CLARO-CMOS performance tests.

The input noise was obtained by measuring the baseline noise at the auxiliary output and referring it to the input of the charge amplifier as an equivalent noise charge $\sigma_Q$. The measured $\sigma_Q$ for an input capacitance of 3.3 pF is 6 ke$^-$ RMS, consistent with equation (2.5), which predicts 5.6 ke$^-$, as already mentioned, once the correct rise and fall time measured at the output of the analog buffer are considered. The importance of low noise is mainly related with timing performance, which will be discussed in the section 2.10.

Figure 2.18 shows the signal at the output of the discriminator when the CLARO-CMOS is operated in “low power” mode. The threshold was set at 800 ke$^-$, and signals from 810 ke$^-$ to 5.6 Me$^-$ were injected at the input. This range of input signals corresponds to the typical single photon response of a photomultiplier in nominal bias conditions, such as those corresponding to the spectra of figure 2.7. As altready mentioned, the output stage of the discriminator is designed to drive a capacitive load of a few pF. In these tests the capacitive load at the output was measured to be 8 pF, contributed by the pads, the QFN48 package, and a short (a few cm) PCB trace to a Texas Instruments LMH6703 fast opamp used as a low impedance driver to the sampling scope. With this load, the 10% to 90% rise time is 2.2 ns, and the 90% to 10% fall time is 9.3 ns. The 50% pulse width depends on the amount of charge injected at the input, ranging from 7.2 ns for the shortest signal in figure 2.18, that is just above threshold, to 21.7 ns for the largest signal in figure 2.18, that is almost a factor of 10 above threshold. The delay between the input charge pulse and the time when the output of the discriminator reaches 50% is 5 ns for signals just above threshold, and lowers to about 2.5 ns for signals well above threshold. The delay is due to the rise time of the charge amplifier pulse at the input of the comparator and to the difference in the speed of the comparator for different levels of overdrive. The difference between the two extreme values, about 2.5 ns in “low power” mode, constitutes the time walk

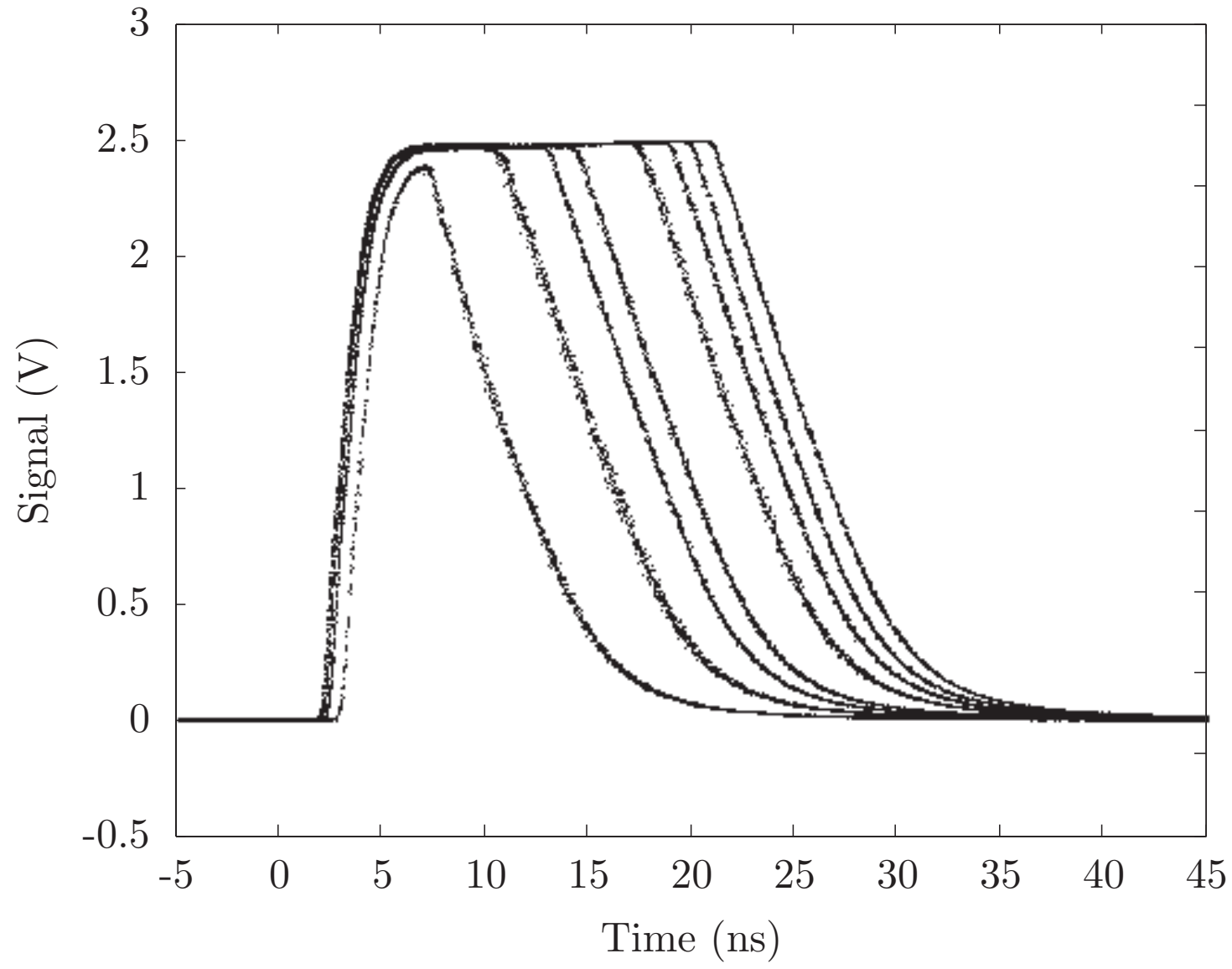

Figure 2.18: Signals at the output of the discriminator (main output) in "low power" mode.

of the discriminator, which is critical for timing performance, to be discussed in section 2.10.

This performance was obtained in "low power" mode, with an overall continuous power dissipation per channel of 0.7 mW. If the discriminator is triggered with a 10 MHz rate, the average power consumption increases to 1.9 mW per channel. It is worth noting that the signals in figures 2.16 and 2.18 are acquired at the output of the sampling scope: the displayed signals are obtained as the superposition of dots from several output signals, while the sampling trigger was synchronized with the step generator. In this way the figure incorporates at a glance also noise and jitter. The output signals shown demonstrate the capability of the CLARO-CMOS to count fast pulses from photomultipliers, from the single photoelectron up to larger gains, with a low noise, very high rate (up to 10 MHz), and a very low power consumption.

When the prototype is operated in "timing" mode, the power consumption is increased to 1.5 mW per channel (becoming 2.3 mW per channel with a 10 MHz rate). The difference in the output signals between "low power" and "timing" modes is small: the different power consumption affects only the output of the charge amplifier, but the difference cannot be directly appreciated on the shape of the buffered signals because of the bandwidth limitation of the auxiliary output buffer. The differences between the two operating modes can be appreciated in the crosstalk and jitter measurements presented in sections 2.9 and 2.10.

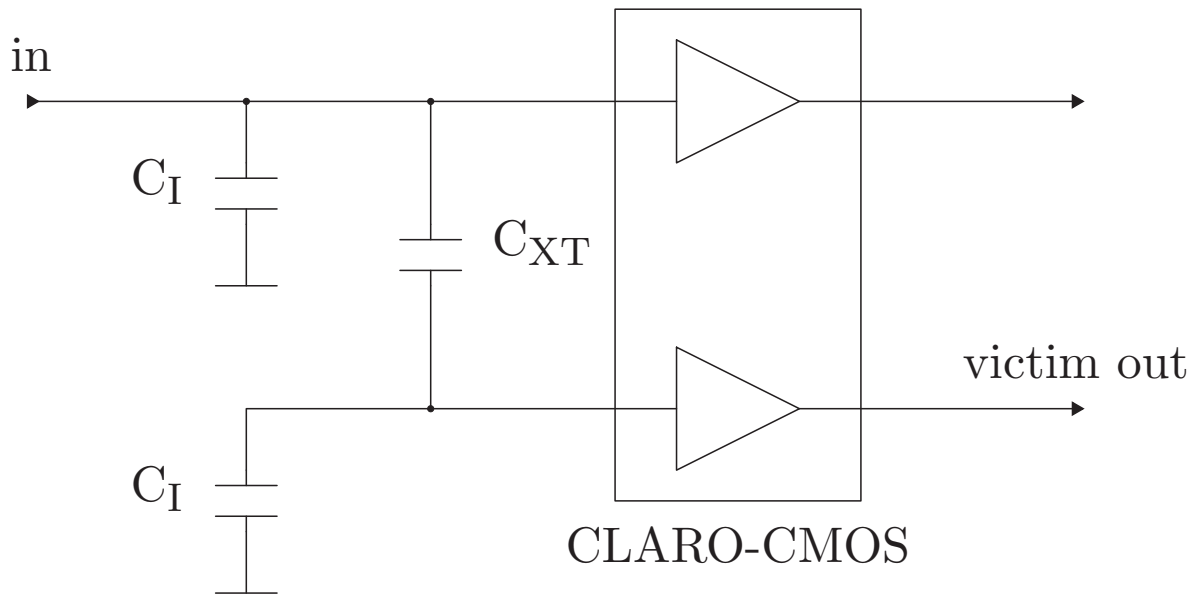

Figure 2.19: Setup for crosstalk measurement. The capacitance $C_{XT}$ represents the stray capacitance between the pixels of the sensor.

## 2.9 Crosstalk

With fast circuits such as the CLARO-CMOS, crosstalk may be critical. Fast signals could be capacitively coupled to neighbouring channels through parasitic capacitances much more easily than with slower circuits. The level of crosstalk between channels was measured as follows. The gain of the victim channel was set to the maximum value and its threshold was set at 300 ke$^-$. No signal was applied at the input of the victim, while large signals were injected at the input of a neighbouring channel. The crosstalk could be estimated from the amplitude of the minimum signal which triggers the discriminator of the victim. To simulate the real case where different pixels of a pixellated photodetector are connected to the inputs of the CLARO, a capacitance $C_{XT}$ was added between the inputs as depicted in figure 2.19. The input capacitance to ground in this measurement was $C_I = 6.5$ pF.

The level of crosstalk was measured with different values of $C_{XT}$ both in "low power" and "timing" modes, and the results are plotted and linearly fitted in figure 2.20. The crosstalk found on chip, that is with $C_{XT} = 0$, is negligible. Signals up to 10 Me$^-$ where injected without triggering the victim. Increasing the value of $C_{XT}$ causes the crosstalk to increase correspondingly. The measured data were fitted with lines, whose intercept value is compatible with zero, confirming that no crosstalk is observed if no capacitance is added outside the ASIC between the inputs. The value of $C_{XT}$ in a given application depends on the type of sensor. For instance, the capacitance between the anodes of a Hamamatsu R7600 Ma-PMT is less than 0.5 pF. This would translate in a crosstalk level below 2% in "low power" mode, and below 1% in "timing" mode. A lower level of crosstalk is obtained in "timing" mode thanks to the lower input impedance, due to the larger loop gain in the CSA as already discussed. For fast readout of pixellated sensors it is mandatory that the parasitic capacitance between neighbouring inputs is kept under control. In the cases where the capacitance $C_{XT}$ cannot be reduced due to the characteristics of the sensor, a larger $C_I$ should be used. This would affect noise and bandwidth, but would help in eliminating crosstalk.

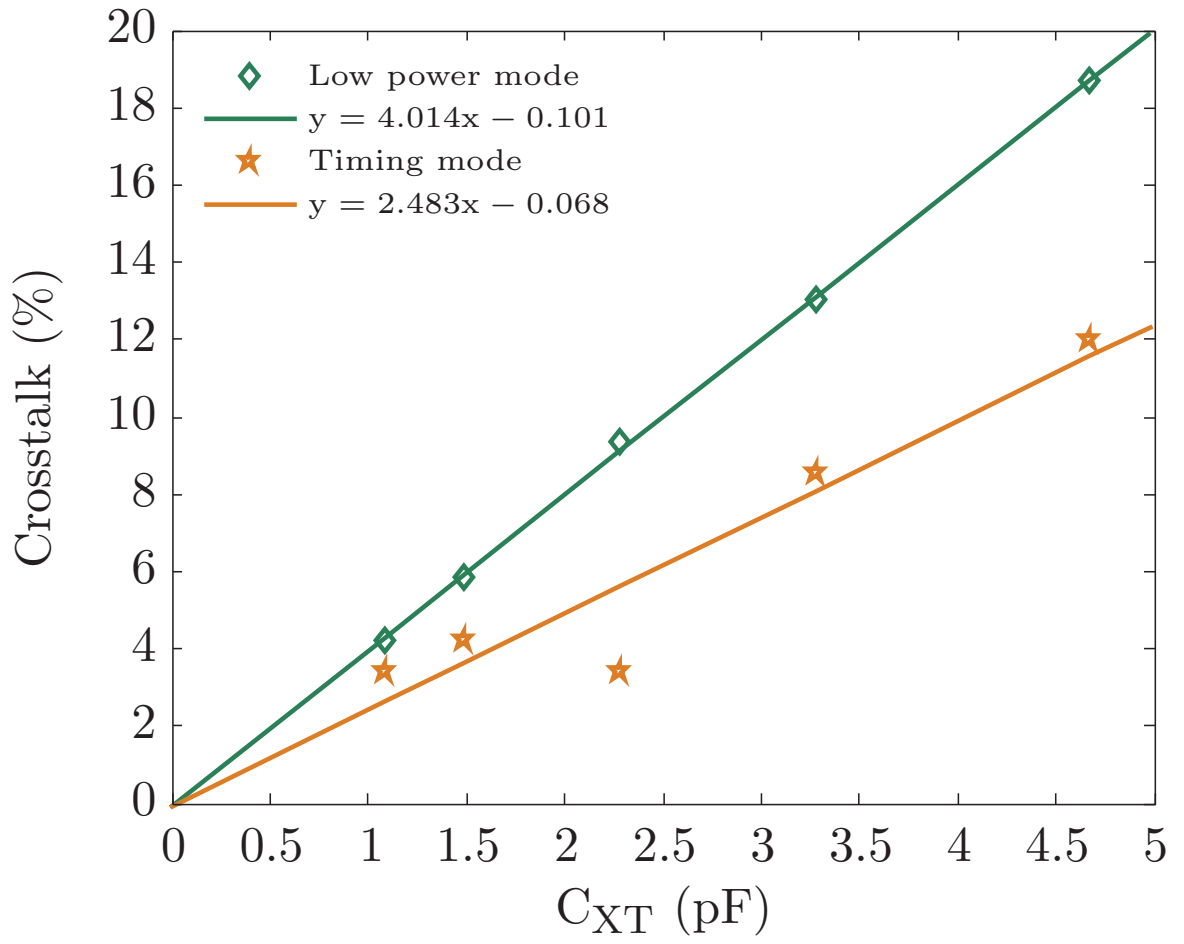

Figure 2.20: Crosstalk versus inter-pixel capacitance $C_{XT}$ in "low power" and "timing" modes.

## 2.10 Timing resolution

To evaluate the timing performance of the CLARO-CMOS prototype the gain of the CSA was set to the maximum value, and the threshold of the discriminator was set at 300 ke$^-$. Since the timing performance is expected to be directly proportional to the signal to noise ratio, the use of small input signals corresponds to a conservative, worst case scenario. The time resolution of the measurement setup was estimated to be 7 ps RMS by directly connecting the Agilent 81130A step generator to the Agilent DCA-X 86100D sampling scope. Some of the measurements presented in the following reach 10 ps: in these cases the result is partially limited by the setup. The setup contribution of 7 ps was subtracted in quadrature from the measurements. Moreover, as already mentioned, the 10% to 90% rise time of the input test signals is 0.6 ns, which is not negligible compared to the rise time predicted at the output of the CSA by equation 2.4 in "timing" mode and with a low input capacitance. As expressed by equation 2.9, the timing resolution on the rising edge of the discriminator signal is limited by the time constant of the first stage of the comparator $\tau_C$ about 5 ns. Thus the contribution of 0.6 ns due to the test signal generator is expected to be negligible in the jitter measurements. It may anyway have some impact on the effectiveness in time over threshold compensation presented in the following.

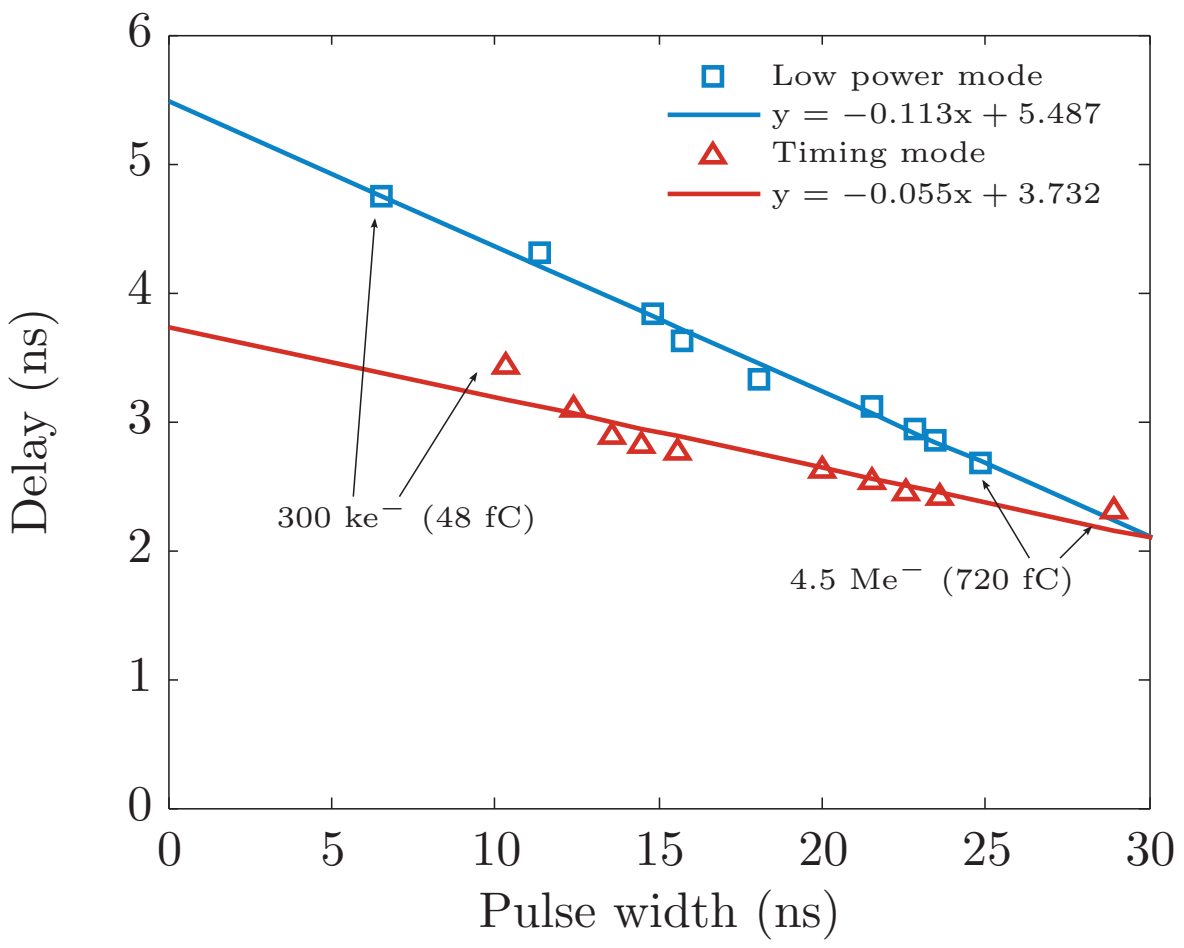

Figure 2.21: Delay versus pulse width in "low power" and "timing" modes.

The overall timing performance of a system composed of a sensor and a low jitter readout circuit depends also on the precision of time walk compensation, otherwise the low jitter would be spoiled by the time walk induced by the amplitude spread of the signals coming from the sensor. Figure 2.21 shows the dependence of the delay on the pulse width, starting from signals just above threshold. The difference in the delay for a given range of input charge is the time walk of the discriminator. This is the fundamental curve on which the time walk compensation based on time over threshold measurement relies. The slope of the fitting lines can be used to estimate the time over threshold effectiveness in compensating time walk. To a first order approximation, the curves of figure 2.21 do not depend on threshold. The measurements were taken both in "low power" mode and "timing" mode. In "low power" mode, as already mentioned, the delay ranges from about 5 ns to 2.5 ns, thus the time walk for this range of input signals, that is the difference between the two, is 2.5 ns. In "timing" mode, as shown in figure 2.21, the time walk of the discriminator reduces by about a factor of 2. Thus, even if the shape of the output signals and the maximum sustainable rate are the same as in "low power" mode, the effectiveness of a time over threshold measurement in compensating time walk is improved by a factor of 2.

The measured RMS jitter versus input charge is displayed in figure 2.22 for the "low power" mode. The plot shows the jitter on the rising edge, that is

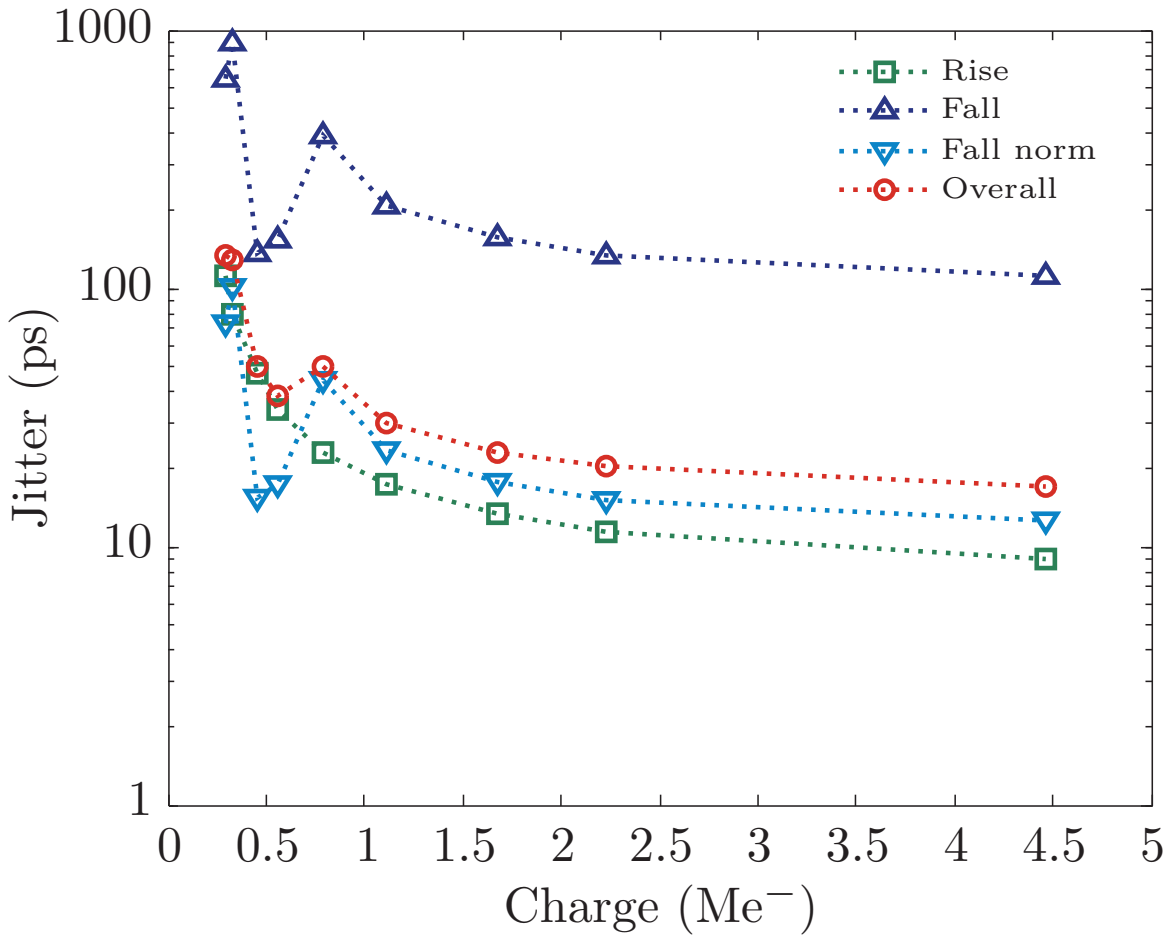

Figure 2.22: Jitter versus input charge in “low power” mode.

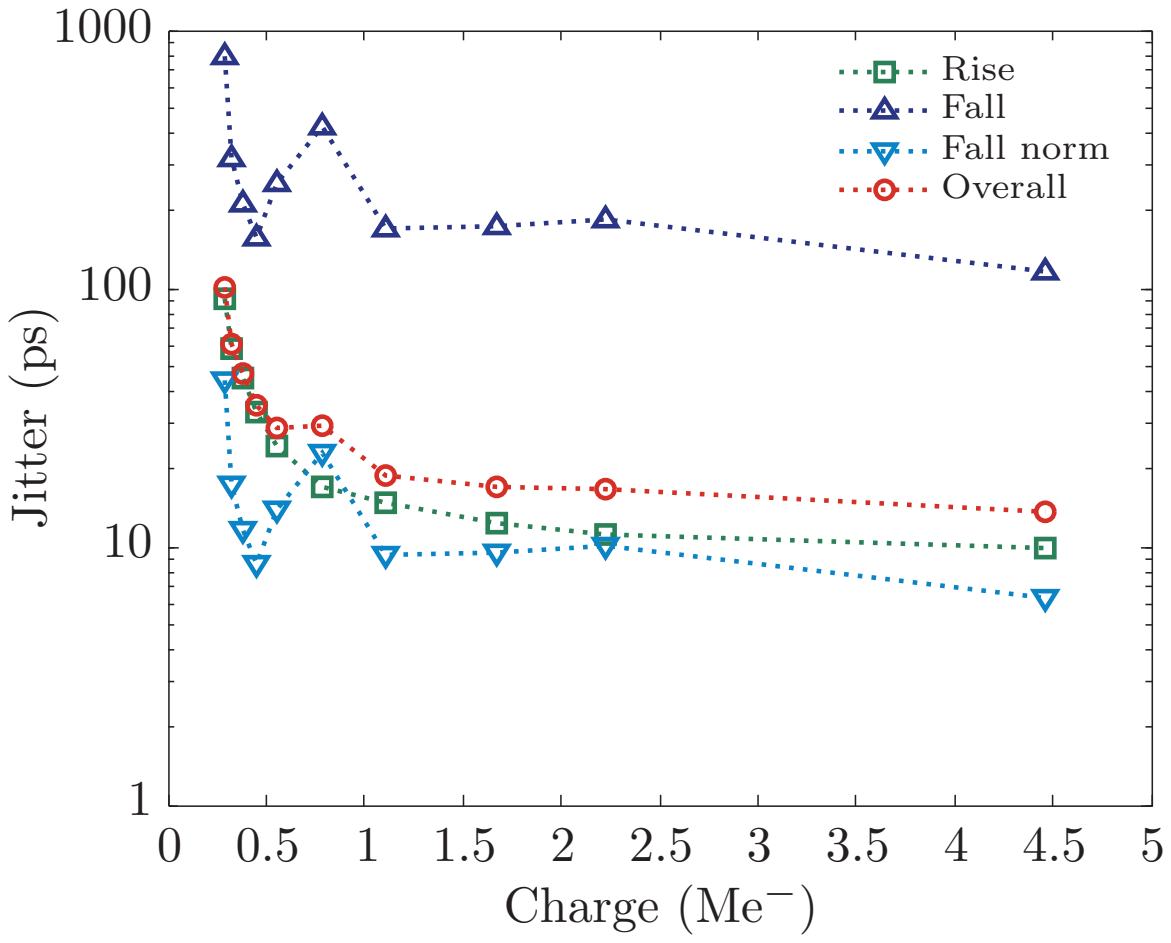

Figure 2.23: Jitter versus input charge in “timing” mode.

113 ps on threshold (about 300 ke$^-$, decreasing to 34 ps for signals of 560 ke$^-$ and then reaching 9 ps for large signals (4.5 Me$^-$. The measured values are in a good match with the values predicted by equation 2.9. For larger pulses, the rising edge jitter stops decreasing and saturates to a constant value.

The jitter on the falling edge is larger because the transition is slower. Moreover, the jitter on the falling edge is affected by a small disturbance which occurs on ground when the discriminator triggers. This explains the non-monotonic behaviour of the falling edge jitter shown in figure 2.22. Anyway, the falling edge is only used to compensate time walk: thus the weight of the falling edge jitter on a timing measurement is given by relation between time walk and pulse width, that is the slope $\gamma$ of the lines used to fit the data in figure 2.21. In other words, the jitter on the falling edge is normalized according to

$$\sigma_{\text{Fall norm}} = \gamma\ \sigma_{\text{Fall}} \tag{2.10}$$

where $\gamma$ is 0.113 in "low power" mode and 0.055 in "timing" mode, as shown in the legend of figure 2.21. The jitter on the falling edge normalized with this weight is shown in the plot, and is about 100 ps just above threshold, decreasing to 13 ps with large signals. The overall timing performance (including time walk compensation) is given by the quadratic sum of the rising edge jitter and the normalized falling edge jitter, and is shown in the red curve of figure 2.22, going from 135 ps just above threshold to 50 ps at 780 $ke^-$, further decreasing to 17 ps with 4.5 $Me^-$ signals.

The same measurements are given in figure 2.23 for the "timing" mode. The RMS jitter on the rising edge goes from 92 ps just above threshold (300 $ke^-$) to 10 ps with large signals (4.5 $Me^-$). Now the rise time $\tau_R$ of the CSA pulse is smaller than in "low power" mode, so the jitter on the rising edge is a bit smaller than in "low power" mode, but since the speed is in any case limited by the first stage of the discriminator the values are still in agreement with the values predicted by equation (2.9). Since now the time walk compensation is twice as effective than before, the normalized jitter on the falling edge goes from 44 ps to 6 ps, becoming almost negligible. The overall timing resolution is thus 102 ps just above threshold, quickly decreasing below 50 ps above 380 $ke^-$, and ultimately reaching 14 ps for 4.5 $Me^-$ signals.

The presented values are comparable with the timing resolution of the fastest photosensors available, that is MCP-PMTs. The CLARO-CMOS can then be used with such sensors for very precise timing applications, such for instance the LHCb TORCH, currently in the R&D phase.

## 2.11 Test with SiPMs

Even if designed for Ma-PMTs, the CLARO-CMOS can also be used to readout another kind of Silicon-based photon sensors which are gaining wide diffusion in recent years thanks to their low cost, that is Silicon Photomultipliers (SiPMs). Each pixel of such sensors is made of many cells connected in parallel. Each cell is a reverse biased photodiode operated in Geiger mode in series with a quenching resistor [31, 32]. An incoming photon triggers an avalanche in the photodiode, and a large number of electrons and holes are liberated and reach the readout electrodes. The gain of these devices is close to $10^6$, as in the case of vacuum-based photomultipliers (Ma-PMTs and MCP-PMTs), but the capacitance is larger, being of the order of tens of pF per $mm^2$ of active area. Upon the detection of a photon, each photosensitive cell gives a binary signal. Single photon signals can be detected, but the large amount of thermally generated dark counts, of the order of tens of KHz per mm$^2$, makes it practically impossible to separate them from the background. But since a pixel is composed of many (up to the order of 1000) microscopic cells hooked in parallel, the analog information allowing to distinguish two or more photons from single photons is recovered. The output of each pixel is then an analog charge signal whose amplitude is proportional to the number of photons hitting the sensor. The dark counts with amplitudes corresponding to signals of two or more photons are negligible if the readout circuit is fast enough to separate the individual dark counts from each other.

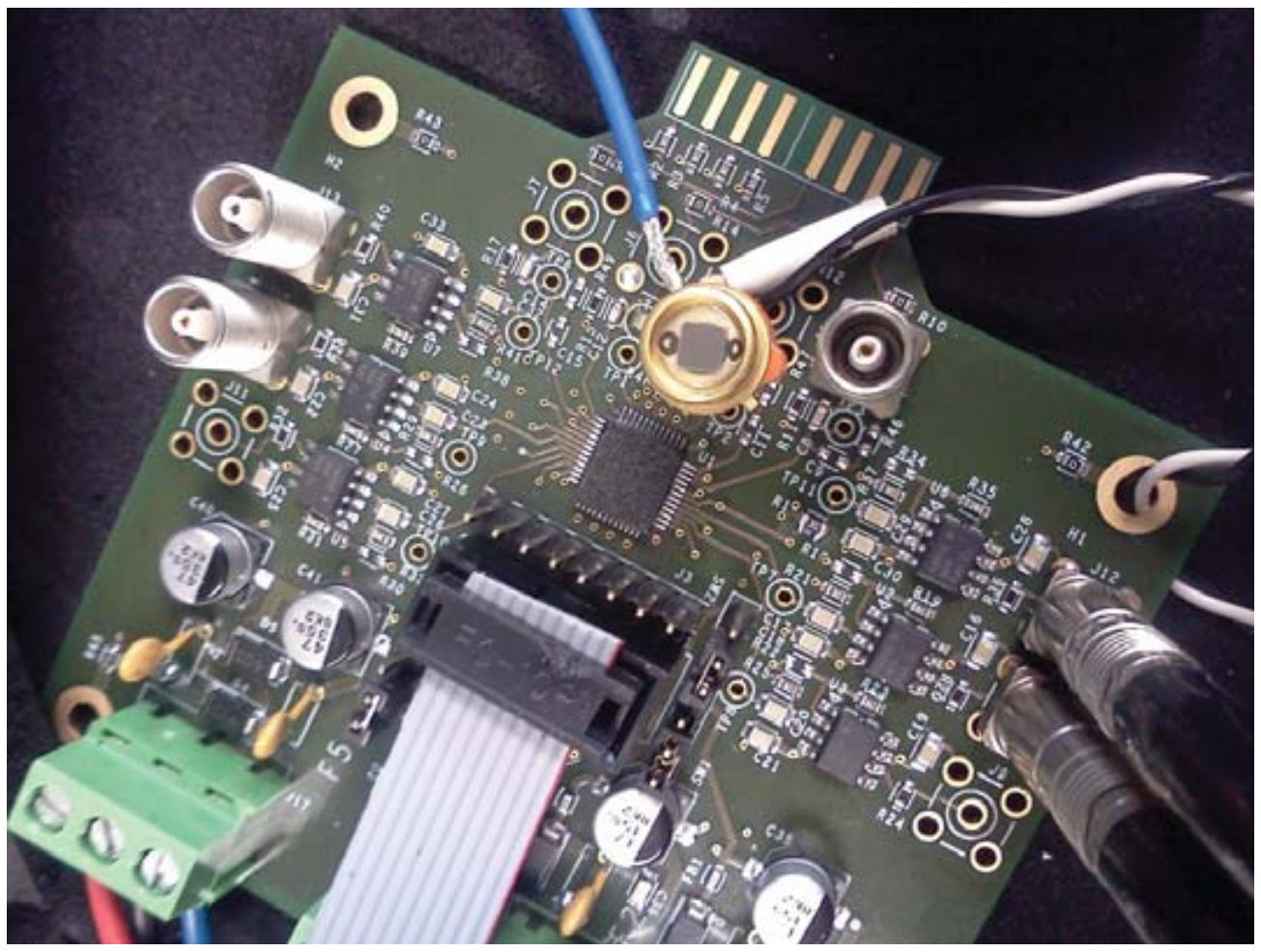

Figure 2.24: A picture of the CLARO-CMOS test board with a SiPM in TO-18 package connected to one input. The CLARO-CMOS is enclosed in a QFN48 package, which can be seen at the center of the test board.

Figure 2.24 shows a picture of the CLARO-CMOS test board with a SiPM in TO-18 package directly connected to the input. The model used for the tests is a SensL MicroSL-10050-X18 with an area of 1x1 mm$^2$. The capacitance of this device in operating conditions is of the order of 10 pF. On the test board the CLARO-CMOS in a QFN48 package can be seen at the center. The signals from the CLARO-CMOS are buffered with fast current feedback operational amplifiers which drive 50 $\Omega$ terminated lines to the oscilloscope, a Rohde&Schwartz RTO 1044.

The SiPM was biased with 29.0 V and was illuminated by a pulsed LED. The voltage to the LED was adjusted by hand to obtain signals of a few photoelectrons from the SiPM. The threshold of the discriminator of the CLARO was set at two photoelectrons to reject the large number of dark single photoelectron pulses from the SiPM. The oscilloscope was triggered on the binary

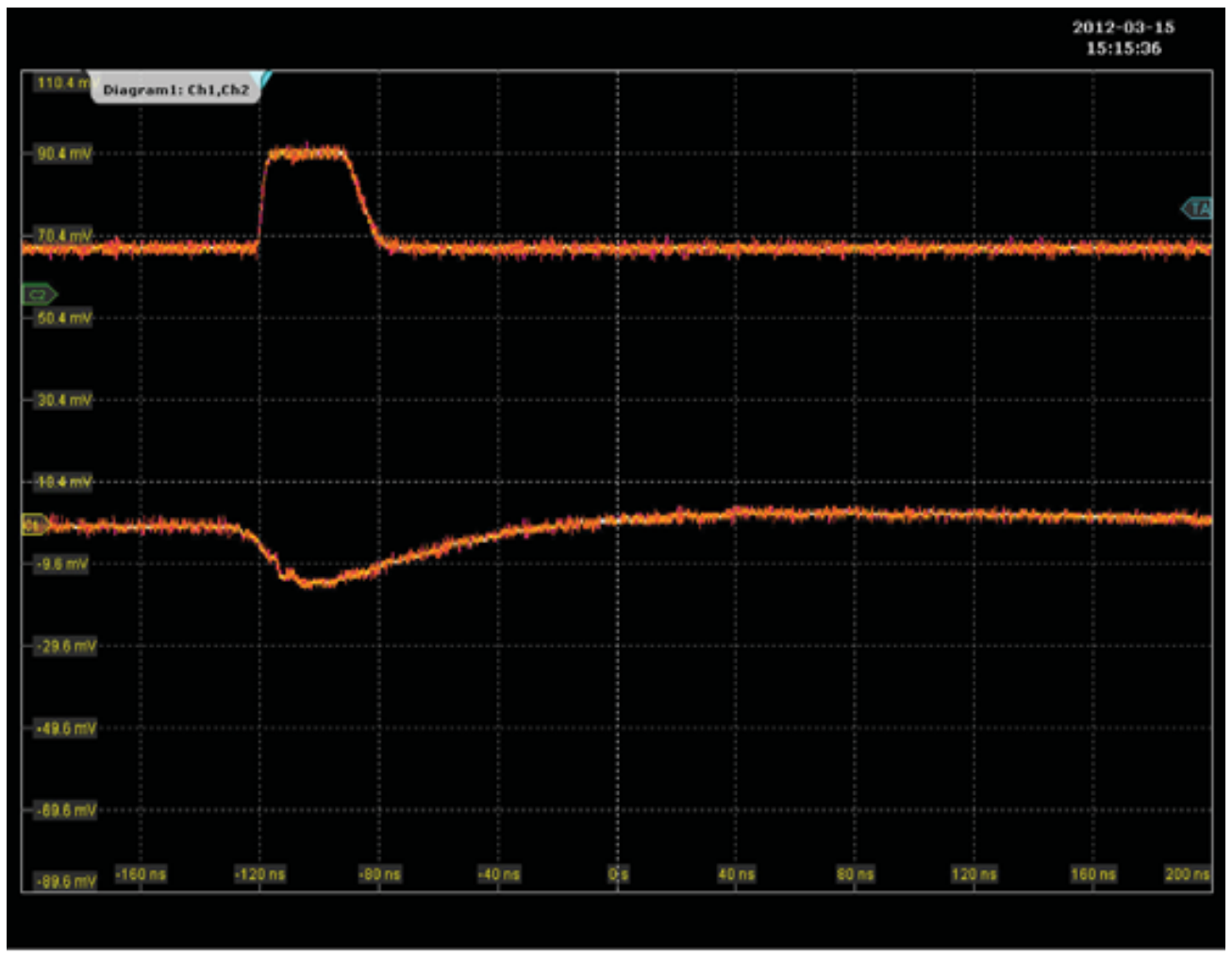

Figure 2.25: Signals at the analog output of a CLARO-CMOS channel with a 1x1 $mm^2$ SiPM directly connected to the input of the ASIC. The upper signal is taken at the binary output, the lower signal is taken at the analog output. This is a two photoelectron signal taken just above threshold.

---

output of the CLARO-CMOS. Figure 2.25 shows the oscilloscope screen when a two photoelectron event just above threshold was triggered. The rise time of the binary signal is 2.7 ns, the fall time about 8 ns. The binary pulse width is close to 30 ns. The analog signal has a leading negative edge of about 10 ns, and a trailing edge of about 50 ns. As already mentioned, the speed of the analog signal is limited by the analog output buffer. The analog signal which feeds the input of the discriminator is faster, and this is reflected in figure 2.25, where the output of the discriminator can be seen to return to the baseline before the analog signal.

Figure 2.26 shows many analog signals acquired in the same conditions. The capability of resolving single photoelectrons can be appreciated with the threshold set to two photoelectrons. The mean pulse from the LED can be seen to be centered at about 6.5 photoelectrons. The amplitude spectrum was also plotted on the left on the oscilloscope, showing a separation between photoelectron peaks of about one FWHM.

Figure 2.27 shows the same measurement, except for the fact that the SiPM was connected to the input through a Environflex EF178 coaxial cable with a capacitance of 95 pF/m. The length of the cable was 120 cm. The total capacitance at the input contributed by the SiPM and the cable is thus close to 120 pF. As can be clearly seen in the figure, the analog signals are slower, as expected from the larger capacitance at the input. The increase in series noise expected from the larger input capacitance is mitigated by the bandwidth reduction. The signals corresponding to different numbers of photoelectrons are still well resolved, with a separation between peaks close to one FWHM.

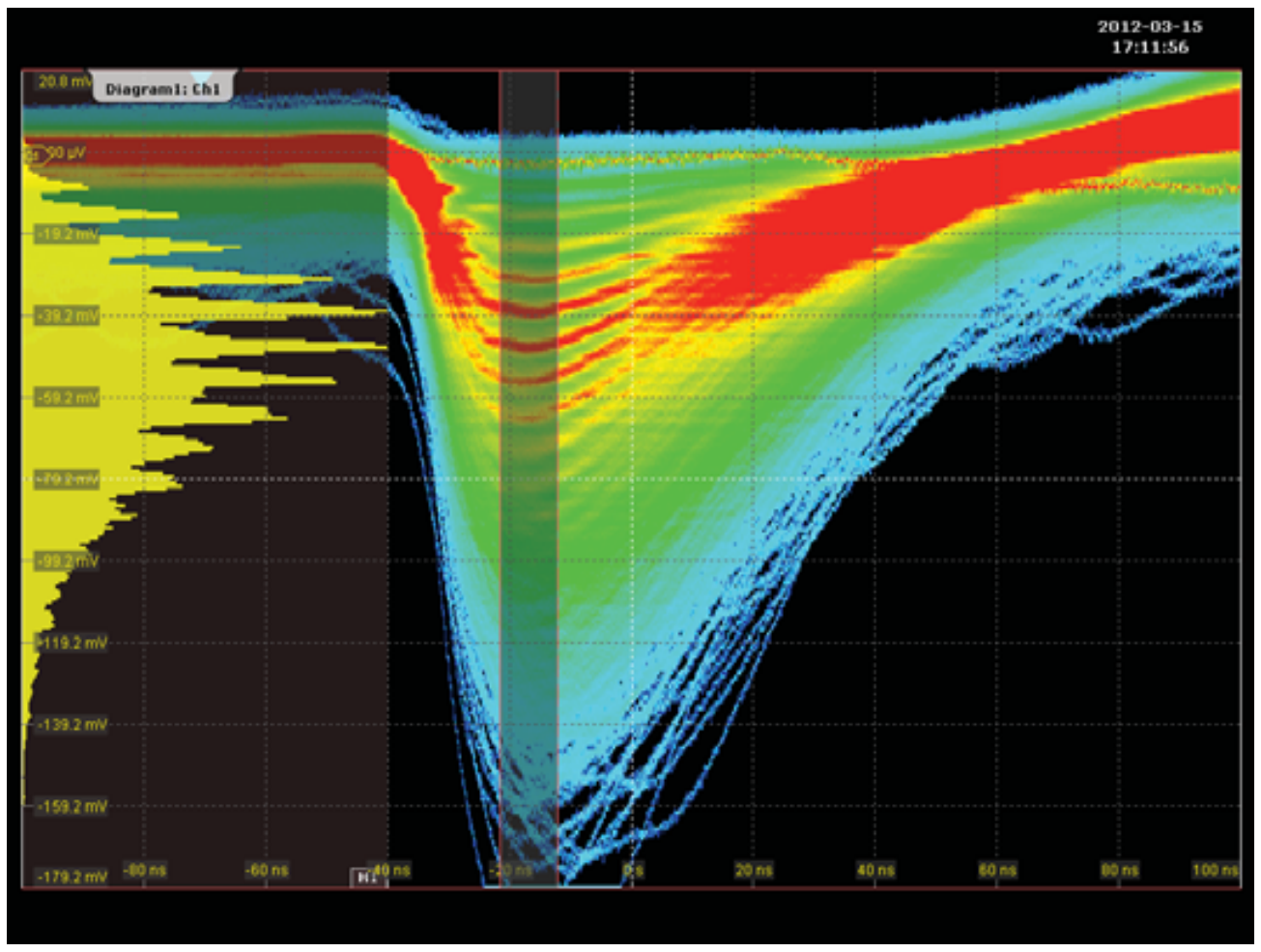

Figure 2.26: Signals at the analog output of a CLARO-CMOS channel with a 1x1 $mm^2$ SiPM directly connected to the input of the ASIC. The amplitude histogram is shown on the left side of the oscilloscope screen.

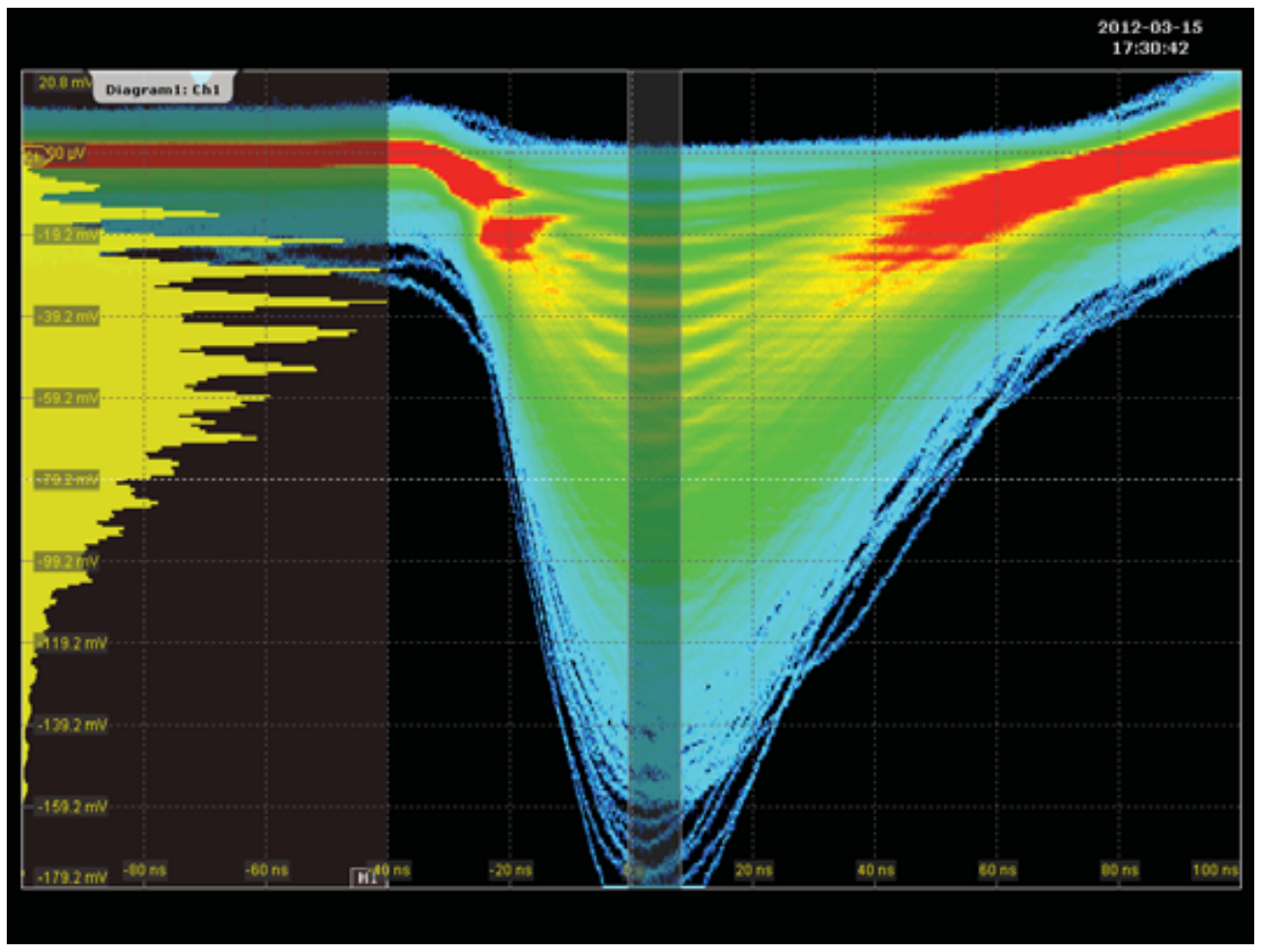

Figure 2.27: Signals at the analog output of a CLARO-CMOS channel with a 1x1 $mm^2$ SiPM connected to the input of the ASIC with a coaxial cable 120 cm long. The amplitude histogram is shown on the left side of the oscilloscope screen.

Similar measurements were also performed with other SiPMs from various manufacturers. With 3x3 mm$^2$ devices the noise is larger due to the larger capacitance. Moreover since the dark count rate is close to $10^7$ counts per second pile-up from the dark single photoelectron events occurs, and the discrete photoelectron peaks cannot be clearly separated. Anyway the threshold in this case can still be set at the level of a few photoelectrons. In the case of Hamamatsu MPPC devices, where holes are meant to be collected at the readout electrode, the $n$ substrate was connected to the input of the CLARO-CMOS in order to get negative signals at the input. For specific use with such devices a dedicated version of the ASIC could be designed, keeping the same topology and changing N-MOS transistors with P-MOS and viceversa, and exchanging the discriminator inputs.

## 2.12 Future work with the CLARO

The first prototype of the CLARO presented in this chapter works well and fully satisfies the design requirements for the LHCb RICH upgrade. In fact it also shows potential for other applications, such as high resolution timing measurements with MCP-PMTs and low power, fast photon counting with SiPMs. Some improvements to the design can still be done and will be implemented in the next versions of the chip, as pointed out in the previous sections. In particular, the area required for each channel can be reduced by changing the design of the DAC for threshold setting. This would allow to fit eight channels in a chip, in order to match the number of pixels in a row of the R11265 Ma-PMT. Moreover, the design of the analog part of the chip could be made differential, improving the rejection of common mode and power supply disturbances. This would also allow to eliminate the AC coupling between the charge amplifier and the discriminator. The design of an upgraded 8-channel version of the chip is forseen for 2013.

Figure 2.28 shows a picture of a R11265 Ma-PMT attached to the readout boards hosting the present 4-channel CLARO-CMOS chip. Figure 2.29 shows the full system designed to readout a pair of R11265. The DAQ board housing the FPGA to count the pulses at the output of the CLARO-CMOS chips and transmit the data to a PC was developed for the LHCb Collaboration by S. Wotton of the University of Cambridge. This system will be tested in laboratory. When the 8-channel version of the CLARO will be available, more R11265 readout modules will be realized. A few of such modules could be already deployed in the peripheral regions of the LHCb RICH2 for tests in the real LHCb environment by the first months of 2015.

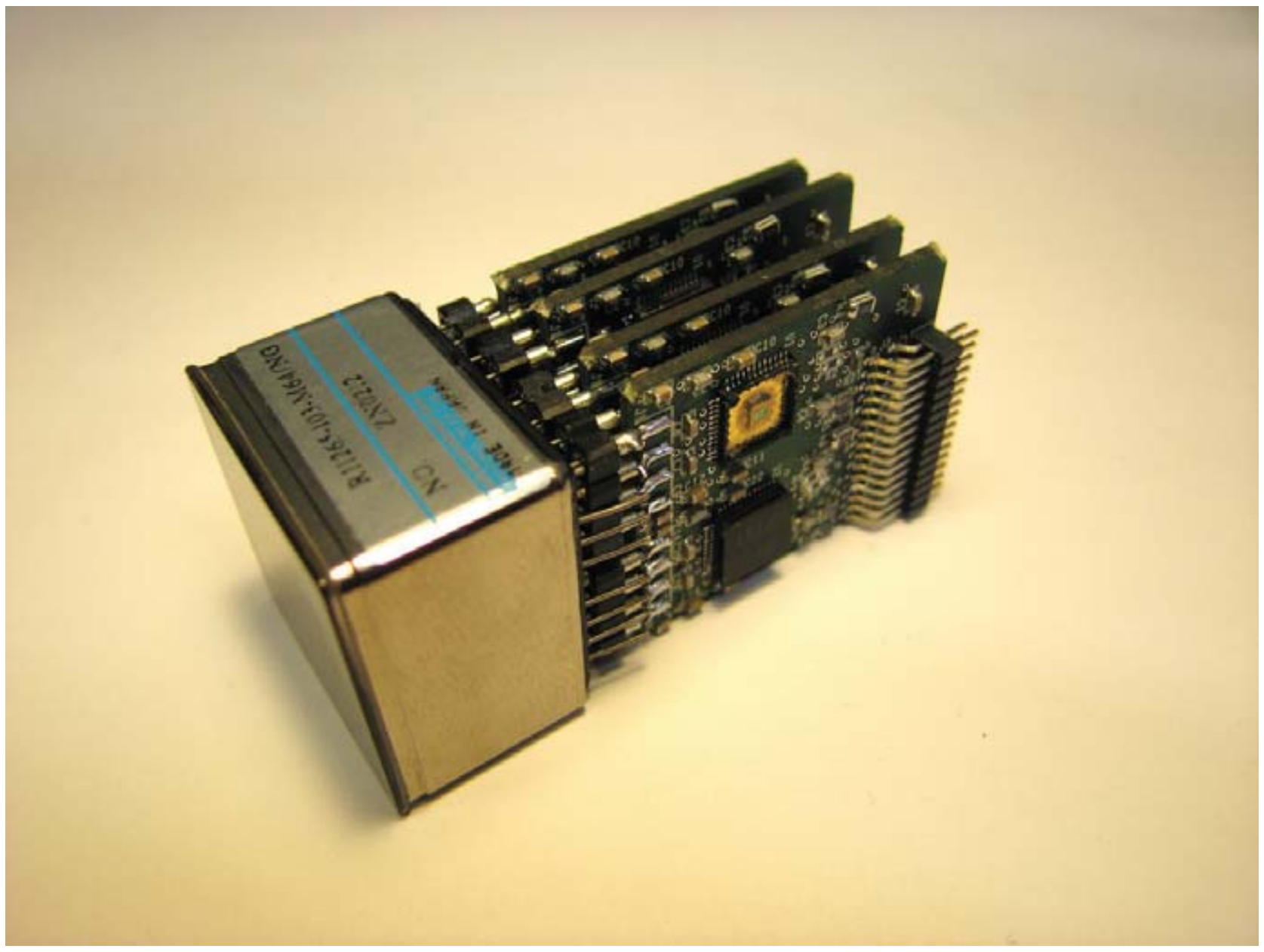

Figure 2.28: Photograph of a R11265 Ma-PMT coupled to front-end PCBs based on the CLARO-CMOS prototype. The lid of a CLARO-CMOS package was opened, showing the CLARO-CMOS chip.

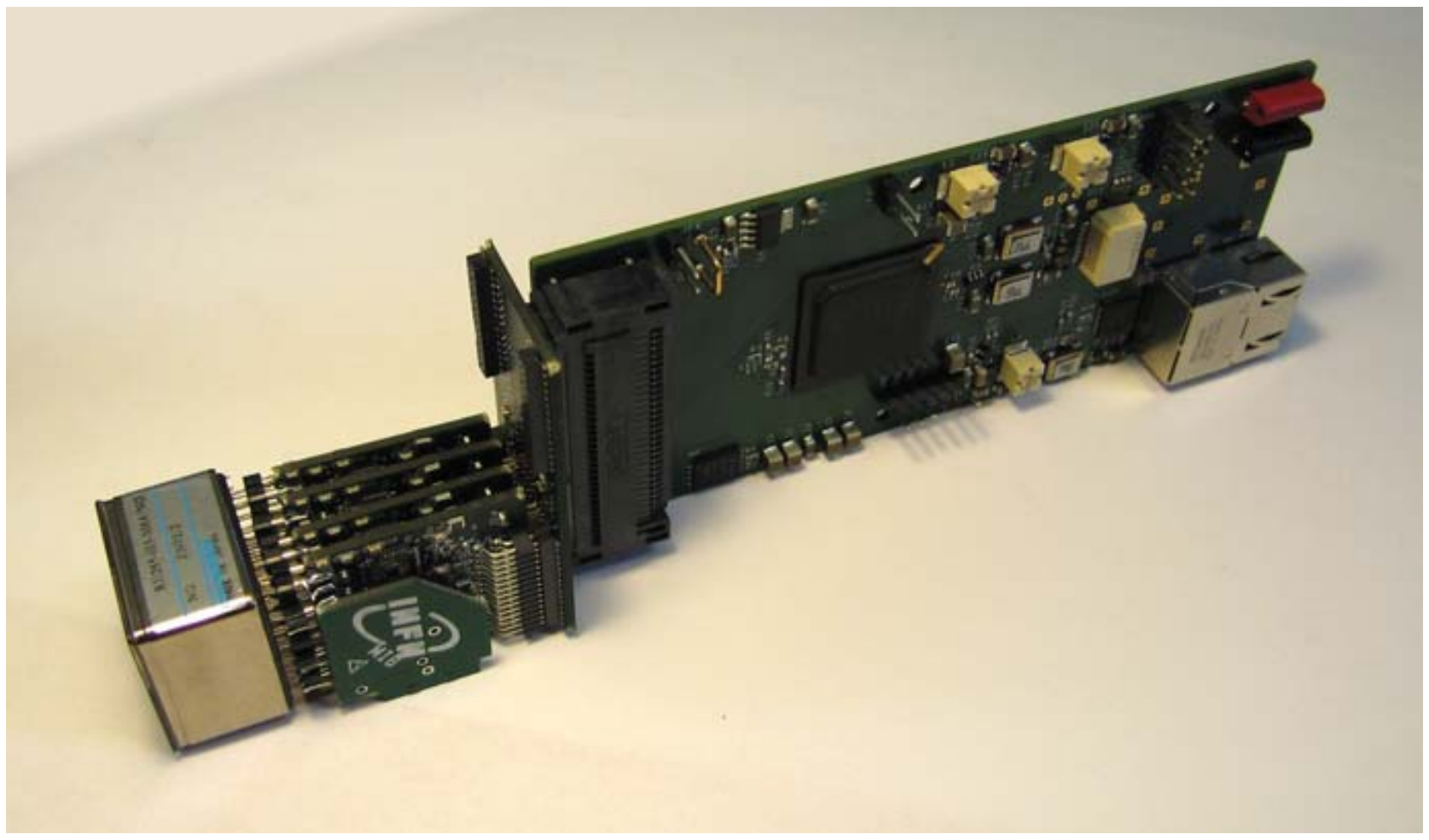

Figure 2.29: Photograph of a R11265 Ma-PMT coupled to the full CLARO-CMOS readout prototype. The high voltage for the Ma-PMT is provided by the smaller PCBs at the sides of the device. On the back, the data acquisition prototype board is shown, developed by S. Wotton of the University of Cambridge.

# 3 The GeFRO circuit for the phase II of GERDA

## 3.1 The GERDA experiment

The GERDA experiment is a search for the neutrinoless double beta decay with $^{76}$Ge as the candidate isotope [33, 34]. The importance of such process is closely related to the properties of neutrinos. Neutrinos are very elusive particles which, according to the Standard Model, have no mass, no electric charge, and interact with matter only through the weak force. In the last few decades experimental evidence was gathered for neutrino oscillations in time between its three flavours, $\nu_e$, $\nu_\mu$ and $\nu_\tau$. Oscillations require neutrinos to have mass, although small, which requires an extension of the Standard Model. Anyway the values of neutrino masses are still unknown, and only upper limits are known up to now.

Among the experimental approaches designed to probe neutrino properties the search for the neutrinoless double beta decay plays a major role. The double beta decay, where two neutrons decay into protons with the emission of two electrons and two antineutrinos, is allowed in the Standard Model. This process was indeed observed in many different isotopes in high sensitivity experiments, obtaining half lives of the order of $10^{19} - 10^{21}$ years. If neutrinos have mass, another decay could be allowed, that is the neutrinoless double beta decay, where two neutrons decay into two protons without emitting neutrinos. This is possible only if neutrinos and antineutrinos are the same particle, which in turn is possible only if neutrinos have mass. The observation of neutrinoless double beta decay would also violate the lepton number conservation, which is postulated to be constant in the Standard Model. The presence of neutrinos is inferred through the measurement of the energy of the two electrons emitted in the decay: whereas the two-neutrinos double beta decay exhibits a continuous spectrum, the neutrinoless double beta decay presents a monochromatic peak at the energy of the endpoint of the process. Energy resolution is thus one of the important factors determining the sensitivity of the search. Others are the reduction of background in the energy region of interest, together with a large mass of the candidate isotope and stable operation over an adequate measurement time.

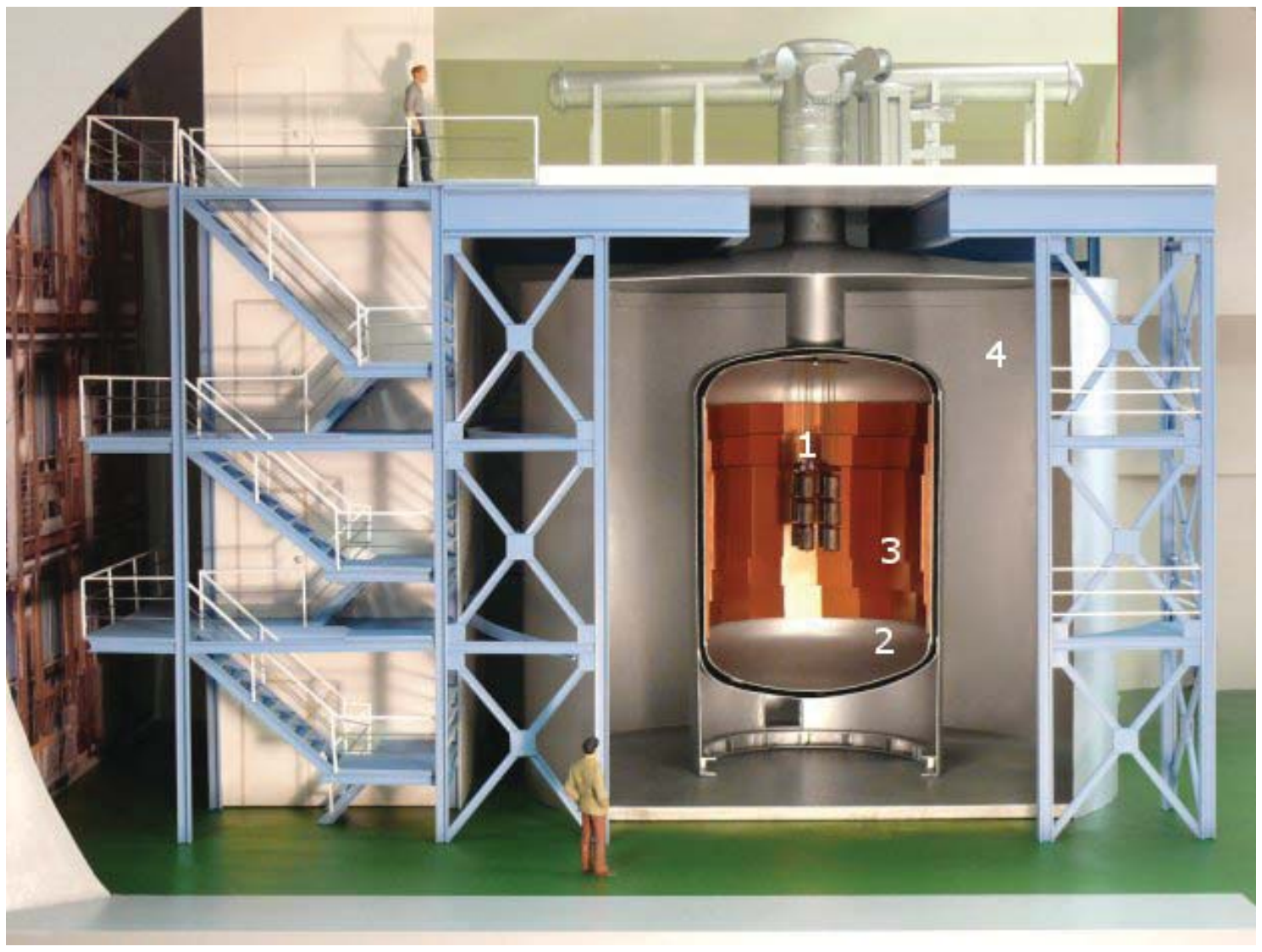

Figure 3.1: Sketch of the GERDA experiment, showing (not to scale) the Germanium detector array (1), the liquid Argon cryostat (2), its inner copper shield (3), and the water tank (4).

A positive observation of the neutrinoless double beta decay would not only provide important qualitative information on the nature of neutrinos, that is the fact that it coincides with its antiparticle, but would also allow to indirectly measure its mass, whose value can be related to the measured half life of the process through the calculation of the nuclear matrix elements for the isotope which undergoes the decay. The experimental effort towards the observation of such decay was boosted by the controversial claim of its observation by a subgroup of the Heidelberg-Moscow collaboration in 2004 [35]. The experiment was based on Germanium detectors enriched in $^{76}$Ge and operated in liquid Nitrogen. The result, not really accepted by the physics community, is that of a half life for the process close to $10^{25}$ years.

The GERDA experiment was started with the main purpose of scrutinizing the claimed observation of the neutrinoless double beta decay in $^{76}$Ge, perfecting the precision of the measurement or ruling it out. A sketch of the whole experiment is shown in figure 3.1. As happens for most double beta decay experiments, the candidate double beta decay emitters coincide with the sensors to obtain the maximum detection efficiency. GERDA employs high purity Germanium diodes, that is solid-state ionization sensors generally used to detect gamma radiation with very high resolution. The natural abundance of $^{76}$Ge is about 8%. In the case of GERDA the sensors are realized with Germanium enriched up to about 86% in $^{76}$Ge, a double beta decaying isotope. Double beta decays in the $^{76}$Ge nuclei composing the diodes are detected as charge signals in the diodes themselves. The phase I of the GERDA experiment employs 8 enriched coaxial Germanium detectors for a total $^{76}$Ge mass of 18 Kg plus some non-enriched detectors for reference. The sensors, coaxial high purity Germanium diodes, are operated bare in liquid Argon at about 87 K, which provides the required cooling with better shielding properties than liquid Nitrogen. The background from natural radioactivity is kept low thanks to the fact that the experiment is located deep underground in Laboratori Nazionali del Gran Sasso (LNGS), where the cosmic ray flux is reduced by several orders of magnitude with respect to the surface. The outer water tank housing the experiment provides further background reduction, since the Cherenkov light produced by external radiation is readout with photomultipliers and provides a veto on the measurements.

GERDA phase I started data taking at the end of 2011, and is expected to provide a check of the above mentioned claim of neutrinoless double beta decay before the end of 2013. After such result an upgrade, named phase II, is planned in order to increase the isotope mass to about 35 Kg. Moreover, passive and active background reduction techniques are to be employed to reduce background by a factor of 10. This is made possible by new point-contact Germanium sensors which allow to discriminate between single-site and multi-site events, in order to separate double beta decay candidates from gamma background. The new sensors are being fabricated with great care for radiopurity, and the scintillation light of liquid Argon is planned to be readout with SiPMs to provide an additional veto, further reducing the background. The properties of phase II sensors and the requirements for their readout circuit are presented in the following section.

A similar and concurring experiment to GERDA is MAJORANA, which shares the same goals and independently develops the technology for neutrinoless double beta decay search in $^{76}$Ge. After GERDA phase II, a common effort from both collaborations towards a joint ton-scale $^{76}$Ge experiment is foreseen, where the best solutions between those found in GERDA and MAJORANA will be employed.

## 3.2 Phase II sensors and readout requirements

The phase II of the GERDA experiment plans to deploy Germanium sensors with point-type anodes. The sensors to be used by GERDA are realized by Canberra, and are commercialized with the name of Broad Energy Germanium (BEGe) sensors. The BEGe sensors to be used in GERDA will be made of Germanium enriched in $^{76}$Ge. Figure 3.2 shows a photograph of a BEGe diode. The cathode is to be biased with a positive high voltage of about 4 kV. The leakage current of the diode can be as low as a few pA if proper care is taken in handling the devices, especially when operated bare as in the GERDA experiment. The values for the leakage current are anyway considered to be acceptable when below a few hundred pA, although this results in a larger noise. Positive charge signals are collected at the anode, which is tied to ground by the readout circuit. The advantages offered by such sensors with respect to standard coaxial diodes are two. First, the smaller anode contact results in a small capacitance at the readout electrode, of the order of one pF, at least one order of magnitude below the coaxial diodes used in GERDA phase I. As was shown in chapter 1, a smaller sensor capacitance results in a lower contribution of the series noise to the energy resolution. Second, the electric field in a BEGe in nominal bias, where the diode is fully depleted, is weaker than in coaxial detectors, resulting in a slower charge collection time which ranges up to about 500 ns. This is crucial for the physics goals of GERDA, as the time-stretching feature of BEGe diodes can be exploited to discriminate between single-site and multi-site interactions [36, 37].

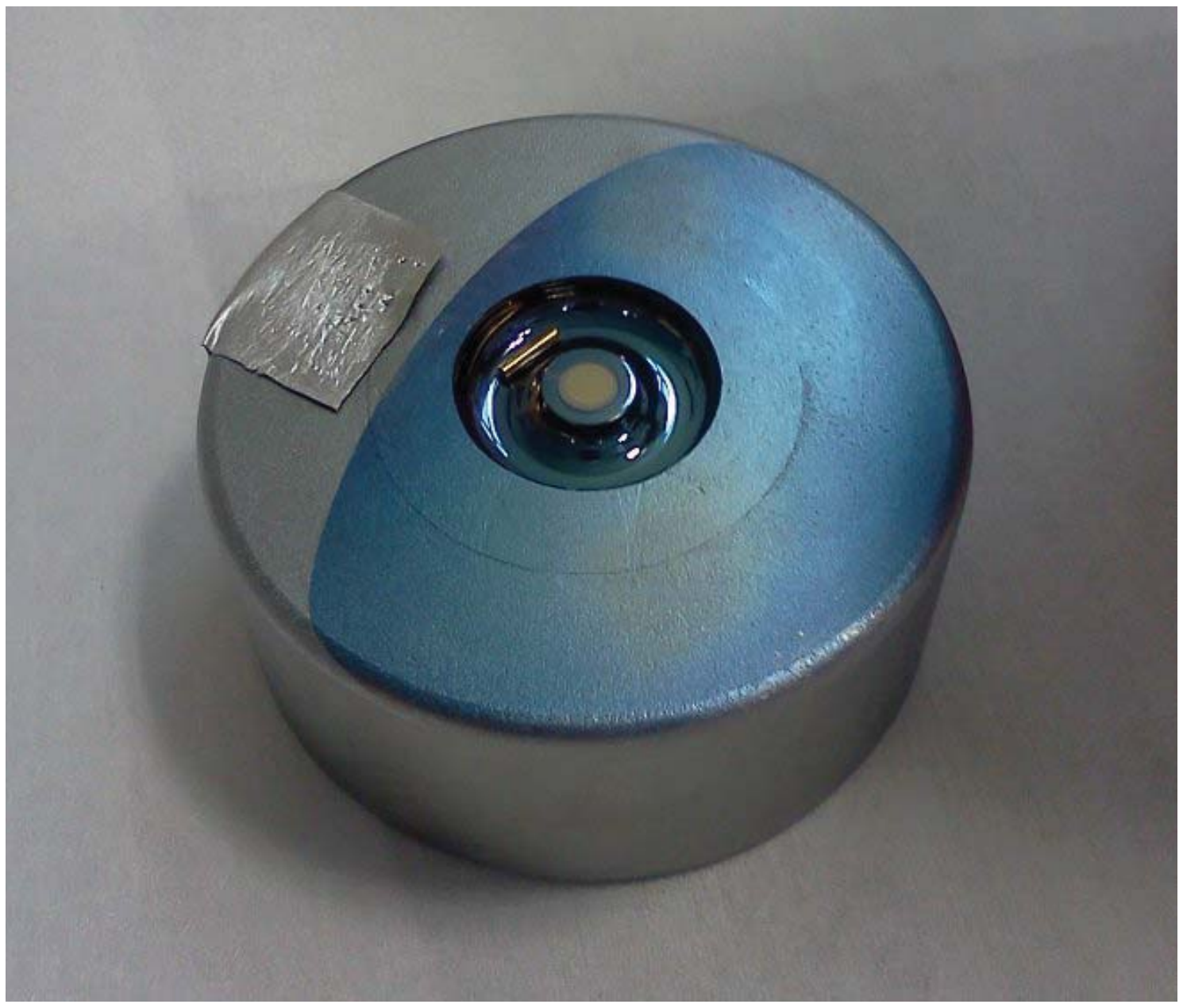

Figure 3.2: A BEGe radiation sensor. The cathode covers all the surface of the diode, except for the small contact in the center, that is the anode. The diameter of the diode is about 8 cm, its height about 3 cm.

The typical single-site and multi-site signals are shown in figure 3.3. As can be seen, the difference is in the rising edge of the integrated signals, which in case of multi-site interaction shows distinguishable steps on the timescale of tens of nanoseconds. Provided that the readout circuit is fast enough to preserve the signal shape, this difference can be used to separate single-site events, to which the double beta decay belongs, from multi-site events which mainly come from the gamma background, reducing it by at least a factor of two. Since the rising edge is of the order of 200 ns, the shape of the rise time is safely preserved with a bandwidth of at least about 20 MHz, which corresponds to a 20 ns rise time in response to a delta-like current pulse.

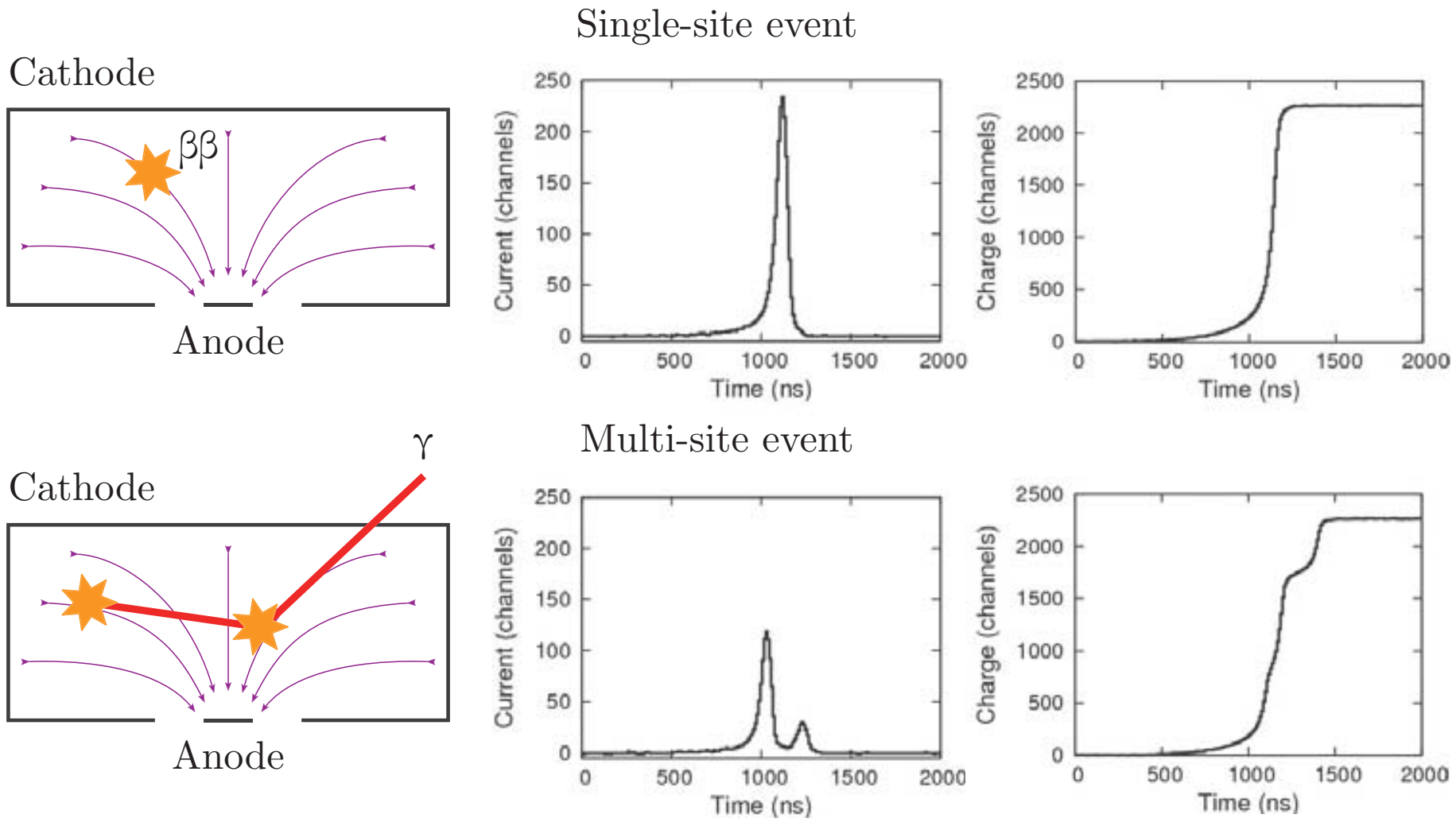

Figure 3.3: Single-site and multi-site event discrimination in BEGe diodes. The figures on the right showing the current and charge signals are taken from [37].

Germanium ionization sensors exhibit a good intrinsic energy resolution [38]. Given an energy E deposited in the sensor by a particle event and converted to ionization, the average number of electrons and holes created in the process is

$$N = \frac{E}{\varepsilon} \tag{3.1}$$

where $\varepsilon$ is the energy required to create an electron-hole pair, which in the case of Germanium is 2.98 eV. For fully depleted high purity sensors, losses due to electron-hole recombination can be neglected and the signal readout at the anode carries a charge which is simply given by

$$Q = Nq = \frac{Eq}{\varepsilon} \tag{3.2}$$

where q is the electron charge. Assuming all the electron-hole creation events to be independent, one may expect the number of ionization pairs to fluctuate following Poisson's statistics, yielding $\sigma_N = \sqrt{N}$. According to more accurate models (supported by experimental evidence) which do not consider the events to be independent, the observed fluctuation is instead lower and is given by

$$\sigma_N = \sqrt{FN} = \sqrt{\frac{FE}{\varepsilon}} \tag{3.3}$$

where F is the Fano factor, which in the case of Germanium is 0.129. The resulting energy resulution can be written as

$$\sigma_E = \frac{E}{N}\sigma_N = \sqrt{\varepsilon F E} \tag{3.4}$$

The resolution is also often expressed in terms of the full width at half maximum (FWHM), that is $\sqrt{8 \ln 2} \times \sigma_E$, with $\sqrt{8 \ln 2} \simeq 2.35$. The neutrinoless double beta decay in $^{76}$Ge is expected to give a monoenergetic peak at an energy of 2039 KeV. The intrinsic resolution of a Germanium diode at such energy as calculated from equation (3.4) is 2.1 KeV FWHM (close to 0.1%). This is the best energy resolution which can be achieved with an ideal Germanium sensor, operated in ideal conditions, assuming a noiseless front-end amplifier. When using a real charge amplifier, its noise must be summed in quadrature to the intrinsic resolution to obtain the resulting energy resolution. The goal for the phase II of the GERDA experiment is to keep the resolution better than 0.2%, that is 4 KeV FWHM. Considering a safe margin, the contribution from the front-end amplifier should be less than 1.5 KeV FWHM, which would provide an overall energy resolution at the neutrinoless double beta decay better than 2.5 KeV FWHM.

Aside from wide bandwidth, needed to allow the necessary timing resolution for pulse shape discrimination, and low noise, needed to preserve the energy resolution at the energy of the neutrinoless double beta decay peak, the third main requirement of the readout amplifier is to be made with pure materials, in order not to contribute to the radioactivity background. The charge amplifier currently used in GERDA phase I, made of a commercial JFET (BF862 from NXP) and an operational amplifier (OPA353 from Texas Instruments) cannot be located close to the sensors [39]. This poses a lower limit to the input capacitance, which is directly proportional to noise, and an upper limit to bandwidth, since the close-loop bandwidth of a charge amplifier results to be inversely proportional to the ratio between the input capacitance and the feedback capacitance. Moreover, various disturbances and crosstalk can be picked up at the input due to the long connecting cable from the sensor to the amplifier. In order to place the front-end circuit closer to the sensors, a higher degree of radiopurity needs to be obtained. This implicitely requires also the minimization of the size of the front-end circuit, since for any given material the background contribution is directly proportional to its mass. One choice might be that of splitting the front-end circuit after the first stage, placing only the input transistor and the feedback elements near the sensor and the rest of the circuit farther apart. Also in this case, bandwidth may prove to be critical, since separating the circuit in this way adds a large capacitance and phase delay in the feedback loop, which affects speed or stability.

In this PhD work, a third way named GeFRO was developed, which allows to overcome the limitations of the above solutions and satisfies the requirements of GERDA phase II. The design choices and the results of tests on

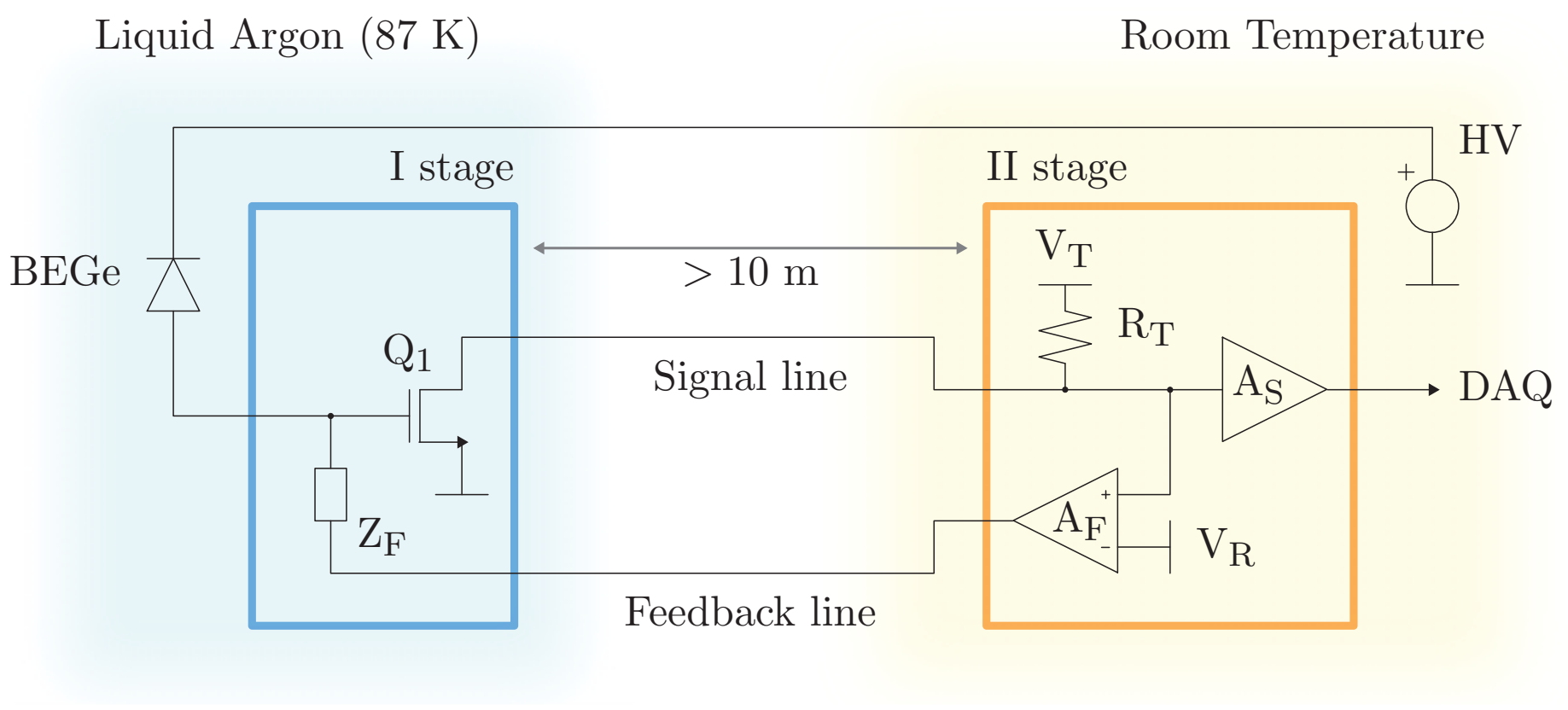

Figure 3.4: Block schematic of the GeFRO circuit.

Germanium detectors carried out both in Milano Bicocca and in LNGS are presented in the following sections.

## 3.3 The GeFRO circuit

The proposed solution, named GeFRO for Germanium front-end, consists of a new approach to the readout of ionization sensors [40,41]. The basic schematic of the circuit is shown in figure 3.4. The circuit is composed of two stages. The first stage employs a minimal number of components and is located near the sensor, submerged in liquid Argon at the temperature of 87 K. The second stage is located at room temperature. The two stages are connected by two transmission lines: one carries the fast signal from the sensor, while the second carries a feedback signal to discharge the input node of the circuit after each event. The figure also shows the high voltage line used to bias the sensor.

As can be seen in the figure, the first stage is based on a JFET operated in common source configuration. The working point of the input transistor $Q_1$ is determined and kept stable by the feedback amplifier $A_F$. At DC the feedback works to keep the drain of $Q_1$ at the same voltage as $V_R$. In this simple scheme the bias current of $Q_1$ can be set by acting on the reference voltage $V_T$ (in the actual prototype other solutions are preferred, as will be shown). The feedback impedance $Z_F$ is a large value resistor, of the order of 1 GΩ, or a diode in reverse bias, which was proven to work as a nonlinear large value resistor.

A signal from the sensor is a current pulse carrying a given amount of charge Q. The current signal is integrated on the input capacitance, that is contributed by the sensor, the JFET, the feedback element $Z_F$ and parasitics, and becomes a voltage step at the input. The JFET converts the voltage step at the input to a current step at its output, which is driven into a properly terminated signal line of characteristic impedance $R_T$ to avoid reflections. The line impedance $R_T$ is expected to be $50 - 100\ \Omega$ depending on the cables which will be chosen for GERDA phase II, which have to satisfy mechanical and radiopurity constraints. The signal across $R_T$ is amplified by the fast voltage amplifier $A_S$ and sent to the data acquisition chain (DAQ). Meanwhile, the feedback amplifier $A_F$, whose bandwidth is purposely limited, modifies its output to force a small current through the feedback element $Z_F$, which discharges the input node and defines the fall time of the signal.

At a few mA of bias current, the JFETs tested so far at low temperature have a transconductance over capacitance ratio of the order of 2 mA/(V pF). For a transconductance of $10 - 20$ mA/V, necessary to drive the signal on the terminated $50 - 100\ \Omega$ transmission line without significant gain loss, the resulting input capacitance from the JFET is $5-10$ pF. The first prototypes of the GeFRO used the BF862 from NXP as the input transistor $Q_1$, featuring a total gate capacitance of 10 pF and a transconductance of about 20 mA/V at a few mA of bias current. The BF862 is the JFET currently used in GERDA phase I. For a higher degree of radiopurity, in the final version of the circuit it will be mandatory to use components in bare die in the cold stage. Thus other JFETs which can be purchased in bare die are currently being tested, and a very promising candidate was already found.

In the first tests with the GeFRO a reverse-biased Schottky diode of small capacitance, below 1 pF, was chosen as the feedback element $Z_F$, namely the BAT17 from NXP. Its current-voltage characteristic curve at low temperature shows a very good performance, being able to sustain a broad range of currents from the sensor, from a few pA to several nA, with a high dynamic impedance. In contrast the reverse current of Silicon diodes of small capacitance at low temperature was found to be negligible, so that such devices cannot be used for this purpose. The presence of the diode is also beneficial to protect $Q_1$ from negative overvoltages, which may happen when the bias voltage of the sensor is decreased. Another more conventional solution is that of employing as the feedback element a standard large value resistor. Resistors in bare die of values up to 150 MΩ were found to be commercially available and are currently being tested. A few of such devices may be used in series to obtain a feedback resistor of a large enough value. The drawback of using only a resistor as the feedback element is that the gate of $Q_1$ would not be protected from negative overvoltage, which may occur when the HV bias of the sensor is decreased too fast. For these reasons, the parallel combination of a large value resistor and a diode may in the end be the best solution for robust operation. In this case, the diode does not need to be a Schottky diode, as any Silicon diode of small capacitance would do, relaxing the requirements on the choice of the device. The final decision will be taken based on the results of radiopurity tests of the components which will be found available. From the point of view of the signals, the feedback diode behaves as a small capacitance $C_F$ which is

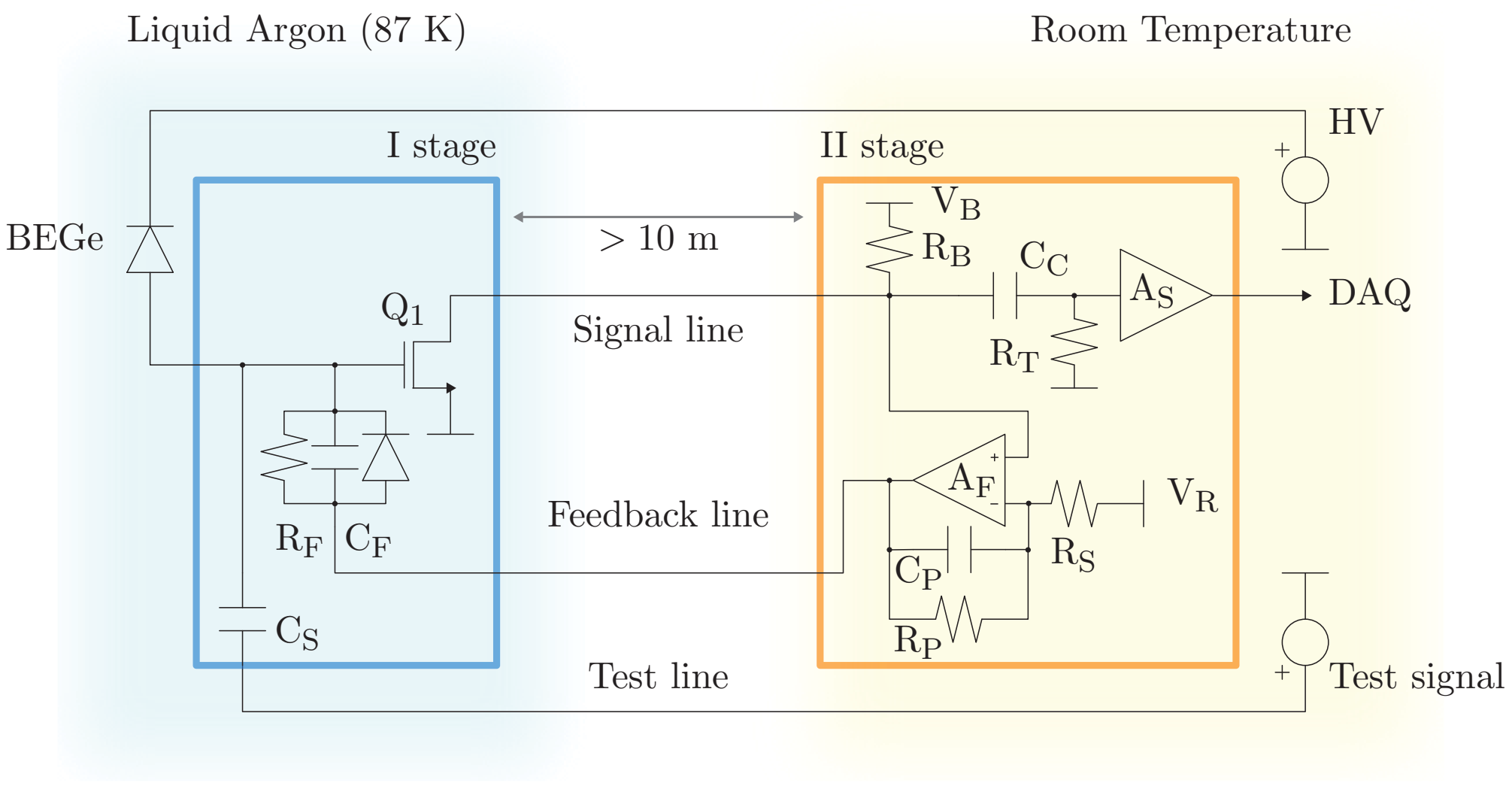

Figure 3.5: Full schematic of the GeFRO circuit.

---

beneficial in compensating the low frequency pole formed at the input by the feedback resistance $R_F$ and the total input capacitance $C_I$, as will be shown. If the diode is not used, a small capacitance of about 1 pF should be used for the same purpose.

Since, as already mentioned, the transconductance $g_m$ of the JFET is expected to be of the order of $10-20$ mA/V and the impedance of the transmission line $R_T$ is likely to be $50-100\ \Omega$, the gain of the first stage given by $g_m R_T$ is close to one. Care must then be taken in the design of the second stage $A_S$ to avoid adding its contribution to the total noise. The current prototype under test employs a LT6230-10 operational amplifier from Linear, featuring a bandwidth of 140 MHz at gain 10 with an input white voltage noise density of 1.1 nV/$\sqrt{\text{Hz}}$, rising to 5 nV/$\sqrt{\text{Hz}}$ at 10 Hz. The LT6230-10 is operated with a gain of 20. The output of the LT6230-10 is amplified and inverted by a pair of AD811 operational amplifiers from Analog Devices with a gain of 10, so that the signal at the output of the second stage can be driven over a differential 50 Ω transmission line. The overall gain of $A_S$ over a differential terminated output pair is then $G_S = 200$ V/V.

A more detailed schematic of the circuit is presented in figure 3.5. The feedback impedance $Z_F$ is the parallel combination of a resistor $R_F$, a capacitance $C_F$ and an ideal protection diode. The bias current for $Q_1$ is set by $R_B$ and $V_B$, and the line termination resistor $R_T$ is now AC coupled through $C_C$. The components $C_P$, $R_P$ and $R_S$, which rule the behaviour of the feedback amplifier $A_F$ are also shown. As the feedback amplifier $A_F$, where wide bandwidth is not required, an OP27 operational amplifier from Analog Devices was used. A baseline restorer circuit, not shown in the schematic, was incorporated into $A_S$ to eliminate the possible undershoot due to the AC coupling. As will be shown in the following calculations the undershoot is already suppressed if a proper choice of $C_P$, $R_P$, $R_S$ and $C_C$ is made. The baseline restorer circuit is then not expected to be necessary in the final version of the circuit. A series resistor of a few kΩ and a diode to ground, not shown in the schematic, were connected to the output of $A_F$ to prevent the possibility of the feedback amplifier to drive a positive signal on the feedback line during the switch-on transients, which may damage the JFET and the feedback diode. The schematic also shows the line used to inject a test signal through the test capacitance $C_S$, whose value was chosen to be 0.5 pF. If the test signal is a voltage step of amplitude $V_S$, a current pulse carrying charge $Q = C_S V_S$ is generated.

Neglecting the slow feedback, the signal at the output of the circuit in response to an instantaneous current pulse carrying a charge Q is given by

$$V_{OL}(s) = -G_S \frac{Q}{sC_I} g_m R_T \left( \frac{sC_C R_B}{1 + sC_C(R_B + R_T)} \right) \tag{3.5}$$

where $C_I$ is the total input capacitance, contributed by the sensor, the JFET, the feedback elements, the test capacitor $C_S$ and parasitics. Equation (3.5) is obtained by considering the drain of the JFET as an ideal current source. Let us note that if the JFET can be considered an ideal current source, then the amplitude of the output signal depends only on $g_m R_T$ and is independent of the series resistance of the cables. This is particularly remarkable, since the cables to be used in GERDA phase II should have a small cross-section to minimize their contribution to the radioactive background, and a series resistance of the order of $1 - 2$ Ω/m is expected at room temperature. In the real case the output impedance of the JFET, contributed by its drain-source resistance and gate-drain capacitance, should be included in the calculations. Its effect on the overall gain is a second order contribution and will not be considered here. It will be considered in the next section, since it gives a small contribution to the gain drift if the series resistance of the cables changes.

For $s \gg 1/C_C R_B$ and $R_T \ll R_B$ equation (3.5) simplifies to

$$V_{OL}(s) = -G_S \frac{Q}{sC_I} g_m R_T \tag{3.6}$$

If this approximation holds, the output signal in time domain neglecting the slow feedback is given by

$$V_{OL}(t) = -G_S \frac{Q}{C_I} g_m R_T \vartheta(t) \tag{3.7}$$

where $\vartheta(t)$ is the step function. Since the gain of the cold stage $g_m R_T$ is close to one, the Miller effect on the gate-drain capacitance of the input JFET is small. The contribution of the JFET to the total input capacitance is then dominated by the gate-source contribution. In the measurements carried out so far, $C_I$ was about 20 pF, contributed in almost equal parts by the JFET and parasitics. The capacitance of the BEGe sensor and of the feedback elements is of the order of 1 pF and can be neglected, together with the test capacitance $C_T$. As already mentioned, the need to keep $g_m R_T \geq 1$ prevents the use of a JFET with a capacitance lower than about 10 pF, but the parasitics at the input can in principle be minimized with respect to the current test setup. A total input capacitance of $10-15$ pF, dominated by the JFET, is thus foreseen for the final versions of the GeFRO.

The discharge of the input node occurs through the feedback resistor $R_F$ and is ruled by the feedback amplifier $A_F$. Let us consider $R_F$ to be 1 GΩ, in parallel with the small capacitance $C_F$, about 1 pF, given by the parasitics of the resistor and of the protection diode. In the domain of the complex frequency the loop gain is given by

$$\begin{aligned} T(s) = -\left(\frac{1+sC_F R_F}{1+sC_I R_F}\right)\left(g_m R_B \frac{1+sC_C R_T}{1+sC_C(R_T+R_B)}\right) \times \\ \times \left(\frac{R_P+R_S}{R_S}\frac{1+sC_P(R_P\|R_S)}{1+sC_P R_P}\right) \end{aligned} \tag{3.8}$$

The first term is due to $R_F$, which forms a pole with the total input capacitance $C_I$, which is compensated with a zero by $C_F$. The second term is due to the gain of the JFET on the total impedance it sees at its output. This term contributes with a pole at $C_C(R_T + R_B)$ and a zero at $C_C R_T$. The last term is due to the transfer function of the feedback amplifier $A_F$. This term contributes with a pole at $R_P C_P$ and a zero at $(R_P\|R_S)C_P$. As already mentioned, $C_I$ is expected to be about 20 pF, while $R_F$ and $C_F$ are expected to be close to 1 GΩ and 1 pF respectively. The value of $R_T$ depends on the characteristic impedance of the transmission line, that is $50 - 100$ Ω. $R_B$ sets the current in the JFET, and if $V_B = 12$ V, $V_R = 2$ V, as in the current prototype, then $R_B = 1$ KΩ gives a current of 10 mA. Let us then consider $R_B \gg R_T$ in the following calculations. The components whose values are not fixed by other constraints are $C_C$, $C_P$, $R_P$ and $R_S$. $R_S$ can be chosen to be equal to $R_B$, and let us assume $R_P \gg R_S$. By approximating for $R_B \gg R_T$ and $R_P \gg R_S$ equation (3.8) becomes

$$T(s) = -\left(\frac{1+sC_F R_F}{1+sC_I R_F}\right)\left(g_m R_B \frac{1+sC_C R_T}{1+sC_C R_B}\right)\left(\frac{R_P}{R_S}\frac{1+sC_P R_S}{1+sC_P R_P}\right) \tag{3.9}$$

Let us now choose $C_C$ so that

$$C_C R_T = C_I R_F \tag{3.10}$$

With the values given above, $C_I R_F \simeq 20$ ms, then if $R_T = 50$ Ω we choose $C_C \simeq 400$ µF. Let us also choose $R_P$ so that

$$C_P R_P = C_F R_F \tag{3.11}$$

which is equal to 1 ms with the vales considered above for $C_F$ and $R_F$. With such choices, equation (3.9) simplifies to

$$T(s) = -g_m R_B \frac{R_P}{R_S} \frac{1 + sC_P R_S}{1 + sC_C R_B} \tag{3.12}$$

The loop gain shows a pole at very low frequency and a zero at higher frequency. Above the frequency of the zero, for $s \gg C_P R_S$, equation (3.12) becomes

$$T(s) \simeq -g_m \frac{C_P R_P}{C_C} \simeq g_m R_T \frac{C_F}{C_I} \tag{3.13}$$

where the conditions (3.10) and (3.11) were used. Since $g_m R_T \simeq 1$ and $C_I \gg C_F$, the loop gain is less than one above the frequency of the zero. The phase shift is then always less than 90° in the frequency range where $|T| > 1$, and stability is assured.

As well known from feedback theory, the open loop signal $V_{OL}$ expressed by equation (3.5) is modified by the presence of the feedback loop according to the relation

$$V_O(s) = \frac{V_{OL}(s)}{1 - T(s)} \tag{3.14}$$

From equation (3.5) approximated for $R_B \gg R_T$, and equation (3.12) which is already an approximation for $R_B \gg R_T$ and $R_P \gg R_S$, we obtain

$$V_O(s) = -G_S \frac{Q}{sC_I} g_m R_T \left( \frac{sC_C R_B}{1 + sC_C R_B} \right) \frac{1}{1 + g_m R_B \frac{R_P}{R_S} \frac{1+sR_S C_P}{1+sC_C R_B}} \tag{3.15}$$

which can be written as

$$V_O(s) = -G_S \frac{Q}{sC_I} g_m R_T \frac{sC_C R_B}{1 + sC_C R_B + g_m R_B \frac{R_P}{R_S} (1 + sC_P R_S)} \tag{3.16}$$

Since $g_m R_T \simeq 1$, and $R_B \gg R_T$, $R_P \gg R_S$, then clearly $g_m R_B R_P / R_S \gg 1$ and equation (3.16) becomes

$$V_O(s) = -G_S \frac{Q}{sC_I} g_m R_T \frac{sC_C R_B}{sC_C R_B + g_m R_B \frac{R_P}{R_S} (1 + sC_P R_S)} \tag{3.17}$$

or, by rearranging the terms,

$$V_O(s) = -G_S \frac{Q}{sC_I} g_m R_T \frac{s \frac{C_C R_S}{g_m R_P}}{1 + s \left( \frac{C_C R_S}{g_m R_P} + C_P R_S \right)} \tag{3.18}$$

We can now define

$$\tau_F = \frac{C_C R_S}{g_m R_P} + C_P R_S = \left( 1 + \frac{C_I}{C_F} \frac{1}{g_m R_T} \right) C_P R_S \tag{3.19}$$

and

$$\tau_F' = \frac{C_C R_S}{g_m R_P} = \frac{C_I}{C_F} \frac{1}{g_m R_T} C_P R_S \qquad (3.20)$$

where the conditions expressed by equations (3.10) and (3.11) were used. Equation (3.18) can then be written as

$$V_O(s) = -G_S \frac{Q}{sC_I} g_m R_T \frac{\tau_F'}{\tau_F} \frac{s\tau_F}{1 + s\tau_F} \qquad (3.21)$$

but since $g_m R_T \simeq 1$ and $C_I \gg C_F$ we have that

$$\frac{\tau_F'}{\tau_F} = \frac{\frac{C_I}{C_F} \frac{1}{g_m R_T}}{1 + \frac{C_I}{C_F} \frac{1}{g_m R_T}} \simeq 1 \qquad (3.22)$$

and equation (3.21) becomes

$$V_O(s) = -G_S \frac{Q}{sC_I} g_m R_T \frac{s\tau_F}{1 + s\tau_F} \qquad (3.23)$$

In time domain this translates to the output signal

$$V_O(t) = -G_S \frac{Q}{C_I} g_m R_T \vartheta(t) e^{-\frac{t}{\tau_F}} \qquad (3.24)$$

that is a voltage step, as expressed by equation (3.7), which discharges with time constant $\tau_F$ given by (3.19). Using the relation (3.11) the fall time constant can also be written as

$$\tau_F = \left(1 + \frac{C_I}{C_F} \frac{1}{g_m R_T}\right) C_P R_S = \left(C_F + \frac{C_I}{g_m R_T}\right) R_F \frac{R_S}{R_P} \qquad (3.25)$$

Assuming $R_S$ to be fixed, the discharge time constant can be controlled by acting on the value of $C_P$, which is conveniently located at room temperature. If $C_P$ is changed then $R_P$ should be adjusted accordingly in order to maintain the relation (3.11) valid. The typical signal shape of a charge amplifier was then recovered, even though the circuit topology was not that of a typical charge amplifier. Most notably, long transmission lines connect the first cold stage and the second stage. The only upper limit to the length of such connecting lines is in the speed of propagation of the feedback signals: for very long lines, if the propagation delay becomes similar to $\tau_F$ the stability of the feedback loop may not be assured. Nevertheless, with a typical value for $\tau_F$ of 100 μs, the propagation delay along the transmission lines is below 1% for connecting lines up to 100 m long, which is more than adequate for the requirements of GERDA, that is 10 m.

The noise of the GeFRO circuit can be evaluated from the equivalent noise charge formula derived in chapter 1, that is

$$\sigma_Q = \sqrt{i_T^2 \beta \tau + A_f C_T^2 \gamma + v_w^2 C_T^2 \frac{\alpha}{\tau}} \tag{3.26}$$

where $i_T$ is the current noise spectrum, considered to be white, $C_T$ is the total input capacitance, $A_f$ is the $1/f$ voltage noise coefficient and $v_w$ is the white voltage noise spectral density. As already discussed in chapter 1 the coefficients $\alpha$, $\beta$ and $\gamma$ in the case of Gaussian shaping take the following values:

$$\begin{aligned} \alpha &= \frac{\sqrt{\pi}}{4} \simeq 0.44 \\ \beta &= \frac{\sqrt{\pi}}{2} \simeq 0.89 \\ \gamma &= \pi \simeq 3.14 \end{aligned} \tag{3.27}$$

In the case of the GeFRO, the current from the BEGe sensor in nominal conditions is expected to be negligible, say of the order of 1 pA, even if in real operating conditions a larger current may arise, and is considered to be acceptable below about 100 pA. The feedback amplifier $A_F$ can also in principle contribute to the parallel noise, but since the bandwidth of the feedback loop is limited the contribution results to be negligible. With a small total current noise $i_T$ the best signal to noise ratio is found at longer shaping times. At 10 µs shaping, for instance, a current of 1 pA contributes to $\sigma_Q$ with about 10 $e^-$ RMS. This contribution rises to 30 $e^-$ RMS and 100 $e^-$ RMS in case the leakage current from the sensor increases to 10 pA or 100 pA respectively. The feedback resistor, assuming a value of 1 GΩ, contributes with about 40 $e^-$ RMS. The $1/f$ voltage noise contribution is dominated by the series noise of the input transistor. At the temperature of liquid Argon (87 K) the $1/f$ noise is expected to be strongly dependent on the quality of a given JFET, since impurities in Silicon can introduce spurious states which result in low frequency noise, as discussed in chapter 1. Several JFETs are being evaluated. The devices found so far have a $1/f$ voltage noise coefficient $A_f$ of about $3 \times 10^5$ nV$^2$ at low temperature. Assuming an input capacitance of 20 pF, this results in a contribution of about 120 $e^-$ RMS to the equivalent noise charge independently of the shaping time chosen. If the gain of the cold stage was much larger than one, as would happen with an input transistor with a transconductance $g_m$ much larger than the reciprocal of the line impedance $R_T$, than the noise from the warm stage could be neglected. In our case $g_m R_T \simeq 1$, so the noise from the termination resistor $R_T$ and the amplifier $A_S$ should be considered, their weight divided by $(g_m R_T)^2$. By a proper choice of the amplifier $A_S$ these contributions can be considered to be purely white, and should be summed to the white noise voltage contribution from the input JFET in the evaluation of the overall white series noise. The overall white voltage noise can be expected to be about 2 $nV/\sqrt{Hz}$. Again considering an input capacitance of 20 pF, at 10 µs shaping time the white voltage noise contribution to the equivalent noise charge results to be about 50 $e^-$ RMS. The overall equivalent noise charge at 10 µs with an input capacitance of 20 pF and a leakage current from the sensor below 100 pA is then expected to be close to 150 $e^-$ RMS. This can be converted to the expected energy resolution by using equation (3.1), obtaining 450 eV RMS, or 1 KeV FWHM. This value is surely remarkable, given the difficult constraints on radiopurity, circuit bandwidth and temperature of operation dictated by the GERDA environment and requirements. Since this value is dominated by series noise it is directly proportional to the input capacitance, and care should then be put in the mechanical design of the sensor holder to avoid to add unnecessary parasitic capacitance at the input. In principle, a reduction of the input capacitance up to 50% with respect to the present values could be attained by minimizing the parasitics. With a total input capacitance of 10 pF, for instance, the equivalent noise charge is expected to drop to less than 100 $e^-$ RMS, or 0.7 KeV FWHM.

## 3.4 Performance of the GeFRO

Figure 3.6 shows the signals at the output of the GeFRO connected to a BEGe radiation sensor. On the time scale where the charge collection can be considered instantaneous, on the left side of the figure, the signals closely follow equation (3.24). The signals at the output of a Ortec 672 Gaussian shaper with pole zero compensation are also shown. On the right site of the figure the rise time of the signals can be apppreciated. The response to the test signal shows a leading edge of about 30 ns limited by the bandwidth of the pulser. The figure also shows the response to single-site and multi-site events in the BEGe sensor at the same energy. Although the shaped signals are identical, the difference in the leading edge of the unshaped signals between single-site and multi-site events can be clearly appreciated, allowing pulse shape discrimination algorithms to be applied to separate the events of interest from background in GERDA.

Several spectra were acquired by facing radioactive sources to the BEGe sensor. The sources used were $^{22}$Na, which emits 511 KeV and 1275 KeV gamma lines, of which only the second is strictly monoenergetic, and $^{228}$Th, which emits several gamma lines, the most prominent being at 583 KeV and 2615 KeV due to $^{208}$Tl in its decay chain. Another peak frequently observed in the test measurements carried out with the BEGe sensor in Milano Bicocca is the 1461 KeV peak of $^{40}$K due to natural radioactivity, since the setup in Milano Bicocca is not optimized in this regard. Figure 3.7 shows a $^{228}$Th and $^{22}$Na spectrum obtained with a BEGe radiation sensor readout with the GeFRO prototype in Milano Bicocca. The spectrum is taken over an hour

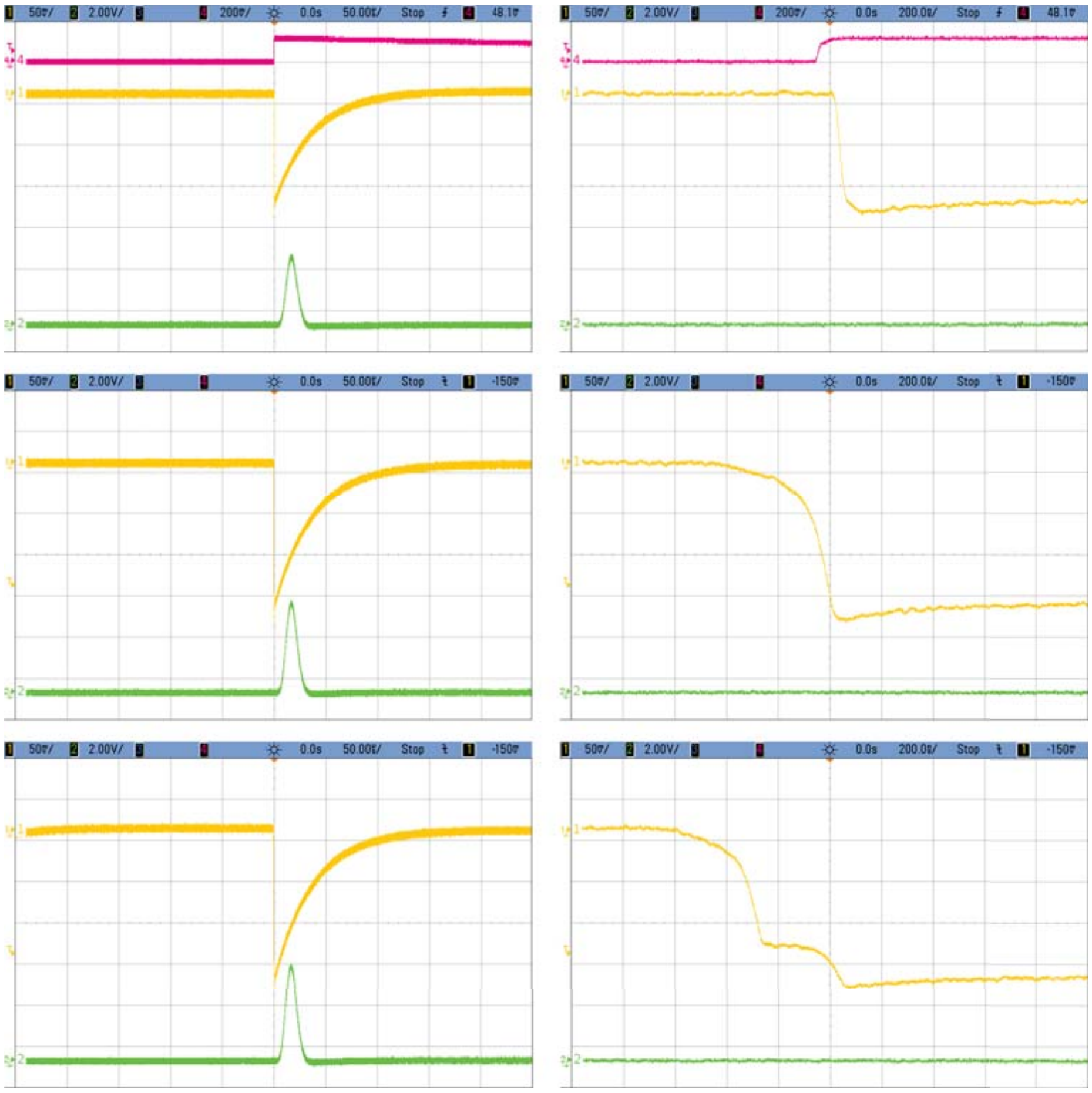

Figure 3.6: Signals at the output of the GeFRO circuit connected to a BEGe radiation sensor. From top to bottom, the oscilloscope shows the response of the circuit to a test pulse, a candidate single-site event and a candidate multi-site event. The output of the GeFRO is shown in yellow, while the output of the shaped signal after a Ortec 672 Gaussian shaper is shown in green. The test signal is shown in purple. On the left, the horizontal scale is 50 µs/div, on the right it is 200 ns/div.

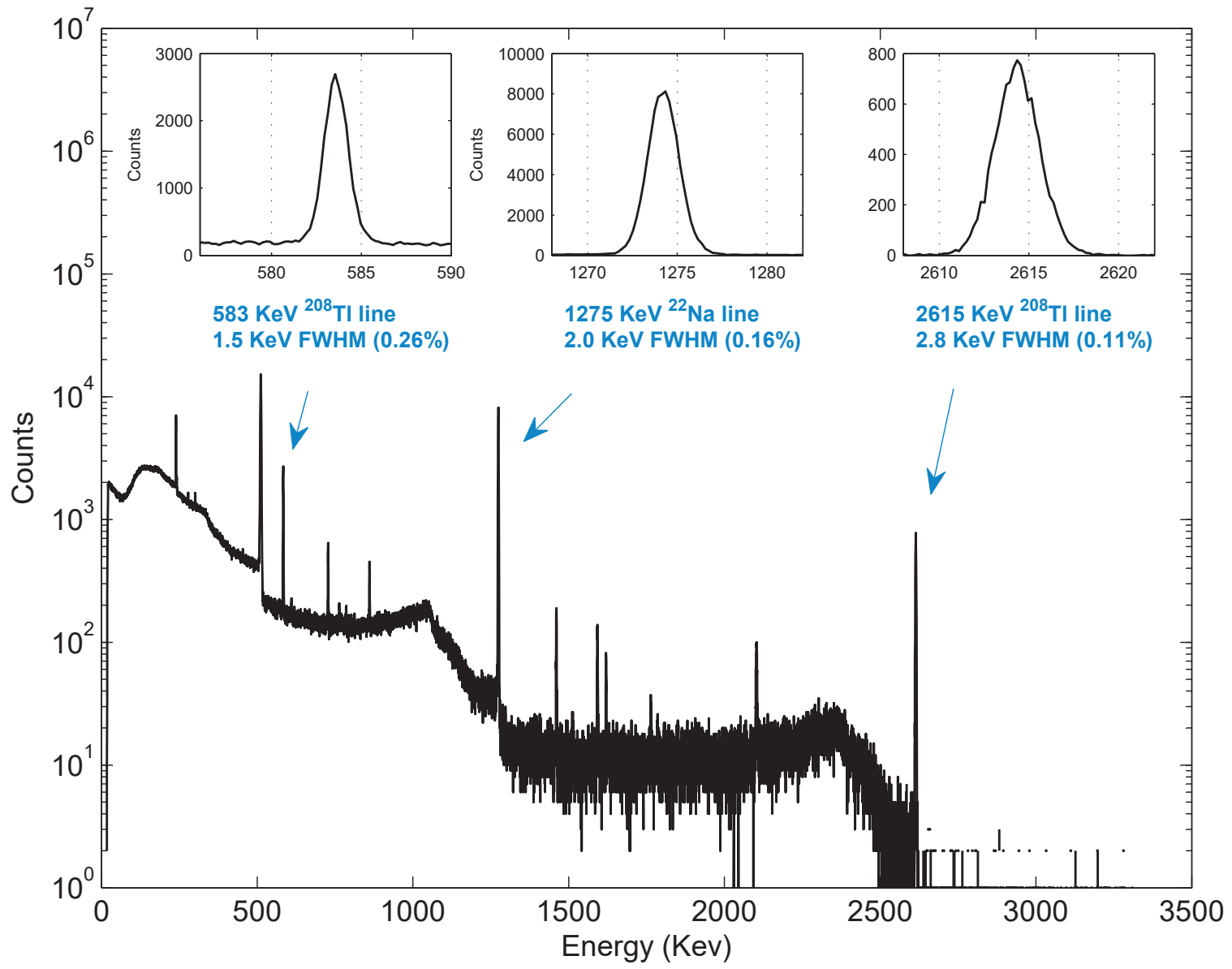

Figure 3.7: Spectrum of $^{228}$Th and $^{22}$Na sources obtained with a BEGe radiation sensor readout with the GeFRO prototype in Milano Bicocca.

---

of measurement. The spectrum was acquired by filtering the GeFRO signals with a Ortec 672 Gaussian shaper at 10 µs shaping time and acquiring the peak amplitudes of the shaped signals with a Ortec multichannel analyzer. The insets show the peaks used to evaluate the energy resolution, which in this measurement are 1.5 KeV FWHM at the 583 KeV line, 2.0 KeV FWHM at the 1275 KeV line, and 2.8 KeV FWHM at the 2615 KeV line. These values are in agreement with those expected with a Germanium sensor, considering a noise contribution of 1 KeV FWHM.

The $^{228}$Th source is used in order to test the pulse shape discrimination capability of the setup, since the double escape peak corresponding to the 2615 KeV line is located at 1593 KeV and is dominated by single-site events, while it also emits gamma radiation at the similar energy of 1620 KeV, due to the presence of $^{212}$Bi, which is instead dominated by multi-site Compton scattering events. The maximum total event rate achievable with our $^{228}$Th source was about 300 cps (counts per second). The $^{22}$Na source was by far more active, and was used in combination with the other to obtain rates up to 1300 cps. Even if GERDA is a rare event search experiment with negligible data rates in operating conditions, radioactive sources are periodically inserted in order to calibrate the setup, and can give rates up to the order of 1000 cps. High resolution and stability in the position of the peaks over

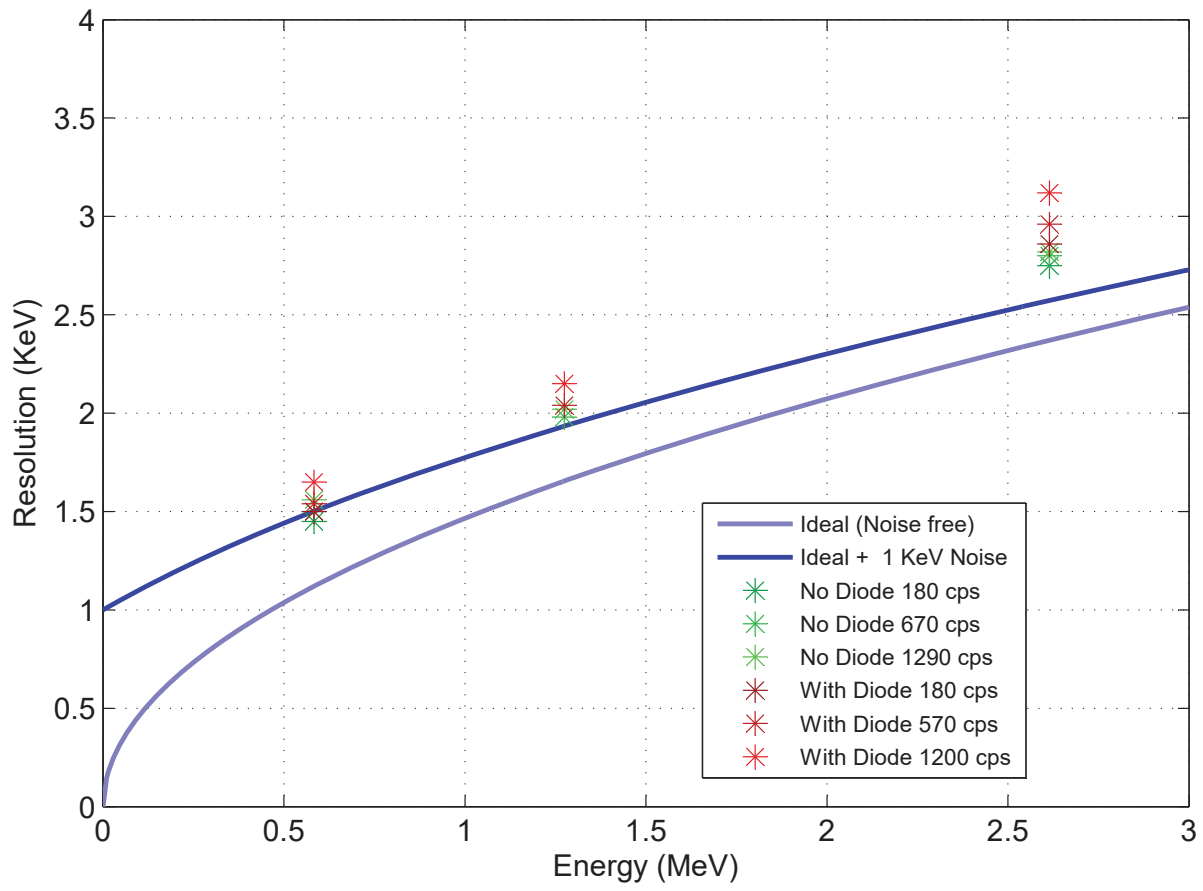

Figure 3.8: Resolution of the GeFRO circuit connected to a BEGe radiation sensor. The light blue curve shows the resolution of an ideal Ge sensor, while the dark blue curve shows the expected resolution with 1 KeV noise from the electronics. The six sets of data, taken at different event rates with and without the protection diode, are shown in the figure.

---

time and at different rates is then required to guarantee the consistency of the measurements. Figure 3.8 shows the resolution versus energy for an ideal Germanium radiation sensor, as expressed by equation (3.4), and the expected resolution if the sensor is readout with a charge amplifier having a FWHM noise of 1 KeV, corresponding to an equivalent noise charge of 150 $e^-$ RMS. The resolution values, measured at different rates with and without the protection diode in the feedback loop, are also shown in the figure. While the match is good at low energy, a rate-dependent deviation can be observed at the 2615 KeV peak. The expected FWHM resolution at the 2615 KeV peak is close to 2.6 KeV. Without the protection diode the measured resolution ranges from 2.7 KeV at low rates to 2.8 KeV at high rates. With the diode the measured resolution ranges from 2.9 KeV to 3.1 KeV. The additional contribution due to the presence of the protection diode is likely to be related with the pile-up of the unshaped signals, causing small fluctuation in the capacitance of the diode, and is being investigated. In these measurements the protection diode was a BAT17 Schottky diode, but a lower dependence on rate is expected to be found with a Silicon diode, having a larger built-in voltage and thus a lower dependence of capacitance with applied voltage. In any case, for all these sets of measurements, the extrapolated energy resolution at the neutrinoless double beta decay energy of 2038 KeV is close to 2.5 KeV FHWM or even better, which is adequate for GERDA phase II.

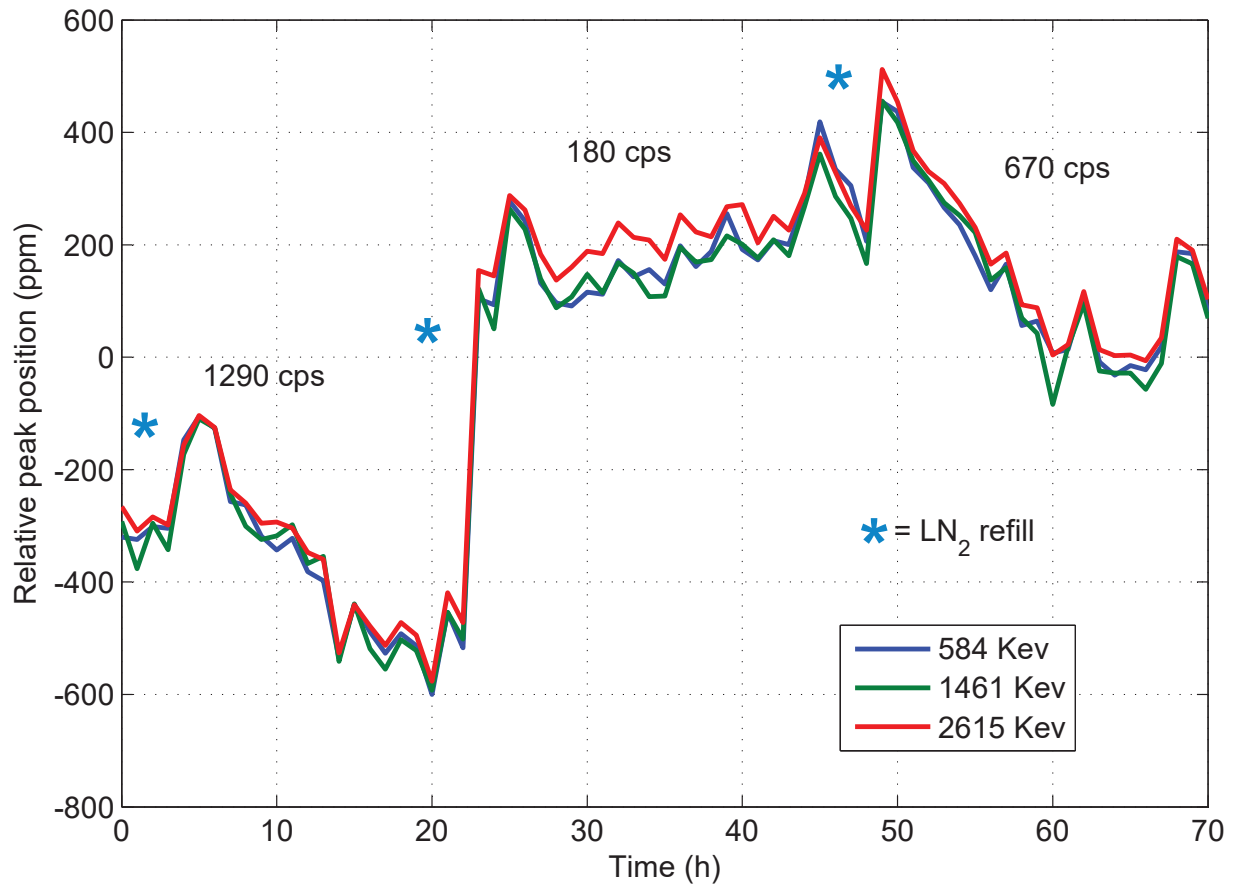

Figure 3.9: Relative peak position versus time over 70 hours of measurement, showing a typical gain drift below 100 ppm/h. The event rate was also changed during the measurement, showing a minimal variation in the position of the peaks.

---

Figure 3.9 shows the relative position of three peaks (583 KeV, 1461 KeV and 2615 KeV) versus time, in a set of measurements where also the rate was changed. The BEGe sensor was held in liquid Nitrogen which was refilled every about 24 hours. On each refill the position of the radioactive sources was changed, varying the overall event rate as shown in the plot. As can be seen, excluding the refill regions the drift is typically below 100 ppm/h. A small variation in the position of the peaks at different rates can also be seen, at the level of 500 ppm for a range of rates from zero up to 1300 cps.

At a fixed event rate, the gain drift was found to depend on three distinct contributions. During the first hours after the high voltage is raised, the capacitance at the input of the GeFRO relaxes towards a constant value. The gain drift in the first hours is dominated by this contribution, which can give drifts as high as several hundred ppm/h. After a few hours this startup transient is over. The gain drift is then due to the change in resistivity of the output cable, due to the evaporation of the cooling medium (liquid Argon or Nitrogen), or to changes in the temperature or humidity of the room, which affects the second stage, at least in the case of the present prototype.

The resistance of the cables to be used in GERDA is of the order of $1-2\ \Omega/\mathrm{m}$ at room temperature. At cold their resistance can be considered to be negligible. The total resistance $R_C$ of the output cable is then proportional to the length L of cable which is at room temperature, and a change in L directly affects a change in $R_C$. In other words,

$$\frac{\Delta R_C}{R_C} = \frac{\Delta L}{L} \tag{3.28}$$

In the setup used in Milano Bicocca, for instance, the cryogenic liquid evaporates at about 1 cm/h, and the length of the cable at room temperature is about 1 m. This gives a relative variation in $R_C$ of about 1%/h, or $10^4$ ppm/h. This is somewhat a worst case evaluation, since the presence of cold vapours above the surface of the cryogenic liquid makes the temperature changes in the cables less significant. If the output cable was driven with an ideal voltage source, and terminated at the far end on $R_T$, the gain would be proportional to $R_T/(R_T + R_C)$, giving a relative gain drift of

$$\frac{\Delta G}{G} = -\frac{\Delta R_C}{R_T + R_C} \simeq -\frac{R_C}{R_T}\frac{\Delta R_C}{R_C} \tag{3.29}$$

Assuming $L = 1$ m, $R_C$ is $1-2\ \Omega$. If $R_T = 50\ \Omega$, then $R_C/R_T$ is about $2-4\%$, and the overall gain drift would then be of the order of $200-400$ ppm/h. On the other hand, if the cable was driven with an ideal current source then the gain would not depend on $R_C$, as was calculated in section 3.3. In this case a change in $R_C$ would not cause any gain drift. The situation of the real GeFRO is somewhat in between since the JFET is not an ideal current source. By considering the effect of the series resistance of the cable $R_C$, and of the gate-drain capacitance $C_{GD}$ and drain-source resistance $R_{DS}$ of the JFET, the gain which was expressed by equation (3.5) becomes

$$\begin{aligned} V_{OL}(s) = -G_S \frac{Q}{sC_I} g_m R_T \left( \frac{sC_C R_B}{1 + sC_C(R_B + R_T)} \right) \times \\ \times \left( \frac{1 - \frac{sC_{GD}}{g_m}}{\eta + \frac{C_{GD}}{C_I}\left(\eta + g_m(R_T + R_C)\right) + sC_{GD}(R_T + R_C)} \right) \end{aligned} \tag{3.30}$$

where

$$\eta = \frac{R_C + R_T + R_{DS}}{R_{DS}} \tag{3.31}$$

Since $R_{DS}$ is expected to be at least a few k$\Omega$, $\eta$ can be approximated to be equal to one. The effect of $R_{DS}$ is then negligible, and the output impedance of the JFET is dominated by $C_{GD}$. By putting $\eta = 1$ into equation (3.30) and differentiating it with respect to $R_C$ we obtain

$$\frac{\Delta V_{OL}}{V_{OL}} = -\frac{\frac{C_{GD}}{C_I} g_m R_C}{1 + \frac{C_{GD}}{C_I}\left(1 + g_m(R_T + R_C)\right) + s\frac{C_{GD}}{C_I}(R_T + R_C)} \frac{\Delta R_C}{R_C} \tag{3.32}$$

This equation gives the relative gain drift of the circuit. The slow feedback was neglected, since as was shown in the previous section it only affects the baseline recovery and does not contribute to the gain term. We can neglect the frequency-dependent term in equation (3.32), obtaining

$$\frac{\Delta \mathrm{G}}{\mathrm{G}} \simeq -\frac{\mathrm{C_{GD} g_m R_C}}{\mathrm{C_I} + \mathrm{C_{GD}}\left(1 + \mathrm{g_m R_T}\right)} \frac{\Delta \mathrm{R_C}}{\mathrm{R_C}} \tag{3.33}$$

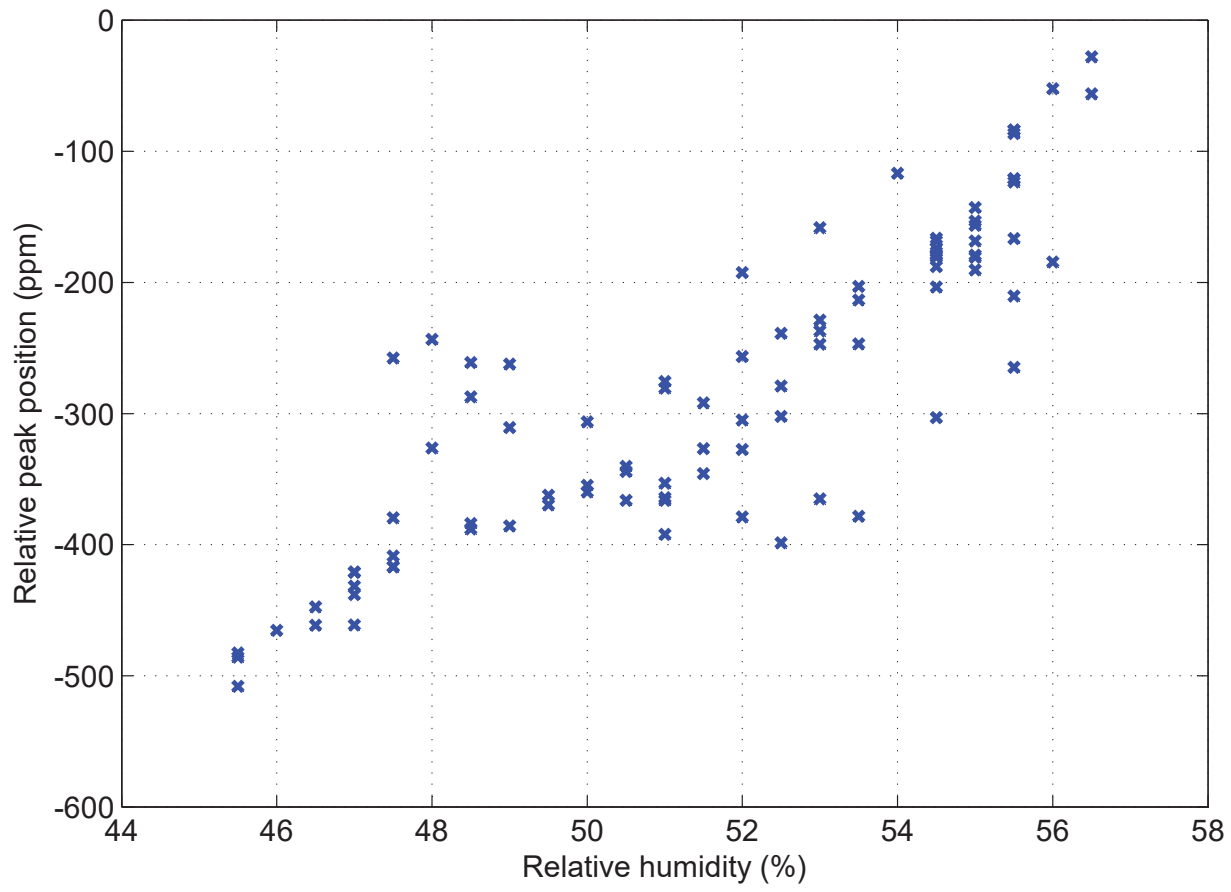

Figure 3.10: Relative peak position versus relative humidity near the second stage over several days of measurement. A correlation can be clearly seen.

Since $\mathrm{g_m R_T} \simeq 1$, and since $\mathrm{C_{GD}}$ is expected to be a few pF at the most, while $\mathrm{C_I}$ is of the order of 10 pF, the expression can be simplified to

$$\frac{\Delta \mathrm{G}}{\mathrm{G}} \simeq -\mathrm{g_m R_T} \frac{\mathrm{C_{GD}}}{\mathrm{C_I}} \frac{\mathrm{R_C}}{\mathrm{R_T}} \frac{\Delta \mathrm{R_C}}{\mathrm{R_C}} \tag{3.34}$$

The factor $\mathrm{g_m R_T}$ is close to one and can be neglected. With respect to the case of the ideal voltage source, equation (3.29), the relative gain drift due to the change in resistivity of the cables is then further reduced by the ratio $\mathrm{C_{GD}/C_I}$. Assuming $\mathrm{C_{GD}/C_I} \simeq 1/5$, the result is of the order of 40−80 ppm/h, consistent with the results shown in figure 3.9.

The results in figure 3.9 are affected also by a contribution coming from the second stage, which could be correlated with a change in the relative humidity of the room. Figure 3.10 shows the correlation between gain variation and relative humidity measured at the second stage, for another run where only the GeFRO circuit (without the BEGe) was operated for several days. A correlation can be clearly seen. The overall drift is in any case below 100 ppm/h.

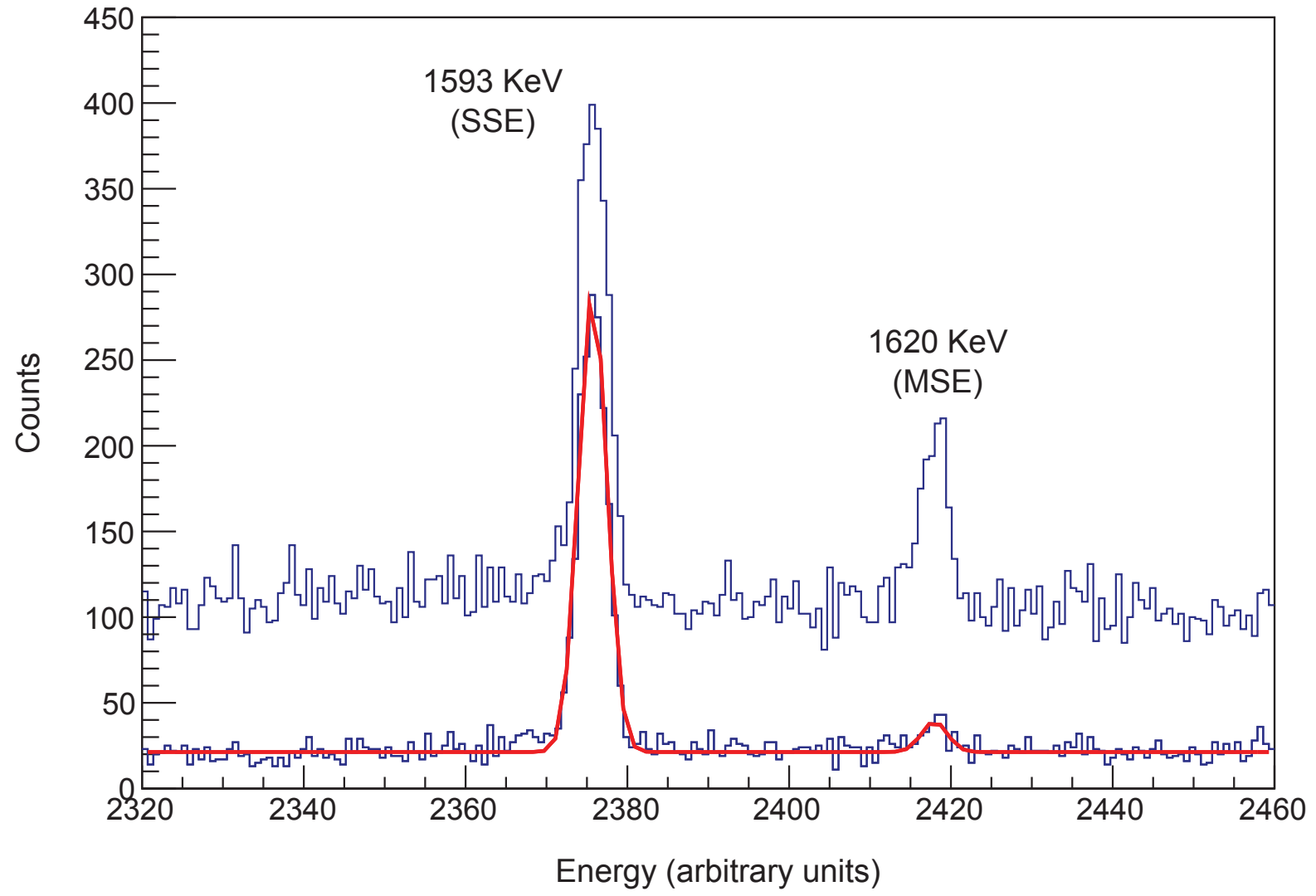

Figure 3.11: Effectiveness of the pulse-shape discrimination in cutting multi-site events (1620 KeV line) while preserving single-site events (1593 KeV line). This plot was produced by processing the data from the GeFRO with the algorithm developed by the GERDA collaboration.

---

It should be noted that the level of liquid Argon in GERDA is kept constant by a dedicated system. No change in the resistance of the cables is expected, and the corresponding gain drift will be negligible. Moreover, the effect of the room humidity and temperature on the gain of the second stage could be to some extent compensated by dedicated circuitry, if necessary. The GeFRO is then expected to provide stable operation in GERDA over the long data taking time required.

## 3.5 Future work with the GeFRO

As illustrated in the previous section, the performance of the GeFRO coupled to a BEGe radiation sensor is well understood and satisfies the requirements of GERDA phase II. The circuit presented is AC coupled, but with a careful compensation of the time constants of the feedback loop the transfer function of an ideal, DC coupled charge amplifier was recovered. An alternative design solution which allows the cold stage to be DC coupled to the warm stage is under study.

All the measurements on the test bench and with a BEGe operated in vacuum in Milano Bicocca indicate good resolution and stable operation. The

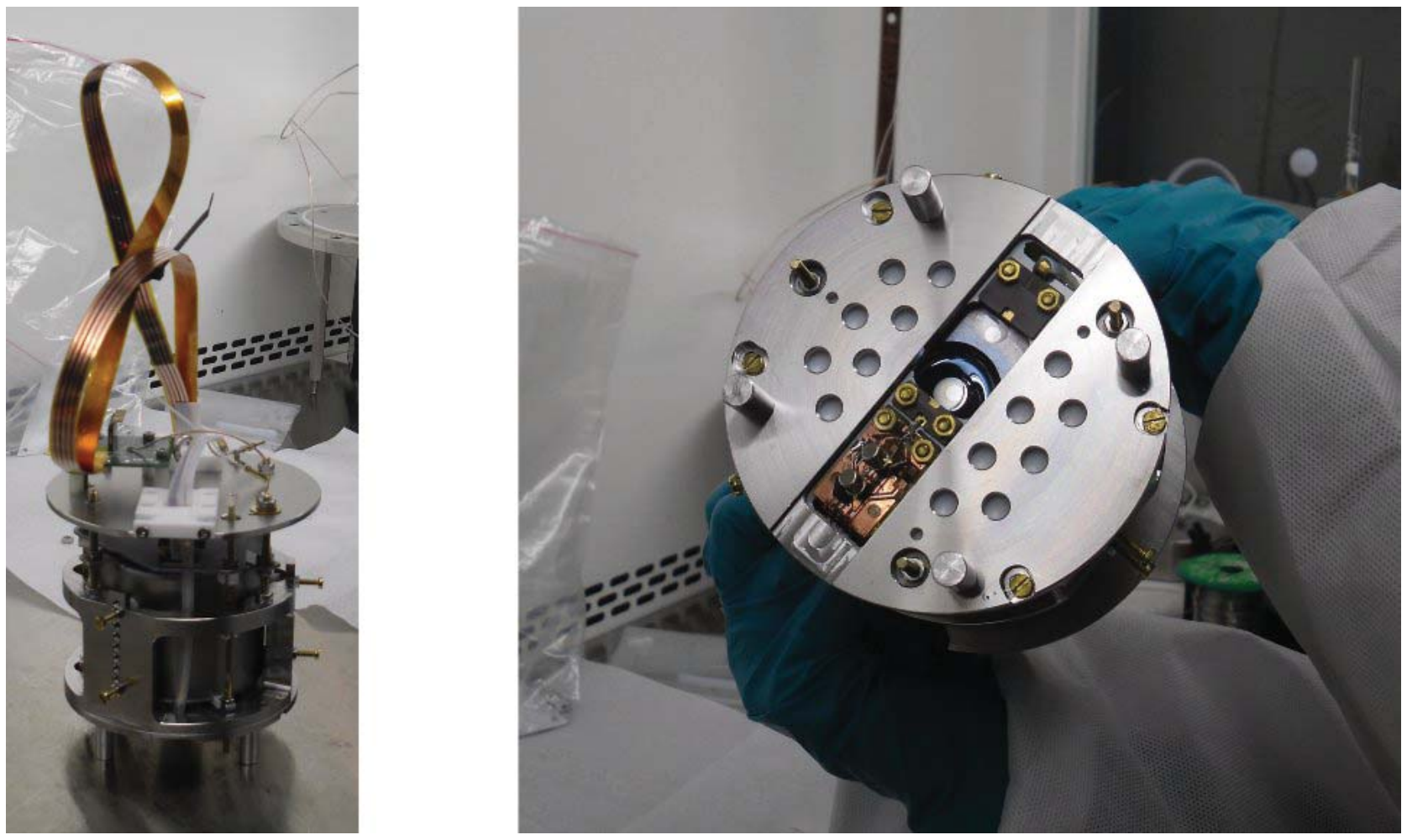

Figure 3.12: On the left side, a prototype holder for GERDA phase II, housing two BEGe sensors and the cold stages of their GeFRO channels. On the right side the contact of the bottom BEGe and the cold stage of the corresponding GeFRO circuit can be seen.

---

resolution at the $^{76}$Ge neutrinoless double beta decay line is better than 0.13%, and stability of operation at constant event rate is maintained within less than 100 ppm/h, including a major contribution related with the variation of the temperature of the output cables, which is expected to be constant in GERDA. The wide bandwidth of the circuit, close to 20 MHz, allows the discrimination of single-site and multi-site events to a level which is adequate for the GERDA experiment. For instance, figure 3.11 shows the performance of the A/E cut algorithm developed by the GERDA collaboration to separate single-site events from multi-site events, applied to the data obtained with the GeFRO coupled to a BEGe sensor in Milano. Although the algorithm is not yet optimized for use with the GeFRO, it shows a survival probability below 15% for multi-site events when the acceptance on single-site events is set to 90%, very close to the requirements of GERDA phase II. Further optimization of the algorithm to match it to the GeFRO will surely prove beneficial. All the components of the cold stage were found in die, and preliminary screening measurements indicate a very low radioactivity background.

Tests should now be carried out in LNGS with BEGe sensors operated bare in liquid Argon. The main concern in this configuration regards gain stability, since the gain of the GeFRO is inversely proportional to the total input capacitance. Figure 3.12 shows the prototype of the low background holder for the BEGe sensors which was designed by the collaboration, coupled to two GeFRO channels. The GeFRO cold stages are mounted directly on the Cuflon tapes which also constitute the first part of the connecting links to the second stage. In the present prototype the components of the cold stage are the packaged versions, for practical reasons. The same components are available in bare die for direct bonding on the Cuflon tapes. Extensive tests on this two-channel setup are needed, and will be performed in the near future.

# 4 Low noise amplifiers for bolometric sensors

## 4.1 The bolometric technique for neutrino physics

Bolometers are thermal sensors, usually made of crystals held at very low temperature. The events of interest release energy in the bolometers, inducing temperature variations which are detected by proper sensing elements which convert the temperature changes into electric signals. Bolometers are used in particle physics experiments since many years, one of their main fields of application being neutrino physics, and in particular the search for the neutrinoless double beta decay and the direct kinematic measurement of neutrino mass.

Among the experiments for double beta decay search which make use of the bolometric technique are CUORE and LUCIFER, both based at Laboratori Nazionali del Gran Sasso (LNGS). In such experiments the bolometers are realized with crystals containing the double beta decay emitting isotopes, so that the decay source coincides with the sensor and the detection efficiency is maximized. Since double beta decay searches require large mass, the bolometers used are large and are usually referred to as macrobolometers. In the case of CUORE the bolometers are crystals of tellurium dioxide ($TeO_2$) of $5 \times 5 \times 5$ cm$^3$ size. Their temperature is readout with semiconductor thermistors glued onto the crystals. CUORE will be an array of 988 of such crystals, and is currently under construction at Laboratori Nazionali del Gran Sasso (LNGS). LUCIFER, currently still in the R&D phase, will employ scintillating crystals to discriminate double beta decays from spurious alpha events due to background, gaining a higher degree of background rejection and thus a higher sensitivity for a given amount of mass. About 50 channels are foreseen, each with a dual readout, since the scintillation light signals need to be acquired together with the thermal signals in order to provide the necessary discrimination capability for alpha events. Among the experiments which instead make use of bolometric sensors for the direct measurement of the neutrino mass is the MARE experiment. The project, currently still in the R&D phase, employs microbolometers of silver perrhenate ($AgReO_4$) glued onto arrays of

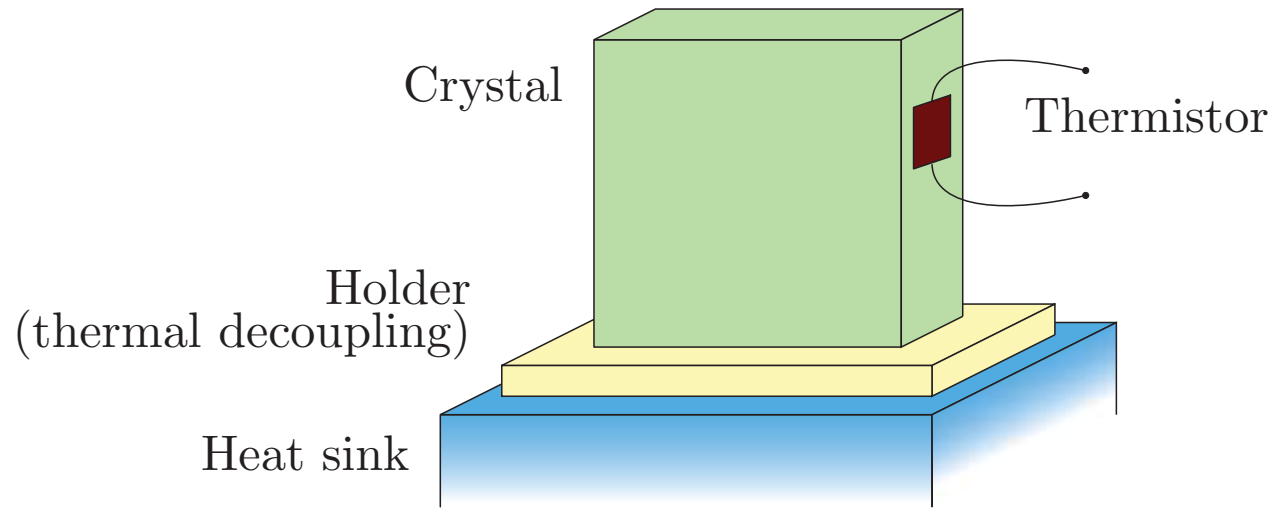

Figure 4.1: Typical setup of a bolometric sensor.

Silicon thermistors to obtain a precise measurement of the single beta decay spectrum of $^{187}$Re, from which a kinetic measurement of neutrino mass can be extracted. The possibility of exploiting the electron capture spectrum of $^{163}$Ho is also being considered. A more detailed description of the experiments just mentioned will be given in the following sections.

The typical setup of a bolometric sensor is sketched in figure 4.1. For a bolometer with a heat capacity $\mathcal{C}$ the temporary temperature variation induced by an energy release E in the bolometer is

$$\Delta \mathrm{T} = \frac{\mathrm{E}}{\mathcal{C}} \tag{4.1}$$

A small heat capacity is required to obtain high sensitivity to small energy events, such as those corresponding to individual particle interactions and nuclear decays. Double beta decay experiments require a large mass of the candidate isotope. The means to obtain a small heat capacity with a substantial sensor mass is to cool down the crystals to a very low temperature. According to the Debye model, the heat capacity of dielectric and diamagnetic solids below a given temperature known as the Debye temperature is proportional to $\mathrm{T}^3$ and thus drops to vanishing levels. In the experiments considered here, the bolometers are operated in cryostats cooled by dilution refrigerators, devices which can reach the lowest known temperatures, down to the order of 10 mK. At such temperature the heat capacity of the CUORE bolometers for instance is of the order of $10^{-9}$ J/K, low enough to enable the detection of events with energy down to a few KeV.

The intrinsic resolution of bolometric sensors is related to the fundamental fluctuations of vibrational modes in the crystals, called phonons, and can be expressed as

$$\sigma_\mathrm{E} = \sqrt{\mathrm{k_B T^2} \mathcal{C}} \tag{4.2}$$

where $\mathrm{k_B}$ is the Boltzmann constant. Equation (4.2) can be considered to be the bolometric counterpart of the similar expression which holds for ionization sensors. In fact it can also be written as

$$\sigma_E = \sqrt{\varepsilon E} \tag{4.3}$$

where $\varepsilon = k_B T$ is the average phonon energy and $E = \mathcal{C}T$ is the energy of the event, as follows from (4.1) if the temperature rise due to the event of energy E can be considered to be much larger than the base temperature of the bolometer. Since as already discussed bolometers are held at very low temperature, $\varepsilon$ can be as low as a few tens of μeV, to be compared with the values for $\varepsilon$ of ionization sensors, of the order of the eV. The intrinsic resolution limit of bolometric sensors is then of many orders of magnitude better than the Fano limited resolution of ionization sensors discussed in chapter 3. The actual energy resolution anyway depends also on the quality and operating conditions of the sensors, which are much more challenging with bolometers than with standard ionization sensors. Resolutions down to 5 eV have been obtained with microbolometers detecting X rays of a few KeV [42, 43]. The best resolutions reported by MIBETA, a pilot experiment for MARE, are below 20 eV FWHM at the $^{187}$Re beta decay endpoint energy of 2.47 KeV. The corresponding intrinsic resolution of an ideal Germanium ionization sensor at the same energy by comparison is about 70 eV FWHM. At higher energy, the best resolutions reported by CUORICINO with macrobolometers in data taking conditions are close to 2.5 KeV at the $^{130}$Te double beta decay endpoint at 2528 KeV, which is very close to the best resolution which would be obtained at the same energy with an ideal Germanium ionization sensor readout with a noiseless amplifier. The real measured resolutions are then in the end of the same order of magnitude as those obtained with ionization sensors, even if the superior resolution of the bolometric technique remains evident at lower energies with microbolometers.

The thermistors used in the above mentioned experiments are obtained from doped Germanium or Silicon. The thermistors used in CUORE are neutron transmutation doped (NTD) Germanium resistors, whose value at low temperature goes as

$$R_T(T) = R_0 \exp\left(\sqrt{\frac{T_0}{T}}\right) \tag{4.4}$$

where $R_0$ and $T_0$ are technological parameters. In the case of CUORE, for instance, the NTD Germanium thermistors have $R_0 \simeq 10\ \Omega$ and $T_0 \simeq 3$ K. Their value at 10 mK is of the order of 1 GΩ. Other devices exist in the field, such as transition edge sensors (TES), magnetic microcalorimeters (MMCs), kinetic inductance devices (KIDs) and others. Some of the above can be naturally arranged for multiplexed readout of the array elements, which would allow to scale the existing setups to larger size without a linear increase in the number of readout wires. Such technologies are thus more naturally apt to be employed in future large scale experiments with respect to semiconductor thermistors. They generally require different readout solutions with respect to the case of thermistors, and will not be considered in this thesis.

The quantities that rule the thermal dynamics of a bolometer are $\mathcal{C}$ and $\mathcal{G}$, respectively the heat capacity of the bolometer and its thermal conductance to the heat sink. The heat sink is maintained at constant temperature. As already discussed, an energy release in the bolometer causes an increase in temperature which is inversely proportional to $\mathcal{C}$. The bolometer then relaxes to base temperature through the thermal conductance $\mathcal{G}$, which for macrobolometers is tipically of the order of $10^{-9}$ W/K. To a first order approximation, which neglects the power dissipated by the bias current in the thermistor, the time constant of the thermal relaxation is given by $\mathcal{C}/\mathcal{G}$, and is tipically of the order of 1 s for macrobolometers. In the following section the thermal dynamics of bolometers will be considered in deeper detail.

## 4.2 The electro-thermal dynamics of bolometers

If a bolometer were held at constant temperature with a good thermal contact to the heat sink, the power dissipated on the thermistor by its bias current would be negligible, and its source impedance would simply be given by the thermistor resistance $R_T$. But the temperature of the bolometer is instead purposely allowed to change by thermally decoupling it from the heat sink, since the thermal conductance $\mathcal{G}$ is chosen to be very small. The power dissipated on the bolometer by the bias current $I_B$ cannot be neglected then, and the impedance of the bolometer results in general to be different from $R_T$ [44–48]. In the following calculations we will consider $\mathcal{G}$ to be constant. In a more realistic case, its dependence on temperature should be considered, but we will neglect it here.

As introduced in chapter 1, the dynamic impedance of the bolometer can be modeled as the parallel combination of the thermistor impedance $R_T$ and an additional impedance, which is given by the series combination of a resistance $R_P$ and an inductance $L_P$. The model is shown in the inset of figure 4.2. The values of $R_P$ and $L_P$ are related to the thermal dynamics of the bolometer. At very low frequency, the inductance can be neglected, and the impedance of the bolometer is given by the parallel combination of $R_T$ and $R_P$. It will be shown that the result can be written as

$$R_S = R_T || R_P = R_T \frac{\mathcal{G} - I_B^2 \alpha}{\mathcal{G} + I_B^2 \alpha} \tag{4.5}$$

where $I_B$ is the bias current in the thermistor and

$$\alpha = -\frac{dR_T(T)}{dT} \tag{4.6}$$

is the thermal coefficient of the thermistor with the sign changed. The value of $\alpha$ depends on the thermistor only, and can be obtained by differentiating equation (4.4). The bolometers considered here are the so-called "negative" thermistors, where the derivative of equation (4.4) is negative, and thus $\alpha > 0$. If $R_T$ follows equation (4.4), then

$$\alpha = \frac{R_T(T)}{\sqrt{4TT_0}} \tag{4.7}$$

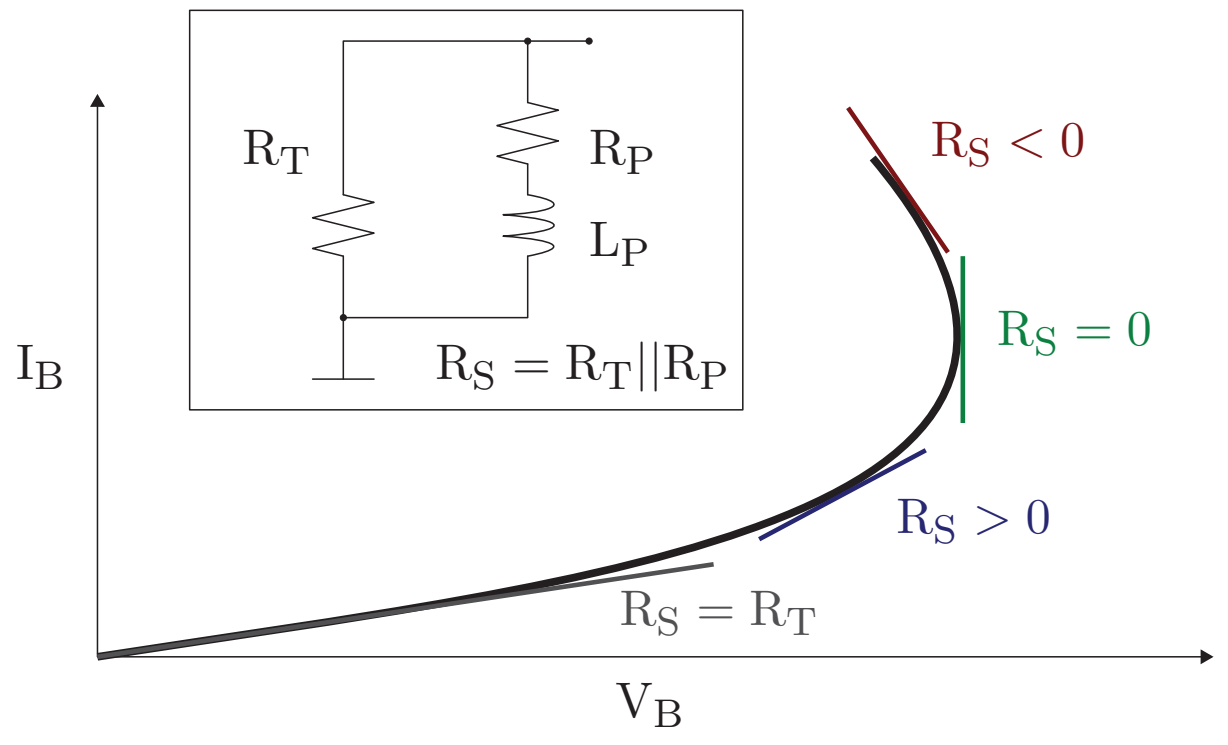

Figure 4.2: The equivalent model of a bolometric sensor, and its typical characteristic curve (or load curve).

For the thermistors of CUORE, $T_0 \simeq 3$ K and $R_T \simeq 1$ G$\Omega$ at 10 mK, thus $\alpha \simeq 1$ M$\Omega$/mK.

The resistance $R_S$ given by equation (4.5) is the dynamic impedance of the bolometer at DC. As can be seen from equation (4.5), if the thermal conductance $\mathcal{G}$ is large, which means that its temperature is forced to be constant and equal to that of the heat sink, then the impedance of the bolometer becomes equal to $R_T$. If instead $\mathcal{G}$ is small, then $R_S$ depends on the bias current $I_B$. In this case $R_S$ is equal to $R_T$ only for small values of $I_B$, where the power dissipated by the bias current is negligible. Otherwise it becomes smaller, reflecting the heating effect of the dissipated power.

Equation (4.5) was obtained under the assumption that the bolometer is biased with a constant current. This is generally implemented with a voltage source in series with a large value load resistor $R_L$, much larger than the impedance of the bolometer. In case the connecting links are long, their parasitic capacitance should be considered in parallel with $R_L$, giving a load impedance $Z_L$. Equation (4.5) can be modified to include the effect of a finite $Z_L$. Anyway, as will be shown, a small $Z_L$ can lead to instability in the working point. Bolometers are then usually operated with large values for $Z_L$, and equation (4.5) holds. This requires the load resistor $R_L$ to have a very large value and the parasitic capacitance across the bolometer to be kept under control.

By varying the bias current $I_B$, the resulting voltage $V_B$ on the bolometer can be measured and plotted in the current versus voltage plane. This is the "load curve" of a bolometer, shown in figure 4.2. For a given bolometer, its DC dynamic impedance $R_S$ at various values of $I_B$ can be obtained as the derivative of the load curve. As expressed by equation (4.5) and shown in figure 4.2, $R_S$ can assume positive, negative and zero values, depending on the values of $\mathcal{G}$, $I_B$ and $\alpha$. The point where $R_S = 0$ is known as the "inversion point".

As already discussed, and implicitly contained in equation (4.5), the temperature of the bolometer changes by changing $I_B$ due to the power dissipation in the thermistor. Changes in temperature cause changes in resistance, which in turn affects power dissipation. This mechanism is known as "electrothermal feedback", and can be positive or negative, leading to instability or stability in the working point. Working points where $R_S$ is positive are thermally stable. An increase of temperature in the bolometer causes its resistance to decrease, reducing the power dissipated by the bias current, and then lowering its self-heating. On the contrary, instability occurs when $R_S$ is negative. Electrothermal feedback makes it convenient to bias the bolometers with a constant current and to readout the voltage signals with voltage amplifiers, instead of doing the opposite.

At high frequency, the inductance $L_P$ hides $R_P$, and the impedance of the bolometer is resistive and equal to $R_T$, that is the electrical resistance of the thermistor at fixed temperature. This reflects the fact that for a high frequency excitation the heat capacity behaves as a thermal short circuit to the heat sink, dominating over the thermal conductance. In between the two regimes, the inductive contribution is noticeable. Its value is related to the thermal dynamics of the bolometer and is given by

$$L_P = R_T \frac{\mathcal{C}}{2 I_B^2 \alpha} \tag{4.8}$$

The lower the heat capacity $\mathcal{C}$ the higher the frequency at which the effect of $L_P$ becomes noticeable. From equation (4.5) the value of $R_P$ can be extracted, and is given by

$$R_P = R_T \frac{\mathcal{G} - I_B^2 \alpha}{2 I_B^2 \alpha} \tag{4.9}$$

Clearly $R_P$ is not a physical resistive element, since its value becomes negative for large values of $I_B$. The overall dynamic impedance of the bolometer can then be expressed as

$$Z_S(s) = R_T || (R_P + sL_P) = R_T \frac{s\mathcal{C} + \mathcal{G} - I_B^2 \alpha}{s\mathcal{C} + \mathcal{G} + I_B^2 \alpha} \tag{4.10}$$

which reduces to equation (4.5) at DC, where $s = 0$. For slow thermal signals, as happens for most of the particle physics experiments described in this thesis, the bolometers can usually be considered to be operated at DC. This is particularly true for macrobolometers. Their dynamic impedance can then to a first order approximation be considered as purely resistive and equal to $R_S$.

Equation (4.10) can be derived from the following evaluations. By denoting the voltage across the thermistor as $V_B = R_T I_B$, the dynamic impedance of the bolometer at temperature T can be calculated as

$$Z_S = \frac{dV_B}{dI_B} = \frac{d\left(R_T I_B\right)}{dI_B} = \frac{dR_T}{dT}\frac{dT}{dI_B} I_B + R_T = -\alpha\frac{dT}{dI_B} I_B + R_T \tag{4.11}$$

The term $dT/dI_B$ can be evaluated from the thermal balance equation. In time domain the thermal balance of the bolometer is given by

$$\mathcal{C}\frac{dT}{dt} = V_B I_B + P(t) - \mathcal{G}\left(T - T_C\right) \tag{4.12}$$

The terms on the right represent respectively the power dissipated by the bias current $I_B$, the power $P(t)$ dissipated by a particle event, and the thermal link to the heat sink, which is at constant temperature $T_C$. By considering the Laplace transform of equation (4.12) and rearranging the terms one obtains

$$\left(s\mathcal{C} + \mathcal{G}\right)\left(T - T_C\right) = V_B I_B + P(s) \tag{4.13}$$

Since we are now considering the equilibrium conditions we can calculate the small temperature variation dT due to a change of bias current $dI_B$. By differentiating with respect to $dI_B$, equation (4.13) becomes

$$\left(s\mathcal{C} + \mathcal{G}\right)\frac{dT}{dI_B} = \frac{dV_B}{dI_B} I_B + V_B = Z_S I_B + V_B \tag{4.14}$$

By substituting the value for $dT/dI_B$ from (4.14) into (4.11) we obtain equation (4.10), demonstrating its validity.

Let us assume the inductance $L_P$ to be negligible, so that we can safely approximate $Z_S$ with $R_S$ at low and moderate frequency. From equation (4.8), this happens if the heat capacity $\mathcal{C}$ is small and $\alpha$ is large, the ideal working conditions for a bolometer. Let us now consider the effect of an external power P, that is a particle event, on the dynamic impedance of the bolometer. Differentiating equation (4.13) with respect to temperature gives the relation between P and the corresponding temperature variation $\Delta T$, that is

$$\left(s\mathcal{C} + \mathcal{G}\right)\Delta T = \frac{dR_T}{dT} I_B^2 \Delta T + P(s) \tag{4.15}$$

which gives

$$\Delta T = \frac{1}{s\mathcal{C} + \mathcal{G} + \alpha I_B^2} P(s) \tag{4.16}$$

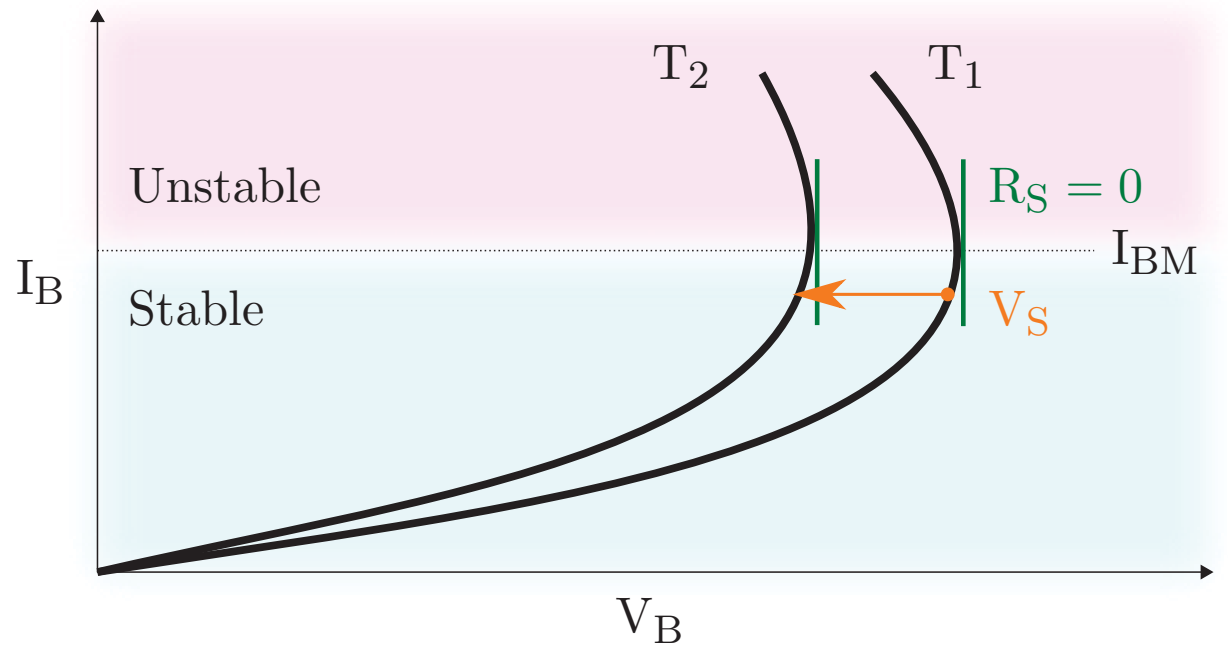

Figure 4.3: The working point of a negative bolometer. When energy is deposited in the bolometer by a particle event, its temperature changes from $T_1$ to $T_2$. The maximum output signal occurs at a bias current $I_{BM}$, which also separates the regions of stable and unstable working points.

The corresponding resistance variation $\Delta R_T$ due to the temperature change can be straightforwardly obtained as

$$\Delta R_T(s) = -\alpha \Delta T = -\frac{\alpha}{s\mathcal{C} + \mathcal{G} + \alpha I_B^2} P(s) \tag{4.17}$$

On the fast transient, that is for large s, the expression can be approximated as

$$\Delta R_T(s) \simeq -\frac{\alpha}{s\mathcal{C}} P(s) \tag{4.18}$$

which in time domain becomes

$$\Delta R_T(t) \simeq -\frac{\alpha}{\mathcal{C}} \int P(t)dt = -\frac{\alpha}{\mathcal{C}} E = \frac{dR_T}{dT} \frac{E}{\mathcal{C}} \tag{4.19}$$

where E is the total energy released, given by the integrated power. This expression is clearly consistent with equation (4.1).

Figure 4.3 shows the load curves at two temperatures $T_1$ and $T_2$, such that $\Delta T = T_2 - T_1$, inducing a resistance change $\Delta R_T$ in the thermistor. Since the bolometer is biased with a constant current, the voltage across the thermistor jumps between the curves following an horizontal line. From equation (4.17) the voltage signal across the thermistor is given by

$$V_S(s) = \Delta R_T(s) I_B = -\frac{\alpha I_B}{s\mathcal{C} + \mathcal{G} + I_B^2 \alpha} P(s) \tag{4.20}$$

It is clear from equation (4.20) that there is an optimum value for $I_B$ which maximizes $V_S$. In fact, by calculating the derivative of $V_S$ with respect to $I_B$ and imposing it to be zero, and neglecting the frequency-dependent terms, a maximum can be found which corresponds to the condition

$$I_{BM} = \sqrt{\frac{\mathcal{G}}{\alpha}} \tag{4.21}$$

By plugging this condition into equation (4.5) it is straightforward to see that the maximum occurs for $R_S = 0$. This is evident also from figure 4.3, where the maximum voltage displacement between the two curves is found for $R_S = 0$. With the values mentioned above for $\mathcal{G}$ and $\alpha$ in the case of CUORE, that is $10^{-9}$ W/K and 1 MΩ/mK respectively, we find the optimum bias current $I_{BM} \simeq 1$ nA.

In this evaluation, the heat capacitance $\mathcal{C}$ was considered constant. But since large values of bias current heat the bolometers, causing the heat capacitance to decrease, the optimum bias current is found at smaller values than $I_{BM}$. Fortunately this is also in agreement with the fact that the electrothermal feedback guarantees a stable operation only for positive values of $R_S$. The bolometers then find their optimum operating conditions at a smaller current than $I_{BM}$, of the order of a few hundred pA in the case of CUORE.

The energy release in the bolometer due to particle events can be considered to be instantaneous. In time domain, the power release can be expressed as

$$P(t) = E\delta(t) \tag{4.22}$$

where $\delta(t)$ is the Dirac delta. Equation (4.20) becomes

$$V_S(s) = -\frac{\alpha I_B}{s\mathcal{C} + \mathcal{G} + I_B^2\alpha}E \tag{4.23}$$

By calculating the inverse Laplace transform, the expression can be translated to the time domain, obtaining

$$V_S(t) = -\alpha I_B \frac{E}{\mathcal{C}}\vartheta(t)e^{-\frac{t}{\tau_S}} \tag{4.24}$$

that is a voltage step which discharges with time constant $\tau_S$, given by

$$\tau_S = \frac{\mathcal{C}}{\mathcal{G} + I_B^2\alpha} \tag{4.25}$$

Equation (4.24) gives the voltage signal across the bolometer biased with a constant current. The absolute value of the peak amplitude of the signal $V_P$ can be calculated from equation (4.24). Assuming a heat capacity of $10^{-9}$ J/K, a bias current of 500 pA, and $\alpha = 1$ MΩ/mK, we obtain

$$\frac{V_P}{E} = -\frac{\alpha I_B}{\mathcal{C}} \simeq 100\ \mu\text{V/MeV} \tag{4.26}$$

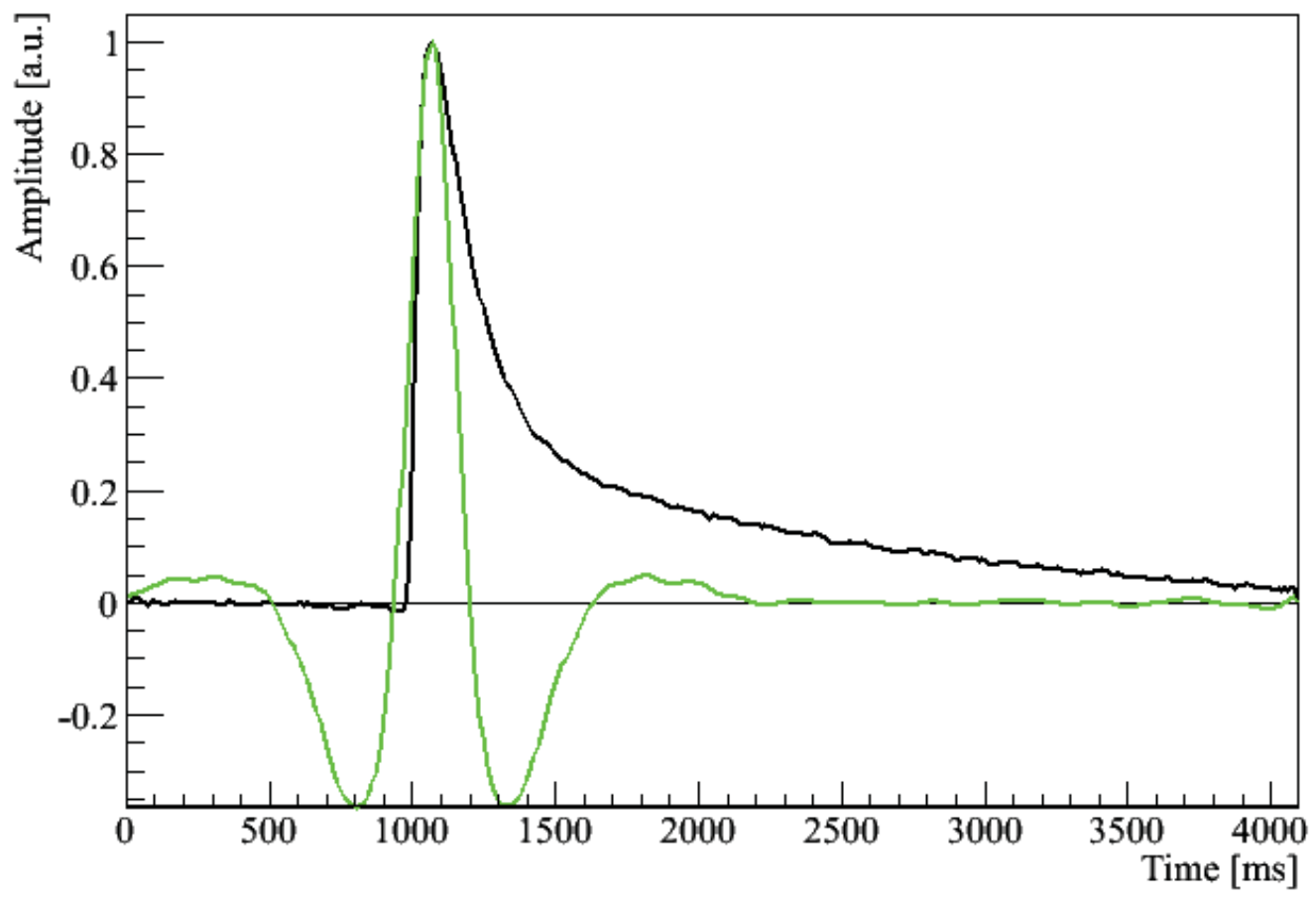

Figure 4.4: Typical signal from a CUORE bolometer. In black, the signal as acquired by the data acquisition system. In green, the same signal after applying the offline optimum filtering algorithm.

This quantity is the energy conversion gain of the bolometer.

The signal $V_S$ expressed by equation (4.24) must be readout with a voltage amplifier with large gain. As already discussed, the opposite case, where the bolometer is biased with a voltage source and the current signal is readout with a current sensitive amplifier, is disfavoured since it changes the sign of the electrothermal feedback and does not allow a stable operation.

Figure 4.4 shows a typical CUORE signal at the output of the readout chain, as acquired by the data acquisition system. In the same figure, the signal after the optimum filter is also shown. As can be clearly seen from equation (4.25) if the bias current is small the fall time is given by $\mathcal{C}/\mathcal{G}$. As the bias current is increased the fall time tends to become faster. The rise time of the signal given by equation (4.23) is considered to be instantaneous. In a more realistic case, the signal is slowed by the heat transfer between the crystal and the thermistor, and can also be limited by the inductive component of the impedance of the bolometer. Moreover, since the bolometers are operated at very low temperature in dilution refrigerators, the readout circuits usually cannot be placed too close. The farther the readout circuits are placed, the more the rise time of the thermal signals is slowed by the parasitic capacitance of the connecting links. If the rise time of the signals does not carry important information and baseline recovery is slow, the parasitic capacitance is not a concern and the readout circuits can be placed outside the cryostat. Otherwise, if the rise time needs to be preserved (to allow pile-up rejection, for instance) then the first stage of the electronics should be placed closer to the sensors inside the cryostat. The design issues and noise calculations of both cases will be treated in the following sections.

The simple model which brought to equation (4.24) gives a rough estimate of the behaviour of the bolometer as a function of the various parameters which rule its thermal and electrical dynamics. As already mentioned, the heat capacity $\mathcal{C}$ and the thermal conductance $\mathcal{G}$ were considered to be constant, but instead they generally depend on temperature, and such dependence should be included in the calculations. Moreover, the model can be refined to take into account the decoupling between the electrons and the lattice of the thermistor, which cause part of the heat to be retained by the lattice, resulting in smaller signals than expected. However the presented equations give anyway a good approximation of the signals observed with bolometric sensors.

Let us finally consider how the electrothermal feedback affects the thermal noise of the thermistor. Without electrothermal feedback, the thermal noise of the thermistor can be expressed as

$$\mathrm{v_{Tw} = i_{Tw} R_T = \sqrt{4 k_B T R_T}} \tag{4.27}$$

where $\mathrm{k_B}$ is the Boltzmann constant. This expression results to be accurate only for $\mathrm{Z_S \simeq R_T}$, that is at high frequency or for low bias currents. If the thermal dynamics of the whole bolometer is considered, then the power dissipation from noise itself should be included in the noise formula. The voltage fluctuation then becomes

$$\mathrm{v_{Sw} = v_{Tw} + I_B \Delta R_T} \tag{4.28}$$

where the first term is due to current fluctuations at fixed temperature, and the second term is due to temperature fluctuations at fixed current. The term $\mathrm{\Delta R_T}$ is the fluctuation in $\mathrm{R_T}$ induced by the temperature fluctuation due to noise. The noise generator $\mathrm{v_{Tw}}$ dissipates a power $\mathrm{I_B v_{Tw}}$. From equation (4.17) the corresponding temperature fluctuation can be calculated and plugged into equation (4.28), obtaining

$$\mathrm{v_{Sw}} = \mathrm{v_{Tw}} \frac{\mathrm{s}\mathcal{C} + \mathcal{G}}{\mathrm{s}\mathcal{C} + \mathcal{G} + \alpha \mathrm{I_B^2}} = \mathrm{v_{Tw}} \frac{\mathrm{Z_S(s) + R_T}}{\mathrm{2R_T}} \tag{4.29}$$

From equation (4.29) we find that at high frequency, where electrothermal feedback is ineffective, $\mathrm{Z_S = R_T}$ and the noise is simply given by the thermal noise of the thermistor. But at low frequency and for large values of $\mathrm{I_B}$ the electrothermal feedback can reduce noise of up to a factor of 2. The lowest value is at low frequency near the inversion point, where $\mathrm{Z_S = 0}$ and $\mathrm{v_{Sw} = v_{Tw}/2}$.

## 4.3 Readout at room temperature

Figure 4.5 shows the schematic of a bolometric sensor with the biasing and readout circuits located at room temperature. The bolometer is represented by its impedance $Z_S$. It is biased with a current $I_B$ generated by applying the bias voltage $V_B$ through the large value load resistors $R_L$. The whole readout scheme is differential, in order to minimize microphonic and electromagnetic disturbances which otherwise may be easily injected. Since the connecting lines are long, their shunt capacitance $C_P$ must be considered in parallel with $Z_S$. The connecting links are a critical item, whose shunt parasitic impedance should be carefully selected and tested [49]. Alternatively, the capacitance could be considered to be referred to ground, without substantial differences in the following evaluations.

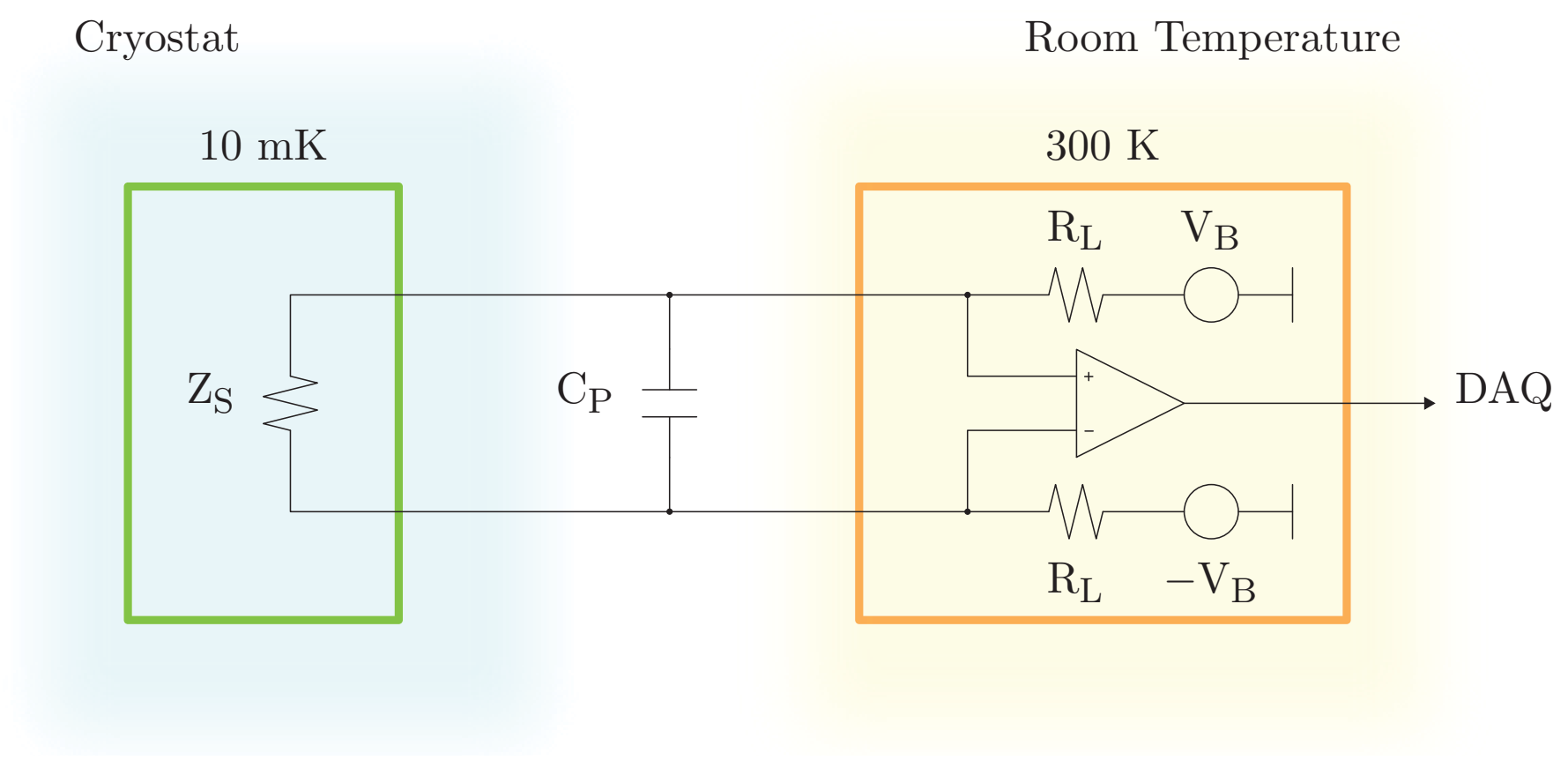

Figure 4.5: Readout of a bolometric sensor at room temperature.

The main advantage of this readout scheme is its simplicity. The bolometer is the only element which is not located at room temperature, and thus every part of the readout circuit is directly accessible without opening the cryostat. Moreover, all of the components are located far from the sensors, which eliminates issues related with their radiopurity in case the experiment requires low background, as happens for double beta decay searches.

The drawbacks are due to the long connecting lines from the sensor to the readout circuit. Even if the disturbances can be kept under control with a proper differential configuration, the large capacitance $C_P$, likely to be of the order of 1 nF, can form a low pass with the source impedance $Z_S$ and limit the rise time of the signals. The large input capacitance may also deteriorate the signal to noise ratio in case the readout circuit is an integrator, as discussed in chapter 1.

Let us now consider the case of a voltage amplifier with flat gain, and let us consider the white noise contributions at the input. Let us assume the load resistors to be much larger than the resistance of the bolometer, and the input impedance of the readout amplifier to be infinite. The thermistor $R_T$ contributes with its thermal noise. From equation (4.29) the resulting noise at low frequency depends on $R_S$, and is given by

$$v_{Sw} = \sqrt{\frac{4k_B T_C (R_S + R_T)^2}{4R_T}} \quad (4.30)$$

where $T_C$ is the temperature inside the cryostat, of the order of 10 mK. The closer the bolometer is to the inversion point, the smaller its series noise. The load resistors contribute with a current noise $i_{Lw}$ which develops a voltage on $R_S$. The total current noise at the input due to the two load resistors is given by

$$i_{Lw} = \frac{\sqrt{2}}{2}\sqrt{\frac{4k_B T_R}{R_L}} = \sqrt{\frac{2k_B T_R}{R_L}} \quad (4.31)$$

The factor of $\sqrt{2}$ comes from the differential configuration, since the two load resistors contribute with uncorrelated current noise sources, while the factor 1/2 comes from the fact that if $R_L$ is much larger than the resistance of the bolometer, only half of the noise current from each load resistor flows through the bolometer. The corresponding voltage noise is simply given by

$$v_{Lw} = i_{Lw} R_S = \sqrt{\frac{2k_B T_R R_S^2}{R_L}} \quad (4.32)$$

where $T_R$ is the room temperature, 300 K. The ratio between the two contributions is

$$\frac{v_{Lw}}{v_{Sw}} = \sqrt{2\frac{T_R}{T_C}\frac{R_T R_S^2}{(R_S + R_T)^2 R_L}} \quad (4.33)$$

It is clear that due to the large difference between $T_C$ and $T_R$ the white noise from the load resistors is generally larger than that from the bolometer, unless $R_L$ is larger than $R_T$ and $R_S$ of several orders of magnitude. Assuming the value of $R_T$ to be 1 G$\Omega$, $R_S$ to be 10 M$\Omega$ and $R_L$ to be 30 G$\Omega$, which are reasonable values for the case of CUORE, noise is contributed almost in equal parts by the thermistor and the load resistors, and is close to 10 nV/$\sqrt{\text{Hz}}$. For larger values of $R_S$ the thermal noise from the load resistors dominates over the noise from the thermistor.

By a proper design of the amplifier at room temperature its white voltage noise contribution can be kept below a few nV/$\sqrt{\text{Hz}}$. Due to the large impedance of the input node, the current noise should also be minimized, and this is obtained by using JFETs in the first stage of the amplifier. Their operating point should also be carefully selected to minimize their gate current. The shot noise $i_A$ due to the input current at the amplifier terminals $I_A$ gives a shot noise $\sqrt{2qI_A}$, where q is the electron charge. Due to the differential configuration only half of that noise current passes through the bolometer, giving a current noise $\sqrt{qI_A/2}$ across the bolometer. Since the amplifier input currents are two, the overall current noise due to the amplifier input currents results to be $\sqrt{qI_A}$. The ratio between the shot noise due to the input current of the amplifier and the thermal noise of the load resistors is given by

$$\frac{i_{Lw}}{i_{Aw}} = \sqrt{\frac{2k_B T_R}{qI_A R_L}} \tag{4.34}$$

The current noise from the amplifier is then negligible if

$$I_A \ll \frac{2k_B T_R}{qR_L} \tag{4.35}$$

which with the value chosen above for the load resistor, that is 30 GΩ, gives $I_A \ll 2$ pA. The choice of the input transistor of the amplifier is then crucial to keep the input current $I_A$ at both its terminals under control.

Since bolometric signals are slow, and bandwidth in this readout scheme is reduced further by the low pass filtering due to the long connecting links, the $^1/_f$ noise should be considered, and may well become the main noise component if care is not taken in the design. The $^1/_f$ noise due to the load resistors is generally proportional to the applied bias, since the mechanisms involved, that is fluctuation of number and mobility of the charge carriers, become more evident when the current is increased. For a given resistor $R_L$, the $^1/_f$ contribution can be modeled as

$$i_{L^1/_f} = \sqrt{\frac{A_L R_L I_B^2}{f}} \tag{4.36}$$

where $A_L$ is a parameter which depends on the geometry and technology of the resistor, and $I_B$ is the bias current [50]. The $^1/_f$ noise from the load resistors is then negligible at low bias currents, and tends to increase at larger $I_B$, even if its effect in this case is mitigated by the fact that $R_S$ decreases at larger $I_B$. The load resistors should be in any case selected in order to use devices with a low value for $A_L$. The need to obtain a low $^1/_f$ from the input amplifier again requires the input JFET transistors to be carefully selected. The $^1/_f$ term may be due both to voltage and current noise. By a proper choice of the input transistor and of its operating point, the overall $^1/_f$ noise coefficient from the amplifier may be kept below a few $nV^2$. Further details will be discussed in section 4.5, which describes the readout solution for CUORE.

## 4.4 Readout at cryogenic temperature

Figure 4.6 shows the sketch of a bolometric sensor readout with a cold stage located inside the cryostat. The cold stage is made of two JFETs in source follower configuration. Their drain is biased with the common voltage $V_D$, while their source is individually biased at room temperature through the resistors $R_J$. Since now the front-end is closer to the sensors, the parasitic capacitance of the connecting wires is reduced, which allows a faster response. Moreover, since the output impedance of the JFETs is low, the signals at their outputs are immune from microphonic and electromagnetic disturbances. For Silicon transistors the optimal noise performance is found around 120 K. Other devices may be used if it is convenient to bring the cold stage at lower temperature. The main drawback of this solution is the larger number of wires needed and the increased complexity.

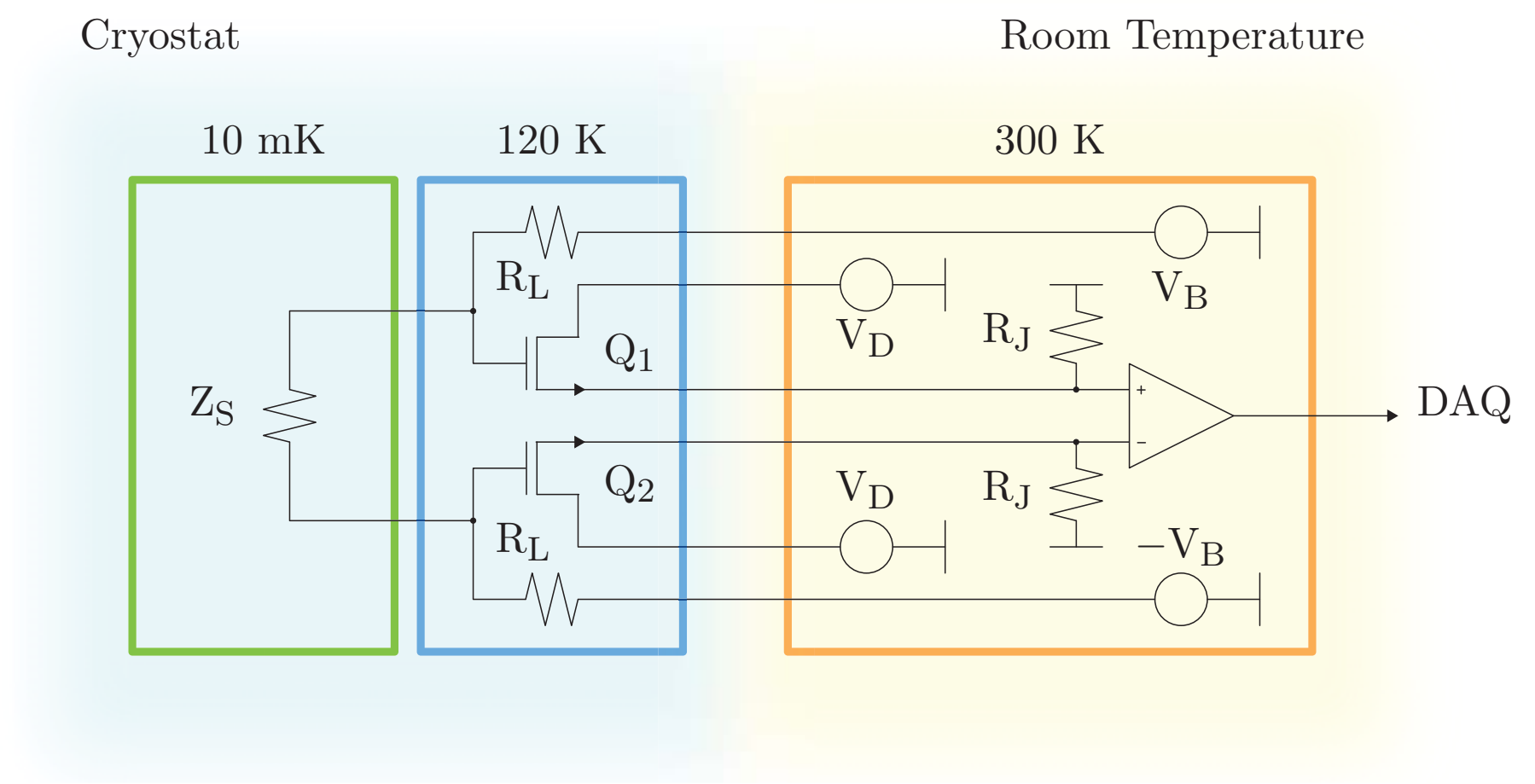

Figure 4.6: Readout of a bolometric sensor at cryogenic temperature.

Aside from the addition of the cold JFETs $Q_1$ and $Q_2$, the main aspects of the readout chain are the same. The load resistors can now be placed inside the cryostat, reducing their thermal noise. In figure 4.6 they were placed at 120 K near the JFETs, but in principle they can also be kept at a lower temperature, provided that they are able to work. On the other side, the contribution to the series noise of the two JFETs $Q_1$ and $Q_2$ must now be considered. And since they are used with unity gain in a source follower configuration, the noise from the second stage cannot be neglected. The source follower configuration is preferred because the gate-source capacitance of the JFETs is bootstrapped and does not limit the bandwidth of the signals. The transistors $Q_1$ and $Q_2$ are critical for noise performance. As introduced in chapter 1, their noise at low temperature depends on the presence of inpurities in the band gap, which can be minimized with a proper choice of materials and processes. Even when a good process is found, the good performance of a given JFET at low temperature cannot be taken for granted, as its noise may also depend on the production batch. Individual testing of each JFET at low temperature is then advisable to obtain optimal performance. For this purpose, an automatic system to characterize the JFETs at low temperature was built and operated in Milano Bicocca. The system is named A/C VISCASY, Ambient to Cryogenic Vibrationless Scan System, and allows to measure the static characteristics and noise of the devices from room temperature down to 77 K or even below, down to 4 K if liquid Helium is used instead of liquid Nitrogen [51,52]. The system allowed to select the transistors of the cold stage of the MARE experiment, which will be described in section 4.7.

## 4.5 The readout of the CUORE experiment

The CUORE experiment is a neutrinoless double beta decay search in $^{130}$Te based on the bolometric technique [53]. The bolometers are $5 \times 5 \times 5$ cm$^3$ crystals of $TeO_2$. Since the natural isotopic abundance of $^{130}$Te is close to 30%, largely above that of most of the other double beta candidate nuclei, the CUORE experiment can be launched without enriched material. The use of enriched crystals can be envisioned for a later stage of the experiment. Each CUORE crystal has a mass of 750 g, of which about 200 g of $^{130}$Te. CUORE will be composed of 988 crystals, arranged in 19 towers of 52 crystals each. The total mass of $^{130}$Te will be close to 200 Kg. The energy of the neutrinoless double beta decay of $^{130}$Te is 2528 KeV, where the average resolution of the bolometric sensors is expected to be about 5 KeV. Since the double beta decay emitters coincide with the sensors, the detection efficiency is expected to be very high, close to 90%. The feasibility of CUORE was demostrated by the pilot experiment CUORICINO, an array of 62 $TeO_2$ crystals succesfully operated at LNGS between 2003 and 2008 which provided a lower limit to the half life of the neutrinoless double beta decay of $^{130}$Te of $2.8 \times 10^{24}$ years [54]. CUORE should provide a sensitivity to the half life of the neutrinoless double beta decay larger than $10^{26}$ years, a goal comparable with that of GERDA phase II.

Figure 4.7 on the left shows a floor of a CUORE tower, made of 4 bolometers. The same figure on the right shows a sketch of the whole setup, with the 19 towers inside the custom dilution cryostat. The CUORE cryostat is currently under construction at LNGS, and the towers are being assembled. As a test for the final assembly materials techniques and procedures to be employed in CUORE, the performance of a single CUORE-like tower housed in the CUORICINO cryostat is being studied. This fundamental test, named CUORE-0, will not only validate the CUORE final setup, but will also constitute a neutrinoless double beta decay search experiment with a sensitivity close to that of GERDA phase I. In a first period, CUORE-0 will use the electronics of CUORICINO. At a later stage the final electronics of CUORE will be employed.

The design concepts of the readout electronics for CUORE are based on the pilot experiment CUORICINO, updated to use modern components and upgraded to incorporate more features [55]. The front-end electronics are located at room temperature, and the connecting links from the sensors to the electronics will be nearly 5 m long. The first part of the connecting links is realized with flexible tapes of Cu-PEN (Copper on a Polyethylene 2.6 Naphthalate substrate) whose radiopurity was tested and found compliant with the requirements [56]. The tapes come in a few different types, following the geometry of the CUORE towers, the longest being about 2.5 m long. A tape is shown in figure 4.8. Each has 29 copper traces. The tapes which carry the signals from the bolometers serve 10 differential channels, with grounded traces interposed between channels to avoid crosstalk. The thermistors are directly bonded to the pads at one end of the tapes. The other ends plug into ZIF connectors mounted on Kapton boards at the first thermalization stage inside the cryostat. This is the most critical part of the connecting links, since it is in direct contact with the sensors and is held at a temperature of 10 mK. The second part of the connecting links is made of twisted NbTi-NOMEX wires, from the first thermalization stage to the multipole connectors at the top of the cryostat. The third and last part are the twisted cables from the top of the cryostat to the front-end crates. Since the source impedance of the bolometers in operating conditions is in the hundreds of M$\Omega$ range, the connecting links must guarantee a negligible parallel parasitic conductance to ground and to neighbouring connecting links at DC.

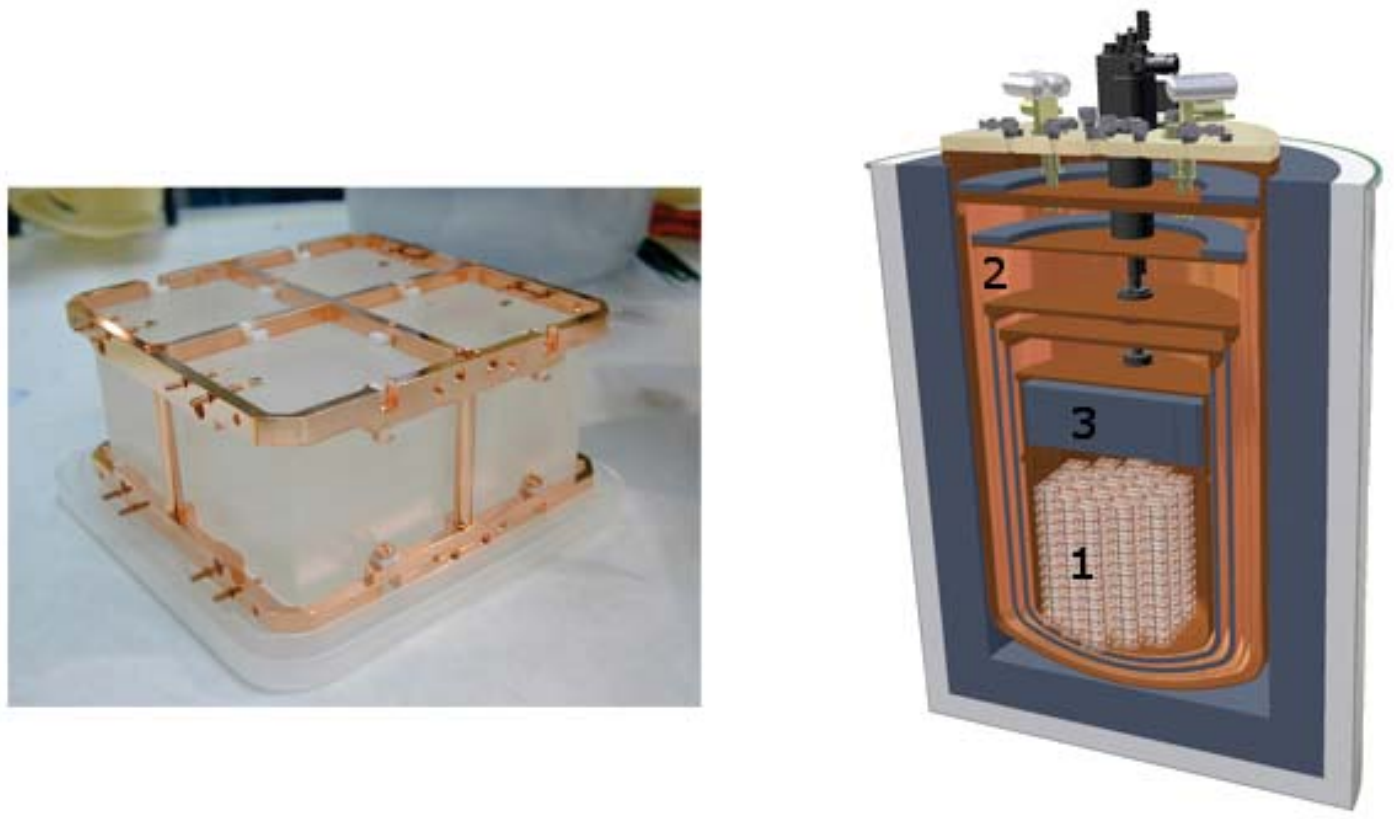

Figure 4.7: On the left side, a floor of a CUORE tower, housing 4 $TeO_2$ crystals. On the right side, a sketch of the whole CUORE experiment, showing the array of 988 $TeO_2$ bolometers (1), the dilution cryostat (2), its inner shield made of low-background Roman lead (3).

During this PhD work a characterization procedure was developed to ascertain the compliance of the tapes with the requirements of the experiment [57]. First the tapes are tested against fabrication defects, which may cause some traces to be shorted or broken. In the first part of the characterization, a Agilent DCA-X 86100D sampling scope is used to study the quality of the tapes with the time domain reflectometry (TDR) technique. A sketch of the setup is shown in figure 4.9. The instrument is connected to the ZIF end of the tape, and sends a differential voltage step of $t_R = 20$ ps rise time down a differential pair of traces, which is seen as a transmission line. The pad end of the tape is left open, untouched, preventing any possible damage to the bonding pads. At the ZIF side, the voltage is sampled with a bandwidth of 18 GHz. From the amount of reflected signal on a good tape, the differential characteristic impedance can be estimated to be $Z_0 = 150\ \Omega$. From this value, and from the value of the propagation delay, which is $t_{pd} = 4.6$ ns/m, the capacitance of the differential pair can be estimated to be

$$C_d = \frac{t_{pd}}{Z_0} = 30 \text{ pF/m} \tag{4.37}$$

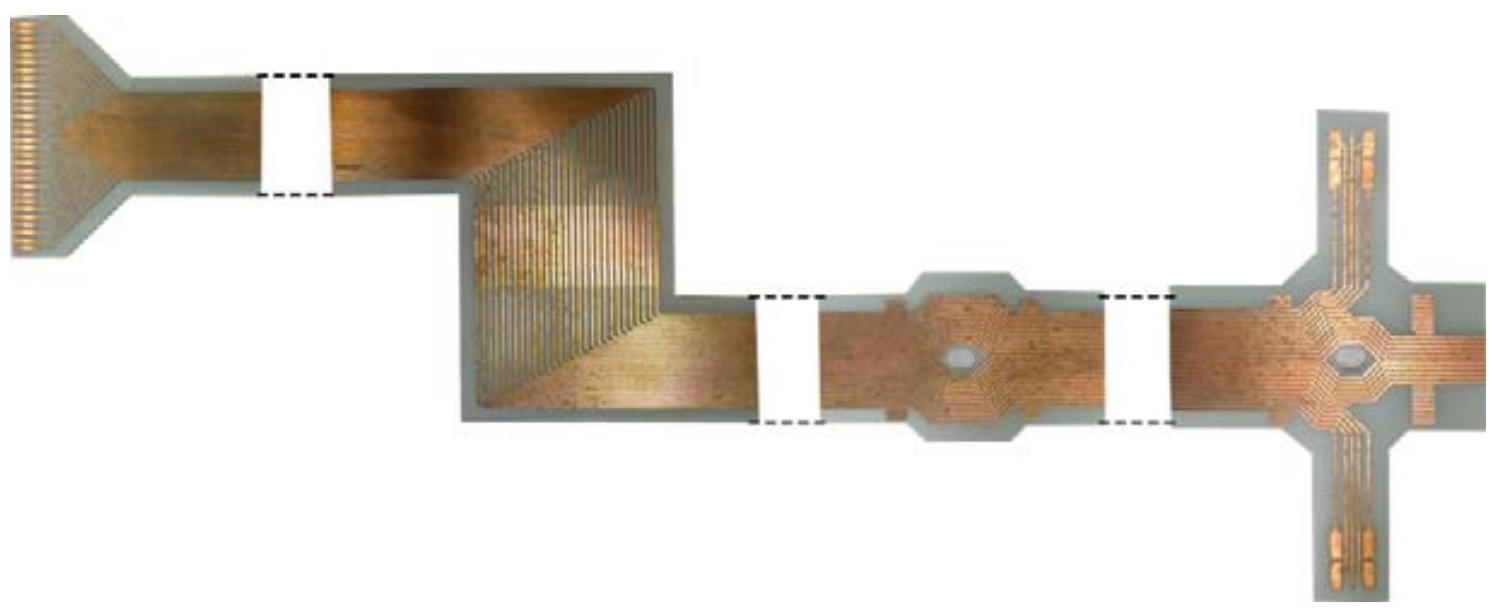

Figure 4.8: The main details of a CUORE Cu-PEN tape. On the left side is the end to be plugged in the ZIF connector. On the right side are the thermistor bonding pads. The double bend in the middle with the larger trace spacing is needed to allow the alignment of the different masks used in the fabrication process.

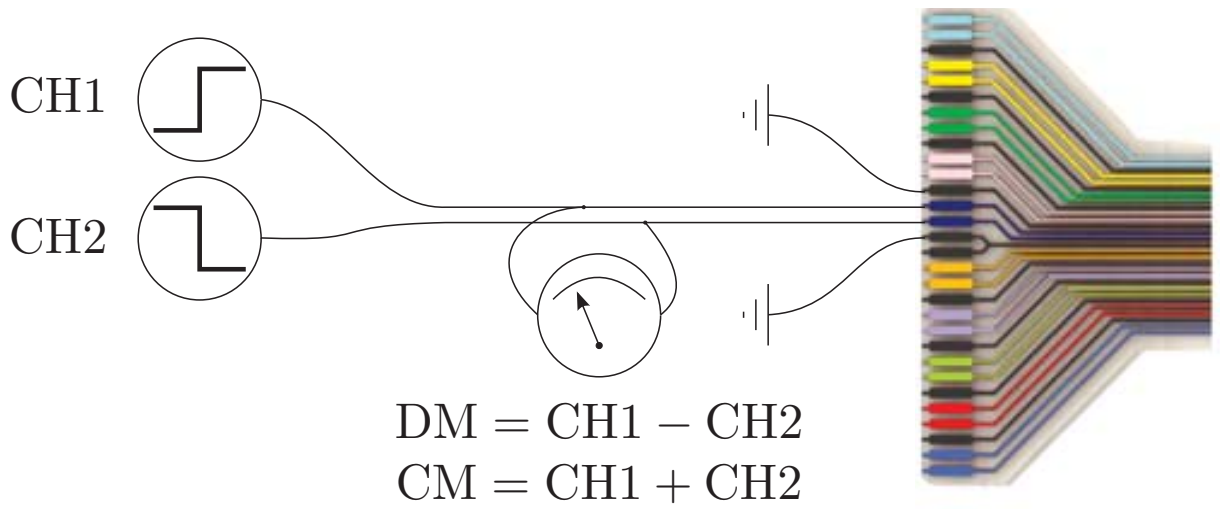

Figure 4.9: Setup for TDR measurement. In this sketch, different colors are used to denote the 10 differential pairs.

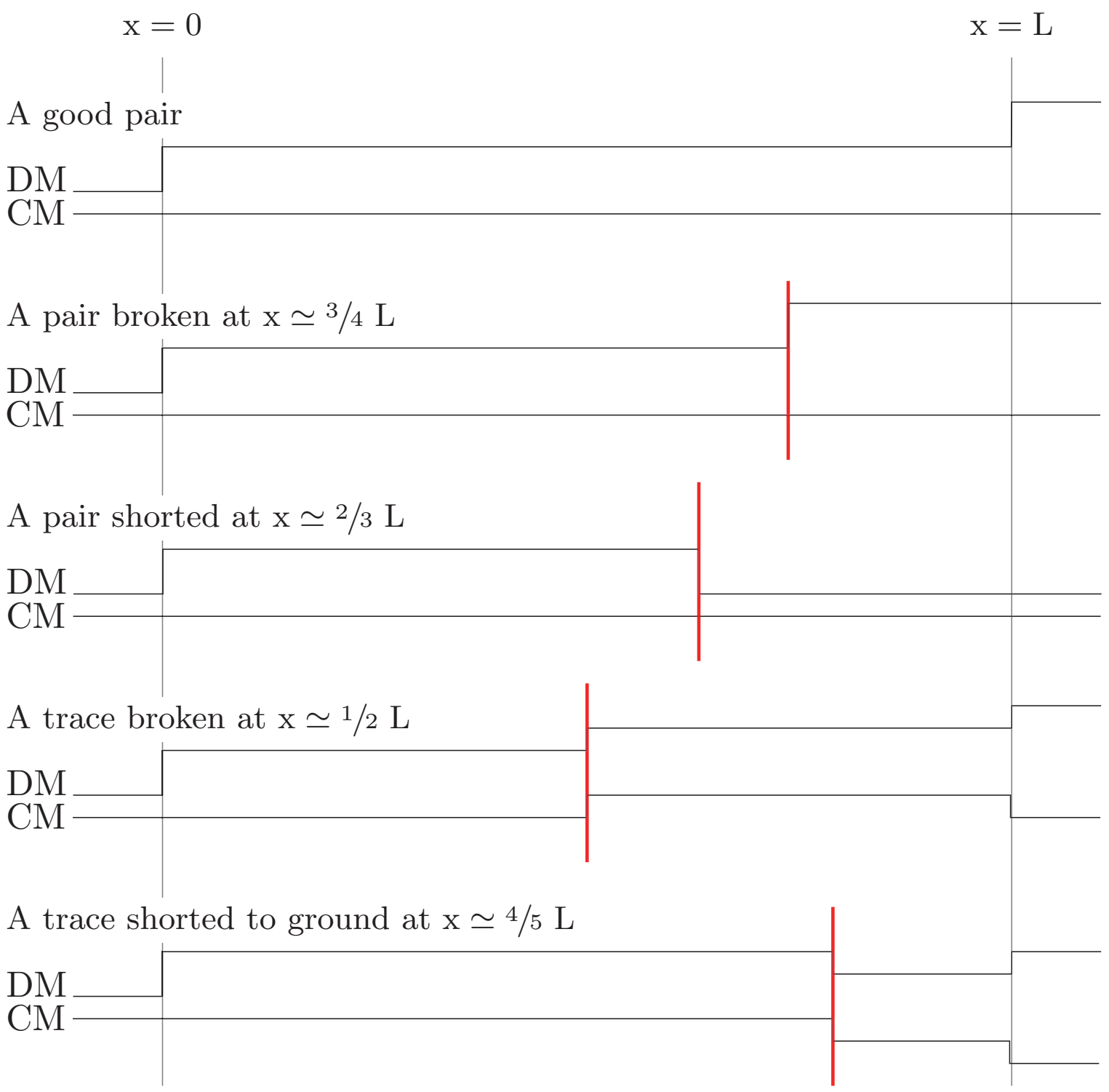

Figure 4.10: Schematic representation of the results of a TDR measurement for a differential pair of traces of length L, showing the differential mode (DM) and the common mode (CM) reflected waveforms for the case of a good pair (top) and of typical defects.

---

The speed of propagation of the voltage step on the tape is $c/\sqrt{\varepsilon}$, where $\varepsilon$ is the relative dielectric constant of the traces. Knowing the tape length, $\varepsilon$ can be measured to be 1.9. This setup allows to locate defects with a spatial resolution of $t_R/t_{pd} \simeq 5$ cm. Figure 4.10 shows the typical TDR results in the case of a good pair and of typical defects. The differential mode (DM) reflects the behavior of the differential pair of traces. The common mode (CM) exposes any possible asymmetry between the two traces in the pair, which would indicate a defect in one of the two traces.

The tapes which are indicated by the TDR to be good can be accepted, and are tested for electrical insulation. To test the electrical insulation between traces on the tape, all the odd and even traces are connected in parallel forming two groups, which are connected to a Keithley 6514 Electrometer with a full scale of 200 GΩ. Thanks to the parallel connection, the maximum measurable impedance per pair is increased by a factor of ten, obtaining a sensitivity of 0.5 pA/V for the mean parasitic conductance of the pairs on the tape. The measurement must be performed in vacuum to avoid parasitic effects due to air humidity. A set of boards equipped with relays were realized to switch the electrometer inputs on different tapes. The relays are remotely controlled from a PC through a I2C protocol. This setup allows to test several tapes held together in vacuum. The tapes which pass also the electrical insulation test can be considered compliant with the electrical requirements of the CUORE experiment.

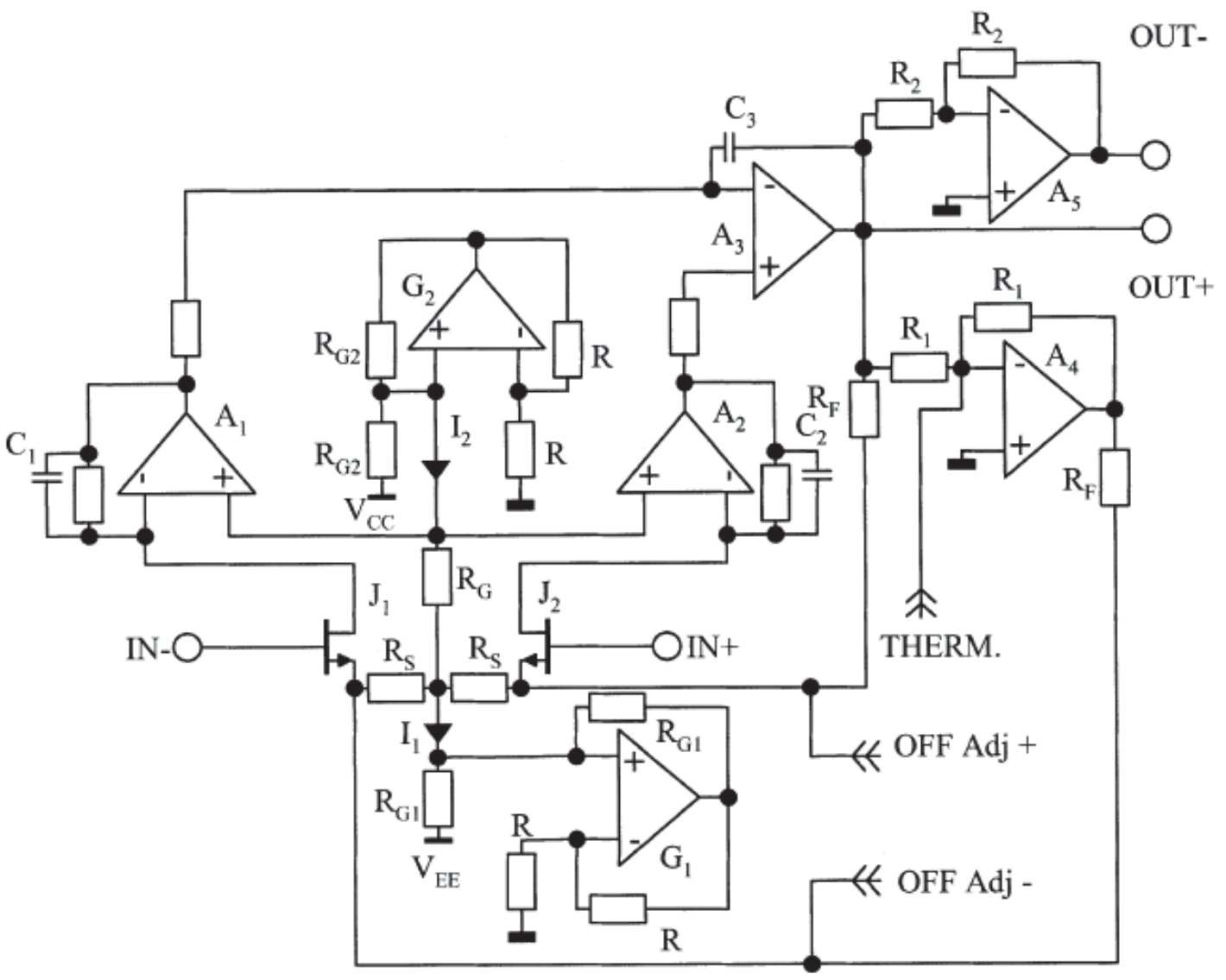

Figure 4.11: Schematic of the room temperature amplifier of the CUORICINO experiment, on which the CUORE amplifiers are based. The schematic was taken from [55].

The front-end is based on voltage sensitive differential amplifiers located at room temperature at the top of the cryostat. As previously discussed, this choice allows to minimize the number of wires entering the cryostat, which is important for CUORE due to the large number of channels. Since the signals are slow, with a bandwidth extending down to 1 Hz, the low pass due to the larger capacitance at the input induces a negligible deterioration of the signal to noise ratio. As discussed in the previous sections, biasing the bolometers with a constant current and reading out the voltage signals assures the stability of the electrothermal feedback below the inversion point. Even if the differential configuration exhibits twice the noise power than an equivalent single ended configuration, it is preferred since its capability to reject common mode signals allows to suppress most of the electromagnetic and microphonic disturbances which may affect the signal to noise ratio. The design is based on the circuit developed for CUORICINO, whose schematic is reported in figure 4.11.

The input stage is based on a matched pair of SNJ132287 JFETs from Interfet, operated at a reduced drain current and drain to source voltage to minimize their noise [58]. The gate current of a JFET is contributed by two sources. One is related with the diffusion of thermally generated minority charge carriers in the channel (holes for a n-channel JFET), which depends on the squared intrinsic concentration $n_i^2$, and is independent of bias. The other is related with the recombination of charge carriers in the depletion region, which depends on the presence of trapping centers and on the size of the depletion region, and is proportional to the intrinsic concentration $n_i$. This contribution depends on bias, since the size of the depletion region depends on the working point of the JFET. Since

$$n_i^2 \propto e^{-\frac{E_G}{k_B T}} \tag{4.38}$$

and

$$n_i \propto e^{-\frac{E_G}{2k_B T}} \tag{4.39}$$

the two contributions are expected to show different behaviour with temperature. An empyrical relation can be written to merge the two contributions and describe the overall gate current of JFETs, and is

$$I_G \simeq I_S(V_{GD}) e^{-\frac{E_G}{mk_B T}} \tag{4.40}$$

where $V_{GD}$ is the largest bias across the junction, $E_G$ is the bandgap of the semiconductor, $k_B$ is the Boltzmann constant and T is the junction temperature. The relation closely resembles the Shockley diode law with the addition of the empyrical terms $I_S$ and m. For the SNJ132287 JFETs to be used in CUORE operated with $V_{GD} = 0.75$ V, the values $I_S = 0.653$ A and $m = 1.355$ were measured. The small value for m indicates that in this case the number of trapping centers is small, due to the good quality of the samples. The small value for $V_{GD} = 0.75$ V is allowed by the fact that the devices were selected to have a small pinch-off voltage. Their gate to source voltage at 5 mA current is close to $-0.15$ V, and their transconductance at such working point is 5 mA/V. The input capacitance is about 19 pF, of which 8 pF are due to the the gate to drain contribution. Thanks to the considerations described above, the gate current could be kept below 0.1 pA. The resulting parallel noise of the amplifier is about 0.1 fA/$\sqrt{\text{Hz}}$, which makes this contribution negligible compared with that of the 30 GΩ resistors, as calculated in section 4.3. Since the gate current is strongly dependent on temperature, the operating temperature of the JFETs should be monitored, and their heating due to the power dissipation in the crates which house the front-end circuits should be kept under control.

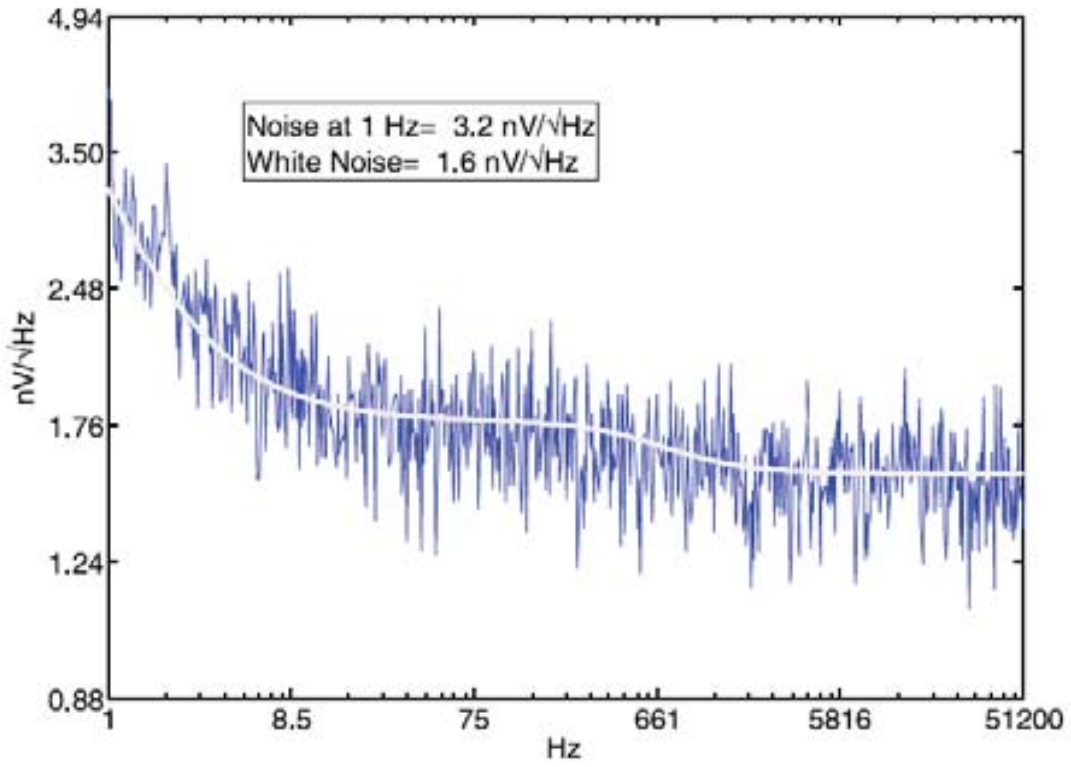

Figure 4.12: Voltage noise of a SNJ132287 JFET from Interfet employed in the front-end of the CUORE experiment. The plot was taken from [58].

The series noise of the JFETs at such operating point is 1.5 nV/$\sqrt{\text{Hz}}$ white, rising to 3 nV/$\sqrt{\text{Hz}}$ at 1 Hz, as shown in figure 4.12. The $1/f$ noise component is low, confirming the very small number of trapping centers inferred from the input current evaluations. The contributions from the two JFETs must be summed in quadrature to obtain the total voltage noise of the amplifier, that is 3 nV/$\sqrt{\text{Hz}}$ white, rising to 4.9 nV/$\sqrt{\text{Hz}}$ at 1 Hz. The front-end circuit is DC coupled to the bolometer, which allows to continuously monitor the baseline. Since the bolometer is biased with a DC current, an offset is present at the inputs of the amplifier and is corrected with a dedicated circuit. With proper design of the offset compensation circuit, the noise performance is not significatively affected by the input offset.

The bias current for the thermistors is provided with a circuit located on the same motherboards which house the front-end amplifiers [59]. The typical bias currents of a few hundred pA required to operate the bolometers are provided through carefully selected large value load resistors with negligible $1/f$ noise contributions.

The sum of all the noise sources, dominated by the thermal noise of the load resistors and the voltage noise of the front-end amplifier, results in an overall noise of about 40 nV RMS in the bandwidth from DC to about 10 Hz. Considering the fact that CUORE bolometers have an expected energy conversion gain of the order of 100 µV/MeV, the overall electronic noise results to be below one KeV FWHM, which is about a factor of 5 smaller than the energy resolution of the macrobolometers employed.

The amplifier has a fixed gain close to 220 V/V [60]. A photograph of the printed circuit board of the CUORE front-end amplifier is shown in figure 4.13. The inputs are on the left side, where the differential JFETs can be seen. The JFETs are matched and housed in the same package. The small apertures on the PCB are used to minimize the parasitic conductance between the inputs, which otherwise may arise due to solder paste residues after the boards are populated. Near the center of the PCB a relay can be seen, which is used to control the gain between the normal value of 220 V/V and a smaller value, close to 20 V/V, expanding the dynamic range which can be covered by the circuit. At the right of the relay, two digital to analog (DAC) circuits are used to control the offset. The DACs are managed by the on-board 8-bit microcontroller LPC925 from NXP which can be seen at the far right. The microcontroller is designed to communicate with the 32-bit microcontroller on the motherboard which houses the amplifiers, which in turn is controlled remotely by a PCB through a CAN interface.

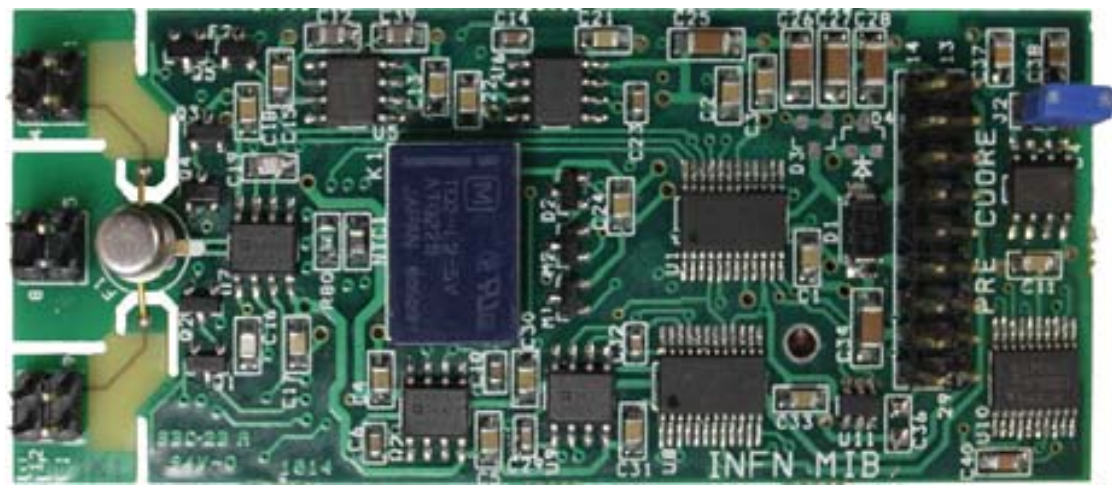

Figure 4.13: Photograph of the front-end amplifier of the CUORE experiment. The PCB is $80 \times 34$ mm$^2$. The input JFETs are on the left side, followed by the operational amplifiers, a relay and the DACs. The on-board microcontroller is on the far right.

The output of the amplifier is fed to a second stage programmable gain amplifier (PGA), which allows to tune the overall gain of each channel from 220 V/V to 5000 V/V. The PGAs are located on the front-end main board, shown in figure 4.14. Each main board houses six amplifiers, together with the bias generator circuits and the large value resistors. The signals are then driven farther away from the cryostat to the data acquisition section. In order to stabilize the overall experimental setup over long data taking conditions, thermal pulses of known energy are injected in the bolometers through dedicated resistors, or heaters, which are glued to each crystal. The known pulses are generated with a dedicated circuit, the pulser, which guarantees a precision of a few ppm. The stability versus temperature of all the circuits mentioned is assured with proper compensation circuitry and was tested at the level of a few ppm/°C. Every aspect of the readout chain is remotely programmable through a CAN bus interface. A dedicated antialiasing filter based on 6-poles Bessel polynomials is used before the acquisition system, composed of commercial devices from National Instruments. The acquired data are filtered offline with optimum filtering algorithms and analyzed.

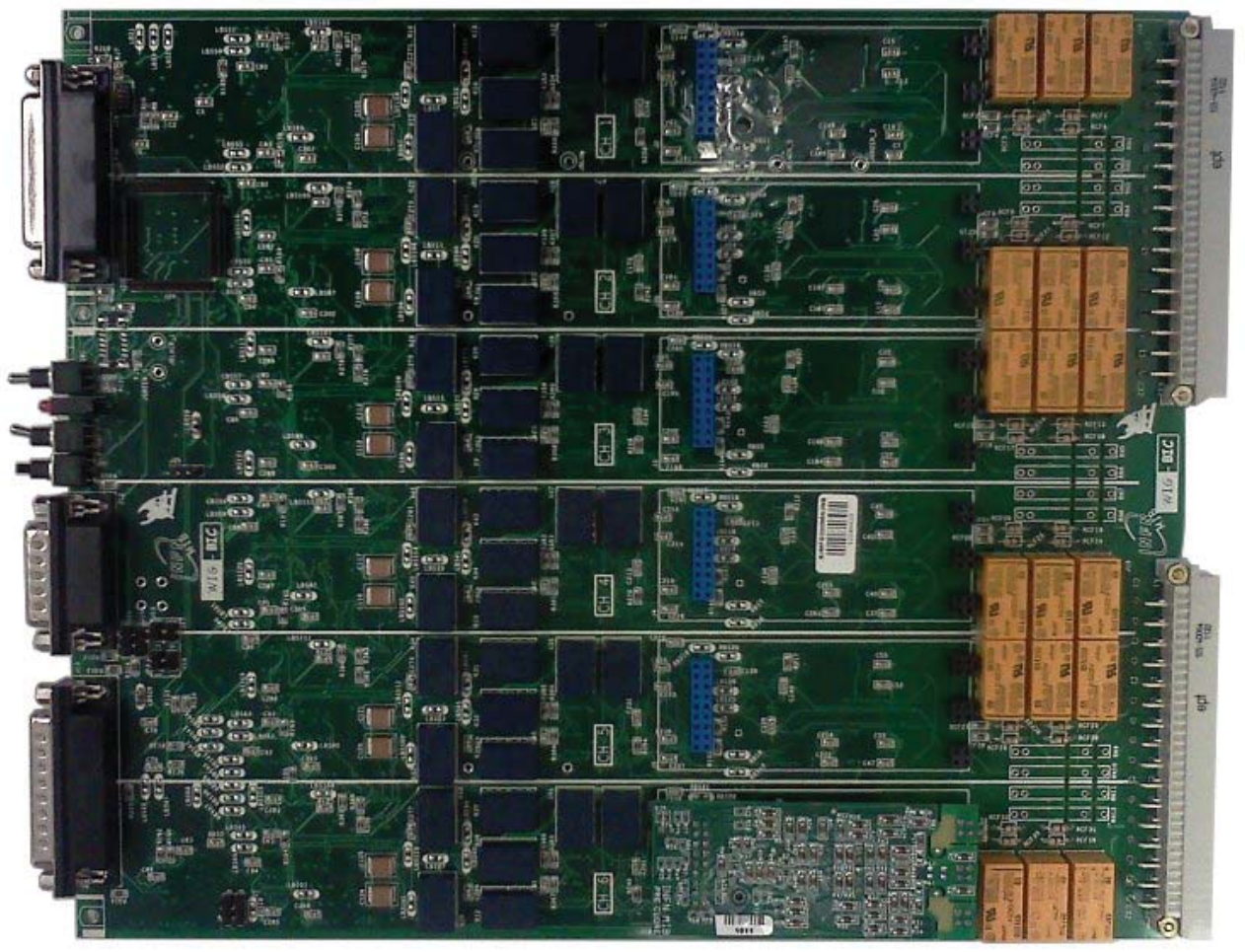

Figure 4.14: Photograph of the front-end main board of the CUORE experiment. Each board houses six amplifiers, together with the bias generators and the PGAs. Only one amplifier is mounted here, and can be seen at the bottom.

## 4.6 The readout of the LUCIFER experiment

The main source of background at the double beta decay endpoint in the CUORICINO experiment was due to alpha particles, mostly due to surface contamination of the copper frames holding the bolometers. In the case of CUORE the cleaning procedures were largely improved, and background is expected to be reduced by about a factor of ten. Nevertheless the dominant background source will still be due to alpha contamination. The capability to actively discriminate alpha events from beta decays in bolometric experiments would be a very welcome feature, and would allow to push the boudaries of double beta decay searches beyond CUORE, towards next generation ton-scale experiments with zero background.

In the case of $TeO_2$ crystals, one way is to readout the Cherenkov radiation produced by beta events in the crystals [61]. Alpha particles do not produce Cherenkov radiation, since they are are much heavier and their velocity is smaller. The main difficulty lies in the fact that Cherenkov photons produced by double beta decay events in $TeO_2$ is relatively small, of the order of one hundred, and the feasibility of this approach is still to be demonstrated.

On the other side, crystals which produce scintillation light proved in general to offer the capability to discriminate between alpha and beta/gamma events. Such property is known, and already exploited in experiments for dark matter searches. The discrimination mechanism relies on the fact that alpha particles release their energy in a much smaller volume than beta/gamma par-ticles, and scintillation light in that case is quenched, resulting in a smaller number of photons produced for a given energy. As an effect of the loss of energy into scintillation, the shape of the thermal signal for scintillating bolometers is also generally different for alpha events compared to beta/gamma events, allowing pulse shape discrimination already in the thermal signal alone. The detection of the scintillation light with a dedicated sensor, readout simultaneously with the thermal signal, is then a very promising technique to provide the required discrimination of alpha particles from double beta decay events. The light sensors used are generally bolometer themselves, made of slabs of Silicon or Germanium. The development of new bolometric sensors which exploit the Neganov-Luke effect or kinetic inductance devices (KIDs) is an active field of research, which may allow a more sensitive and faster readout of thermal and light signals in future experiments.

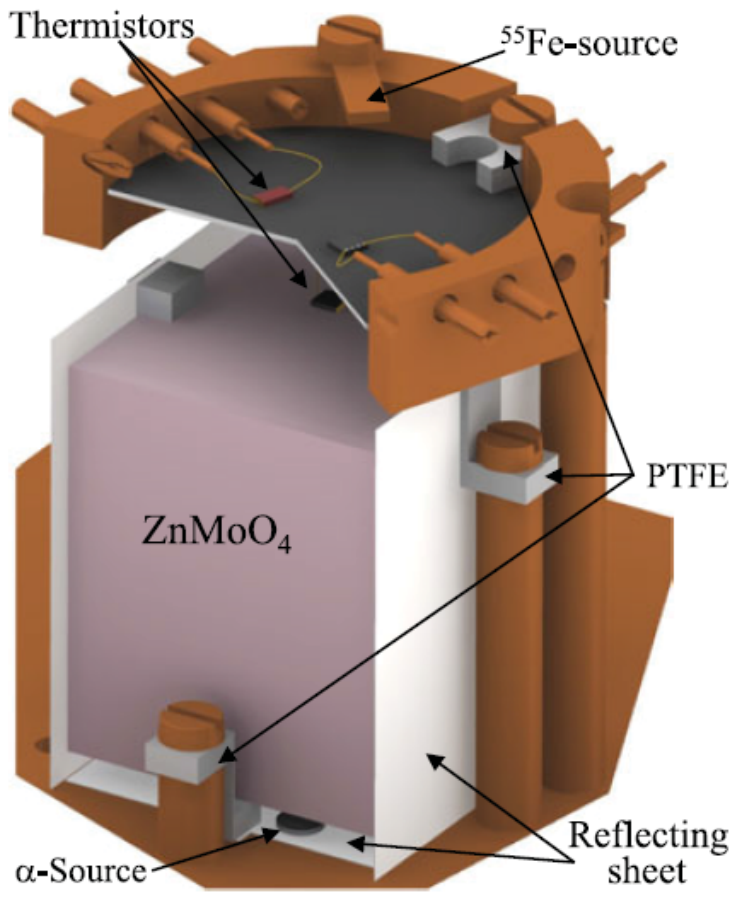

Figure 4.15: Sketch of the setup of a scintillating bolometer, faced to a light sensor which provides the light readout. Taken from [63].

The LUCIFER experiment was proposed in order to extend the application of the scintillating bolometer technique to the neutrinoless double beta decay search [62]. The experiment is mainly intended as a R&D activity towards the next generation of double beta decay searches with bolometric sensors. A large part of the work is related with the selection of the crystals containing double beta emitting isotopes. Among the scintillating crystals which may be employed, those based on Molybdenum, Selenium and Cadmium seem the most promising to be grown in large size. The candidate double beta decay emitting isotopes which can be incorporated are $^{100}$Mo, $^{82}$Se and $^{116}$Cd. The double beta decay endpoints for all these candidates is above the 2615 KeV gamma peak of $^{208}$Tl, which makes the energy region of interest almost free from spurious gamma events, most of which lie below 2615 KeV. The drawback in all the above cases is the need for enriched materials, since the natural abundance of all the double beta decay emitters is below 10%. Among crystals grown with Cadmium, $CdWO_4$ was proven to work as a scintillating bolometer for double beta decay search, but the element is difficult and costly to enrich in $^{116}$Cd. Its use in a future large scale experiment seems then disfavoured for this reason. Among crystals grown with Molibdenum, $ZnMoO_4$ was recently reported to work as a scintillating bolometer [63]. A sketch of the setup used is shown in figure 4.15, showing the $ZnMoO_4$ bolometer in its frame, with the thin light sensor on top. Two thermistors are used for redundancy both on the $ZnMoO_4$ crystal and in the light sensor. A X ray $^{55}$Fe source is used to calibrate the light sensor, while an Uranium source is used to test the discrimination performance of the setup with alpha particles. The energy resolution reported is close to 6 KeV FWHM, competing with the $TeO_2$ crystals of CUORE. The results are shown in figure 4.16, demonstrating that alpha events and beta/gamma events can be clearly separated by the use of the dual readout. A third solution is to use crystals of ZnSe [64]. In this case the quenching factor was surprisingly found to be larger than one, which means that alpha events produce more scintillation light than beta/gamma events. Nevertheless, alpha and beta/gamma discrimination can be performed. An advantage over $ZnMoO_4$ is in the light yield, which is larger than in $ZnMoO_4$ by a factor of 5, making light detection easier. A drawback in this case is energy resolution, which could not be reduced below about 13 KeV FWHM.

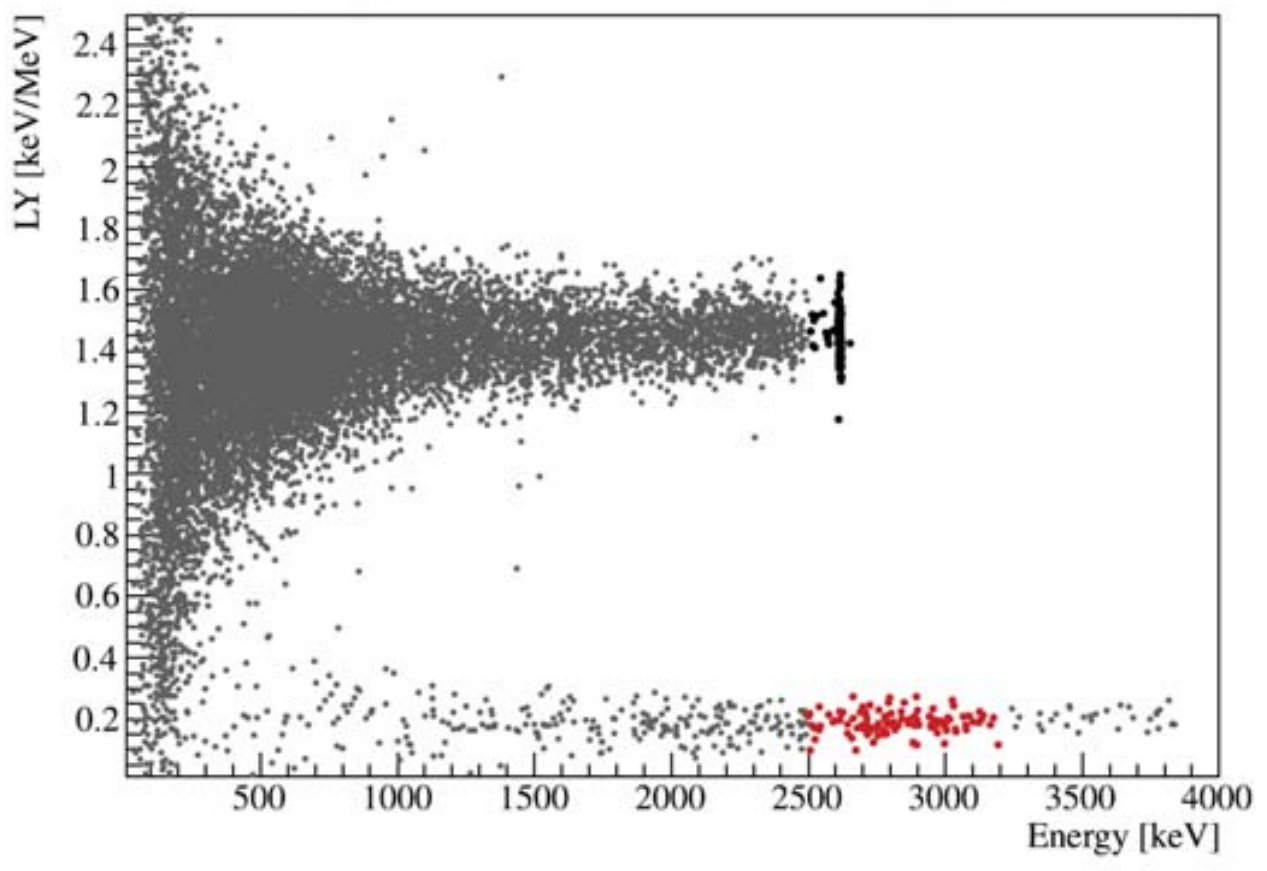

Figure 4.16: Light yield versus thermal signal with a scintillating bolometer, demonstrating the capability of the scintillating bolometer to discriminate between alpha events (the lower band) and beta/gamma events (the upper band). The energy is detected by the thermal signals in the bolometer. The light yield is readout by the light sensor faced to the bolometer. Taken from [63].

The baseline crystal for LUCIFER is ZnSe. An array of 30 to 50 crystals (depending on the size of the cryostat which will be used) is planned to be deployed by 2015 and start data taking. Such measurement will prove the experimental technique and will constitute a neutrinoless double beta decay search experiment in $^{82}$Se. The readout electronics for LUCIFER at this point are not yet clearly defined, since some aspects of the experiment may still change. As a baseline option, a room-temperature solution such as that of CUORE may be employed to readout both the thermal and the light signals. Anyway, if the discrimination capability will be found to require the integrity of the higher frequency components of the signals, which cannot be saved with a room temperature solution, then a first cold stage could be employed.

## 4.7 The readout of the MARE experiment

Unlike the experiments described in the previous sections, which exploit macrobolometers to search for the neutrinoless double beta decay, the MARE experiment uses microbolometers and aims at a kinematic measurement of the neutrino mass through the precise measurement of the single beta decay spectrum. The main advantage of this approach is the fact that being a purely kinematic measurement it is fully model independent. The value of the electronic neutrino mass is extracted from the shape of the spectrum near the endpoint of the decay, where a distortion is expected due to the finite mass of the neutrino. The tighter upper bound to the neutrino mass obtained up to now with a kinematic measurement is 2.2 eV, measured with a spectrometer on the beta decay of Tritium by the Mainz and Troitzk experiments. A further refinement of this technique led to the KATRIN experiment, currently under contruction, which is expected to improve the sensitivity by a factor of ten.

The MARE experiment provides a different approach, which is in principle scalable to large arrays to reach a similar sensitivity [65]. The first phase of the experiment, named MARE-1, aims at a 2 eV sensitivity. The experimental technique was already proven in the past with the MIBETA experiment in Milano, obtaining an upper bound of 15 eV on the neutrino mass. MARE-1 is designed to improve on the previous result and to probe the scalability of the technique to larger arrays. It should reach its goal sensitivity of about 2 eV with an array of 300 elements operated for a few years.

The MARE bolometers are small crystals of Silver Perrhenate ($AgReO_4$) of 500 µg of $600 \times 600 \times 250$ µm$^3$ size. The crystals are glued onto a arrays of $6 \times 6$ Silicon thermistors of $300 \times 300 \times 1.5$ µm$^3$ size. The expected energy resolution at the endpoint of the beta decay under study, that is at 2.47 KeV, is about 25 eV FWHM. The possibility of using transition edge sensor (TES) as the sensing elements is also being pursued, but will not be considered here. As in the case of double beta decay experiments, the beta emitter, that is $^{187}$Re, is contained in the bolometers to provide the maximum detection efficiency. The possibility of performing the same measurement on the electron capture spectrum of $^{163}$Ho is also being considered as a future development [66]. The events of interest are only those closer to the spectrum endpoint, where the electron carries away almost all the energy. A large number of events needs to be recorded to form a spectrum with the required sensitivity. Anyway with the bolometric technique there is no way to select only the decays of interest, and the lower energy events may produce pile-up which can constitute a major background source. The speed of response of the sensors is then a critical factor. The pulses have to be fast to allow to discriminate pile-up signals while collecting the required statistics in a reasonable amount of time. This is a particularly challenging requirement with thermal sensors, which are inherently slow. The timing resolution which can be achieved in MARE-1 coincides with the rise time of the pulses, and it is expected by design to be close to 200 μs, requiring a readout bandwidth which extends to a few tens of KHz.

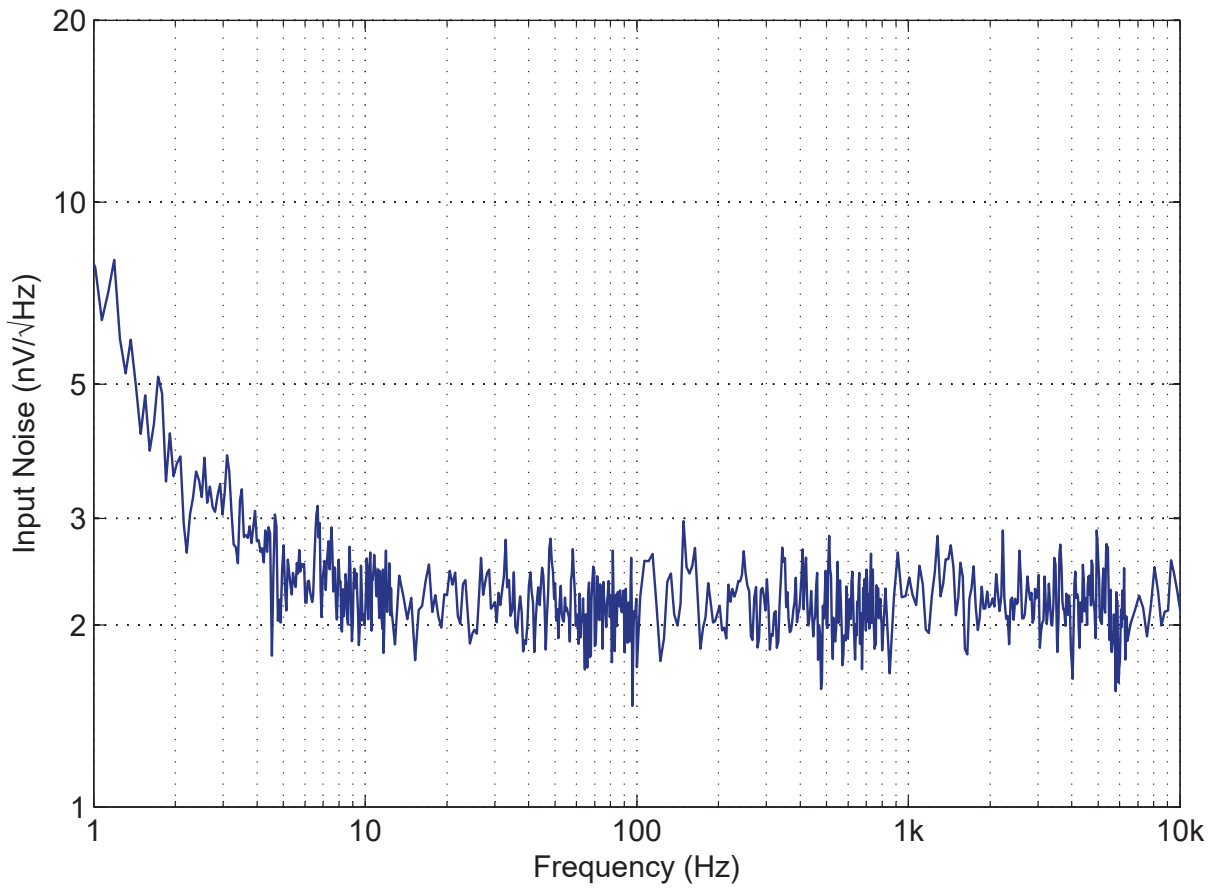

Figure 4.17: Series noise of the JFET transistors used in MARE-1, measured at 120 K with A/C VISCASY.

The thermistors have resistance values of the order of $1-10$ MΩ at 50 mK. From the point of view of the front-end electronics the requirement for a fast readout forces to use a cold stage inside the cryostat, to avoid the parasitic capacitance of the connecting links to the outside. The cold stage is realized with SNJ450 Silicon JFETs from Interfet, operated at about 15 cm from the sensors at the temperature of $120-130$ K. The 50 MΩ load resistors are also located at cold, near the sensors. Since their temperature is the range of tens of mK, the same as the sensors, the thermal noise from the load resistors is negligible. The noise is then dominated by the sensors, giving a few nV/$\sqrt{\text{Hz}}$ of thermal noise, and by the cold JFETs. The JFETs were selected and individually tested with A/C VISCASY, the characterization system mentioned in section 4.4. Figure 4.17 shows the typical noise spectrum of a MARE-1 JFET at 120 K. The white noise is about 2 nV/$\sqrt{\text{Hz}}$. The 1/f component gives about 7 nV/$\sqrt{\text{Hz}}$ at 1 Hz.

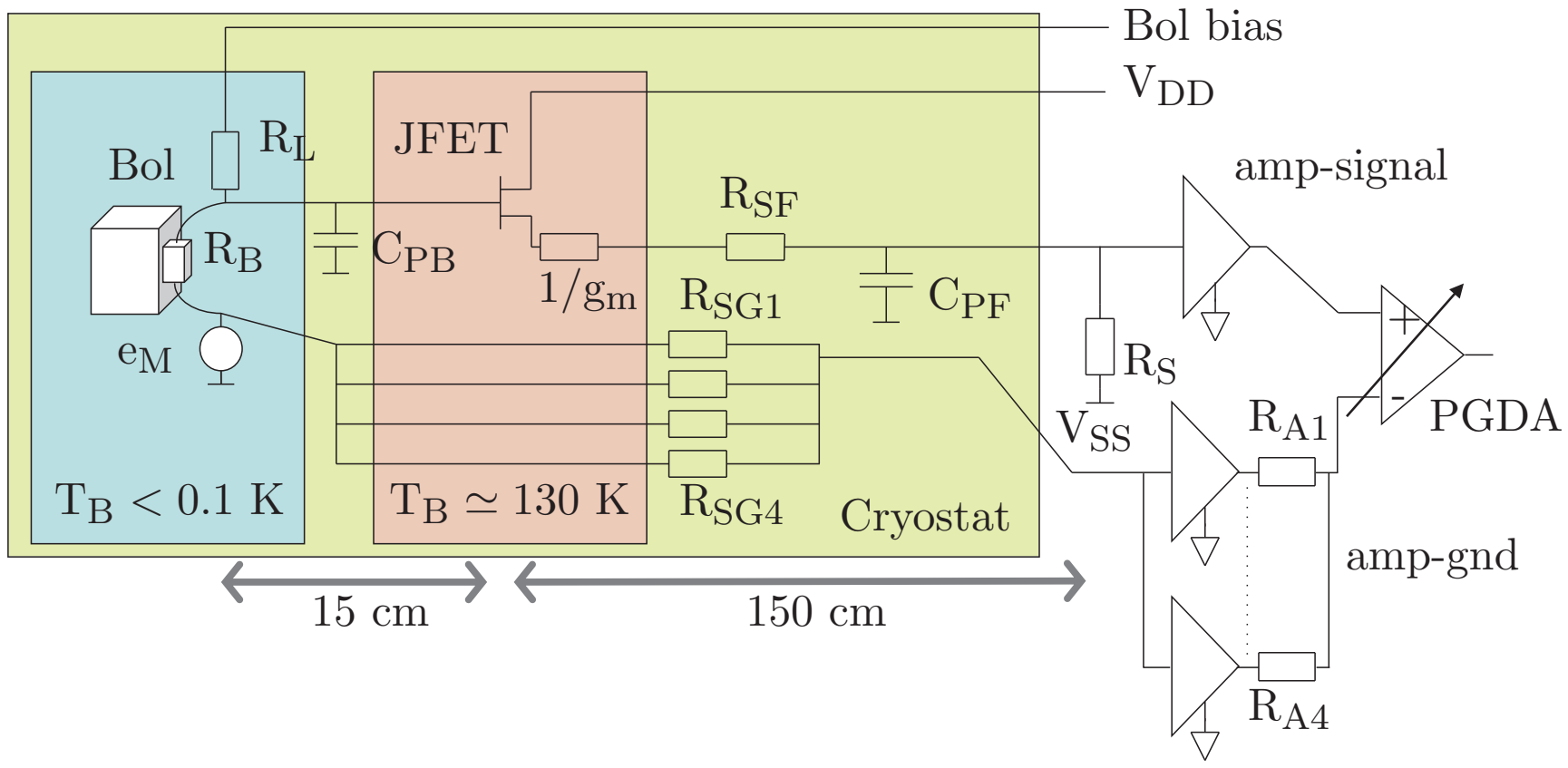

Figure 4.18: Readout scheme of the MARE-1 experiment in Milano.

Instead of a fully differential readout, such as that presented in section 4.4, another solution was implemented which allows to reduce the total number of wires entering the cryostat, while maintaining the common mode rejection capability of a fully differential configuration [67]. Moreover, the noise power with such solution is not twice the noise of a single ended readout, as happens with a fully differential configuration, but only about 25% larger. The schematic of the readout solution is shown in figure 4.18. As already discussed, each sensor is buffered with a cold stage JFET held at $120-130$ K. The parasitic capacitance $C_{PB}$ at the JFET input is of the order of 15 pF. The signal at the source of the JFET is read at room temperature by the very low noise single sided input preamplifier "amp signal" [68]. The second stage amplifier has unity gain at DC, and a differential gain close to 700 V/V at AC. The AC gain of amp-signal can be disabled to allow the DC characterization of the bolometers. The connecting cables from the cold to the warm stage contribute with about 200 pF of parasitic capacitance $C_{PF}$. The input stage of amp-signal matches, at moderate frequency, the impedance seen at its input, that is $1/g_m + R_{SF}$, between 200 and 300 Ω, where $g_m$ is the transconductance of the JFET and $R_{SF}$ is the parasitic series resistance of the connecting wires.

The suppression of the common mode disturbances $e_M$ is done with the help of "amp-gnd", that reads $e_M$ with the same gain of amp-signal. This time the link is implemented with more wires in parallel and its impedance $R_{SG1}$ to $R_{SG4}$ results in a few tens of Ω. The matching of such impedance can be done with an input stage having a very wide area. To achieve this, four amp-signal were put in parallel to form amp-gnd, and their average output taken with the resistors $R_{A1}$ to $R_{A4}$ forms the reference ground signal. Since the series input noise is inversely proportional to the area of the input transistor, the resulting noise power is now only 25% larger than that of a single sided amplifier. Since the values of the source impedance seen by the second stage is small, the parallel noise due to amp-signal and amp-gnd, that is proportional to the transistor area, has a negligible effect. A Programmable Gain Differential Amplifier, PGDA, subtracts the signals of amp gnd from amp signal, rejecting the spurious disturbances due to $e_M$ at the output of the chain. The effectiveness of this approach in cancelling the common mode disturbances is shown in figure 4.19, where the baseline fluctuations of four channels are plotted versus time before (left) and after (right) the subtraction of common mode disturbances from amp-gnd. The voltage gain of the PGDA is settable with 4 configuration bits. The signals at the outputs of the PGDA are shaped with 4-poles Bessel filters, whose frequencies can be remotely selected between 5, 15, 50 and 62 KHz. Their outputs are directly sent to the data acquisition system (DAQ).

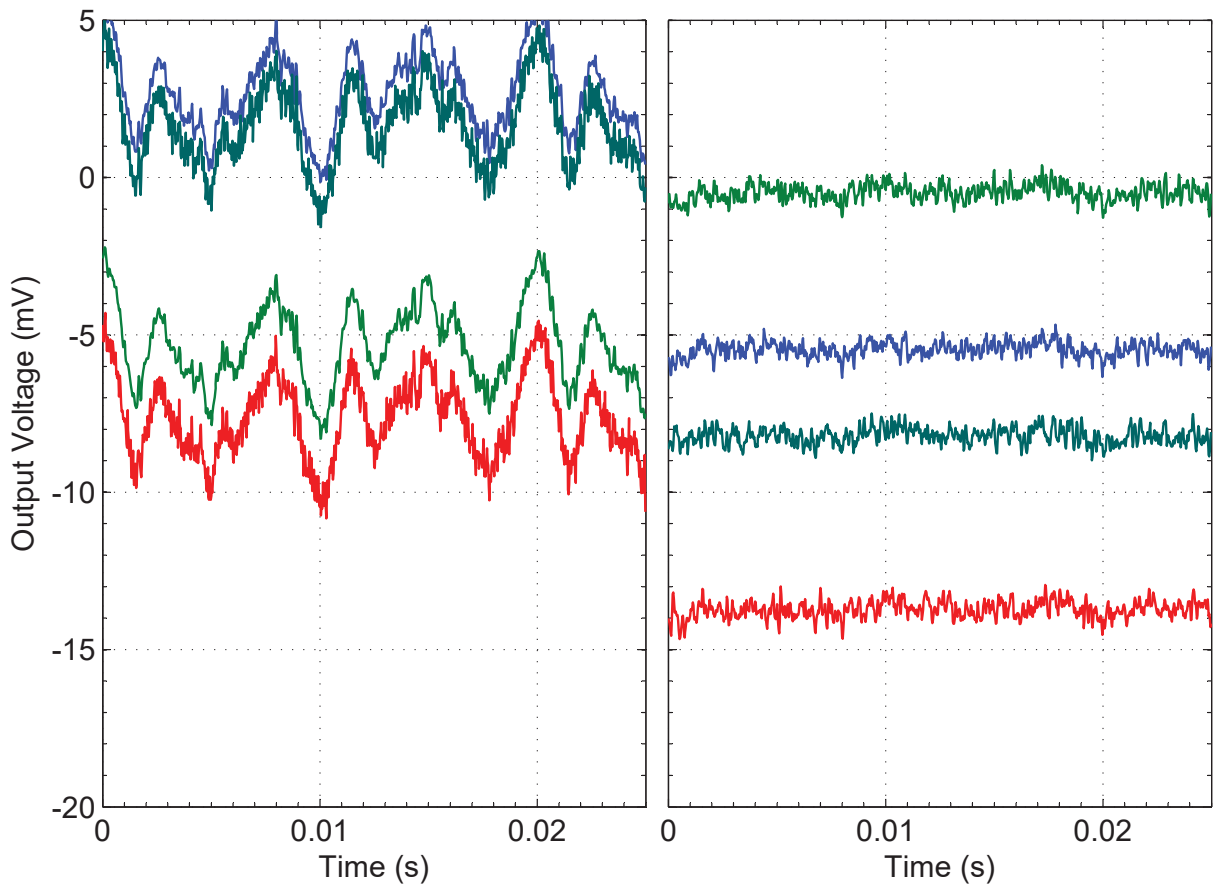

Figure 4.19: Baseline fluctuations versus time on four channels, before (left) and after (right) the subtraction of the common mode signals performed in the PGDA.

The detectors are small and very close together. The ground node is found therefore at the same potential for all channels and only a few amp-gnd can be used, each one serving as the reference for many detectors. In the MARE-1 setup four amplifiers are housed on each main board. One amp-gnd is used every 20 readout channels. Since there is no distinction between main boards

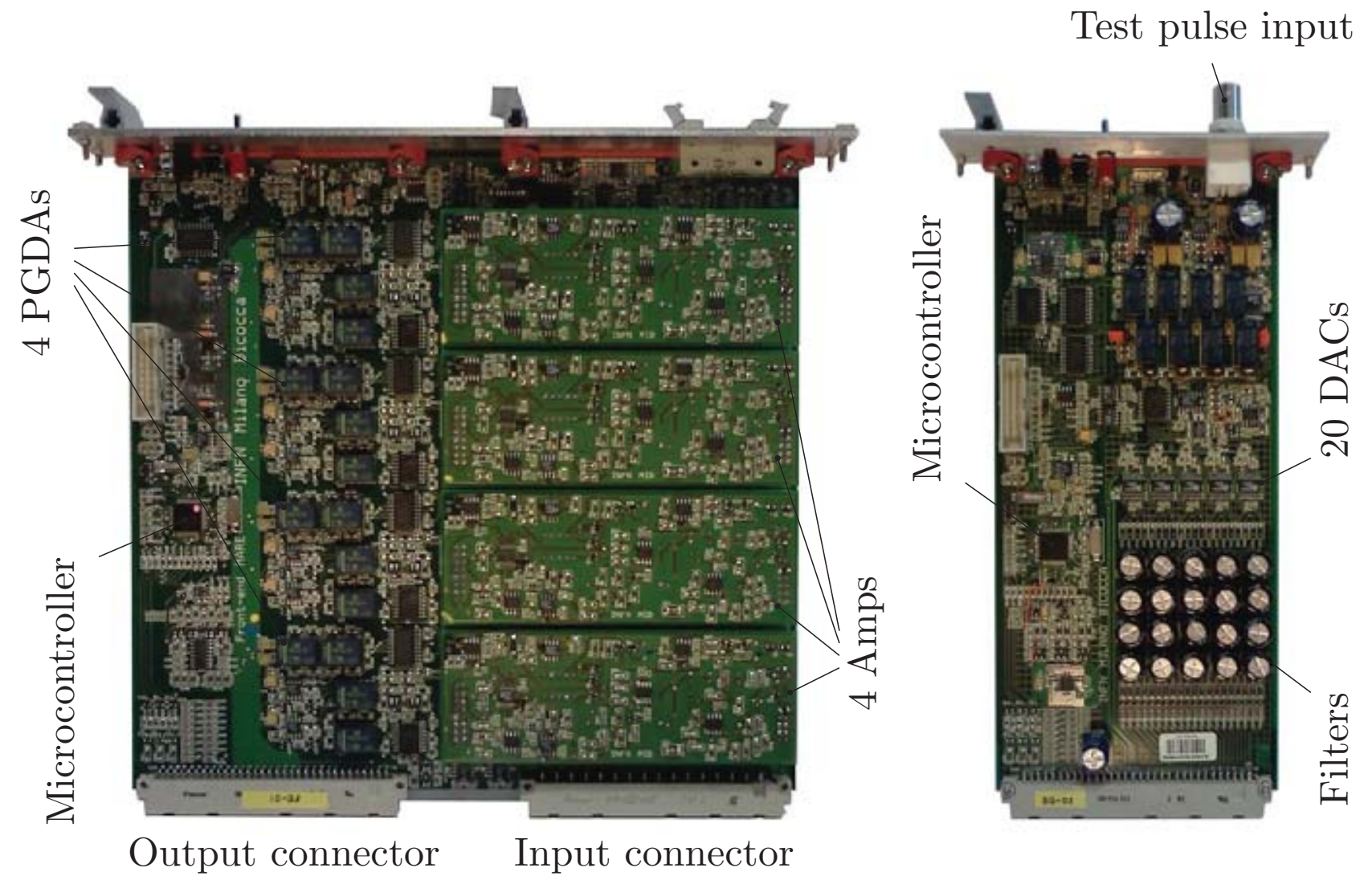

Figure 4.20: Photos of the front-end boards (left) and bias boards (right) of the MARE-1 setup in Milano.

---

with amp-signals or amp-gnd, only one type of PCB was designed. The only difference is that in the amp-gnd boards the PGDA is not used, so it was not populated. The purpose of each amplifier is set by the back panels of the racks, where the input/output connectors are sorted on purpose. After five slots where the amplification of 20 sensors is done, the sixth has the outputs of its 4 preamplifiers averaged with 4 resistors on the back panel (connected as in figure 4.18) and buffered to drive the 20 channels. On this sixth slot the input of each preamplifier is the ground signal from inside the cryostat. On every main board the amp-gnd is routed from the input connector to a digital trimmer and then subtracted from the amp signal at the PGDA differential amplifier. The digital trimmer allows to add a small attenuation on the amp gnd path to compensate for the gain of the cold stage that is slightly smaller than one. The pattern repeats every six main board slots.

The biasing of the bolometric sensors is provided by a different set of circuit boards, each one able to manage 20 bolometers. The input voltage to the bias circuit can be derived from the supply of the rest of the system or from batteries, in order to suppress ground loops. Digital trimmers are present, which allow to tune the optimal biasing for each individual channel. A known pulse, enabled with relays, can be injected in series with the bias and applied across the thermistor in order to stabilize the measurements.

A picture of the front-end and bias boards described so far is shown in figure 4.20. Both the front-end main boards and the bias boards house micro-controllers and are remotely programmable. A graphical user interface (GUI) was implemented in Matlab to control the bias, gain and filtering of each channel remotely from a PC. The PC communicates with the boards through optical fibers to prevent ground loops. A total of 80 channels are ready for MARE-1.

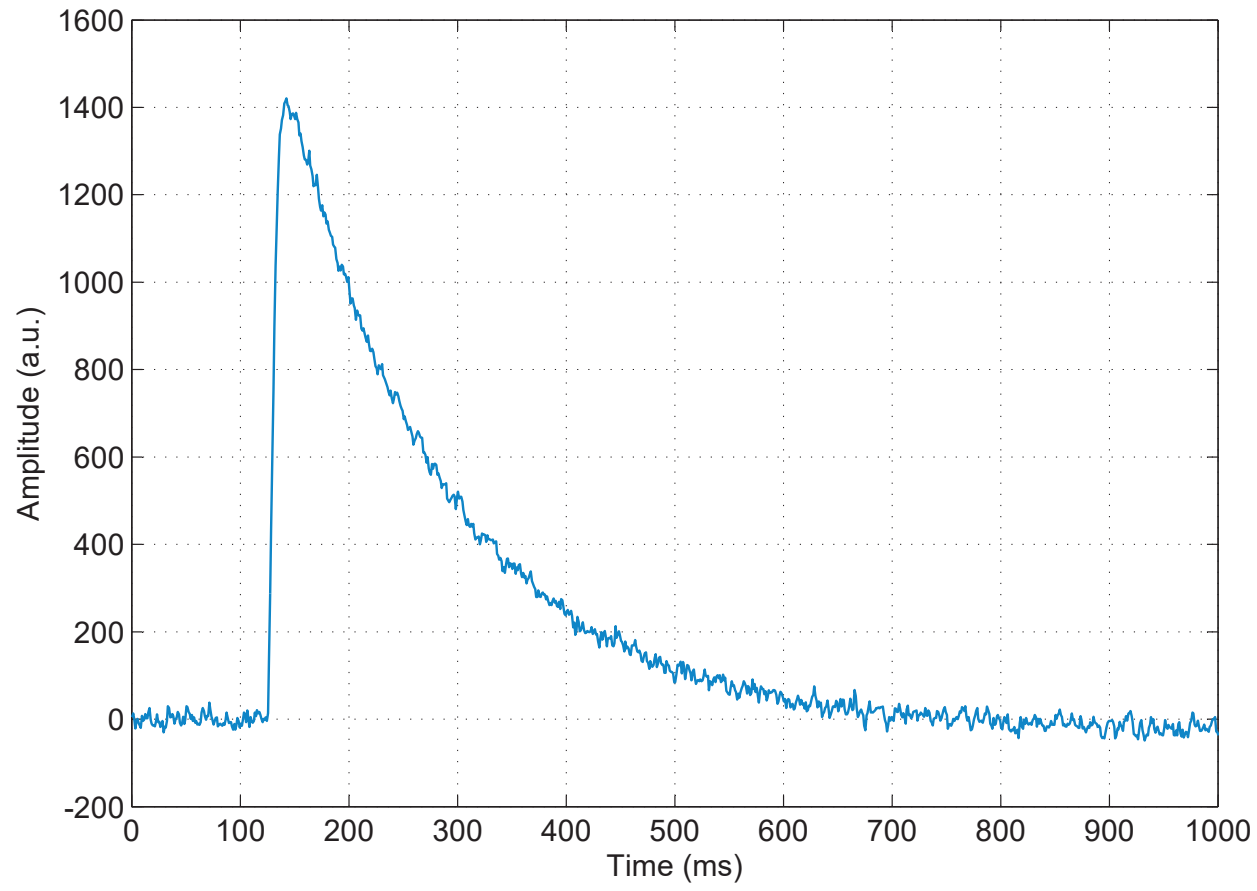

Figure 4.21: A 6 KeV signal from a calibration source.

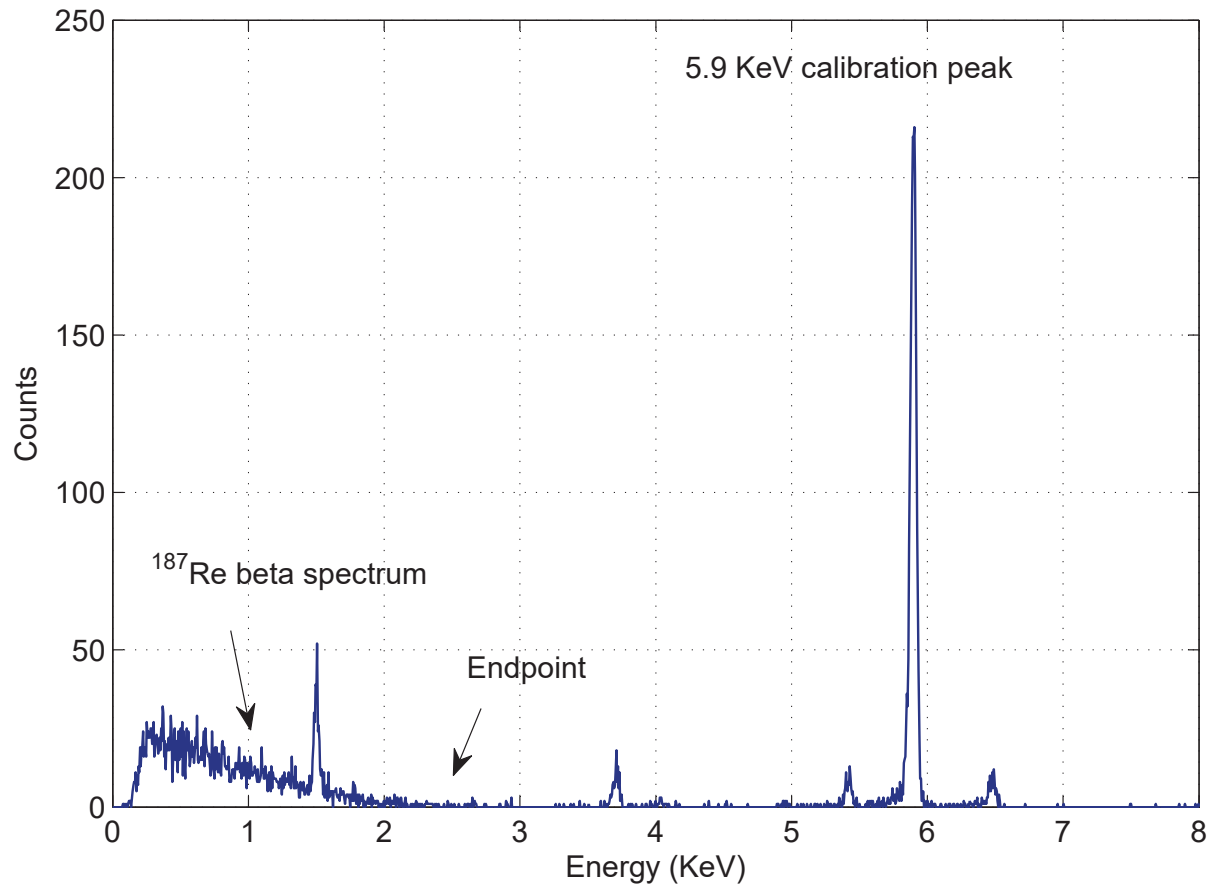

Figure 4.22: A $^{187}$Re beta spectrum recently measured with MARE-1.

Figure 4.21 shows a 6 KeV signal from a calibration source, while figure 4.22 shows a spectrum recently measured in Milano Bicocca with MARE-1. The $^{187}$Re beta spectrum can be clearly recognized, its endpoint at 2.47 KeV. Some peaks from the calibration sources can also be seen, the most prominent being at 5.9 KeV. The spectrum is binned in 5 eV steps. The resolution after optimal filtering is close to the design value of 25 eV FWHM. Optimization of the setup is ongoing. Data taking with two arrays (72 microbolometers) is expected to start in 2013.